\documentclass[10pt]{article}
\pdfoutput=1

\usepackage{jheppub}

\usepackage{todonotes}
\presetkeys{todonotes}{inline}{}
\usepackage{amsmath}
\usepackage[T1]{fontenc}
\usepackage{textcomp}
\usepackage{amssymb}
\usepackage{commath}
\usepackage{array}
\usepackage{calc}
\usepackage{longtable}
\usepackage{multirow}
\usepackage{pstricks}
\usepackage{graphicx}
\graphicspath{{figures/}}
\usepackage{xspace}
\usepackage{listings}
\usepackage{
  pgf,
  tikz}
\usetikzlibrary{
  patterns,
  shapes.multipart,
  arrows,
  trees,
  scopes,
  decorations.pathreplacing,
  decorations.pathmorphing,
  decorations.markings,
  decorations.text,
  positioning,
  calc
}
\usepackage{mciteplus}

\newcommand{\Sherpa}{S\protect\scalebox{0.8}{HERPA}\xspace}
\newcommand{\Pythia}{P\protect\scalebox{0.8}{YTHIA}\xspace}
\newcommand{\Herwig}{H\protect\scalebox{0.8}{ERWIG}\xspace}

\newcommand{\Comix}{C\protect\scalebox{0.8}{OMIX}\xspace}

\newcommand{\Caesar}{C\protect\scalebox{0.8}{AESAR}\xspace}

\newcommand{\MEPSatNLO}{MEPS\protect\scalebox{0.8}{@}NLO\xspace}
\newcommand{\Rivet}{R\protect\scalebox{0.8}{IVET}\xspace}
\newcommand{\fastjet}{F\protect\scalebox{0.8}{AST}J\protect\scalebox{0.8}{ET}\xspace}
\newcommand{\OpenLoops}{O\protect\scalebox{0.8}{PEN}L\protect\scalebox{0.8}{OOPS}\xspace}
\newcommand{\Collier}{C\protect\scalebox{0.8}{OLLIER}\xspace}

\newcommand{\muR}{\ensuremath{\mu_{\text{R}}}}
\newcommand{\muF}{\ensuremath{\mu_{\text{F}}}}

\newcommand{\zcut}{\ensuremath{z_{\text{cut}}}}
\newcommand{\tauPerp}{\ensuremath{\tau_\perp}\xspace}
\newcommand{\tauSD}{\ensuremath{\tauPerp^{\text{SD}}}\xspace}
\newcommand{\alphaS}{\alpha_\text{s}\xspace}
\newcommand{\LO}{\text{LO}\xspace}
\newcommand{\NLO}{\text{NLO}\xspace}    
\newcommand{\NLL}{\text{NLL}\xspace}
\newcommand{\NLLp}{\ensuremath{\text{NLL}^\prime}\xspace}
\newcommand{\NLOpNLL}{\ensuremath{\NLO+\NLL}\xspace}
\newcommand{\LOpNLL}{\ensuremath{\LO+\NLL}\xspace}    
\newcommand{\NLOpNLLp}{\ensuremath{\NLOpNLL^\prime}\xspace}
\newcommand{\LOpNLLp}{\ensuremath{\LOpNLL^\prime}\xspace}
\newcommand{\BSZ}{BSZ\xspace}

\usepackage{placeins}
\usepackage{graphicx}

 \hypersetup{
  pdfauthor={Jeremy Baron, Daniel Reichelt, Steffen Schumann, Niklas Schwanemann, Vincent Theeuwes}, 
  pdftitle={Soft-drop grooming for hadronic event shapes}
}

\preprint{MCNET-20-26}

\title{Soft-drop grooming for hadronic event shapes}

\author[1,2]{Jeremy Baron,}
\author[2]{Daniel Reichelt,}
\author[2]{Steffen Schumann,}
\author[2]{Niklas Schwanemann,}
\author[2]{Vincent Theeuwes}

\affiliation[1]{University at Buffalo, The State University of New York, Buffalo, NY 14260-1500, USA}

\affiliation[2]{Institut f{\"u}r Theoretische Physik,
  Georg-August-Universit{\"a}t G{\"o}ttingen, D-37077 G{\"o}ttingen, Germany}

\emailAdd{jfbaron@buffalo.edu}
\emailAdd{daniel.reichelt@uni-goettingen.de}
\emailAdd{steffen.schumann@phys.uni-goettingen.de}
\emailAdd{niklas.schwanemann@stud.uni-goetingen.de}
\emailAdd{vtheeuwe@gmail.com}

\abstract{
  Soft-drop grooming of hadron-collision final states has the
  potential to significantly reduce the impact of non-perturbative
  corrections, and in particular the underlying-event contribution.
  This eventually will enable a more direct comparison of accurate
  perturbative predictions with experimental measurements. In this
  study we consider soft-drop groomed dijet event shapes. We derive
  general results needed to perform the resummation of suitable
  event-shape variables to next-to-leading logarithmic (\NLL) accuracy
  matched to exact next-to-leading order (\NLO) QCD matrix elements.
  We compile predictions for the transverse-thrust shape accurate
  to \NLOpNLLp using the implementation of the \Caesar\ formalism
  in the \Sherpa event generator framework. We complement this by
  state-of-the-art parton- and hadron-level predictions based on
  \NLO QCD matrix elements matched with parton showers. We explore
  the potential to mitigate non-perturbative corrections for
  particle-level and track-based measurements of transverse thrust
  by considering a wide range of soft-drop parameters. We find
  that soft-drop grooming indeed is very efficient in removing the
  underlying event. This motivates future experimental measurements
  to be compared to precise QCD predictions and employed to constrain
  non-perturbative models in Monte-Carlo simulations. 
}		

\begin{document}
\maketitle

\section{Introduction}

Event-shape variables attribute simple real numbers to a scattering event, determined
by the momenta of the final-state particles, that characterise geometric properties of the
event. Their distributions offer a wide range of potential applications in collider
phenomenology. This includes precision QCD studies, \emph{e.g.}\ extractions of the strong coupling,
the discrimination of hypothetical new physics from Standard Model expectations, as well as stress-tests,
validation and tuning of Monte Carlo event generators. Aside from event generators, theoretical
predictions for event shapes can be obtained from fixed-order and resummation calculations.
However, in general, event-shape observables are rather susceptible to non-perturbative
corrections, \emph{i.e.}\ hadronisation effects and the underlying event. These need
to be taken into account when comparing high-precision calculations with experimental
data, \emph{e.g.}\ through the evaluation of power
corrections~\cite{Dokshitzer:1995zt,Dokshitzer:1995qm,Dokshitzer:1997ew}, or, via
phenomenological models, as done in event generators~\cite{Buckley:2011ms}. Their sensitivity
to non-perturbative phenomena makes event-shape observables very valuable for the tuning
of Monte Carlo generators. 

There exists a vast amount of event-shape studies for $e^+e^-$
colliders~\cite{Heister:2003aj,Abbiendi:2004qz,Abdallah:2003xz,Achard:2004sv} for
DIS~\cite{Aktas:2005tz,Chekanov:2006hv},
and corresponding higher-order and higher-logarithmic perturbative QCD predictions,
see for instance~\cite{Abbate:2010xh, Abbate:2012jh, Hoang:2014wka,
  Banfi:2014sua, Tulipant:2017ybb, Bell:2018gce, Kang:2013lga, Becher:2015gsa,
  Becher:2015lmy, Dixon:2018qgp, Gao:2019ojf, Dixon:2019uzg, Gehrmann:2019hwf}. 
Event shapes such as thrust can be consistently defined also for hadron
colliders~\cite{Banfi:2004nk}, but despite their great potential, they have received
comparably little attention from experiments at hadron colliders to date.

Resummed predictions for event shapes, \emph{e.g.}\ at next-to-leading logarithmic (NLL) accuracy,
can be derived and matched to next-to-leading order (NLO) calculations. Ref.~\cite{Banfi:2010xy}
presented an extensive phenomenological study of a variety of event shapes in
dijet production under Tevatron and LHC conditions, based on \NLOpNLLp predictions
as well as Monte Carlo simulations. 
The smaller number of experimental studies of event-shape observables in hadronic collisions
is certainly related to their pronounced sensitivity to non-perturbative corrections~\cite{Banfi:2010xy},
which complicates the interpretation of measurements in terms of perturbative predictions.
Additional complications arise from constraints on the measurement acceptance region,
\emph{e.g.}\ a maximum rapidity range, $|y|\leq y_{\text{max}}$, or, a non-vanishing particle (track)
transverse momentum cut, $p^{\text{track}}_{T,\text{min}}$, that hinder the direct comparison
between idealised theoretical predictions and experiment. Some recent LHC measurements of
event-shape variables have been based on reconstructed jets, rather than
particles, as input for the observable
calculation~\cite{Aad:2012np,Khachatryan:2014ika,Sirunyan:2018adt,Aad:2020fch}. While
this eases the evaluation of systematic uncertainties and acceptance
corrections, it makes it extremely difficult to address them beyond a
fixed-order calculation or Monte Carlo simulations.

With this work we follow the original approach of using particles as inputs to the
observable. However, we consider grooming the event using Soft Drop~\cite{Larkoski:2014wba},\
and only use the surviving constituent particles as inputs to the event-shape calculation.
This provides the potential to significantly reduce the impact
of non-perturbative effects while retaining the ability to analytically address these observables. 
As a concrete example we focus on the transverse-thrust shape. Through variations of the grooming
parameters its underlying-event sensitivity can be regulated, providing additional means to tune
the corresponding phenomenological models. Furthermore, the impact of using a finite maximum
input-particle rapidity and minimum transverse momentum can be diminished by grooming. In fact, for
sufficiently hard grooming a rather direct correspondence between perturbative predictions and
hadron-level results is found. In Refs.~\cite{Baron:2018nfz,Marzani:2019evv} similar observations
have been made for soft-drop thrust in $e^+e^-$ collisions, where a reduced sensitivity to
hadronisation effects was observed. 
Many other jet-substructure techniques for underlying-event mitigation exist, for
example~\cite{Butterworth:2008iy,Ellis:2009su,Krohn:2009th,Dasgupta:2013ihk,Dreyer:2018tjj}
with a detailed overview given in~\cite{Marzani:2019hun}. However, these approaches are
restricted to grooming the constituents of a given jet only. Soft-drop grooming
individual jets at hadron colliders is also intensely studied theoretically,
see \emph{e.g.} \cite{Frye:2016aiz, Marzani:2017kqd, Kang:2018jwa, Kang:2018vgn, Bell:2018vaa, Kang:2019prh,Hoang:2019ceu,
  Kardos:2020ppl, Kardos:2020gty, Anderle:2020mxj} for recent results. With this
work we suggest to extend those methods to grooming of the entire event,
thereby generalising the work in \cite{Baron:2018nfz,Marzani:2019evv} to
global event shapes at hadron colliders.

We derive resummed predictions at \NLOpNLLp accuracy for soft-drop transverse thrust by employing
the implementation of the \Caesar resummation formalism~\cite{Banfi:2004yd} in the \Sherpa event
generator framework~\cite{Gerwick:2014gya,Gleisberg:2008ta,Bothmann:2019yzt}. We
compare the results against parton-level shower Monte Carlo predictions before
focusing on the mitigation of underlying-event effects through soft-drop grooming. 

This paper is organised as follows: in Sec.~\ref{sec:softdrop} we give a
prescription to apply soft-drop to global event shape observables for dijet
events in hadronic collisions. The \NLOpNLLp resummation and matching calculation are
discussed in Sec.~\ref{sec:res}. In Sec.~\ref{sec:pheno} we present predictions for soft-drop
thrust in dijet-event final states at the LHC. We then compare resummed results and parton-shower
simulations, and study the sensitivity to non-perturbative corrections as well as experimentally
motivated acceptance cuts. Our conclusions are presented in Sec.~\ref{sec:conclusions}.

\clearpage
\section{Soft drop for hadronic event shapes}\label{sec:softdrop}

With all final-state particles contributing to the observable calculation, event-shape
variables can be particularly sensitive to non-perturbative
corrections. This we illustrate here for the
transverse-thrust observable, which we will use throughout the paper as concrete example.
This hadron-collider variant of thrust is defined
as\footnote{As commonly done, we prefer to work with an observable that vanishes
  in the soft limit, \emph{i.e.}\ $\tauPerp = 1-T_\perp$.}  
\begin{equation}\label{eq:thrust_def}
\tauPerp \equiv 1-\max_{\vec{n}_\perp} \left(\frac{\sum_i |\vec{p}_{T,i}\cdot
    \vec{n}_\perp|}{p_{T,\text{tot}}}\right)\;,\quad p_{T,\text{tot}}= \sum\limits_i p_{T,i}\,,
\end{equation}
with the sum extending over all final-state particles, and $\vec{p}_{T,i}$ the
respective two-component transverse-momentum vector with length $p_{T,i} =
|\vec{p}_{T,i}|$. The unit vector $\vec{n}_\perp$ that maximises the sum in
Eq.~\eqref{eq:thrust_def} defines the transverse-thrust axis. As a three-jet observable,
transverse thrust quantifies the deviation from the back-to-back event
configuration. 

\begin{figure}
	\begin{center}
		\includegraphics[width=0.45\textwidth]{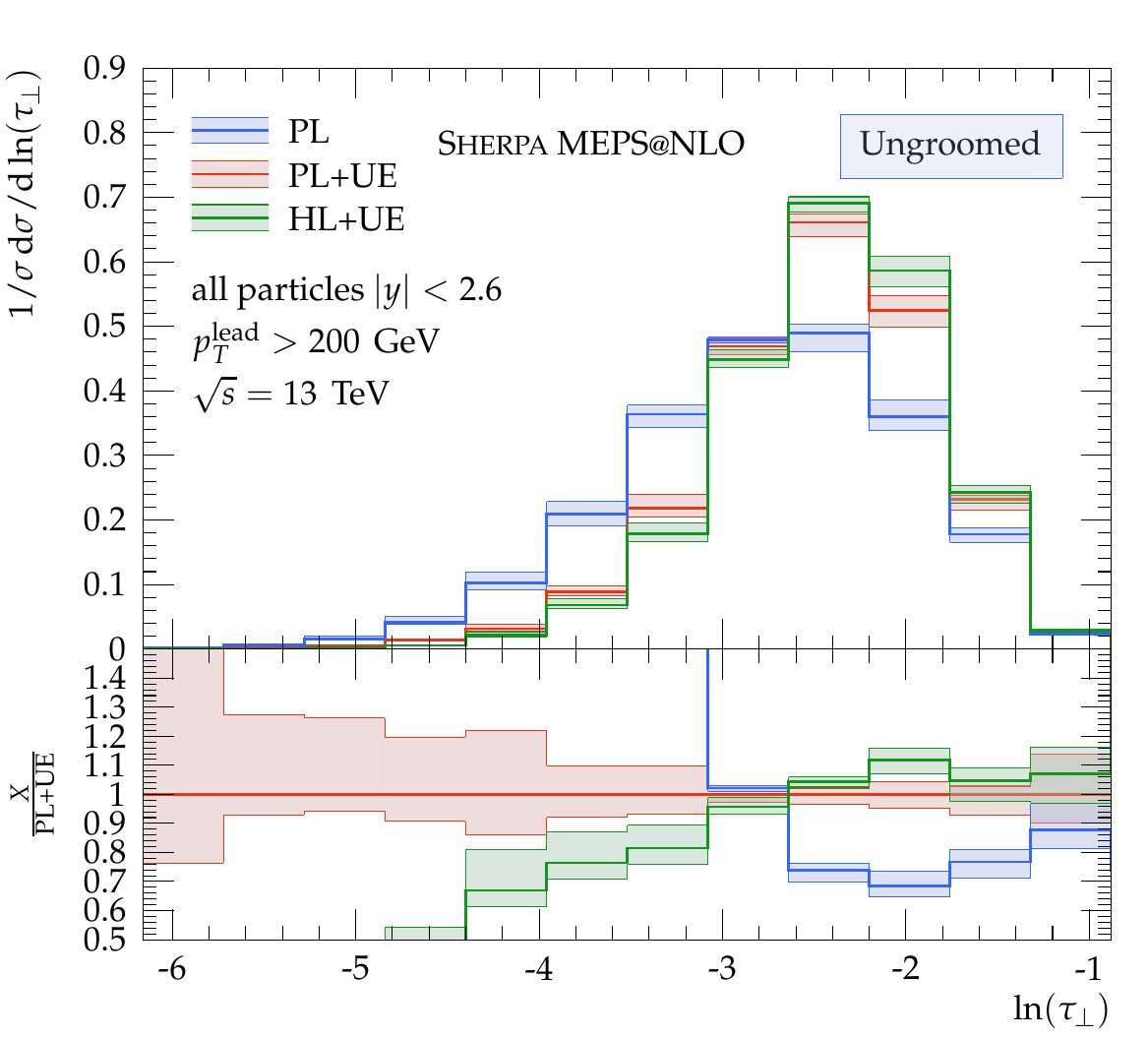}\qquad
		\includegraphics[width=0.45\textwidth]{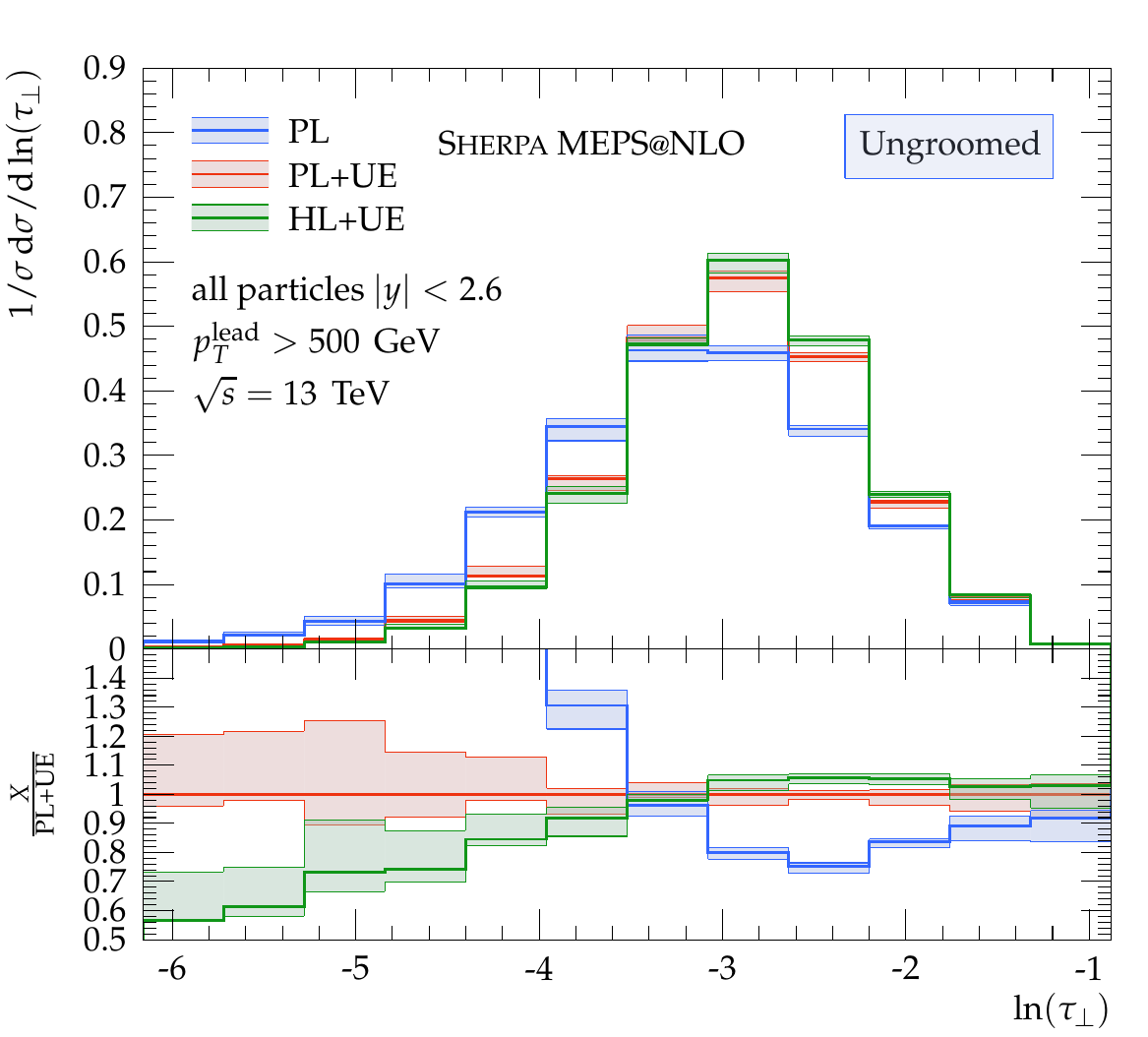}
	\end{center}
	\caption{The transverse-thrust distributions for events with a leading-jet transverse
          momentum $p^{\text{lead}}_T > 200\;\text{GeV}$ (left) and $p^{\text{lead}}_T > 500\;\text{GeV}$ (right).
          Further details on the event-selection cuts are given in Sec.~\ref{sec:eventselections}.
          Shown are \MEPSatNLO\ predictions obtained with \Sherpa at parton level (PL), with the underlying event
          included (PL+UE), and at full hadron level (HL+UE). The lower panels show the ratios with respect to
          the PL+UE prediction.}
	\label{fig:NP_ungroomed}
\end{figure}

In Fig.~\ref{fig:NP_ungroomed} we present results obtained with the
\Sherpa generator at different stages of the event evolution, \emph{i.e.}\
after parton showering but without underlying event (PL), with the underlying-event
contribution included (PL+UE), and fully hadronised (HL+UE). We consider
dijet-production at $\sqrt{s}=13\;\text{TeV}$ with the leading-jet transverse
momentum above $200\;\text{GeV}$ and $500\;\text{GeV}$, respectively. Further details
on the event selections and generator settings can be found in Secs.~\ref{sec:eventselections}
and \ref{sec:MCsims}. We can observe that the inclusion of the underlying event
significantly shifts the distribution, corresponding to up to 40\% corrections in the peak
region for events with a jet above $200\;\text{GeV}$. Even for the higher transverse-momentum
criterion ($500\;\text{GeV}$) the underlying event still retains a similar impact. Hadronisation
corrections are comparably smaller. Qualitatively hadronisation pushes events to somewhat
higher values of transverse thrust, resulting in corrections of order 10\% in the peak
region and even more sizeable in the low-$\tau_\perp$ tail, \emph{i.e.}\ $\ln(\tau_\perp)\lesssim -3$.
This strong susceptibility of the observable to non-perturbative effects over its whole
range makes the comparison of experimental measurements with purely perturbative
calculations rather indirect and plagued by significant modelling uncertainties.

Transverse thrust has been measured by the Tevatron~\cite{Aaltonen:2011et} and
LHC
experiments~\cite{Aad:2012np,Khachatryan:2014ika,Sirunyan:2018adt,Aad:2020fch}.  
However, recent measurements are based on reconstructed jets as inputs to the
observable calculation, which simplifies dealing with the large underlying event
contributions but prevents a direct comparison to perturbative
predictions, see for example the discussion in~\cite{Banfi:2010xy}. Those have so
far been presented at \NLO QCD~\cite{Nagy:2003tz} and resummed to \NLOpNLLp level
in the \Caesar framework~\cite{Banfi:2003je}, and more recently the resummation
has been extended to NNLL accuracy in the context of soft-collinear effective field
theory~\cite{Becher:2015gsa,Becher:2015lmy}. In this work we consider full
particle-level final states, \emph{i.e.}\ charged tracks, as input to the
observable calculation. To mitigate the impact of non-perturbative effects, in particular the underlying
event, we suggest soft-drop grooming the event prior to the observable evaluation.
We argue that the developed method should be applicable to the standard set of
observables studied by the experiments, and the derived formulae are given for a 
general observable within the limitations of the \Caesar approach to
resummation. For concreteness, we focus on transverse thrust, as the "standard
candle" event-shape variable, in our phenomenological studies.

\subsection{Definition groomed event shapes}\label{sec:obs_def}

Soft drop has a wide range of applications, including being used in boosted-particle tagging
or for pile-up mitigation. The method introduced in~\cite{Larkoski:2014wba} has
originally been designed as a jet-substructure technique to remove soft
wide-angle radiation from a jet. For a jet of radius $R_{\text{SD}}$ the
constituents get reclustered using the Cambridge--Aachen (C/A)
algorithm~\cite{Dokshitzer:1997in,Wobisch:1998wt}. Then the last C/A clustering
step is undone and the resulting two subjets are subjected to the soft-drop
criterion
\begin{equation}
\frac{\min(p_{T,i},p_{T,j})}{p_{T,i} + p_{T,j}}\ge \zcut\left(\frac{\Delta R_{ij}}{R_{\text{SD}}}\right)^\beta\,, \label{eq:SD}
\end{equation}
with $p_{T,i}$ the transverse momenta of the two constituent subjets with
respect to the beam, and $\Delta R_{ij}^2 = (y_i-y_j)^2+(\phi_i-\phi_j)^2$ their
separation in the rapidity-azimuth plane. If the condition is satisfied
grooming ends and the jet is the combination of these two subjets, otherwise the
subjet with smaller transverse momentum is removed from the jet and the
procedure is continued for the harder subjet. The two relevant parameters are the
threshold $\zcut$ and the angular exponent $\beta \geq 0$. It should be noted that for
$\beta=0$, which corresponds to the modified Mass-Drop Tagger
(mMDT)~\cite{Butterworth:2008iy,Dasgupta:2013ihk}, all soft emissions at \LO
accuracy are groomed. This alters the logarithmic structure of the resummation
leading to only single-logarithmic enhancement of the observable which is of
collinear origin.

Here, we build up on work in \cite{Baron:2018nfz,Kardos:2018kth}, where soft
drop has been applied to the thrust variable in $e^+e^-$
collisions~\cite{Farhi:1977sg}, among other event shapes. We stress that the main
motivation for this is to study observables close to traditional event shapes
but with reduced impact of non-perturbative effects, even though soft drop is a more
general tool that could suggest alternative observable definitions by itself. Correspondingly, the
event is divided into two hemispheres based on the thrust axis, and the
soft-drop condition is applied to both of them. We will generalise
this idea and define soft-drop groomed versions of hadronic event shapes. To
this end we employ the transverse-thrust axis $\vec{n}_\perp$ to separate a
given event into azimuthal hemispheres, $\mathcal{H}_R$ containing all particles
with $\vec{n}_\perp\cdot\vec{p}_{T,i}\geq 0$, and $\mathcal{H}_L$ containing all
particles with $\vec{n}_\perp\cdot\vec{p}_{T,i}<0$. For each hemisphere
separately, we can then apply the standard soft-drop procedure, \emph{i.e.}
recluster all particles in the hemisphere into a C/A jet, then subsequently undo
clustering steps and check the soft-drop criterion Eq.~\eqref{eq:SD} as described
above for jets. The remaining particles of both hemispheres, which we will refer
to as $\mathcal{H}_L^\prime$ and $\mathcal{H}_R^\prime$, then constitute the
groomed-event final state. The hemispheres do not have an auxiliary radius
associated to them, so we can make a choice $R_{\text{SD}}=1$ in what
follows. The result of this application of soft-drop is quite different from the
usual approach of grooming jets. In general, $\beta>0$ suppresses grooming for
radiation at small angles. For a jet $R_{\text{SD}}$ is typically chosen equal 
to the jet radius and this suppression is happening for all radiation inside the
jet. However, for a hemisphere there will be cases for which $\Delta
R_{ij}>R_{\text{SD}}$. For these particles, grooming will instead be even
stronger. This changes the role of the parameter $\beta$, as values $\beta>0$
can result in more significant grooming. This in particular has the potential to
suppress contributions from multiple-parton interactions, \emph{i.e.}\ the
underlying event, which are largely uncorrelated in angle with respect to the
hard process.

We now want to calculate a given event shape with the particles
that survived this grooming procedure. The exact definition of the groomed
event shape for arbitrary, not necessarily soft and/or collinear configurations
might however not be uniquely fixed by this prescription, and care has to be
taken not to introduce issues regarding collinear safety, \emph{cf.}\ Appendix A
of \cite{Marzani:2019evv} for a detailed discussion. As we are going to focus on
hadronic thrust in our phenomenological studies, we give an explicit definition
of  its groomed variant:
\begin{align}\label{eq:sd_thrust_def}
\tauSD \equiv \left(1 - \frac{\sum_{i\in\mathcal{H}_L^\prime}
    |\vec{p}_{T,i}\cdot
    \vec{n}_{\perp,L}^\prime|}{\left(p_{T,\text{tot}}\right)^{\text{groomed}}} -
  \frac{\sum_{i\in\mathcal{H}_R^\prime} |\vec{p}_{T,i}\cdot
    \vec{n}_{\perp,R}^\prime|}{\left(p_{T,\text{tot}}\right)^{\text{groomed}}}\right)
\frac{\left(p_{T,\text{tot}}\right)^{\text{groomed}}}{\left(p_{T,\text{tot}}\right)^{\text{all}}}\\
\text{with}\quad \vec{n}_{\perp,X}^\prime = 
\frac{\sum_{i\in\mathcal{H}_X^\prime}\vec{p}_{T,i}}{\left|\sum_{i\in\mathcal{H}_X^\prime}\vec{p}_{T,i}\right|},\quad\text{and}\quad
\left(p_{T,\text{tot}}\right)^{\text{groomed}} =  \sum_{i\in\mathcal{H}_L^\prime}
\left|\vec{p}_{T,i}\right| + \sum_{i\in\mathcal{H}_R^\prime}
\left|\vec{p}_{T,i}\right|,
\end{align}
where as indicated all sums run over the particles that survive grooming in the
respective hemispheres, and $\left(p_{T,\text{tot}}\right)^\text{all}$ denotes the scalar 
sum of the transverse momenta of all particles, whether or not affected by
grooming. Multiplying by the ratio of the total groomed and ungroomed transverse
momentum guarantees collinear safety as mentioned above, in full analogy with
the $e^+e^-$ case~\cite{Baron:2018nfz,Marzani:2019evv}.

\subsubsection*{Experimental considerations}

In addition to the theoretical considerations when defining an observable, it
should be viable to be measured experimentally in a setup as close to the
calculation as possible. The resummation and fixed-order consideration are based
on the distribution of partons. After applying a correction according to some
hadronisation model, or convincing ourselves that those corrections are rather
small, they might be taken as a prediction for the observable as defined on all hadrons
in the final state. However, those are not readily available in general in hadron-collider
experiments, though for example the \emph{particle-flow method} used by the CMS
experiment~\cite{Sirunyan:2017ulk} gets rather close to the particle level. However,
for our final phenomenological studies, we here define the observables based on
detectable charged-particle tracks, accessible with conventional tracking techniques.
Those are assumed to resemble the overall distribution of hadrons in the event,
allowing us to relate to the analytic calculation. However, the use of charged tracks
in practice results in several limitations.

The first experimental restriction is the rapidity range where tracks can be
measured reliably. This results in a maximum rapidity $y_{\text{max}}$
within which particles are considered to contribute to the observable calculation.
For transverse thrust this cut-off alters the logarithmic structure for the
resummation. In \cite{Banfi:2010xy} this was addressed by suggesting certain
changes to the observable definitions. Here, we are going to argue that our
modification, \emph{i.e.}\ soft-drop grooming, is already sufficient. The
reason for this is essentially that particles contributing to this difference
in logarithmic structure will have low transverse momentum and therefore be
prone to grooming. This allows us to ignore the rapidity cut-off in the resummed
calculation and take it into account in the final matched distributions by including
it in the fixed-order calculation. 

In addition to the spatial restriction, a track-based measurement can only be
performed based on charged particles with a transverse momentum above some
threshold $p^{\text{track}}_{T,\text{min}}$. This conversion from all particles
to charged tracks can not easily be  consistently included in either resummation
or fixed-order computations. However, we can make use of particle-level Monte
Carlo simulations, including the parton-to-hadron transition process, to
estimate the impact of this restriction. This is studied in detail along with
the parton-shower results in Sec.~\ref{sec:pheno}. There we will also verify
our assumption on the correspondence between measurements based on charged and
all particles in hadronic final states. 

\subsection{Event selection and phase-space constraints}\label{sec:eventselections}

For completeness, we will define the full phase space we consider for both
the analytic calculation and our Monte Carlo studies. We want
to study dijet events in proton--proton collisions at $\sqrt{s}=13\;\text{TeV}$
centre-of-mass energy.  We require events to contain at least two $R=0.4$
anti-$k_t$ jets~\cite{Cacciari:2008gp}, rather central in rapidity, and satisfying
an asymmetric cut on their transverse momenta:
\begin{eqnarray}
 |y_{j}|<1\,,\;\;\text{and}\qquad p_{T}^{\text{lead}} \geq p_{T,\text{min}}\,,\quad   p_{T}^{2\text{nd}} \geq \frac{p_{T,\text{min}}}{2}\,.\label{eq:jetcuts}
\end{eqnarray}
In what follows we consider the two choices $p_{T,\text{min}}= 200\;\text{GeV}$
and $p_{T,\text{min}}= 500\;\text{GeV}$. Note, these jet requirements serve
as triggers only; the observable calculation is based on final-state particles,
\emph{not} on jets. For validation, we produce Monte Carlo samples at parton
level as well as hadron level taking charged and neutral final states into account.
To address the restricted acceptance for particle tracks mentioned above, only
particles with $|y|\leq y_{\text{max}}=2.6$ are considered for both jet reconstruction
and observable calculation. All cuts mentioned so far are implemented in the
fixed-order calculations as well. For our final "experiment-level" Monte-Carlo
prediction, we in addition only take into account charged particles in the hadronic
final state, with a transverse momentum of
\begin{equation}
  p^{\text{track}}_{T}\geq p^{\text{track}}_{T,\text{min}}= 500\;\text{MeV}\,.
\end{equation}
To study the impact of soft-drop grooming, we investigate a variety of parameter
choices, \emph{i.e.}\ $\zcut\in \{ 0.05, 0.1, 0.2, 0.3\}$ with $\beta\in \{0, 1,
2\}$, while keeping $R_{\text{SD}}=1$ fixed. For the higher jet
transverse-momentum selection ($p_{T,\text{min}}= 500\;\text{GeV}$) we consider
in addition smaller grooming thresholds, \emph{i.e.}\ $\zcut=0.01$ and
$\zcut=0.02$. We make use of the \fastjet implementation of the soft-drop
procedure~\cite{Cacciari:2011ma}.  

\subsubsection*{An event display}

\begin{figure}[t!]
        \centering
        \includegraphics[width=0.89\textwidth]{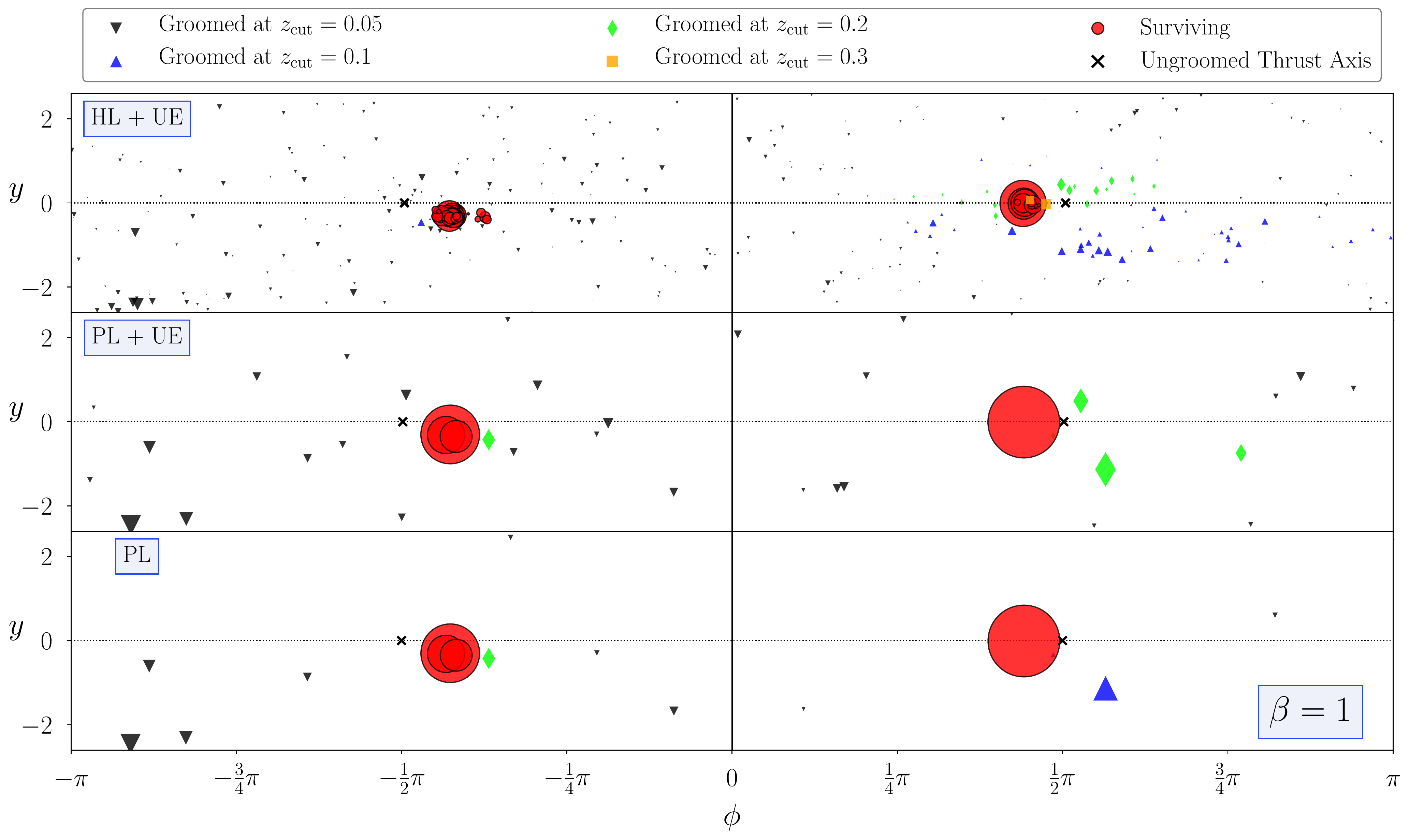}
        \caption{Event-display view of a typical two-jet event from proton--proton
          collision at $\sqrt{s}=13\;\text{TeV}$ at different stages of the event
          evolution. The respective final states are shown for various levels of
          soft-drop grooming, \emph{i.e.}\ without grooming,
          and $\zcut=\{0.05,0.1,0.2,0.3\}$  with $\beta=1$.}\label{fig:eventdisplay}
\end{figure}

To illustrate the potential of soft-drop grooming to mitigate the impact in
particular of the underlying event, we provide in Fig.~\ref{fig:eventdisplay} an
event-display view of a typical event simulated with the \Sherpa generator.
We use $\beta=1$ and show the effect of several $\zcut$ values, indicated by
different colours. As done for Fig.~\ref{fig:NP_ungroomed}, we consider the
event at different levels of its evolution, \emph{i.e.}\ at parton level (lower
panel), with underlying event included (middle panel) and at hadron level (upper
panel). At all stages we take into account all particles in the acceptance region
and do not apply a $p_{T,\text{min}}^{\text{track}}$ cut.
 
Each particle in the event display is scaled linearly in size according to
its transverse momentum, with larger points corresponding to larger values of $p_T$.
The event's two leading jets have transverse momenta of
$p_{T,j}^{\text{lead}}\approx 215\;\text{GeV}$ and $p_{T,j}^{\text{2nd}}\approx 185\;\text{GeV}$,
respectively. The hemispheres of the event are separated by a vertical line,
which we choose to be at $\phi=0$, with $\phi < 0$ and $\phi \ge 0$ for the left
and right hemisphere, correspondingly. It is visible that the particles
surviving the hardest grooming modes can cluster significantly away from the
original thrust axis. For this isolated event this is due to the presence
of semi-hard emissions sufficiently separated from the hardest parton of the PL
event. A majority of the underlying-event activity is already groomed away with
$\zcut \leq 0.1$ (black and blue) while larger values of $\zcut$ (green and
orange) probe into the hard process. In this particular event we can observe
that, in the right hemisphere for $\zcut=0.3$, grooming probes into the
fragmentation products of the hardest PL parton. Thus, by grooming ever more
aggressive, sizeable difference between observable values at parton and hadron
level can emerge despite most hadronisation corrections already being removed at
softer choices of grooming.

\FloatBarrier
\clearpage

\section{\NLL resummation and matching to \NLO QCD}\label{sec:res}

Having established the procedure for the type of observable we wish to study, we can detail the strategy
for the resummation of this school of observables. In this section we will start
from the general \Caesar framework for resummation and show the alterations
needed for groomed event shapes. Finally, we confirm the logarithmic structure by comparing
the expansion of the resummation to fixed order for groomed transverse thrust,
and present matched \NLOpNLLp results. 

\subsection{\Caesar in a nutshell}

We base our calculation on the well known \Caesar formalism for soft-gluon resummation~\cite{Banfi:2003je,Banfi:2004nk}.
Starting from a well separated, hard Born configuration $\mathcal{B}$, it is possible to write
the cumulative distribution of a given observable $v$ --- resummed to \NLL accuracy, with $L\equiv \ln\left(1/v\right)$ being the relevant
logarithm --- in a rather generic master formula 
\begin{equation}\label{eq:CAESAR}
  \begin{split}
    \Sigma_\mathrm{res}(v) &= \sum_\delta \Sigma^\delta(v)\,,\,\,\text{where} \\  
    \Sigma_\mathrm{res}^\delta(v) &= \int d\mathcal{B_\delta}
    \frac{\mathop{d\sigma_\delta}}{\mathop{d\mathcal{B_\delta}}} \exp\left[-\sum_{l\in\delta}
      R_l^\mathcal{B_\delta}(L)\right]\mathcal{P}^{\mathcal{B}_\delta}(L)\mathcal{S}^\mathcal{B_\delta}(L)\mathcal{F}^\mathcal{B_\delta}(L)\mathcal{H}^{\delta}(\mathcal{B_\delta})\,,
  \end{split}
\end{equation}
and \NLL is defined as systematically exponentiating all contributions of the type
$\alphaS^n L^n$ and higher logarithmic powers.\footnote{Despite the resummation structure of soft drop for $\beta=0$ starting at $\alphaS L$ at the lowest order, we stick to a naming scheme independent of $\beta$.} This formulation is
applicable to a wide range of observables. The sum extends over different 
partonic channels $\delta$; we will drop this label in the following if not explicitly
needed. The main ingredients are the hard function $\mathcal{H}$ representing the kinematic cuts
on the Born kinematics $\mathcal{B}$, the function $\mathcal{F}$ accounting for the effect
of multiple emissions, the soft function $\mathcal{S}$ implementing the non-trivial colour evolution, the collinear radiators $R_l$ for all hard legs $l$,
and the ratio of parton-distribution-functions (PDFs) $\mathcal{P}$ to take into account the true initial-state collinear scale,
\begin{equation}
  \mathcal{P}^{\mathcal{B}_\delta}(L) = \prod_{l=1}^2q_l\left(x_l^{\mathcal{B_\delta}},e^{-L/(a_l+b_l)}\mu_F\right)/q_l\left(x_l^{\mathcal{B_\delta}},\mu_F\right).
\end{equation} 
We refer to the original literature on the
\Caesar formalism, in particular the review \cite{Banfi:2004yd}, for a detailed discussion on the construction
and the applicability of the approach. It has been used to resum a number of event shapes
in hadron-hadron collisions \cite{Banfi:2010xy}. These can in principle all be modified and
studied as soft-drop groomed variants. We will present the general formalism to do so, we
focus on groomed transverse thrust for concrete results. In what follows we make use of
the implementation of the \Caesar formalism in the \Sherpa framework presented
originally presented in~\cite{Gerwick:2014gya} which we recently also applied to
obtain resummed predictions for soft-drop thrust~\cite{Marzani:2019evv} and
multijet resolution scales~\cite{Baberuxki:2019ifp} in electron--positron
collisions. 

The building blocks of the \Caesar resummation formula can be calculated for 
observables $V$ (vanishing at Born level) that have a specific scaling behaviour when
assuming an additional soft gluon with momentum $k$, collinear to a leg $l\in\mathcal{B}$, \emph{i.e.}\
\begin{equation}\label{eq:CAESAR_param}
  V(k)=\left(\frac{k_{t}^{\left(l\right)}}{\mu_Q}\right)^{a_{l}}e^{-b_{l}\eta^{\left(l\right)}}d_{l}\left(\theta\right)g_{l}\left(\phi\right)\,.
\end{equation}
Here $k_{t}^{\left(l\right)}$ and $\eta^{\left(l\right)}$ are the emission's transverse
momentum and pseudorapidity relative to leg $l$, respectively. Furthermore, $\phi$ labels the azimuthal angle of
the emission, while $\theta$ is the hard-process scattering angle in the centre-of-mass
frame. Finally, $\mu_Q$ denotes the hard scale, or resummation scale, of the problem.
The main goal for the remainder of this section is to recompute the building blocks including
the effect of soft-drop grooming as described in Sec.~\ref{sec:obs_def}. 

\subsection{\NLL resummation for soft-drop groomed event shapes}\label{sec:SD-NLL}

Due to some of the complexities involved in the treatment of initial-state emissions we
will compute the full resummation in the strict $v\ll \zcut\ll 1$ limit. This allows us to
ensure that the logarithms of the observables are taken into account up to \NLL
accuracy. We stress that we are neglecting any contributions not associated with
at least next-to-leading order in $L$, logarithms of the soft-drop
parameter $\zcut$ are only taken into account if they appear in such a contribution.
In practice the grooming logarithms are not with respect to $\zcut$ but some
related $\zcut^\prime$ absorbing factors dependent on the exact hard kinematics,
which should be taken into account for the appropriate limit instead. We will
specify $\zcut^\prime$ later and just note here that, as long as
$\zcut^\prime/\zcut={\cal O}(1)$, the two limits coincide. The region
$v\sim\zcut$ will also not be treated and therefore the resummation will not
shift into the ungroomed contribution beyond the transition point. Those effects
will hence only be accounted for to the accuracy of the fixed-order
calculation. 

The behaviour of the observable in the presence of an additional soft gluon that
remains ungroomed is still given by Eq.~\eqref{eq:CAESAR_param}, however,
grooming imposes additional phase-space constraints given by
Eq.~\eqref{eq:SD}. Per construction, the C/A algorithm will always cluster the
emission to one of the final-state legs, independent of the parton it is
actually radiated off. We therefore need to treat initial- and final-state
emissions separately.

In the $v\ll \zcut\ll 1$ limit only radiation from the final-state legs will
potentially not be groomed and can result in logarithms of the
observable. Collinear initial-state and the associated PDF contributions will
always be groomed away and result in  logarithms of $\zcut$ or related
variables, but not be enhanced by $L$. The same is
true for wide-angle soft emissions. In practice, we can hence set $R_1 = R_2 =
0$ and $\mathcal{P}=1,\, \mathcal{S}=1$ in the groomed case. Further discussion
of those  contributions, relevant away from the strict limit we are working in,
is presented in App.~\ref{sec:logzc}. We finally note that this
in particular includes the type of emissions with $\Delta R_{ij} \gg
R_\mathbf{SD}$ with respect to the relevant final-state leg, which we discussed
earlier due to their special nature in this approach to grooming events.
Those will hence not need any special consideration in the resummed
calculation at this accuracy.

Finally, in this limit the multiple-emission nature of
the observable is not altered. In particular for the typical case where the
multiple-emission function only depends on the logarithmic derivative of the
radiator, $\mathcal{F}(L) = \mathcal{F}(R^\prime(L))$, we can use the same
functional form as in the ungroomed case, \emph{i.e.}\ for additive
observables like transverse thrust $\mathcal{F}(R^\prime) = e^{-\gamma_E
  R^\prime}/\Gamma(1+R^\prime)$, simply with a different radiator
argument. Including the limit $v\sim\zcut$ would change the $\mathcal{F}$
function, as discussed in detail in Appendix B.2 of~\cite{Marzani:2019evv}.

\begin{figure}
	\begin{center}
		\includegraphics[width=0.42\textwidth]{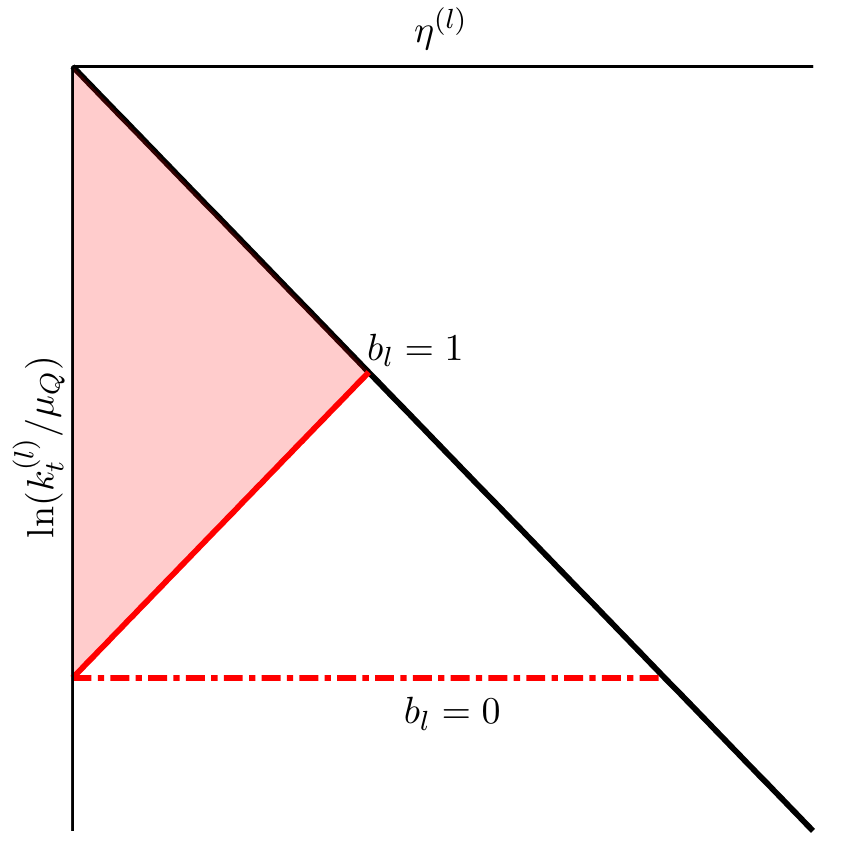}~
		\includegraphics[width=0.42\textwidth]{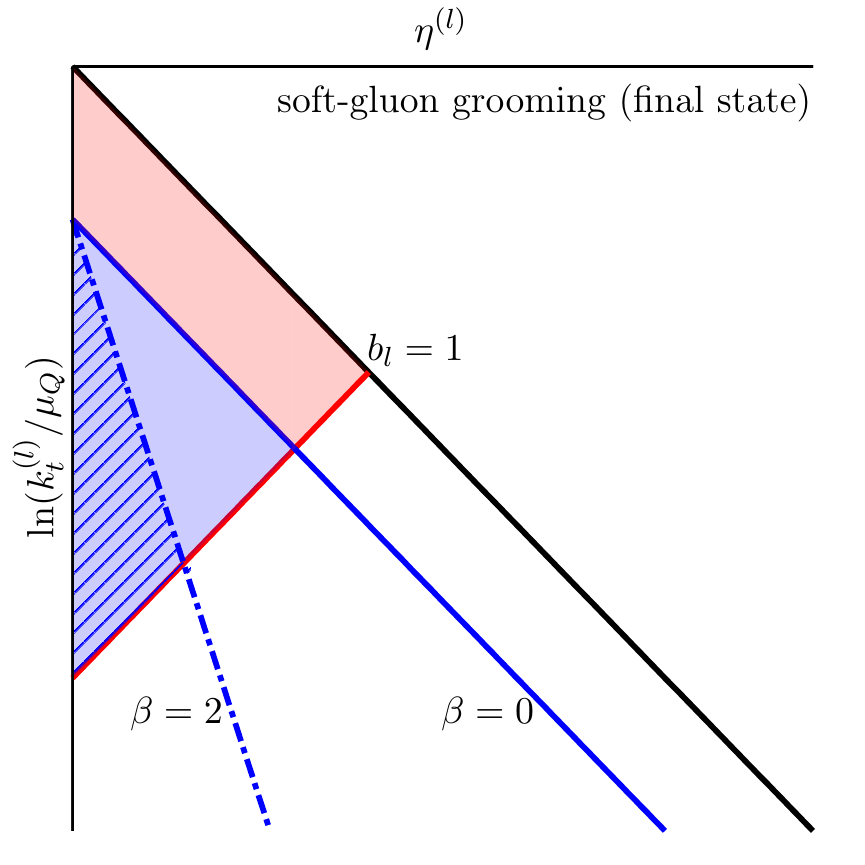}
	\end{center}
	\caption{The emission phase space in the Lund plane illustrating the kinematic
          constraints without (left) and with grooming (right). In the left figure we
          consider the \Caesar parameters $a_l=1$ with $b_l=0$ and $b_l=1$ (valid for thrust).
          The blue areas in the right panel mark the phase-space region removed by
          grooming with boundaries corresponding to $\beta=0$ (solid line and area)
          and $\beta=2$ (dashed line, hatched area). }
	\label{fig:Lund}
\end{figure}

In the remainder of this section we will recompute the final-state leg radiators
$R_l$.  The general phase-space constraints for emissions off leg $l$ are given by:
\begin{enumerate}
\item[](i) $\eta^{(l)}>\ln(2E_l/Q)$ (with $Q$ the mass of the radiating dipole) for the emission to be collinear to $l$,
\item[](ii) the limit given by collinear momentum conservation, \emph{i.e.}\ $\eta^{(l)} <
  \ln(2E_l/k_t^{(l)})$, and,
  \item[](iii) the condition $v>V(k)$, \emph{cf.}\ left hand
    side of Fig.~\ref{fig:Lund} for the resulting Lund-plane diagram.
\end{enumerate}

Soft-drop grooming now imposes an additional constraint:
\begin{enumerate}
\item[](iv) an emission only contributes if it is not groomed, implying\\
  \begin{equation*}
    \frac{k_{t}^{(l)}e^{\left(1+\beta\right)\eta^{\left(l\right)}}}{2E_{l}}\geq\zcut\left(\frac{R_{\text{SD}}\sin\theta}{2}\right)^{-\beta}\equiv\zcut^{\prime}\,.
  \end{equation*}
\end{enumerate}
To derive this last condition, we rewrote the soft-drop criterion
Eq.~\eqref{eq:SD} in the soft/collinear region in terms of $k^{(l)}_{t}$ and
$\eta^{\left(l\right)}$, using
\begin{eqnarray*}
p_{T,k} =k^{(l)}_{t}\frac{e^{\eta^{\left(l\right)}}}{2}\sin\theta\,,&& p_{T,l}  =E_{l}\sin\theta\,,\\
\left|\Delta y\right| =\frac{2e^{-\eta^{\left(l\right)}}\cos\phi}{\sin\theta}\,,&&
\left|\Delta\phi\right| =\frac{2e^{-\eta^{\left(l\right)}}\sin\phi}{\sin\theta}\,.
\end{eqnarray*}
As mentioned earlier, the relevant argument of the logarithms is $\zcut^\prime$,
as we now formally introduced it. It differs from $\zcut$ by a factor
$\left(R_{\text{SD}}\mathop{\sin\theta} / 2\right)^{-\beta}$.
Since we are interested here in the collinear limit relative to centralised
final-state particles, \emph{i.e.}\ $\sin\theta \sim\mathcal{O}(1)$, and choose
$R_{\textbf{SD}}\sim\mathcal{O}(1)$ we indeed have $\zcut^\prime/\zcut \sim\mathcal{O}(1)$.
This implies that the limits $v\ll \zcut^\prime\ll 1$ and $v\ll \zcut\ll 1$ correspond to one
another. In our following numerical analysis we use $R_{\textbf{SD}} = 1$ and require Born-level
events with $|y_j|<1$, \emph{cf.} Eq.~\eqref{eq:jetcuts}, what corresponds to $\sin\theta>0.648$.
Note, the limit $\zcut^\prime\ll 1$ is challenged when considering grooming parameters
$\zcut\geq 0.1$ and $\beta=2$. Despite this, we still probe such choices in our
phenomenological analyses. However, corrections proportional to $\zcut^\prime$ beyond the ones
considered here might become numerically sizeable for these cases. In any case,
they will be taken into account to NLO after matching.

\FloatBarrier
With the above general considerations, radiation in the blue area in the right hand
side of Fig.~\ref{fig:Lund} does not contribute to the groomed observable, and the
integral over this area should be subtracted from the ungroomed case. This results in
the radiator function for the groomed observable being given by

\begin{align}
R\left(v,\zcut\right) =\sum_{l}C_{l}&\left[\vphantom{\frac{\frac12}{\frac{\frac{1\frac12}{2}1}{\frac12}}}\int_{\mu_Q^{2}v^{\frac{2}{a_{l}+b_{l}}}}^{\mu_Q^{2}}\frac{dk_{t}^{2}}{k_{t}^{2}}\frac{\alphaS\left(k^2_{t}\right)}{\pi}\left(\ln\left(\frac{Q}{k_{t}}\right)+B_{l}\right)\right.\nonumber \\
 +&\int_{\mu_Q^{2}v^{\frac{2}{a_{l}}}}^{\mu_Q^{2}v^{\frac{2}{a_{l}+b_{l}}}}\frac{dk_{t}^{2}}{k_{t}^{2}}\frac{\alphaS\left(k^2_{t}\right)}{\pi}\left(\ln\left(\frac{Q}{2E_{l}}\right)+\frac{1}{b_{l}}\ln\left[\left(\frac{k_{t}}{\mu_Q}\right)^{a_l}\frac{d_{l}g_{l}}{v}\right]\right)\nonumber \\
-&\int_{\mu_Q^{2}f\left(v,\zcut^{\prime}\right)}^{\mu_Q^{2}\left(\zcut^{\prime}\right)^{2}}\frac{dk_{t}^{2}}{k_{t}^{2}}\frac{\alphaS\left(k^2_{t}\right)}{\pi}\left(\ln\left(\frac{Q}{k_{t}}\right)-\frac{\beta}{1+\beta}\ln\left(\frac{2E_{l}}{k_{t}}\right)+\frac{1}{1+\beta}\ln\left( \zcut^{\prime}\right)\right)\nonumber \\
-&\left.\int_{\mu_Q^{2}v^{\frac{2}{a_{l}}}}^{\mu_Q^{2}f\left(v,\zcut^{\prime}\right)}\frac{dk_{t}^{2}}{k_{t}^{2}}\frac{\alphaS\left(k^2_{t}\right)}{\pi}\left(\ln\left(\frac{Q}{2E_{l}}\right)+\frac{1}{b_{l}}\ln\left[\left(\frac{k_{t}}{\mu_Q}\right)^{a_l}\frac{d_{l}g_{l}}{v}\right]\right)\right]\,,\label{eq:Radiator}
\end{align}
with the soft-collinear corner parameterised as
\begin{equation}
f\left(v,\zcut^{\prime}\right)=v^{\frac{2\left(1+\beta\right)}{b_{l}+a_{l}\left(1+\beta\right)}}\left(\zcut^{\prime}\right)^{\frac{2b_{l}}{b_{l}+a_{l}\left(1+\beta\right)}}=v^{2p_{l}^{\left(v\right)}}\left(\zcut^{\prime}\right)^{2p_{l}^{\left(z\right)}}\,,
\end{equation}
where in the second equality we defined
\begin{equation}
  p_l^{(v)} \equiv \frac{1+\beta}{b_{l}+a_{l}\left(1+\beta\right)}\quad \text{and}\quad p_l^{(z)} \equiv \frac{b_{l}}{b_{l}+a_{l}\left(1+\beta\right)}\,.
\end{equation}
The first two lines of Eq.~\eqref{eq:Radiator}
correspond to the original ungroomed triangle contribution. Using the usual \Caesar convention the coefficient $B_{l}$
accounts for hard-collinear splittings. The last two lines in Eq.~\eqref{eq:Radiator} subtract off emissions that
are groomed. The difference in the overall scales in the $k_t$-integration is beyond \NLL accuracy and $\mu_Q$ can be used for
all.  Note that, as already indicated by our notation, we are free to use a
unique scale $Q$ to represent the maximal kinematic energy available for
emissions from leg $l$, although it generally differs for the contributions of
the various dipoles $l$ is a part of. These differences, however, can be
captured at \NLL accuracy by the soft function $\mathcal{S}$, which for groomed
event shapes only results in logarithms of $\zcut$. The radiator function can be
brought to a form similar to the ungroomed case, see for instance Eq.~\textup{(5)} in
Ref.~\cite{Banfi:2003je},

\begin{align}
R\left(v,\zcut\right)  =\sum_{l}C_{l}&\left[r_{l}\left(L,\thinspace L_{z}\right)+r_{l}^{\prime}\left(L,\thinspace L_{z}\right)\left(\ln\left(\bar{d}_{l}\right)-b_{l}\ln\left(\frac{2E_{l}}{\mu_Q}\right)\right)\right.\nonumber \\
& \left.+\beta\,\dot{r}_{l}\left(L,\thinspace
    L_{z}\right)\ln\left(\frac{2E_{l}}{\mu_Q}\right)+B_{l}T\left(\frac{\alphaS\beta_0
      L}{a_{l}+b_{l}}\right)+T\left(\alphaS\beta_0 L_{z}\right)\ln\left(\frac{Q}{\mu_Q}\right)\right]\,.
\end{align}
Here we have introduced an additional logarithm $L_{z}=\ln(1/\zcut^{\prime})$. The function $T$ is given by
\begin{equation}
T(X) = \int^{\mu_Q^2}_{\mu_Q^2e^{-\frac{2X}{\alphaS\beta_0}}}
\frac{dk_{t}^{2}}{k_{t}^{2}}\frac{\alphaS\left(k^2_{t}\right)}{\pi} = \frac{-\ln(1-2X)}{\pi\beta_0}\,,\label{eq:T}
\end{equation}
with $\beta_0=(11 C_A -2 n_f)/(12\pi)$ and $\alphaS=\alphaS(\muR^2)$. Furthermore, $\ln(\bar{d}_{l})$
is given by the sum of $\ln(d_l)$ and the azimuthally averaged $\ln(g_l)$ contribution, \emph{i.e.}\
\begin{equation}
 \ln\left(\bar{d}_{l}(\theta)\right) =\ln \left(d_{l}(\theta)\right)+\int_{0}^{2\pi}\frac{d\phi}{2\pi}\ln\left( g_{l}(\phi)\right)\,.
\end{equation}
The appearing resummation functions for the groomed event shapes follow the same structure as in the ungroomed case,
however, the additional function $\dot{r}_{l}$ appears, given by the derivative of $r_l$ with respect to $L_z$, rather
than $L$, as for $r_l^\prime$. Their explicit form reads:
\begin{align}
  r_{l}\left(L,\thinspace L_{z}\right) & =\int_{\mu_Q^{2}e^{\frac{-2L}{a_{l}+b_{l}}}}^{\mu_Q^{2}}\frac{dk_{t}^{2}}{k_{t}^{2}}\frac{\alphaS\left(k^2_{t}\right)}{\pi}\ln\left(\frac{\mu_Q}{k_{t}}\right)\nonumber\\
  &+\int_{\mu_Q^{2}e^{-2\left(p_{l}^{\left(v\right)}L+p_{l}^{\left(z\right)}L_{z}\right)}}^{\mu_Q^{2}e^{\frac{-2L}{a_{l}+b_{l}}}}\frac{dk_{t}^{2}}{k_{t}^{2}}\frac{\alphaS\left(k^2_{t}\right)}{\pi}\left(\frac{L}{b_{l}}+\frac{a_{l}}{b_{l}}\ln\left(\frac{k_{t}}{\mu_Q}\right)\right)\nonumber \\
& +\int_{\mu_Q^{2}e^{-2\left(p_{l}^{\left(v\right)}L+p_{l}^{\left(z\right)}L_{z}\right)}}^{\mu_Q^{2}e^{-2L_{z}}}\frac{dk_{t}^{2}}{k_{t}^{2}}\frac{\alphaS\left(k^2_{t}\right)}{\pi}\frac{1}{1+\beta}\left(L_{z}-\ln\left(\frac{\mu_Q}{k_{t}}\right)\right)\nonumber\\
&=\frac{1}{\alphaS}r_{1,l}\left(\alphaS\beta_0L,\thinspace \alphaS\beta_0L_{z}\right)+r_{2,l}\left(\alphaS\beta_0L,\thinspace \alphaS\beta_0L_{z}\right)\,,\label{eq:rl}\\
  r_{l}^{\prime}\left(L,\thinspace L_{z}\right) & =\frac{1}{b_{l}}\int_{\mu_Q^{2}e^{-2\left(p_{l}^{\left(v\right)}L+p_{l}^{\left(z\right)}L_{z}\right)}}^{\mu_Q^{2}e^{\frac{-2L}{a_{l}+b_{l}}}}\frac{dk_{t}^{2}}{k_{t}^{2}}\frac{\alphaS\left(k^2_{t}\right)}{\pi}\nonumber\\
  &=\frac{1}{b_{l}}\left[T\left(p_{l}^{\left(v\right)}\alphaS\beta_0L+p_{l}^{\left(z\right)}\alphaS\beta_0L_{z}\right)-T\left(\frac{\alphaS\beta_0L}{a_{l}+b_{l}}\right)\right]\,,\\
\dot{r}_{l}\left(L,\thinspace L_{z}\right) & =\frac{1}{1+\beta}\int_{\mu_Q^{2}e^{-2\left(p_{l}^{\left(v\right)}L+p_{l}^{\left(z\right)}L_{z}\right)}}^{\mu_Q^{2}e^{-2L_{z}}}\frac{dk_{t}^{2}}{k_{t}^{2}}\frac{\alphaS\left(k^2_{t}\right)}{\pi}\nonumber\\&=\frac{1}{1+\beta}\left[T\left(p_{l}^{\left(v\right)}\alphaS\beta_0L+p_{l}^{\left(z\right)}\alphaS\beta_0L_{z}\right)-T\left(\alphaS\beta_0L_{z}\right)\right]\,.
\end{align}
The LL and \NLL terms $r_{1,l}$ and $r_{2,l}$ appearing in Eq.~\eqref{eq:rl} are given by
\begin{flalign}
r_{1,l}\left(\lambda,\thinspace \lambda_{z}\right)=\frac{-1}{2\pi\beta_0^2}&\left[\frac{a_l+b_l}{b_l}\left(1-\frac{2\lambda}{a_l+b_l}\right)\ln\left(1-\frac{2\lambda}{a_l+b_l}\right)+\frac{1}{1+\beta}\left(1-2\lambda_{z}\right)\ln\left(1-2\lambda_{z}\right)\right.\nonumber&&\\
&-\left.\frac{a_l(1+\beta)+b_l}{b_l(1+\beta)}\left(1-2(p_{l}^{\left(v\right)}\lambda+p_{l}^{\left(z\right)}\lambda_{z})\right)\ln\left(1-2(p_{l}^{\left(v\right)}\lambda+p_{l}^{\left(z\right)}\lambda_{z})\right)\right]\,,\label{eq:r1}&&\\
r_{2,l}\left(\lambda,\thinspace \lambda_{z}\right)=\frac{K}{4\pi^2\beta_0^2}&\left[\frac{a_l+b_l}{b_l}\ln\left(1-\frac{2\lambda}{a_l+b_l}\right)+\frac{1}{1+\beta}\ln\left(1-2\lambda_{z}\right)\right.\nonumber&&\\
&\left.- \frac{a_l(1+\beta)+b_l}{b_l(1+\beta)}\ln\left(1-2(p_{l}^{\left(v\right)}\lambda+p_{l}^{\left(z\right)}\lambda_{z})\right)\right] \nonumber&&\\
-\frac{\beta_1}{2\pi\beta_0^3}&\left[\frac{a_l+b_l}{b_l}\left(\frac12\ln^2\left(1-\frac{2\lambda}{a_l+b_l}\right)+\ln\left(1-\frac{2\lambda}{a_l+b_l}\right)\right)\right.\label{eq:r2}&&\\
  &\left.+\frac{1}{1+\beta}\left(\frac12\ln^2\left(1-2\lambda_{z}\right)+\ln\left(1-2\lambda_{z}\right)\right)\right.\nonumber&&\\
&\left. - \frac{a_l(1+\beta)+b_l}{b_l(1+\beta)}\left(\frac12\ln^2\left(1-2(p_{l}^{\left(v\right)}\lambda+p_{l}^{\left(z\right)}\lambda_{z})\right)+\ln\left(1-2(p_{l}^{\left(v\right)}\lambda+p_{l}^{\left(z\right)}\lambda_{z})\right)\right)\right]\,,\nonumber&&
\end{flalign}
with the coefficient $K$ in the $\overline{\text{MS}}$ renormalisation scheme
given by $K=\left(\frac{67}{18}-\frac{\pi^2}{6}\right)C_A-\frac{5}{9}n_f$. In
order to perform the actual calculation of the resummation we extended the
\Sherpa resummation plugin~\cite{Gerwick:2014gya} to include the soft-drop 
groomed version of the final-state radiators as presented above. 

\clearpage
\subsection{The non-perturbative realm}\label{sec:np_scales}

As is the case in the original \Caesar formalism, \emph{cf.} also \cite{Luisoni:2015xha}, the formulae presented here
exhibit logarithmic branch cuts as a result of integrating the strong-coupling
constant over the Landau pole. The positions of these branch cuts can be used
to estimated the observable scales at which non-perturbative physics, \emph{i.e.}\
hadronisation effects, dominate. From Eqs.~\eqref{eq:r1} and \eqref{eq:r2} we can
read off their locations as 
\begin{align}
  2\lambda &= a_l+b_l\,, \\
  2 p_{l}^{\left(v\right)}\lambda &= 1-2 p_{l}^{\left(z\right)} \lambda_z\,.  
\end{align}
Here we assumed $\lambda_z < 1/2$. These can easily be expressed in terms of the
observable $v = e^{-\lambda/\alphaS\beta_0}$, which yields 
\begin{align}
  v_\text{Had, collinear} &= \left(e^{-1/2\alphaS\beta_0}\right)^{a_l+b_l} =
  \left(\frac{\Lambda_\text{QCD}}{\mu_R}\right)^{a_l+b_l}\,, \\ 
  v_\text{Had, wide-angle} &= \left(e^{-1/2\alphaS\beta_0}\right)^{a_l+b_l/(1+\beta)}
  \left(e^{L_z}\right)^{b_l/(1+\beta)} =
  \left(\frac{\Lambda_\text{QCD}}{\mu_R}\right)^{a_l}
  \left(\frac{\Lambda_\text{QCD}}{\mu_R \zcut^\prime}\right)^{b_l/(1+\beta)}\,.
\end{align}
For the last equality we introduced $\Lambda_\text{QCD}^2 = \mu_R^2
e^{-1/\alphaS\beta_0}$. The notation reflects that the first solution
corresponds to the collinear limit, whereas the second one is approached for
the softest wide-angle emissions allowed by grooming (\emph{i.e.}\ the intersection of
the blue and red line in Fig.~\ref{fig:Lund}). The condition on $\lambda_z$
translates to $\zcut^\prime > \Lambda_\text{QCD}/\mu_R$, which if violated puts
the full groomed part of the distribution into the non-perturbative
region. Hadronisation corrections
start to dominate when $v$ approaches the larger of the two solutions, such
that, keepint in mind $\beta \geq 0$, we have
\begin{align}
  v_\text{Had} &= \left(\frac{\Lambda_\text{QCD}}{\mu_R}\right)^{a_l+b_l}\,,\qquad
  \text{for }~b_l\leq 0\,,\label{eq:vNP_groomed_blneg}\\
  v_\text{Had} &= \left(\frac{\Lambda_\text{QCD}}{\mu_R}\right)^{a_l}
  \left(\frac{\Lambda_\text{QCD}}{\mu_R \zcut^\prime}\right)^{b_l/(1+\beta)}\,,\qquad \text{for
  }~b_l>0\,.\label{eq:vNP_groomed}
\end{align}
This agrees with the estimates given for example in
Refs.~\cite{Frye:2016aiz,Caletti:2021oor} for the energy-energy correlations
$e_2^{(\alpha)}$ and jet angularities $\lambda^1_\alpha$, where the relevant
parameters are $a_l = 1$, $b_l=\alpha-1$ (see the latter reference for a
detailed treatment using the \Caesar parametrisation). For reference, we note
that, without grooming (\emph{i.e.} $\beta \to \infty$), we would obtain
\begin{equation}
  v^{\text{ungroomed}}_\text{Had} = \left(\frac{\Lambda_\text{QCD}}{\mu_R}\right)^{a_l}\,,\qquad\text{for
  }~b_l>0\,,\label{eq:vNP_ungroomed}
\end{equation}
while the $b_l\leq 0 $ case remains unchanged since here the dominant
contribution is from the collinear limit. By comparing Eqs.~\eqref{eq:vNP_groomed}
and \eqref{eq:vNP_ungroomed} it is apparent that grooming reduces the impact
of hadronisation corrections. 

\begin{figure}
	\begin{center}
		\includegraphics[width=0.8\textwidth]{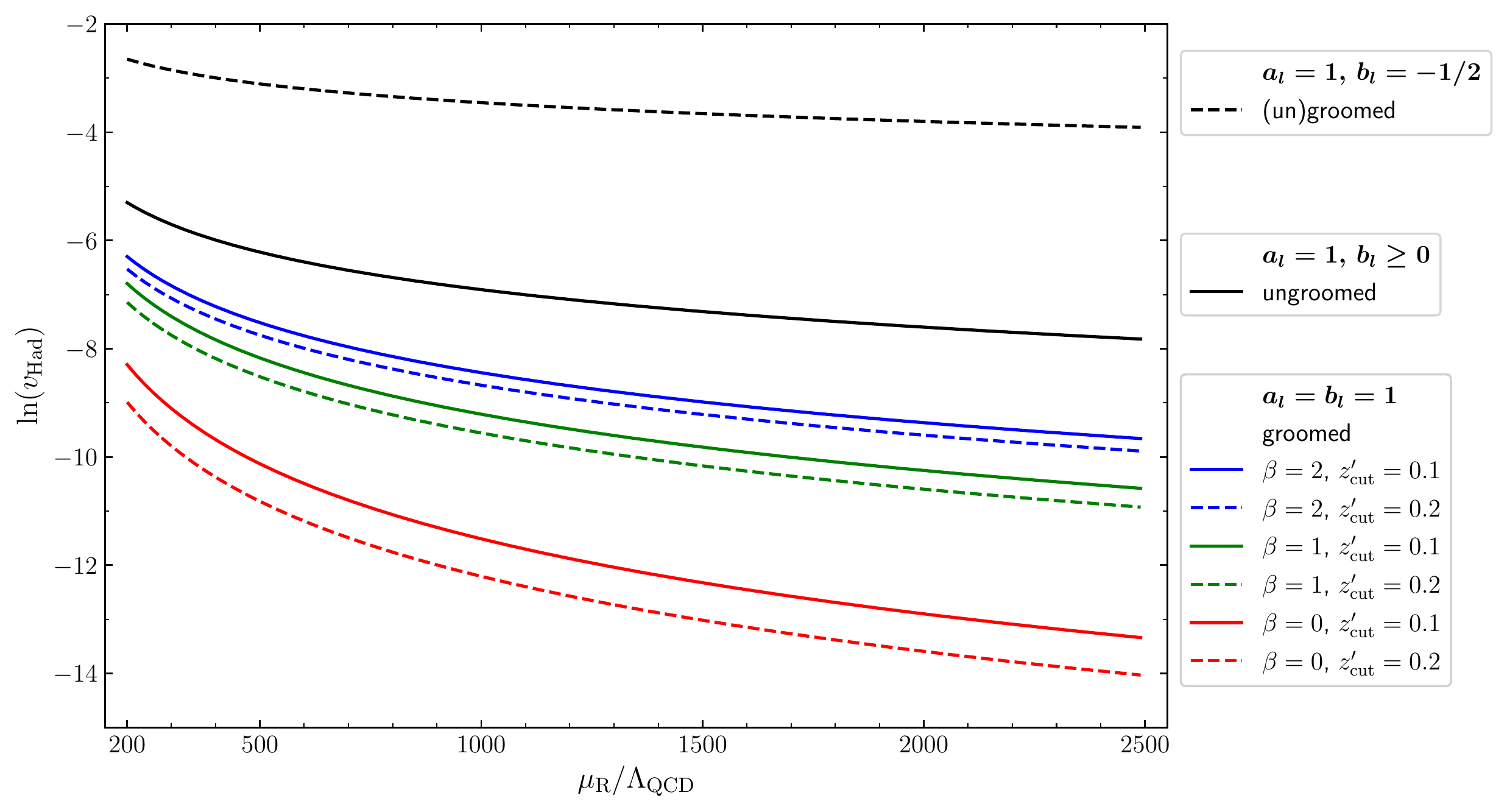}
	\end{center}
	\caption{Observable value $v_\text{Had}$ where to expect dominance of non-perturbative corrections,
          \emph{i.e.}\ hadronisation effects, as a function of the dimensionless quantity $\muR/\Lambda_\text{QCD}$.
          Results are given for different combinations of \Caesar
          parameters. For the case $a_l=b_l=1$, relevant for example for the
          thrust observable, we present results for different grooming parameter combinations.}
	\label{fig:vNP}
\end{figure}

In Fig.~\ref{fig:vNP} we present concrete results for $v_\text{Had}$ for different combinations
of \Caesar observable parameters in dependence on the dimensionless ratio $\muR/\Lambda_\text{QCD}$.
The black solid line corresponds to the case $a_l=1$ and $b_l=-1/2$, representing, for example,
the so-called LHA jet angularity $\lambda^1_{1/2}$. As stated above, for observables with $b_l<0$
the onset of hadronisation dominance is, independent of grooming, given by
Eq.~\eqref{eq:vNP_groomed_blneg}. As also noted in Ref.~\cite{Caletti:2021oor}, this observable
is very susceptible to non-perturbative corrections, even for rather high values of $\muR$.
The black dashed line represents the class of observables parametrised by $a_l=1$ and $b_l\geq0$
for the case of no grooming. Compared to the LHA angularity a somewhat reduced sensitivity to
hadronisation effects is evident. However, this can be significantly reduced through soft-drop
grooming. This is exemplified for the parameter set $a_l=b_l=1$, corresponding to the thrust
observable we are going to study in what follows. We consider here grooming with $\beta\in\{0,1,2\}$
and two representative $\zcut^\prime$ values, \emph{i.e.}\ $\zcut^\prime\in \{0.1,0.2\}$. For
the considered range of $\muR/\Lambda_\text{QCD}$ grooming with $\beta=2$ reduces the transition
point $v_\text{Had}$ by about $1-2$ $e$-folds. However, the dependence on $\zcut^\prime$ is rather
mild. As evident from Eq.~\eqref{eq:vNP_groomed}, the strongest suppression is obtained for $\beta=0$.
For $\zcut^\prime=0.1$ $v_\text{Had}$ gets reduced by about $4$ $e$-folds at $\muR/\Lambda_\text{QCD}=1000$.
By increasing $\zcut^\prime$ to $0.2$ we roughly gain another factor of $e$.

While these estimates are certainly helpful in getting a feeling for the
ultimate breakdown of the perturbative approach, we note that they do not
provide information on the scales related to the underlying event. In
Sec.~\ref{sec:pheno} such effects are addressed in a phenomenological study.

\subsection{Matching to \NLO and achieving \NLLp accuracy}\label{sec:matching_nllp}
We improve the pure resummation by matching to a fixed-order calculation
at \NLO QCD, \emph{i.e.}\ at ${\cal{O}}(\alphaS^2)$ relative to the Born process.
We combine our \NLL resummation and the \NLO in a way that allows us to effectively
achieve \NLLp accuracy. In what follows we detail our approach and discuss
subtleties arising due to soft-drop grooming.

\subsubsection*{Preliminaries}
Before performing the actual calculation it is common to exploit the ambiguities
present at \NLL to ease the matching to a fixed-order result. The natural argument
of the logarithms $L_z$ is the value of the transition point for these types of
emissions. Naturally, grooming vanishes as $v\to\left(\zcut^\prime\right)^{a_l}$,
at which point the observable transitions back to the ungroomed case. However,
this is based on the \NLL approximation of the kinematics. Formally, in the Lund
plane the LO transition point is reached when the intersection of the
$\eta^{(l)}=\ln(2E_l/Q)$ line and the soft-drop line is at the same value
of $k_{t}$ as that with the observable line. When this occurs the soft-drop
line no longer results in a kinematic boundary on the emissions and instead
the observable value is the only constraint. This happens when: 
\begin{equation}
\zcut^{\prime}Q^{1+\beta}\left(2E_{l}\right)^{-\beta} = \left(\frac{v_{\text{trans}}}{d_{l}g_{l}}\right)^{1/a_{l}}\left(\frac{2E_{l}}{Q}\right)^{b_{l}/a_{l}}\mu_Q\,,
\end{equation}
resulting in the transition point
\begin{equation}
v_{\text{trans}}=\left(\zcut^{\prime}\right)^{a_{l}}\left(\frac{2E_{l}}{\mu_Q}\right)^{a_{l}}\left(\frac{Q}{2E_{l}}\right)^{a_l\left(1+\beta\right)+b_{l}}d_{l}g_{l}=\tilde{z}_{\text{cut}}^{a_{l}}\,.
\end{equation}
At fixed logarithmic accuracy we can always shift the argument of the logarithms by 
multiplying $\zcut^\prime$ with a constant and include an additional
contribution to compensate for the leading terms in $L_z$. This way we can make
use of the actual LO transition point for this type of emission as an argument
of the logarithm, by comparing the full relations instead of the \NLL approximation. 
Accordingly, we shift the argument of the logarithm from
$L_{z}=-\ln(\zcut^{\prime})\to-\ln(\tilde{z}_{\text{cut}})$, and  the additional
contribution that needs to be taken into account reads: 
\begin{equation}
R\to R+\frac{1}{a_{l}}\dot{r}_{l}\left(L,\thinspace L_{z}\right)\left[a_{l}\ln\left(\frac{2E_{l}}{\mu_Q}\right)+\left(a_l\left(1+\beta\right)+b_{l}\right)\ln\left(\frac{Q}{2E_{l}}\right)+\ln\left(\bar{d}_{l}\right)\right]\,.
\end{equation}

Next, we include end-point corrections~\cite{Catani:1992ua,Jones:2003yv} to
ensure the cumulative distribution of the resummation approaches unity
and its derivative vanishes at the kinematic end-point of the fixed-order
distribution. Two types of logarithms are included, the one of the observable,
which can be altered to vanish at the end-point, and logarithms of $\zcut$,
which will not vanish there. The observable logarithms are modified
in the usual manner
\begin{equation}
\ln\left(\frac{1}{x_{L}v}\right)\to\frac{1}{p}\ln\left(\frac{1}{\left(x_{L}v\right)^p}-\frac{1}{\left(x_{L}v^{\mathrm{max}}\right)^p}+1\right)=L\,,
\end{equation}
where $x_L$ is chosen to be
\begin{equation}
  \ln(x_L)=-\sum_{l\in f}\ln(\bar{d}_l)/n\,,
\end{equation}
see Ref.~\cite{Baberuxki:2019ifp}, averaged over the final-state particles only,
here $n=2$. Per default we will assume $p=1$.  The argument of logarithms of $\zcut$
we multiply by the same factor, \emph{i.e.}\ $L_z=-\ln(x_L \tilde{z}_{\text{cut}}^{a_l})/a_l$.
Finally the exponential is modified according to  
\begin{equation}
\exp\left[\tilde{R}\left(L,\,L_{z}\right)\right]\to \exp\left[\tilde{R}\left(L,\,L_{z}\right)-\tilde{R}\left(0,\,L_{z}\right)-\left(\frac{v}{v^{\text{max}}}\right)^p \tilde{R}^{\prime}\left(0,\,L_{z}\right)L\right],
\end{equation}
where $\tilde{R}$ includes the multiple-emission function $\mathcal{F}$, and $\tilde{R}^\prime$
is its derivative with respect to $L$. Here we subtract $\tilde{R}\left(0,\,L_{z}\right)$ to consistently remove all pure $L_z$ type of contributions.

To remind the reader, for the practical implementation we focus on the resummation of logarithms
in the observable in the strict limit $v\ll \zcut\ll 1$. Therefore we shall neglect
the transition point in the resummation and expansion. We will only perform the exponentiation
of the logarithms of the groomed observable, which requires the inclusion of final-state
emissions only in the exponential. Other contributions result in logarithms of $\zcut$ and
will be included through means of matching to fixed order, \emph{i.e.}\ \NLO QCD.

\subsubsection*{Definition of the matched distributions}

To match the resummed predictions to the fixed-order calculation, we
follow the strategy presented using our conventions in~\cite{Baberuxki:2019ifp}. 
We adopt the same notation, \emph{i.e.}\ for every partonic channel $\delta$ we
introduce the cumulant distributions  
\begin{equation}\label{eq:barSigma}
  \Sigma^\delta(v) = \int_0^v \mathop{d\sigma^\delta}\quad\text{and}\quad \overline{\Sigma}^\delta(v)
  = \int_v^1 \mathop{d\sigma^\delta}\,,
\end{equation}
both, for the resummed (res) and the fixed-order (fo) calculations. They can be expanded
in $\alphaS$ relative to the Born configuration, which itself is ${\cal{O}}(\alphaS^2)$, 
\begin{equation}
  \Sigma^\delta(v) = \sigma^{\delta,(0)}+\Sigma^{\delta,(1)}(v) + \Sigma^{\delta,(2)}(v) + \dots~,\;\;\text{where}\;\; \Sigma^{\delta,(n)}(v)
  \propto \alphaS^{n+2}~,
\end{equation}
and $\sigma^{\delta,(0)} = \Sigma^{\delta,(0)}(1)$.
We then define (taking the argument of the cumulants as implicit) the multiplicatively
matched cumulant distribution for channel $\delta$
\begin{align}\label{eq:matching}
  \Sigma^{\delta}_\mathrm{mult} &= \Sigma^{\delta}_\mathrm{res}\left[1 + \frac{\Sigma^{\delta,(1)}_\mathrm{fo}-\Sigma^{\delta,(1)}_\mathrm{res}}{\sigma^{\delta,(0)}}
    + \frac{1}{\sigma^{\delta,(0)}}
    \left(-\overline{\Sigma}^{\delta,(2)}_\mathrm{fo}-\Sigma^{\delta,(2)}_\mathrm{res} -
    \frac{\Sigma^{\delta,(1)}_\mathrm{res}}{\sigma^{\delta,(0)}}
    \left(\Sigma^{\delta,(1)}_\mathrm{fo}-\Sigma^{\delta,(1)}_\mathrm{res}\right)\right)\right]\,,
\end{align}
which reproduces the fixed-order result when expanded to ${\cal{O}}(\alphaS^2)$.
In the limit $v\to 0$ it reduces to 
\begin{equation}
  \Sigma_\mathrm{mult}^\delta \to \left(1+\frac{\alphaS}{2\pi}C_1^\delta
  + {\cal{O}}(\alphaS^2)\right)\Sigma_\mathrm{res}^\delta~,\;\;\text{where}\qquad  \frac{\alphaS}{2\pi}C_1^\delta \equiv \lim_{v\to 0} \frac{\Sigma^{\delta,(1)}_\mathrm{fo}-\Sigma^{\delta,(1)}_\mathrm{res}}{\sigma^{\delta,(0)}}\,.
\end{equation}
We later also refer to matched distributions at LO, which corresponds to
including in Eq.~\eqref{eq:matching} only the first two terms in the square brackets.
Finally, the full cumulative distribution is given by the sum over partonic channels,
\emph{i.e.}\
\begin{equation}\label{eq:match_final}
  \Sigma_\mathrm{match} = \sum_{\delta\in\mathcal{B}} \Sigma_\mathrm{match}^{\delta} + \sum_{\delta\notin\mathcal{B}} \Sigma^{\delta}_\mathrm{fo}\,,
\end{equation}
where the second sum takes into account channels vanishing in the soft limit and
hence not part of the resummation.

\subsubsection*{Channel separation}
\begin{figure}
	\begin{center}
		\includegraphics[width=0.6\textwidth]{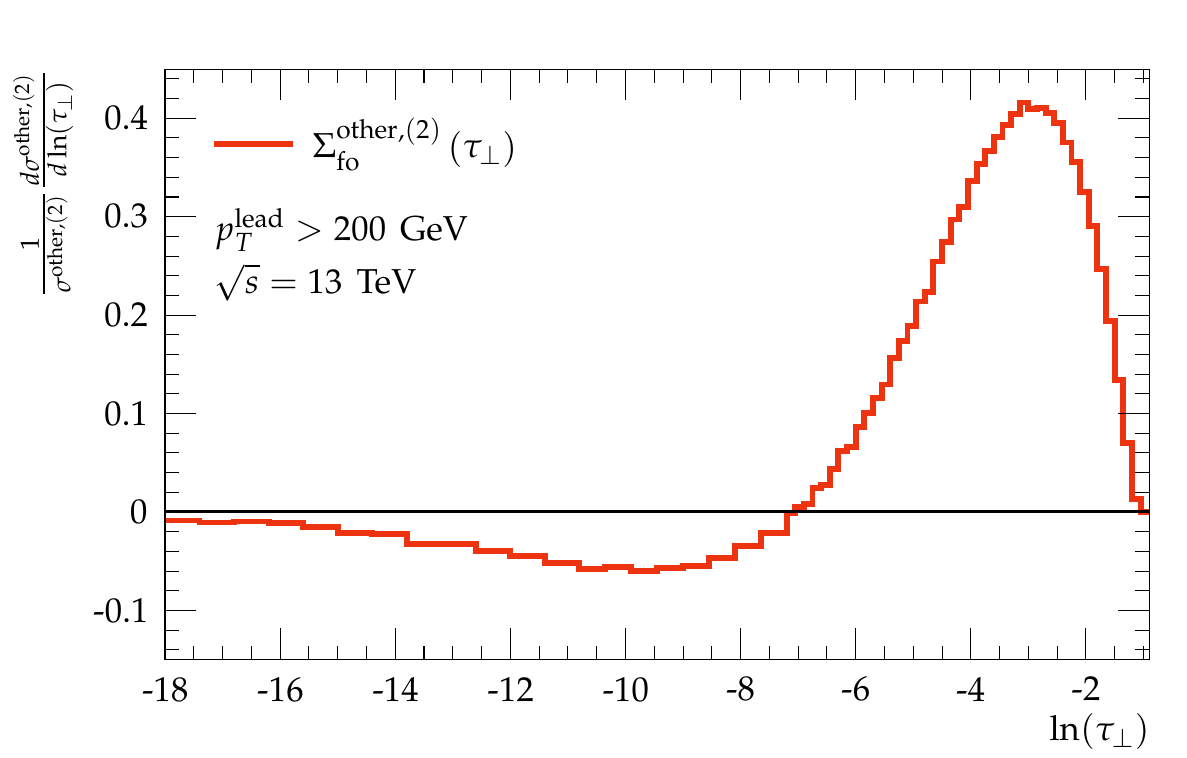}
	\end{center}
	\caption{The ${\cal{O}}(\alphaS^2)$ contribution to the `other' channel in the
          limit $\tauPerp\to0$.}
	\label{fig:other-ch}
\end{figure}

In order to separate the fixed-order phase space into the different channels
$\delta$, we employ the `flavour-$k_t$' algorithm defined in~\cite{Banfi:2006hf},
called \BSZ\ algorithm in the following. This is the approach taken in the resummation
of plain event shapes at hadron colliders in~\cite{Banfi:2010xy}. We implemented and
used the lepton-collider variant of this algorithm in our framework in earlier
work~\cite{Baberuxki:2019ifp}. Here
we will need to consider the hadron-collider version. We obtain the fixed-order
matrix elements from the \Comix generator~\cite{Gleisberg:2008fv}, that provides all the required information
about the flavour assignment of initial- and final-state partons in an automated fashion.
The \BSZ\ algorithm constitutes a sequential recombination algorithm, where the particle
$i$ with the smallest distance measure
\begin{align}
  d_{ij} &= \begin{cases}\min\left(p^2_{T,i},p^2_{T,j}\right)\Delta R^2_{ij}\,, & \text{if softer of
      $i,j$ is a gluon} \\ \max\left(p^2_{T,i},p^2_{T,j}\right)\Delta R^2_{ij}\,, & \text{if softer of
      $i,j$ is a quark}\end{cases},~\\
  d_{iB} &= \begin{cases}\min\left(p^2_{T,i},p^2_{T,B}(y_i)\right)\,, & \text{if $i$ is a gluon} \\
    \max\left(p^2_{T,i},p^2_{T,B}(y_i)\right)\,, & \text{if $i$ is a quark}\end{cases}\,,\quad\\
  d_{i\overline{B}} &= \begin{cases}\min\left(p^2_{T,i},p^2_{T,\overline{B}}(y_i)\right)\,, & \text{if $i$ is a gluon} \\
    \max\left(p^2_{T,i},p^2_{T,\overline{B}}(y_i)\right)\,, & \text{if $i$ is a quark}\end{cases},~\\
  \text{where}&\quad
  \begin{array}{l}
    p_{T,B}(y) = \sum_j
    p_{T,j}\left(\Theta\left(y_j-y\right)+\Theta\left(y-y_j\right)e^{y_j-y}\right)\\
    p_{T,\overline{B}}(y) = \sum_j
    p_{T,j}\left(\Theta\left(y-y_j\right)+\Theta\left(y_j-y\right)e^{y-y_j}\right)
  \end{array}\,,
\end{align}
is combined with either final-state particle $j$ or the forward ($B$) or
backward ($\overline{B}$) beam. The flavour of the combined object is determined by
the sum of the flavours of the two clustered entities. For this purpose, any object
with (anti-)quark flavour is called a quark, a gluon is an object without any net flavour
(including the case of same-flavour quark-antiquark pairs). We run this
algorithm until only two final-state objects are left. Together with the initial
state, their flavours define the channel $\delta$ to which we assign the event. The
algorithm as described so far can lead to channels that are not present as Born
channels, \emph{e.g.}\ including objects that have multiple (anti-)quarks or anti-quarks
and quarks of different flavour associated to them. In our matching scheme they
are taken into account in the sum over $\delta \notin\mathcal{B}$ in
Eq.~\eqref{eq:match_final}. We collect all those configurations in a common
channel, denoted as `other'. To test the infrared safety of the \BSZ\ algorithm and
our implementation we need to check that this channel vanishes in the
soft limit, \emph{i.e.}\ if all but two particles become soft and/or collinear
to one of the two remaining particles. This validation is presented in
Fig.~\ref{fig:other-ch}. We use transverse thrust $\tauPerp$ as a measure for the
hardness of the event. Infrared safety then implies that the differential
cross section for the `other' channel approaches zero in the limit $\tauPerp\to0$.
For the results we present in this work, we will use what is referred to as
`bland' version of the \BSZ\ algorithm in~\cite{Banfi:2006hf}, \emph{i.e.}, we
veto any clustering that would lead to such not Born like jets, by effectively
setting the measure $d_{ij}$ (and likewise the beam distance measures) to
infinity for those cases.

\begin{figure}[ht!]
	\begin{center}
		\includegraphics[width=0.45\textwidth]{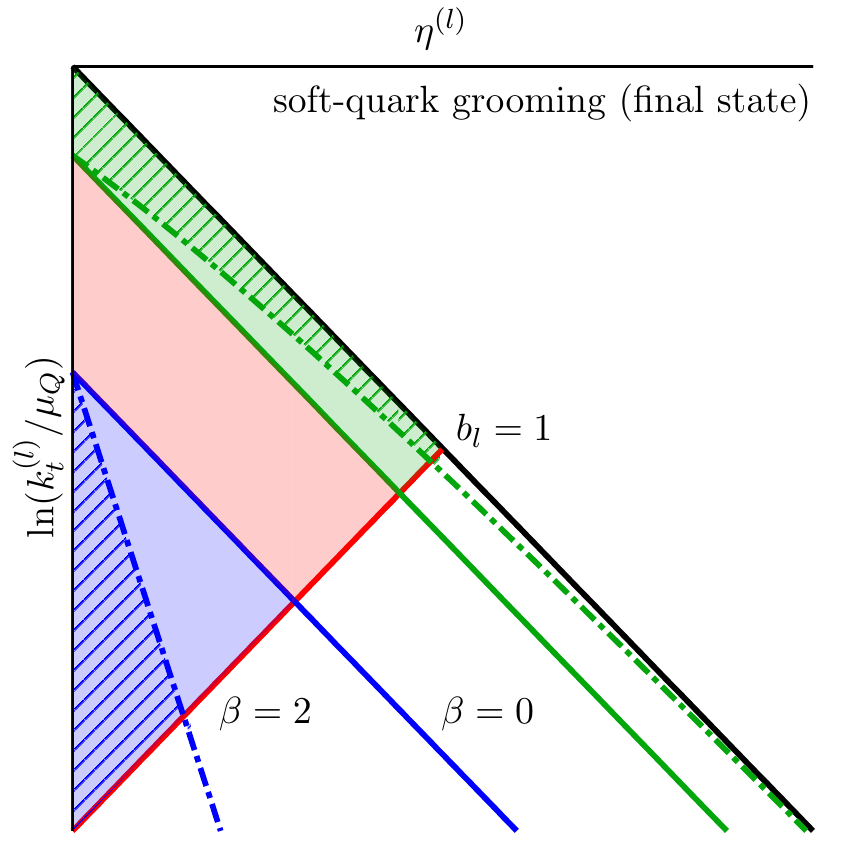}
		\includegraphics[width=0.5\textwidth]{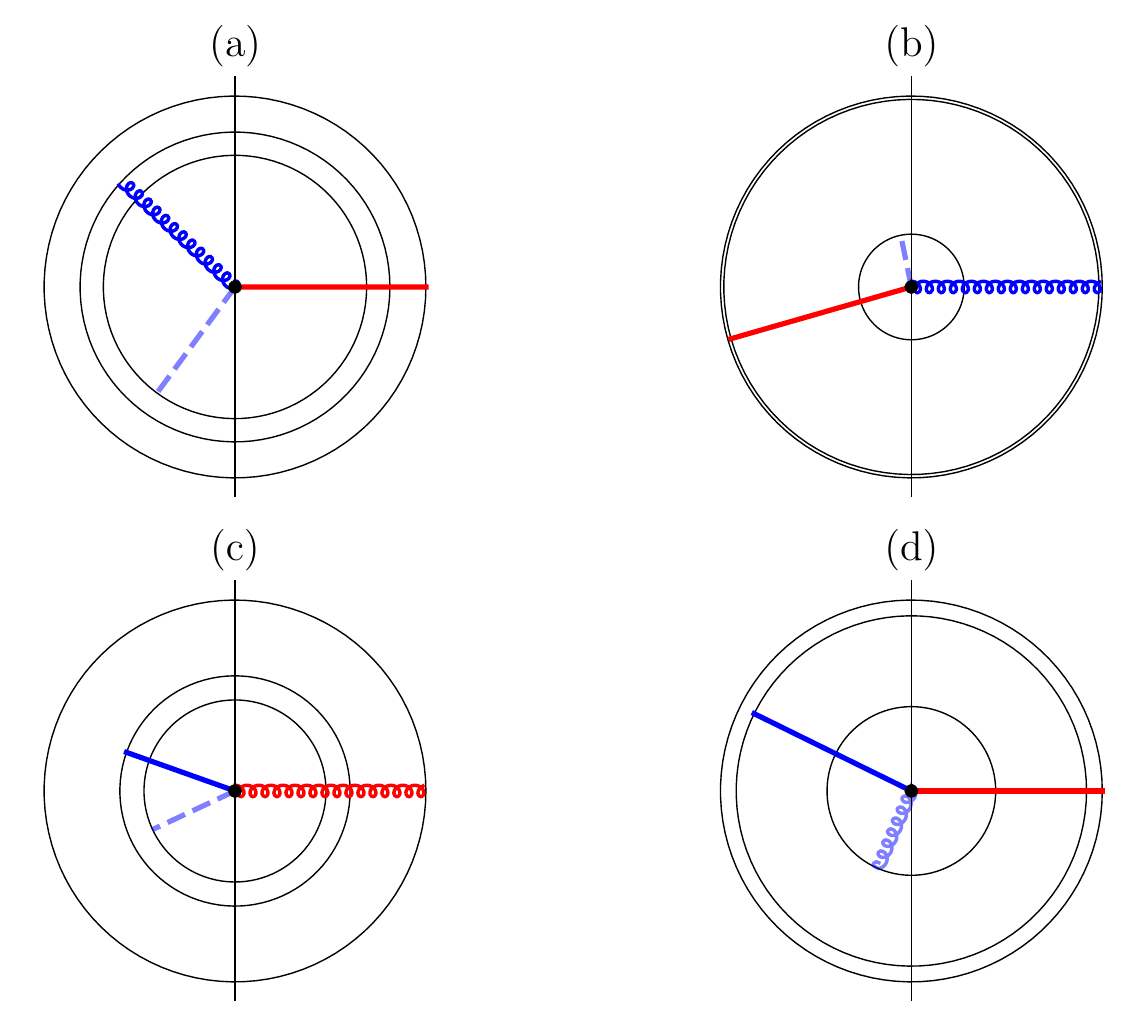}
	\end{center}
	\caption{Left: The emission phase space in the Lund plane with the \Caesar
          parameters $a_l=1$ and $b_l=1$ (valid for thrust).
          The green area marks the phase-space region where the emitting
          particle will be removed by grooming with boundaries corresponding to
          $\beta=0$ (solid line and area) and $\beta=2$ (dashed line, hatched
          area). The blue area marks the region where the emitted gluon is
          removed, \emph{cf.}\ Fig.~\ref{fig:Lund}. Right: Displays of the transverse
          plane of several LO final-state configurations. The colour codes
          indicate jet association according to the \BSZ\ algorithm, groomed
          particles are in light shade and dashed in the case of
          quarks. (a) Sample event from the green area, where the quark is
          groomed instead of the gluon. (b), (c) Sample events with two quarks
          in one hemisphere, with the softer one getting groomed, for different jet
          clusterings. (d) Sample event from the blue area, a soft gluon is
          groomed. }
	\label{fig:Lund_Power}
\end{figure}

\subsubsection*{$C_1^\delta$ and \NLLp accuracy for soft-drop event shapes}

The matching scheme we use in principle provides the correct constant $C_1^{\delta}$
up to terms beyond \NLLp due to implicit averaging over different Born
kinematics, see also the discussion in \cite{Banfi:2010xy, Baberuxki:2019ifp}.  
However, some subtleties arise, as the application of soft-drop
grooming implies that the limit $v^\text{SD}\to0$ does not uniquely impose the soft
limit for all particles beyond the Born multiplicity. Hence
$\Sigma^{\delta,(1)}_\mathrm{fo}$ in this limit consists of several
pieces. As usual, there is the constant contribution of the virtual-real
corrections at $v^\text{SD}=0$ and, for the real correction, there is a part of phase
space where nothing is groomed, contributing a finite remainder between the
integral of this phase space and $\Sigma^{\delta,(1)}_\mathrm{res}$. The pieces
additionally present due to grooming are real corrections that result in $v^\text{SD}=0$
due to one particle being groomed away. In principle those should be multiplied
by the appropriate Sudakov factors for the corresponding $(n+1)$-particle final
state. However, to achieve \NLLp accuracy those are needed at LL only,
\emph{i.e.}\ only in the limit of radiation that is simultaneously soft and
collinear. This results in the same factor as for the $2\to2$ configuration obtained
by ignoring the groomed parton. 

In practice, we sort all events according to the \BSZ\
cluster algorithm described above. A groomed gluon in a LO event will
always also be clustered first, since it will have the smallest transverse
momentum overall and the remaining two particles can not be collinear due to
momentum conservation in the transverse plane. A configuration like this is
shown in Fig.~\ref{fig:Lund_Power} in panel (d). For a gluon, the flavour
assignment from this algorithm is hence the same as if it was simply
discarded. This is still the case if a groomed quark is clustered to the beam,
since our soft-drop observables are insensitive to initial-state radiation
at LL. However, this will no longer be the case if a quark is groomed away
but clustered to another final-state parton. This can happen if two quarks are
in the same hemisphere. This is illustrated in panels (b) and (c) of
Fig.~\ref{fig:Lund_Power}, which would be classified as quark-quark (b) and
gluon-gluon (c) like final states. Finally, a quark can be in a hemisphere 
together with a gluon, but be softer. The corresponding configuration in panel
(a) of Fig.~\ref{fig:Lund_Power} would be attributed to the quark-quark
channel. However, all those cases are suppressed with powers of
$\zcut$. Since we work in the limit $\zcut \ll 1$,
those terms are beyond our accuracy target. The phase-space region where this
happens is illustrated in the left part of Fig.~\ref{fig:Lund_Power}. For
$\beta=0$, these power corrections are logarithmically enhanced and need to be
taken into account at \NLL accuracy if finite $\zcut$ effects are important,
\emph{cf.}\ \cite{Dasgupta:2013ihk, Marzani:2017mva}. With $\beta>0$, they still
provide a constant contribution which would enter at \NLLp accuracy. We
note that, despite being formally irrelevant for us, these phase-space regions
might still be problematic from a practical point of view when matching \NLO
and \NLL resummation as we do here. This is due to flavour
assignments that have no corresponding Born process and would thus be assigned
to the `other' channel. These are not guaranteed to vanish in the $v^\text{SD}\to0$
limit (infrared safety of the flavour algorithm of course mandates that this
happens in the $v\to0$ limit, as we demonstrated above for $v=\tauPerp$). This
could lead to an unphysical behaviour where the matched distribution approaches
a finite constant rather than zero in the soft limit. This could be solved by adjusting
details of the matching scheme. However, we do not encounter this problem as we
use the `bland' variant of the \BSZ\ algorithm.

With the above discussion at hand, we can conclude that our calculations achieve
\NLOpNLLp accuracy, in the $\zcut \ll 1$ limit, for
histograms of differential distributions, scaled by the overall cross section at
the corresponding accuracy, \emph{i.e.}\ 
\begin{equation}\label{eq:normalisation}
  \frac{1}{\sigma}\frac{\mathop{d\sigma}}{\mathop{d\ln(v^\text{SD})}} \equiv \frac{1}{\Sigma_\mathrm{match}(1)}
  \frac{\mathop{d\Sigma_\mathrm{match}(v^\text{SD})}}{\mathop{d\ln(v^\text{SD})}}\,.
\end{equation}

\subsection{Results for soft-drop transverse thrust}

Here we apply our resummation for soft-drop groomed hadronic event shapes for the
transverse-thrust observable introduced in Eq.~\eqref{eq:sd_thrust_def}. The
parameters of its \Caesar representation, \emph{cf.}\ Eq.~\eqref{eq:CAESAR_param}, are collected
in Table~\ref{tab:param}. In addition, in Table~\ref{tab:End-point} we list the
kinematic end-point values $v^{\text{max}}$, extracted from fixed-order
calculations, for the ungroomed observable and those cases where grooming actually limits the
observable range.

\begin{table}[ht!]
  \begin{center}
    \begin{tabular}{|c|c|c|c|c|}
      \hline 
      $\tauPerp$ $\vphantom{\frac{\frac{12}1}{2}}$& $a_{l}$ & $b_{l}$ & $d_{l}(\theta)$ & $g_{l}(\phi)$ \\ 
      \hline 
      $l \in$~initial state $\vphantom{\frac{\frac{12}1}{2}}$& 1  & 0  & $\frac{1}{\sin\theta}\left(\frac{\mu_Q}{Q_{12}}\right)$ & $1-\left|\cos\phi\right|$  \\ 
      \hline 
      $l \in$~final state $\vphantom{\frac{\frac{12}1}{2}}$& 1 & 1 & $\frac{1}{\sin^{2}\theta}\left(\frac{\mu_Q}{Q_{12}}\right)$ & $\sin^{2}\phi$ \\ 
      \hline 
    \end{tabular}
    \caption{The \Caesar parametrisation valid for both plain and soft-drop groomed transverse thrust,
      given for initial- and final-state hard-process legs.}
    \label{tab:param}
  \end{center}
\end{table}

\begin{table}[ht!]
	\begin{center}
		\begin{tabular}{|c|c|c|c|}
			\hline
			$\zcut\vphantom{\frac{\frac{12}1}{2}}$ & $\beta$ & $v^{\text{max}}_\LO$ & $v^{\text{max}}_\NLO$\\
			\hline
			Ungroomed $\vphantom{\frac{\frac{12}1}{2}}$& & $1/3$ & 0.3406\\
			\hline
			0.2 $\vphantom{\frac{\frac{12}1}{2}}$& 2 & 0.2927 & 0.3067\\
			\hline
			0.3 $\vphantom{\frac{\frac{12}1}{2}}$& 1 & 0.2929 & 0.3067\\
			\hline
			0.3 $\vphantom{\frac{\frac{12}1}{2}}$& 2 & 0.1994 & 0.2174\\
			\hline
		\end{tabular}
	\end{center}
	\caption{Table of transverse thrust end-point values determined
          numerically at LO and \NLO. The kinematical end-points for the
          $\beta, \zcut$ values not shown are unaltered by grooming.}\label{tab:End-point} 
\end{table}

All phase-space integrals are performed using
\Sherpa~\cite{Gleisberg:2008ta,Bothmann:2019yzt}. This includes the integration
over the Born phase space in Eq.~\eqref{eq:CAESAR}, and, over the three- and
four-particle phase space needed to evaluate the cumulants at LO and \NLO QCD. For
the LO, \emph{i.e.}\ the calculation of $\Sigma_\mathrm{fo}^{(1)}(v^\text{SD})$, we 
need to add the constant contribution from the virtual correction to the Born
$2\to 2$ parton scattering process and the real correction integrated up to
$\tauSD = v^\text{SD}$. Both parts are regularised using Catani--Seymour dipole
subtraction~\cite{Catani:1996vz,Gleisberg:2007md}. We obtain the required
one-loop virtual matrix elements from \OpenLoops~\cite{Cascioli:2011va}, using
the \Collier library~\cite{Denner:2016kdg} for the evaluation of tensor and
scalar integrals. For
the \NLO calculation we regulate double real-correction divergences in the
infrared by requiring $\tauPerp > \tau^{\text{cut}}_\perp$. This implies that we do not
need to calculate the two-loop virtual corrections, for which $\tauPerp = 0$.
In the remaining phase space, the $2\to3$ parton matrix element is finite (though
possibly numerically large close to the cutoff), and we can evaluate the
one-loop virtual and real corrections, including one real emission that might
be arbitrarily soft/collinear, using the same tools as in the LO case. Since
our matching formula Eq.~\eqref{eq:matching}  only depends on
$\overline{\Sigma}_\mathrm{fo}^{(2)}(v^\text{SD})$, which is an integral from $v^{\text{SD}}$ to $1$,
see Eq.~\eqref{eq:barSigma}, our final results are independent of the infrared
cut for $v^\text{SD} > \tau^{\text{cut}}_\perp$. Using the fact that the value of groomed
transverse thrust is strictly smaller than the corresponding value of ungroomed
thrust, we can generate all results by setting the cut in $\tauPerp$ to the
lowest observable value we are interested in. In practice, we use
$\tau^\mathrm{cut}_\perp = e^{-11} \approx 1.7\times 10^{-5}$.

\begin{figure}[t!]
	\begin{center}
		\includegraphics[width=0.32\textwidth]{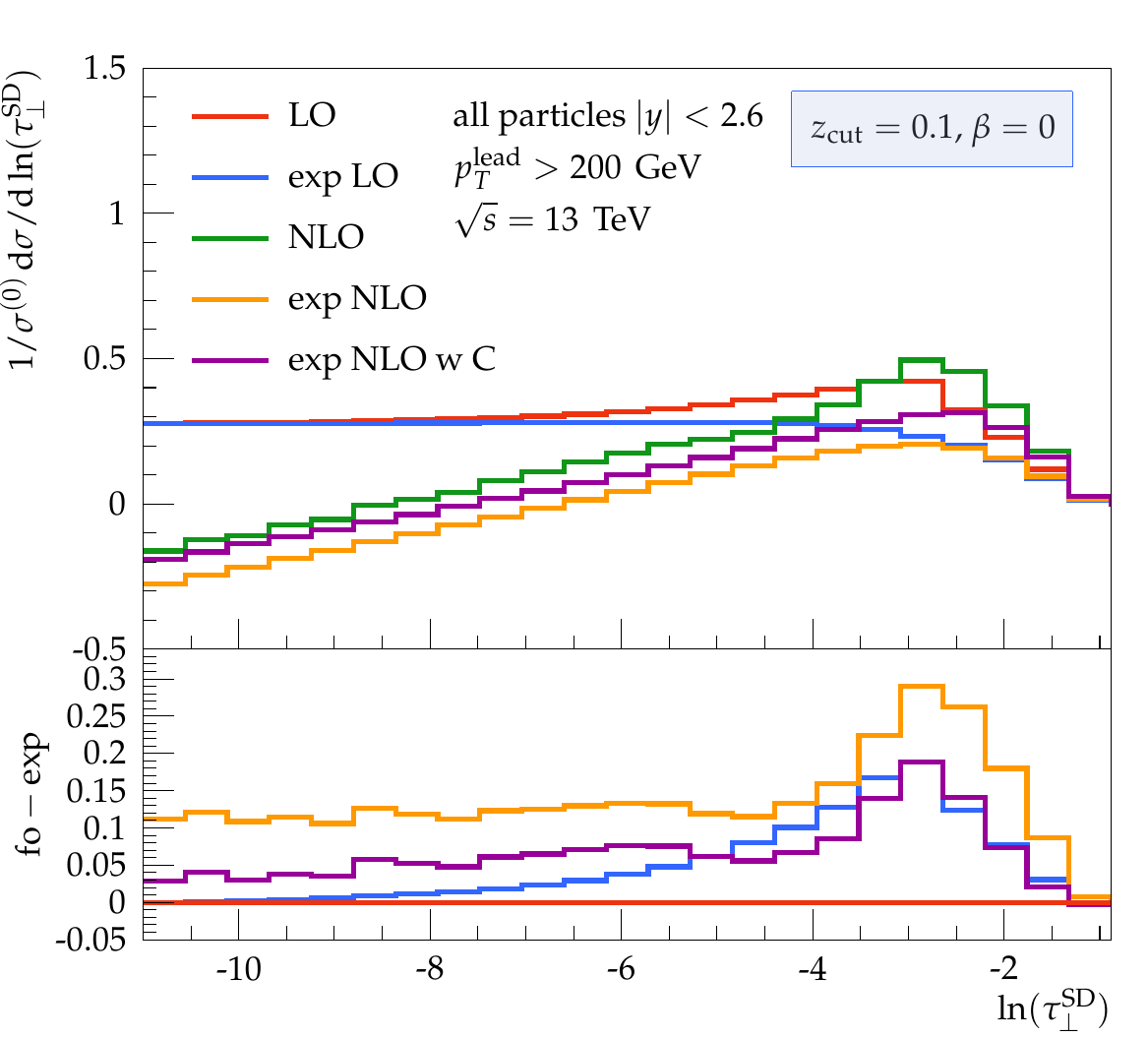}~
		\includegraphics[width=0.32\textwidth]{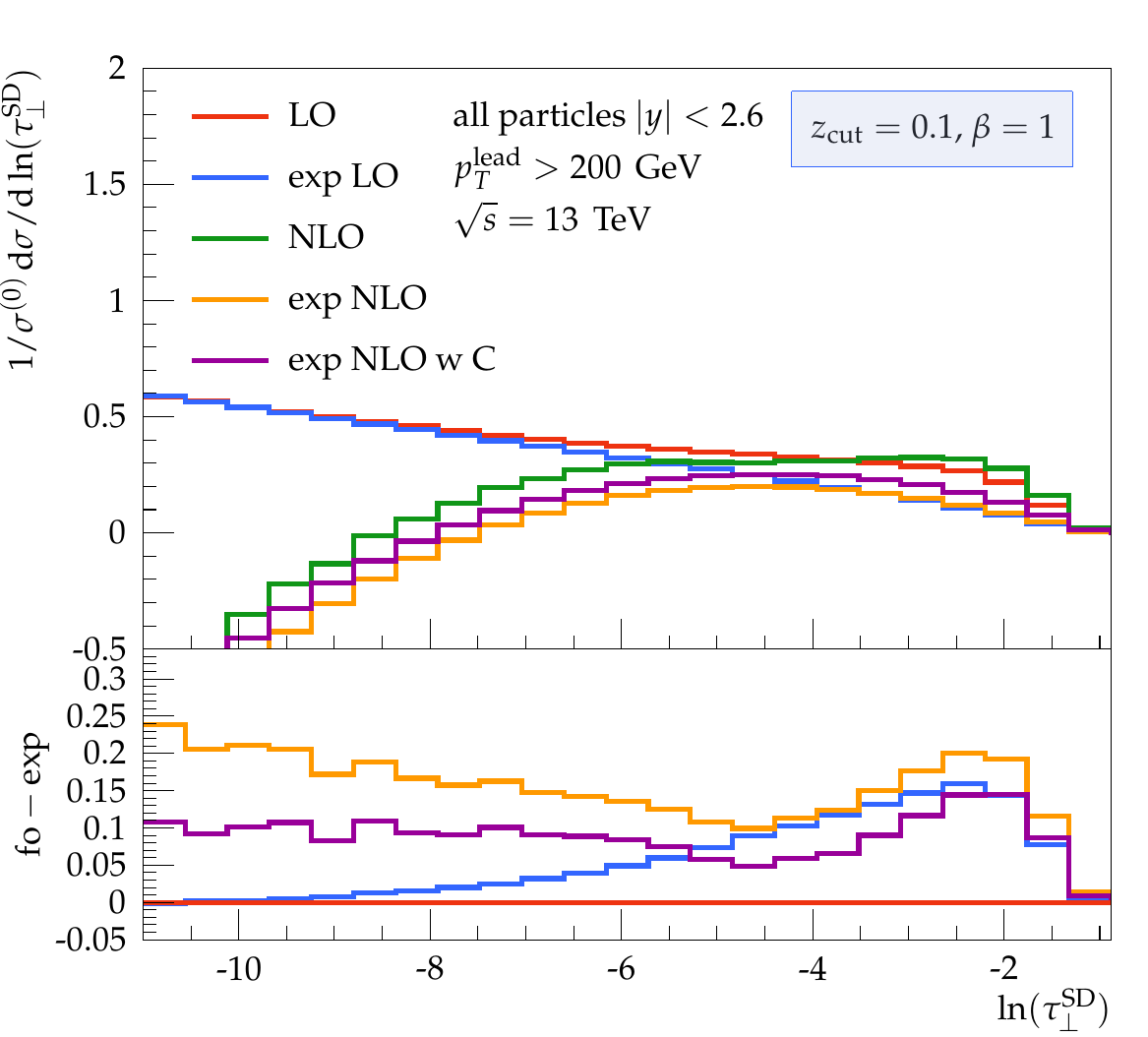}~
		\includegraphics[width=0.32\textwidth]{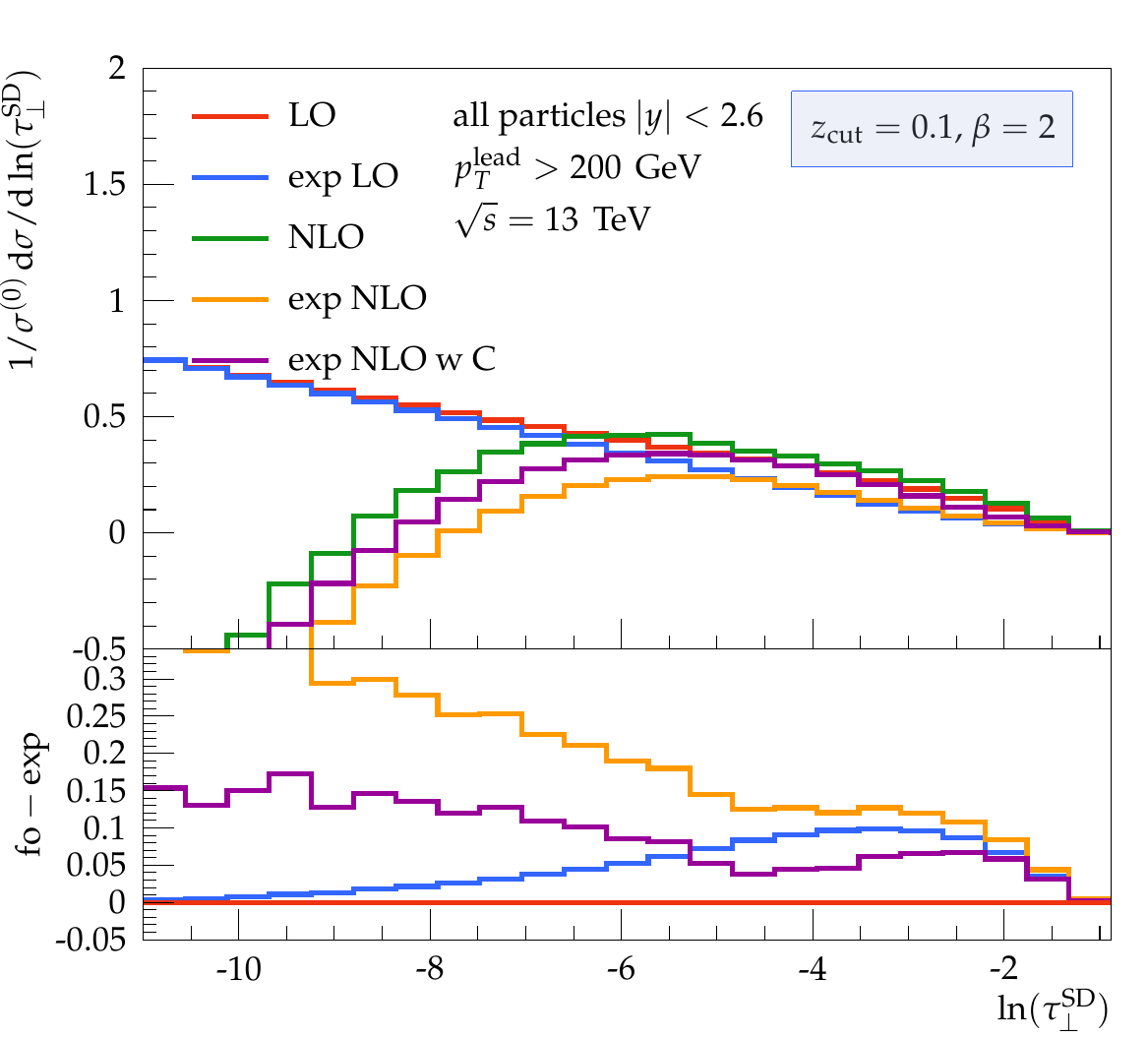}\\
		\includegraphics[width=0.32\textwidth]{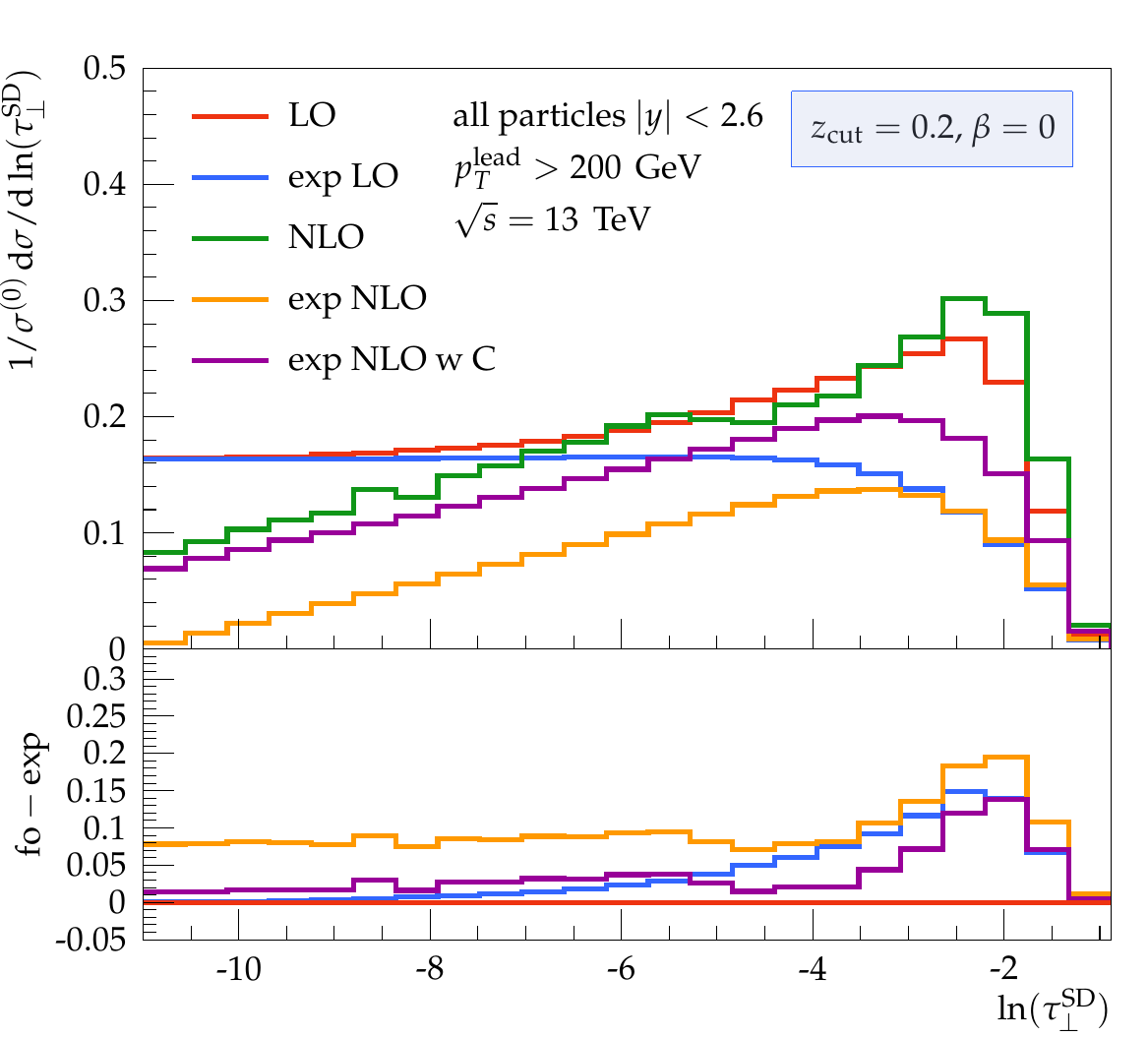}~
		\includegraphics[width=0.32\textwidth]{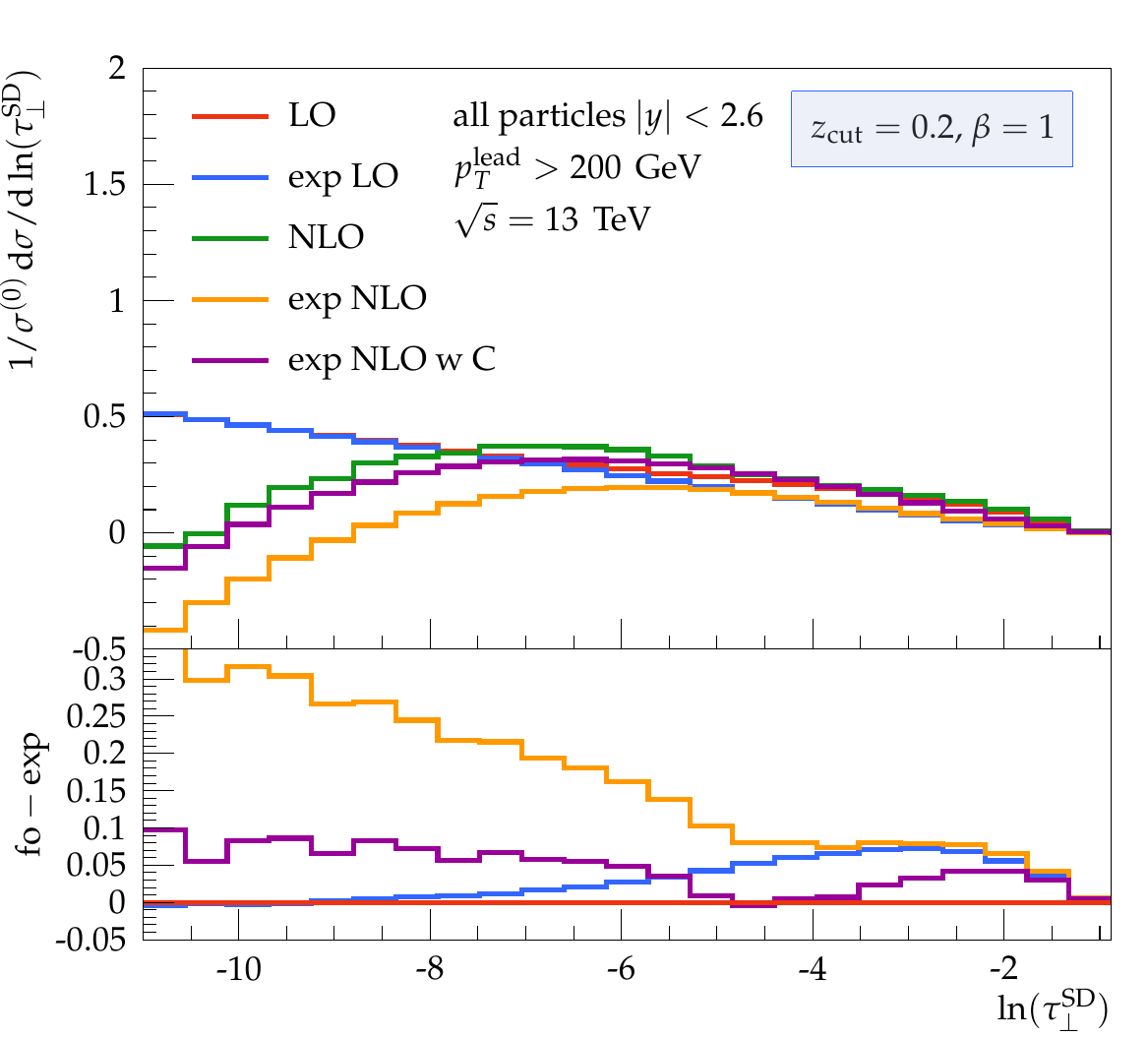}~
		\includegraphics[width=0.32\textwidth]{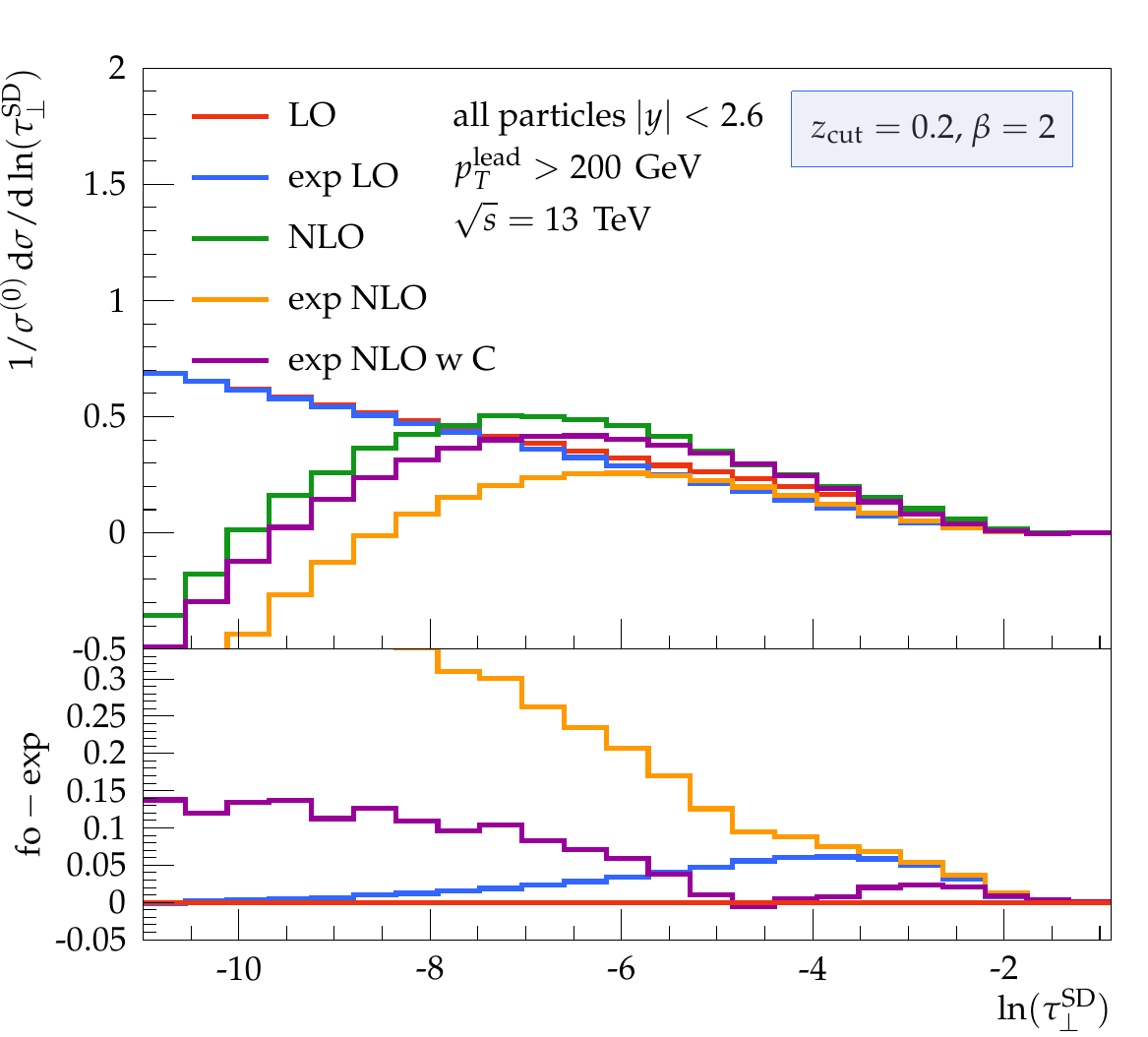}\\
		\includegraphics[width=0.32\textwidth]{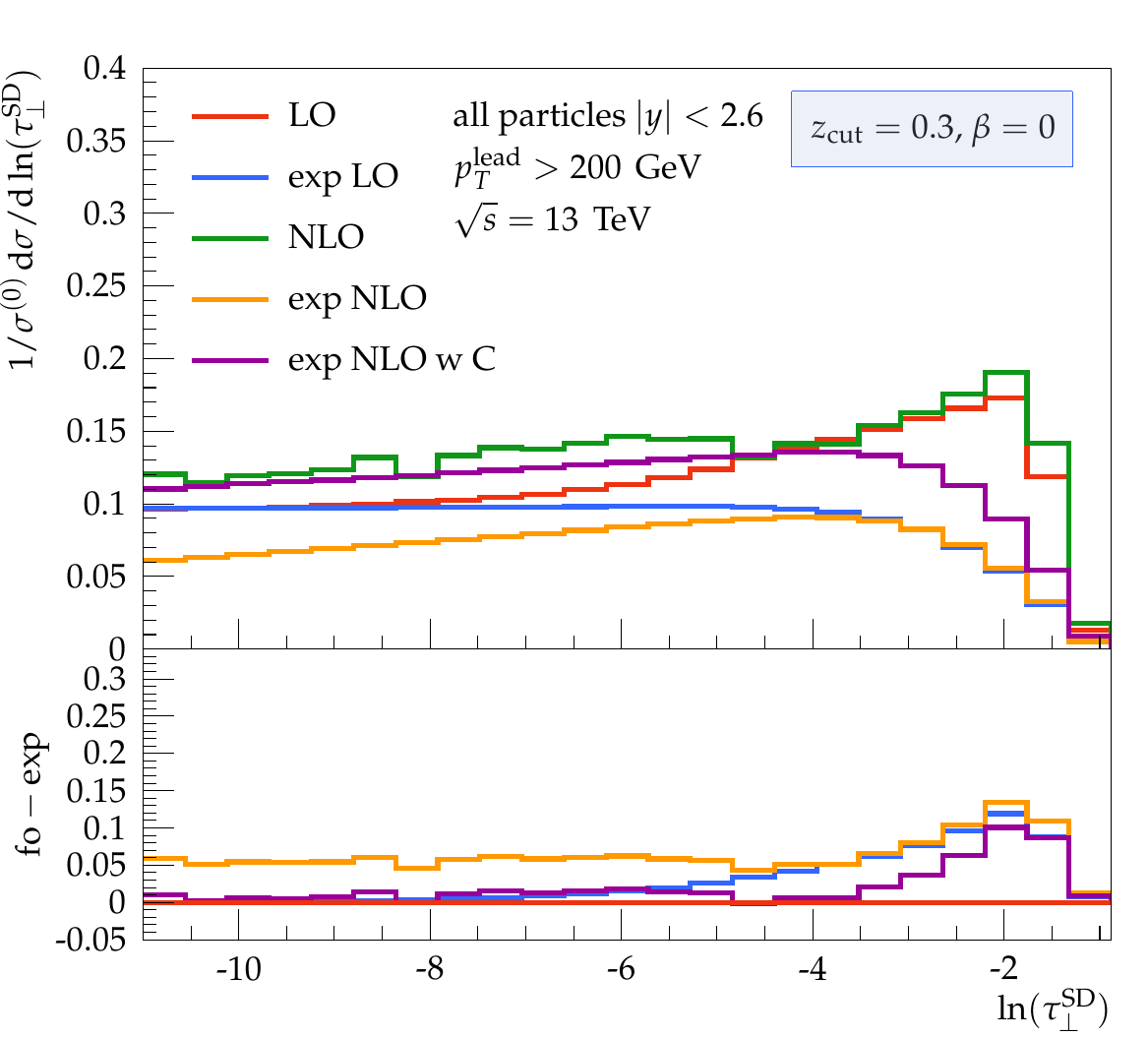}~
		\includegraphics[width=0.32\textwidth]{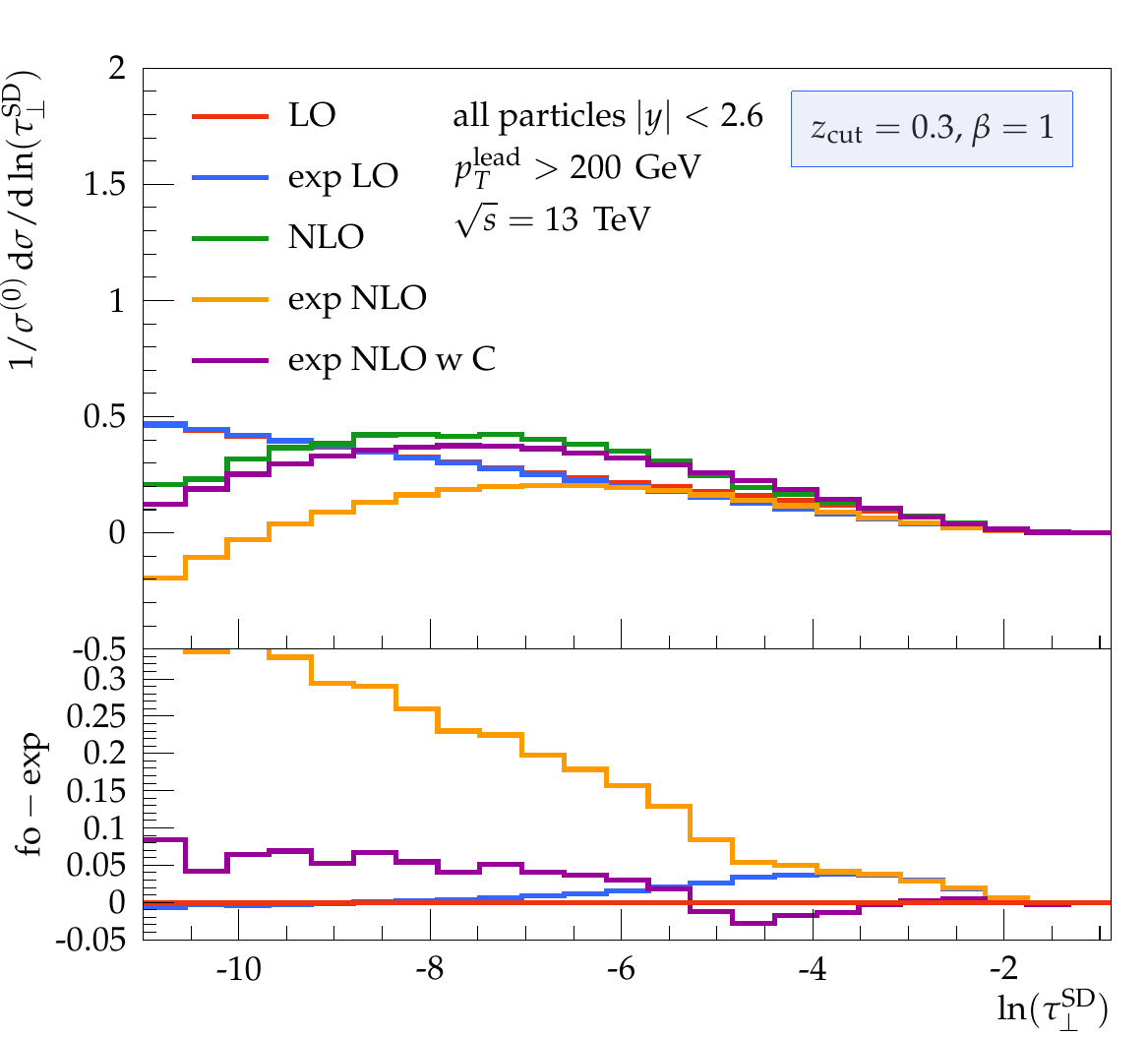}~
		\includegraphics[width=0.32\textwidth]{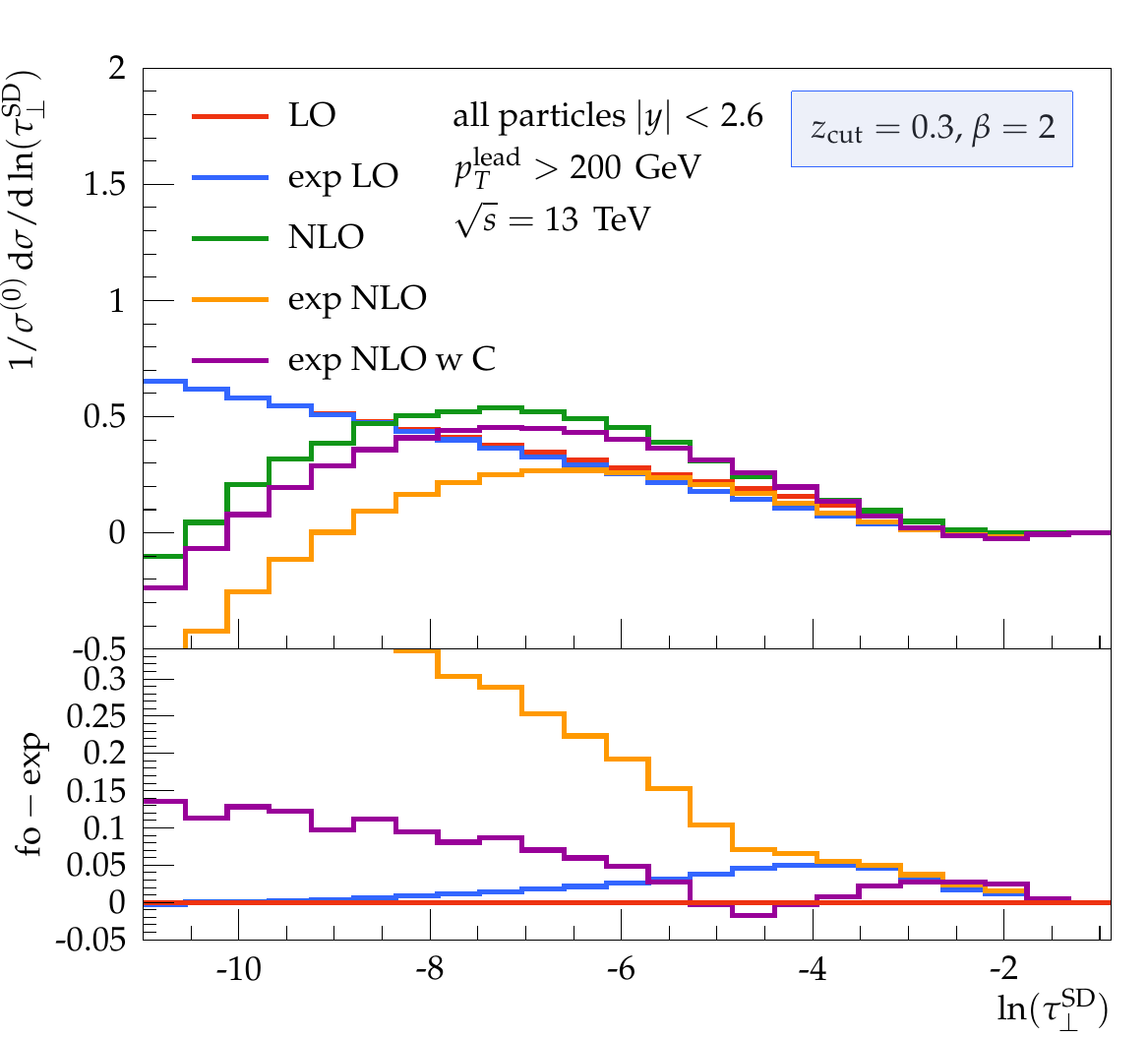}
	\end{center}
	\caption{Comparison of the expansion of the \NLL resummation and fixed-order results
          for soft-drop groomed thrust at LO and \NLO accuracy. The lower panels show the
          difference between expansion and fixed order at LO and \NLO. }
	\label{fig:exp}
\end{figure}

For our calculation, the factorisation, renormalisation and resummation scale
are chosen identical and equal to half the scalar sum of the partonic
transverse momenta, \emph{i.e.}\ 
\begin{equation}
  \mu_{\text{R}}=\mu_{\text{F}}=\mu_{\text{Q}} = \frac12 H_T\,.
\end{equation}
We make use of the NNPDF-3.0 NNLO PDF \cite{Ball:2014uwa} and evaluate
$\alphaS(\muR^2)$ at two-loop starting from $\alphaS(M^2_Z)=0.118$, with
a fixed number of $n_f=5$ active flavours. 

In Fig.~\ref{fig:exp} we compile fixed-order results for the
$p_{T,\text{min}}= 200\;\text{GeV}$ event selection for different combinations of
grooming parameters. Alongside we show the expansions of our resummation
formulae to LO and \NLO in $\alphaS$. Note, our \NLO results are in fact
independent of the infrared regulator $\tau^\mathrm{cut}_\perp$ since
$\frac{\mathop{d\Sigma}(v^\text{SD})}{\mathop{dv^\text{SD}}} = -\frac{\mathop{d\overline{\Sigma}}(v^\text{SD})}{\mathop{dv^\text{SD}}}$,
\emph{cf.}\ Eq.~\eqref{eq:barSigma}. In the notation of the previous section, we
compare the derivatives $\frac{\mathop{d\Sigma}}{\mathop{d\ln \tauSD}}$  of the
cumulative distributions at LO 
\begin{align}
  \Sigma_\text{LO}(\tauSD) &= \sum_\delta \left(\sigma^{\delta,(0)} +
    \Sigma_\mathrm{fo}^{\delta,(1)}(\tauSD) \right)\,, \\
  \Sigma_\text{exp LO}(\tauSD) &= \sum_\delta \left(\sigma^{\delta,(0)} +
    \Sigma_\mathrm{res}^{\delta,(1)}(\tauSD) \right)\,,
\end{align}
and at \NLO
\begin{align}
  \Sigma_\text{NLO}(\tauSD) &= \sum_\delta \left(\sigma^{\delta,(0)} +
    \Sigma_\mathrm{fo}^{\delta,(1)}(\tauSD) -
    \overline{\Sigma}^{\delta,(2)}_\mathrm{fo}(\tauSD) \right)\,,\\
  \Sigma_\text{exp \NLO}(\tauSD) &= \sum_\delta \left(\sigma^{\delta,(0)} +
  \Sigma_\mathrm{res}^{\delta,(1)}(\tauSD) +
  \Sigma^{\delta,(2)}_\mathrm{res}(\tauSD) \right)\,,\\
  \Sigma_\text{exp \NLO}^{C}(\tauSD) &= \sum_\delta \left(\sigma^{\delta,(0)} +
  \left(1+\frac{\Sigma_\mathrm{fo}^{\delta,(1)}(\tauSD)-\Sigma_\mathrm{res}^{\delta,(1)}(\tauSD)}{\sigma^{\delta,(0)}}\right)
  \Sigma_\mathrm{res}^{\delta,(1)}(\tauSD) + \Sigma^{\delta,(2)}_\mathrm{res}(\tauSD) \right)\,,
\end{align}
where the last definition corresponds to the inclusion of $C_1^\delta$ in the
expansion in the limit $\tauSD\to 0$.
For all considered grooming parameters, we observe that the subtractions of the
LO results, \emph{i.e.}\ $\frac{\mathop{d}}{\mathop{d\ln
    \tauSD}}\left(\Sigma_\text{LO}-\Sigma_\text{exp LO}\right)$, tend to zero as 
$\tauSD\to 0$ (blue lines), confirming that all logarithmically enhanced terms
are correctly captured by the expansion. In the difference between the exact \NLO
results and the naive expansion, $\Sigma_\text{NLO}-\Sigma_\text{exp \NLO}$, for $\beta>0$ we
expect and observe residual single-logarithmic contributions $\propto
{\cal{O}}(1)\alphaS^2 L^2$, \emph{i.e.}\ a linear rising difference in the
derivative $\frac{\mathop{d}}{\mathop{d\ln
    \tauSD}}\left(\Sigma_\text{NLO}-\Sigma_\text{exp \NLO}\right)$ (orange
lines). This contribution is a cross term between the $C^\delta_1$ coefficients and
the LL $\alphaS L^2$ contribution. However, as the LO structure for $\beta=0$
starts at $\alphaS L$, this cross term does not exist for this case and the first
order at which there is a difference is ${\cal{O}}(1)\alphaS^2 L$. In addition there
are finite $\zcut$ contributions which, for $\beta=0$, already contribute
$\propto {\cal{O}}(\zcut)\alphaS L$, however these effects cannot be extracted
with any numerical significance.

Upon including the $C^\delta_1$ coefficients (purple lines), 
the NLO is expected to be captured up to a contribution $\propto {\cal{O}}(\zcut)
\alphaS^2 L^2$. The slope of the remaining difference and the dependence on $\zcut$
are indeed very small, to an extent that we are not able to
significantly detect them numerically. We instead observe an almost constant
difference between the derivatives of NLO and expansion, indicating
missing terms of ${\cal{O}}(\alphaS^2 L)$ in the cumulant $\Sigma_\text{exp
  \NLO}^{C}$, \emph{i.e.}, terms beyond \NLLp accuracy. 
Up to these higher-order terms the logarithmic structure of the resummation thus fully
matches the fixed order, in spite of the fact that the latter include a rapidity cut
$|y|\leq y_{\text{max}}$ for the particles contributing to the observable. Emissions far
out in rapidity will be groomed as they originate from soft wide-angle or initial-state
radiation, for details see the discussion in App.~\ref{sec:logzc}. This however does
not hold for ungroomed thrust, for which the logarithmic structure of initial-state
emissions is altered by such cut. 

With the logarithmic structure of the expansion confirmed, we can consistently
match the full \NLL resummation to the fixed-order results, using multiplicative matching
as presented in Eq.~\eqref{eq:matching}.

\begin{figure}[t!]
	\begin{center}
		\includegraphics[width=0.32\textwidth]{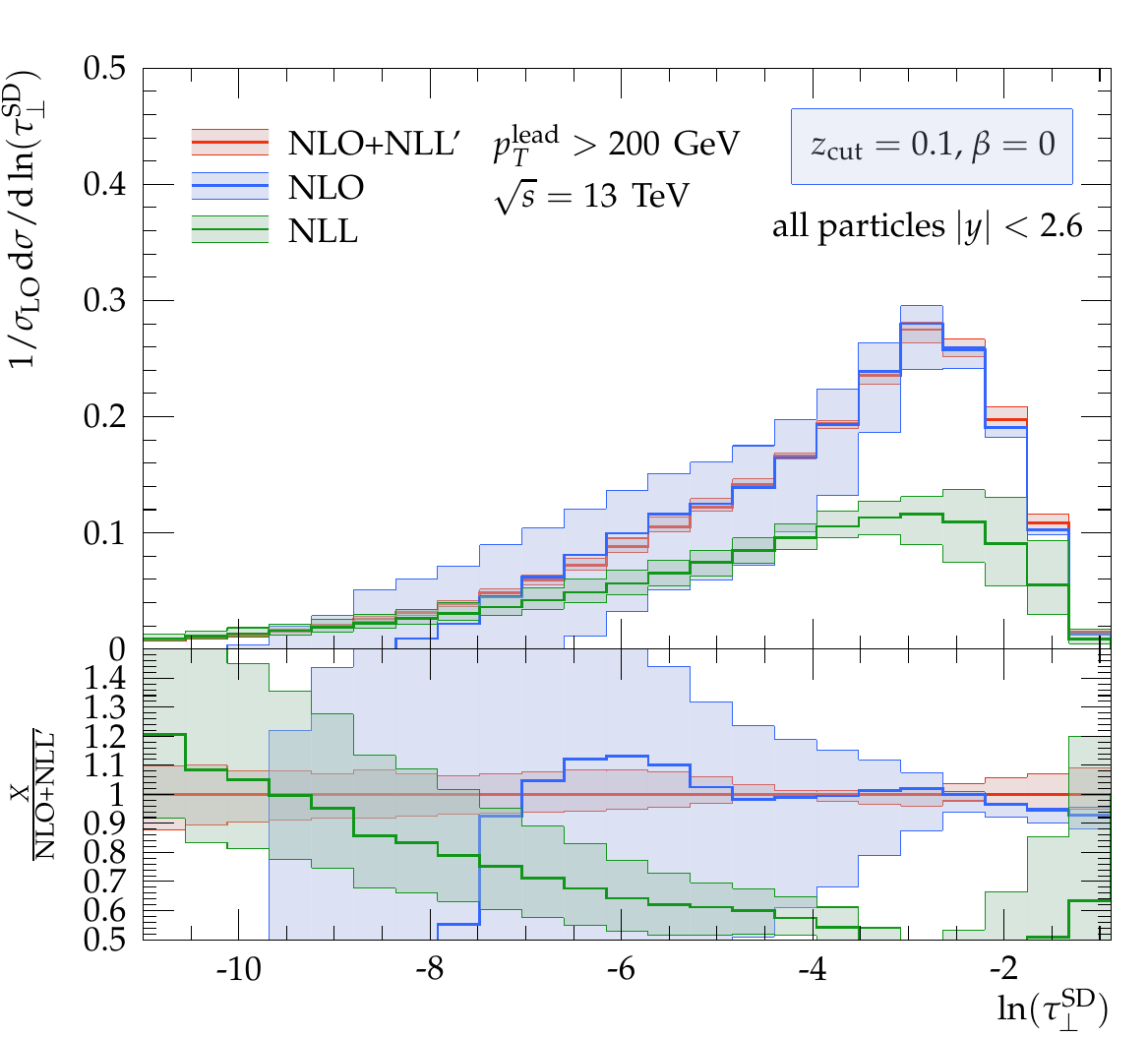}~
		\includegraphics[width=0.32\textwidth]{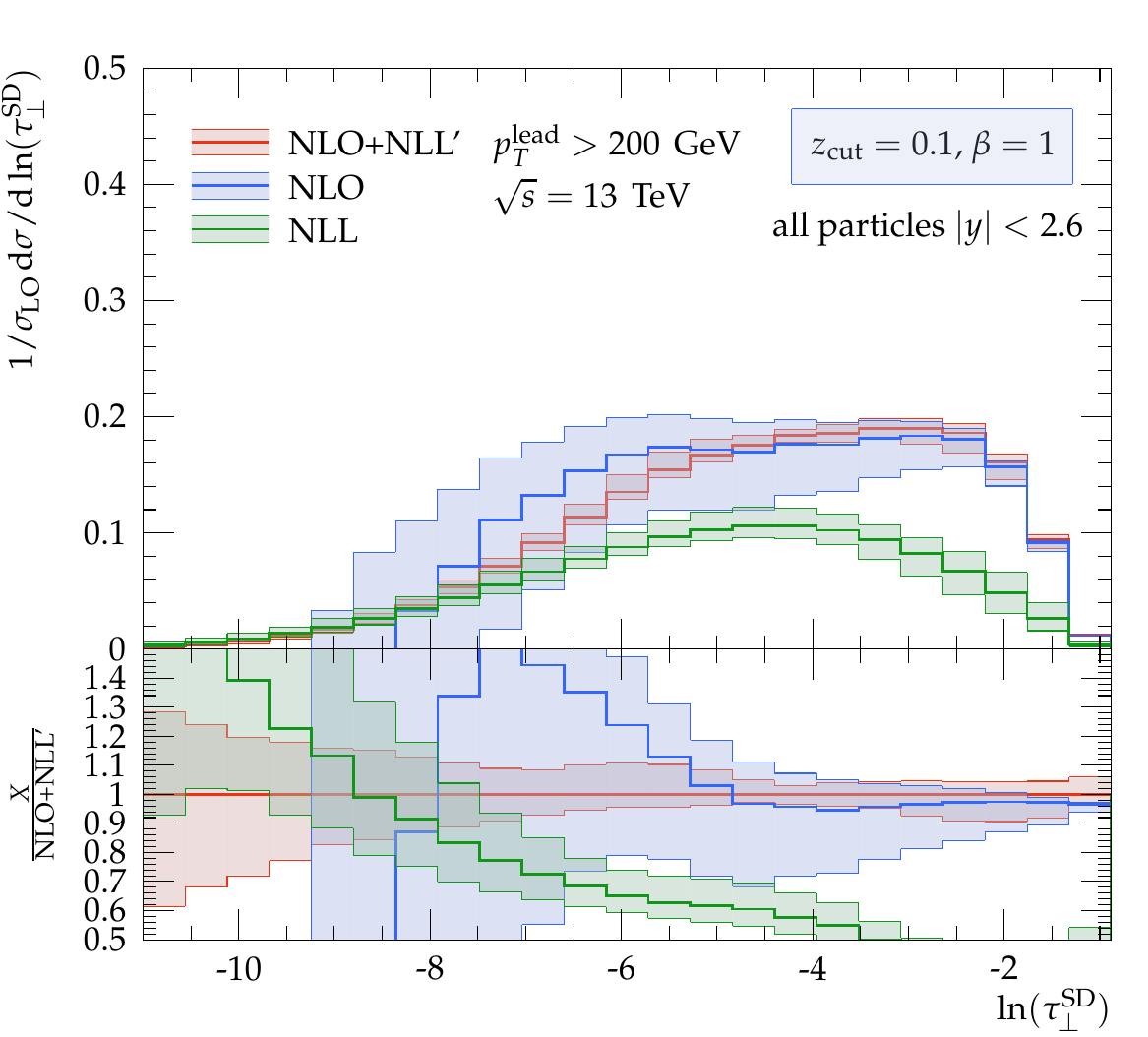}~
		\includegraphics[width=0.32\textwidth]{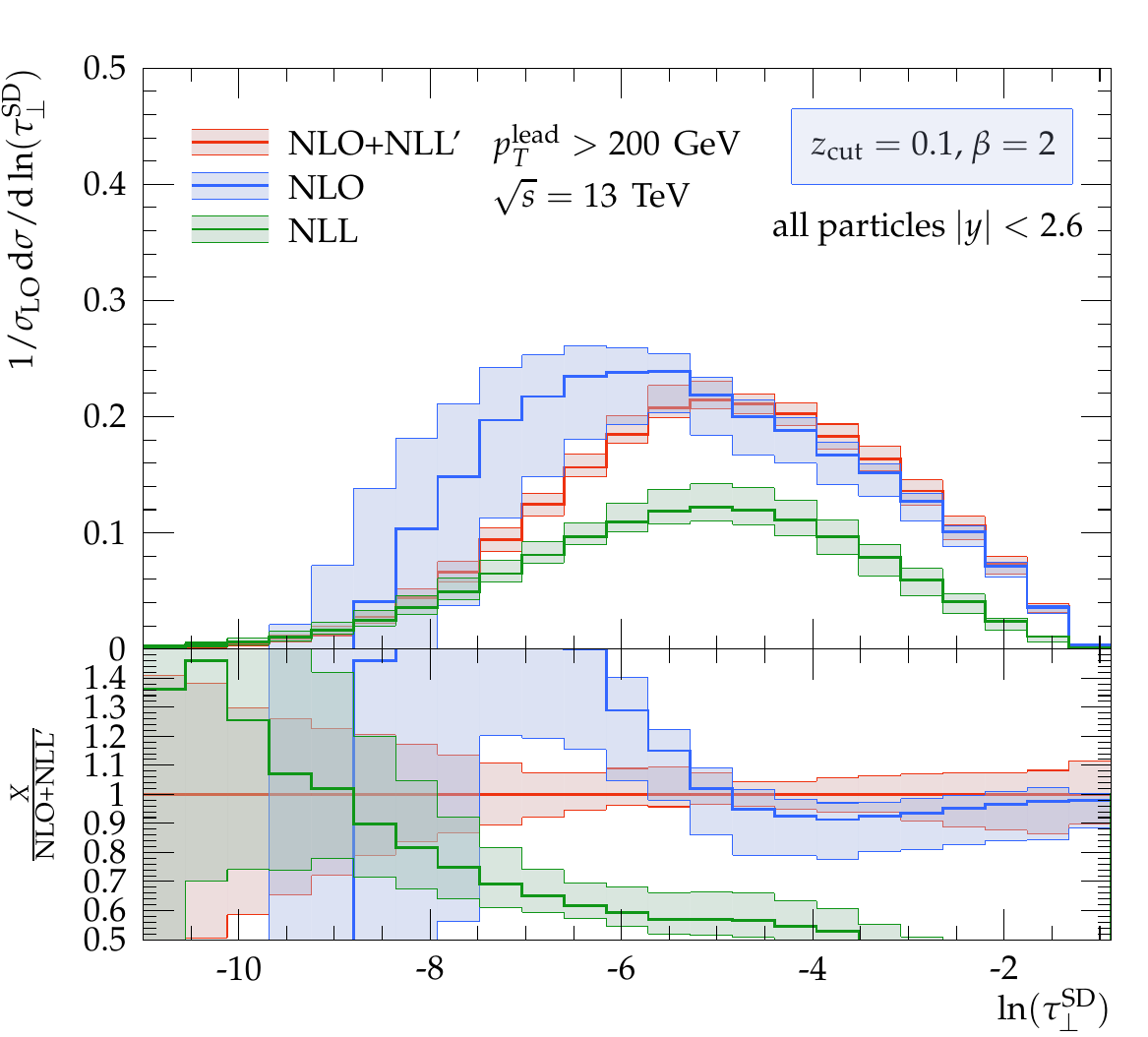}\\
		\includegraphics[width=0.32\textwidth]{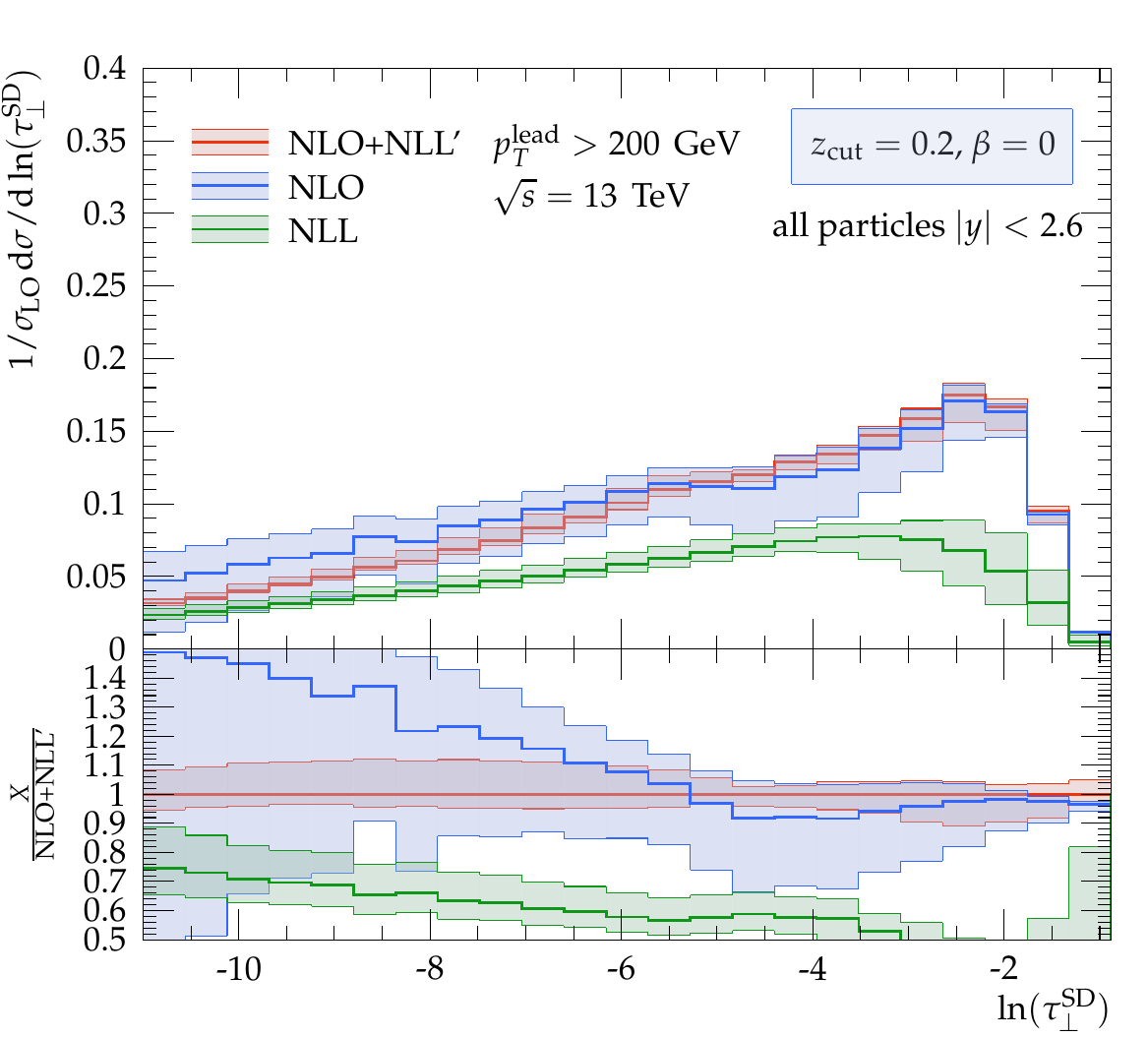}~
		\includegraphics[width=0.32\textwidth]{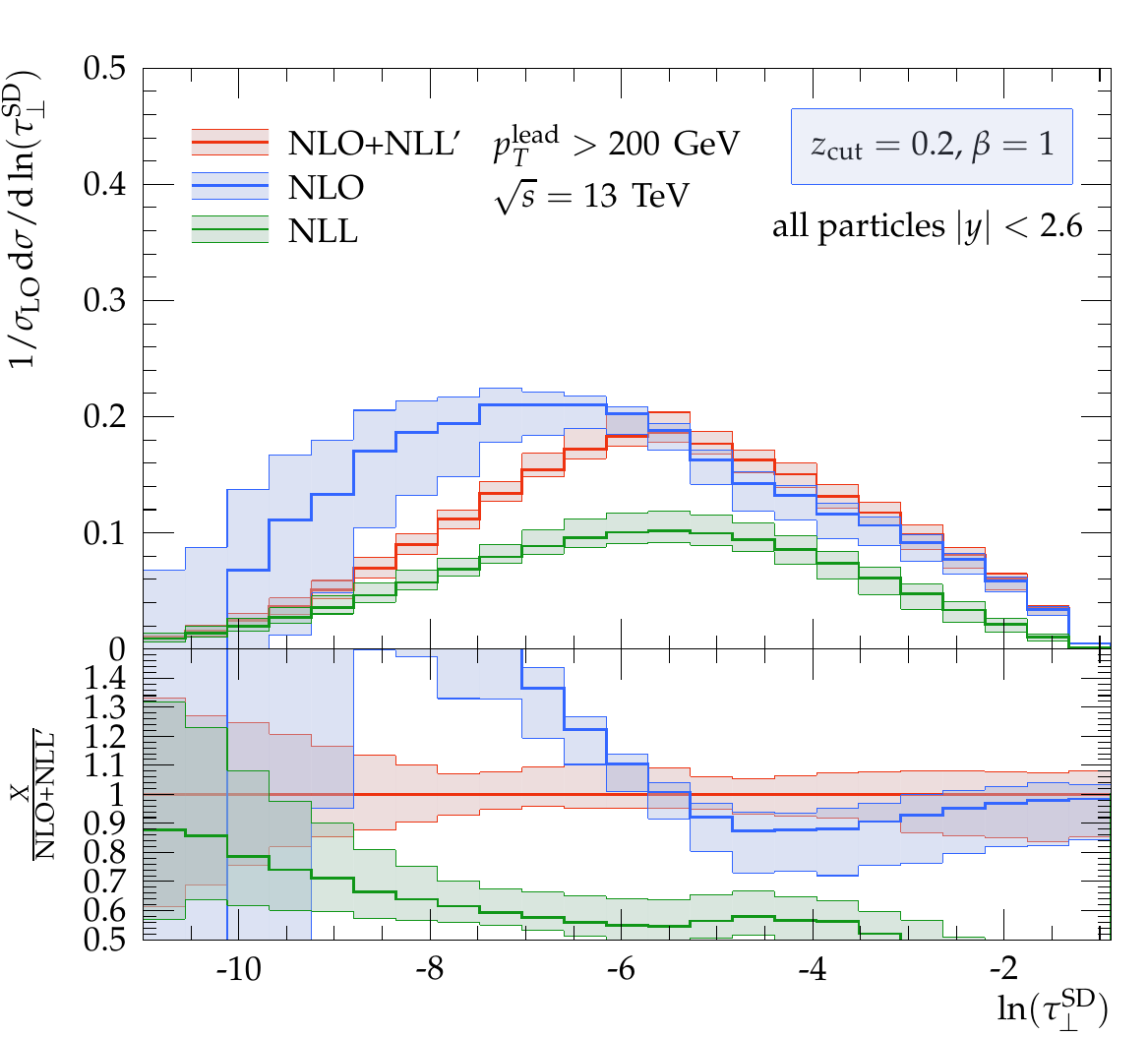}~
		\includegraphics[width=0.32\textwidth]{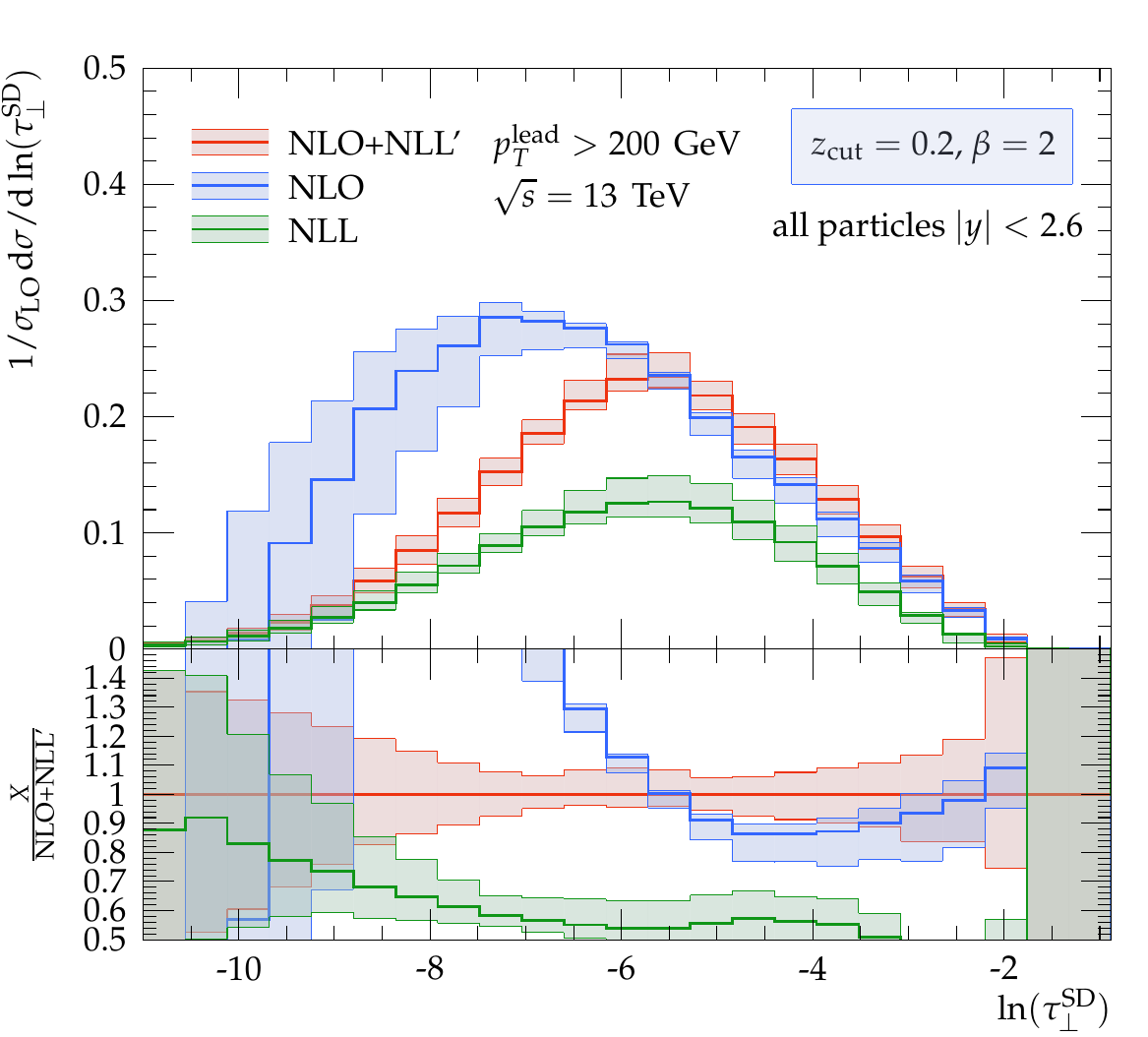}\\
		\includegraphics[width=0.32\textwidth]{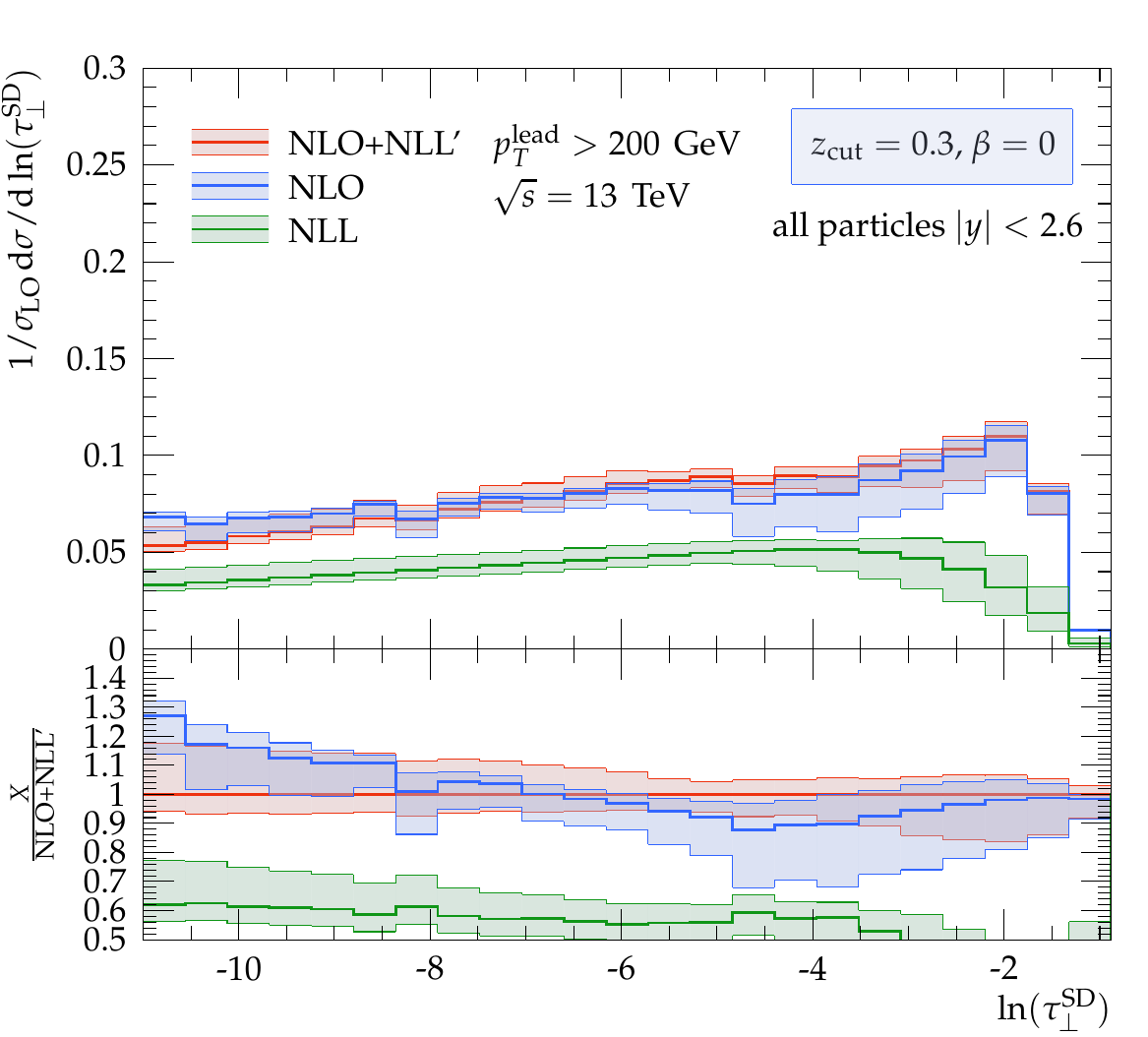}~
		\includegraphics[width=0.32\textwidth]{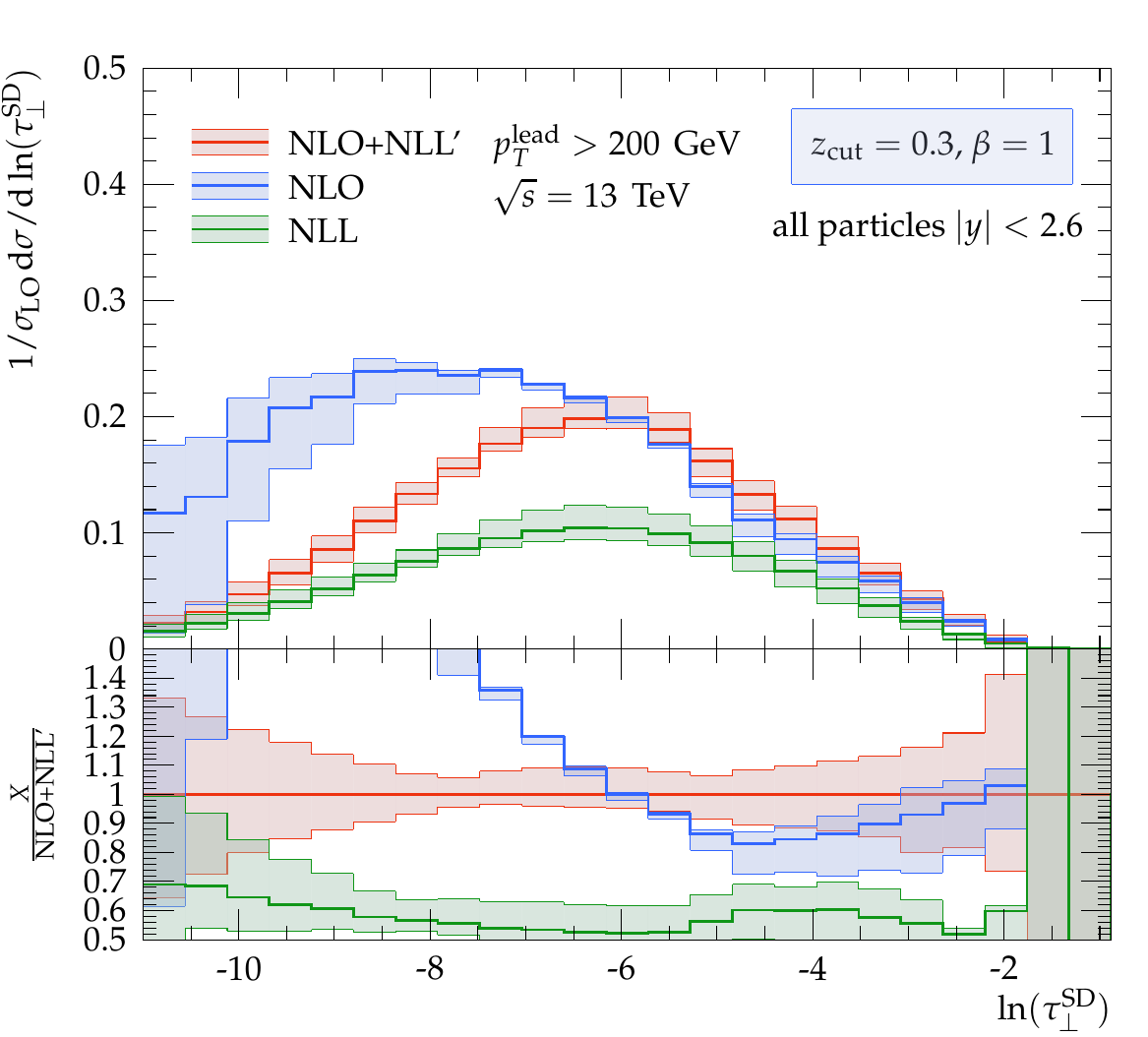}~
		\includegraphics[width=0.32\textwidth]{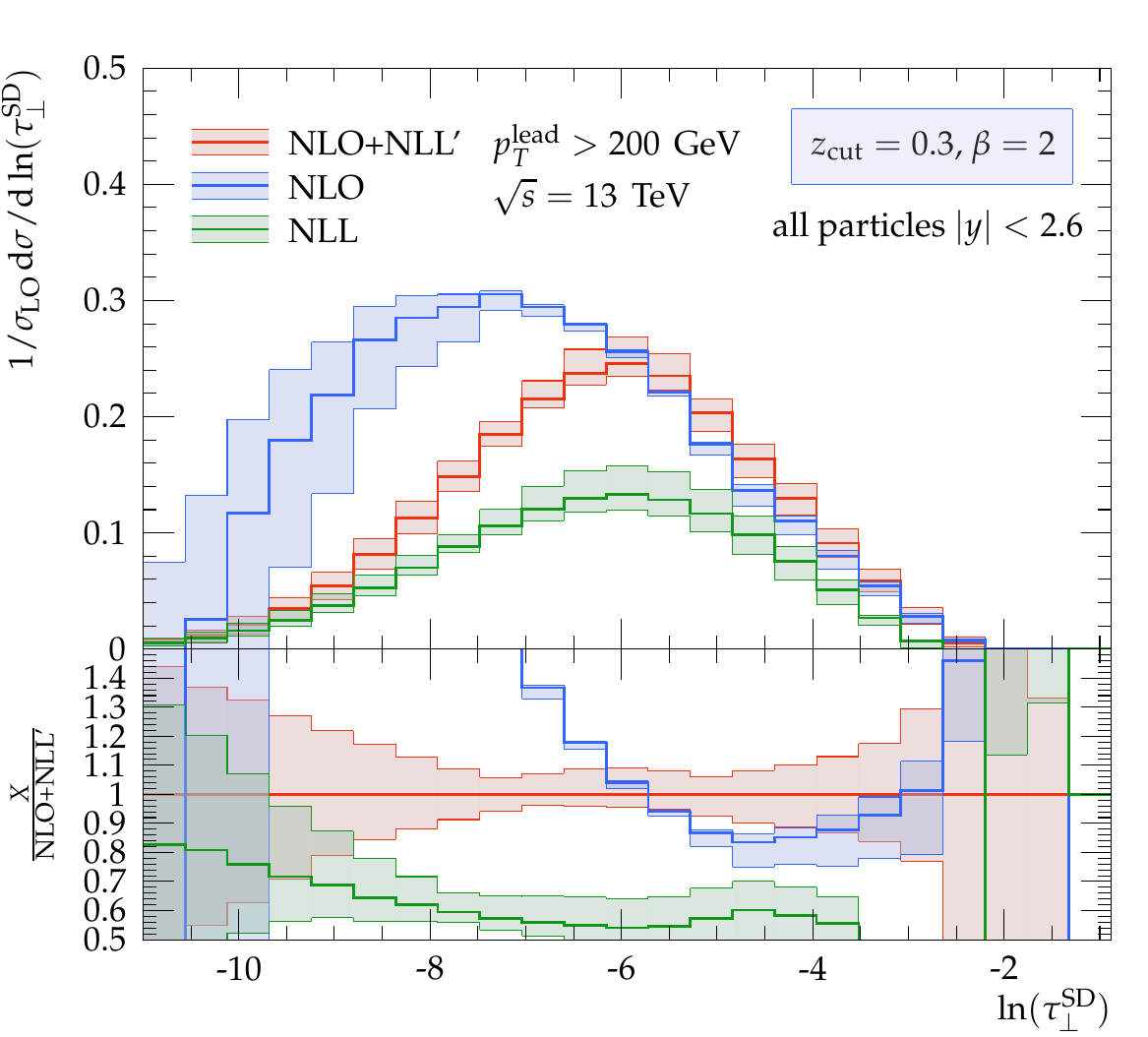}
	\end{center}
	\caption{\NLOpNLLp predictions for groomed transverse thrust for $\beta\in\{0,1,2\}$
          (columns) and $\zcut\in\{0.1,0.2,0.3\}$ (rows) for the $p_{T,\text{min}}=200\;\text{GeV}$
          event selection in comparison to the NLO result and the pure NLL
          resummation.}
	\label{fig:Resum}
\end{figure}

In Fig.~\ref{fig:Resum} we present our results for the matched \NLOpNLLp distributions
for the $p_{T,\text{min}}=200\;\text{GeV}$ event selection, along with the corresponding
\NLO and pure \NLL predictions. The uncertainty bands shown envelop the $7-$point scale
variations for $\mu_{\text{R}}$ and $\mu_{\text{F}}$, 
\emph{i.e.}\ $\{(\tfrac{1}{2}\muR,
\tfrac{1}{2}\muF)$, $(\tfrac{1}{2}\muR,\muF)$, $(\muR,\tfrac{1}{2}\muF)$, $(\muR,\muF)$,
$(\muR,2\muF)$, $(2\muR,\muF)$, $(2\muR,2\muF)\}$, 
and, for the resummation, separate
variations of $\mu_{\text{Q}}$ by factors of $0.5$ and $2$, and the alternative
end-point parameter choice $p=2$. The main result, the \NLOpNLLp prediction, is
shown in red, with its fixed-order ingredient (NLO) shown in blue, and the
resummation contribution (NLL) in green.  \NLOpNLLp and NLO
are normalised to unity including underflow, \emph{cf.}\ Eq.~\eqref{eq:normalisation}.
The \NLL here is also rescaled by $\sigma_\text{LO} = \sigma^{(0)} + \sigma^{(1)}$ instead of
just $\sigma^{(0)}$ in order to reflect the limit for the matched distribution when
the resummed logarithms dominate. As before we consider $\beta\in \{0,1,2\}$ and
$\zcut\in\{0.1,0.2,0.3\}$. In addition the ratio with respect to the \NLOpNLLp
result is included.

Grooming an event will reduce the value of transverse thrust, resulting in final states
that exhibit a more pencil-like topology. This is evident if we increase the threshold
$\zcut$. In consequence the cross section increases for smaller values of
$\tau^\text{SD}_\perp$, including the situation that the event ends up outside the
considered plot range. For $\beta>0$ even the peak position is significantly shifted to
lower values. When comparing the matched predictions to the pure resummation, we
can note that the latter dominates in the soft region, \emph{i.e.}\ for $\ln(\tauSD)<-7$
for all $\beta$. However in the ratios it can be seen that there are still significant
corrections originating from the $C^\delta_1$ coefficients and missing \NLO logarithms.
In particular for $\beta=0$, where the logarithmic enhancement is reduced, effects
from the matching can reach up to $40\%$ even in the logarithmic region. In contrast,
the fixed-order result dominates in the hard region, \emph{i.e.}\
large $\tauPerp^{\text{SD}}$, \emph{i.e.} for $\ln(\tauSD)>-5$.

The \NLOpNLLp prediction smoothly combines fixed order and resummation, thereby
inheriting the strengths of both approaches. It is worthwhile to note that in particular
for the case $\beta=0$ the \NLO results remain close to the matched prediction even in
the logarithmically dominated region. In particular for $\zcut>0.1$ results are found to
agree within the \NLO uncertainty estimate, however for $\zcut=0.1$ it can be read off from
the ratios that the logarithms do still result in significant deviations for $\ln(\tauSD)<-7$.
This agreement between \NLOpNLLp and \NLO, however, does not hold in general and for
$\beta>0$,  below $\ln(\tauSD)<-6$, the fixed-order results deviate very
strongly from the resummed predictions. This applies in particular for $\zcut>0.1$,
while in the $\zcut=0.1$ case the deviation is somewhat smaller and the
uncertainties tend to be larger, rendering this statement more ambiguous. 

Independent of the considered set of grooming parameters, the \NLOpNLLp results offer
a significant reduction of scale-variation uncertainties, when comparing to NLO and pure NLL,
respectively. For the fixed order and the NLL these become rather sizeable away from their
natural habitat, \emph{i.e.}\ in the soft region for the NLO and towards the hard region
for the NLL. For the matched predictions scale variations amount to roughly $10\%$ changes
in the peak region and for $\beta>0$ somewhat increase towards the hard and soft end of
the spectrum where the cross section is significantly reduced.

\begin{figure}[t!]
	\begin{center}
		\includegraphics[width=0.32\textwidth]{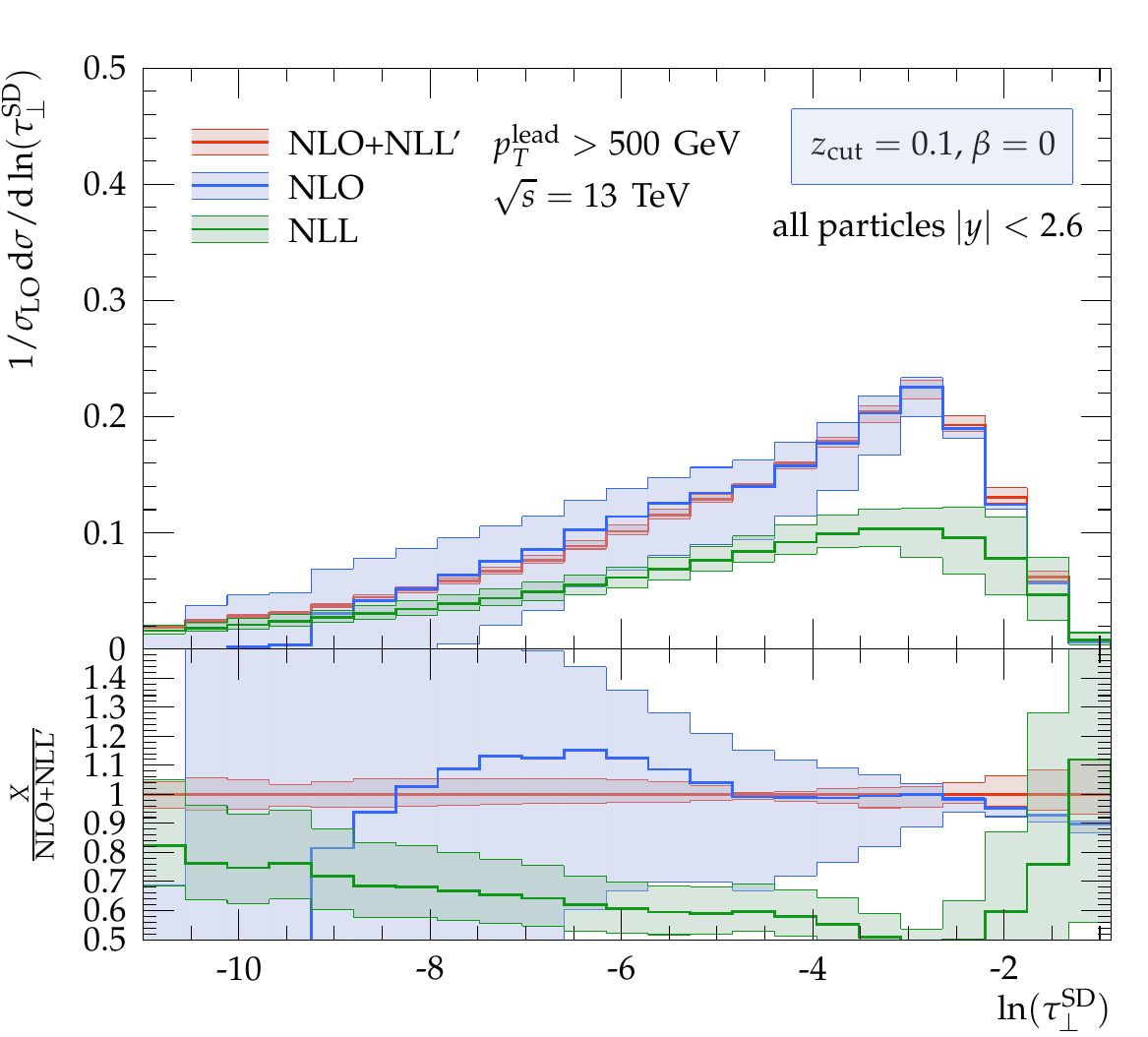}~
		\includegraphics[width=0.32\textwidth]{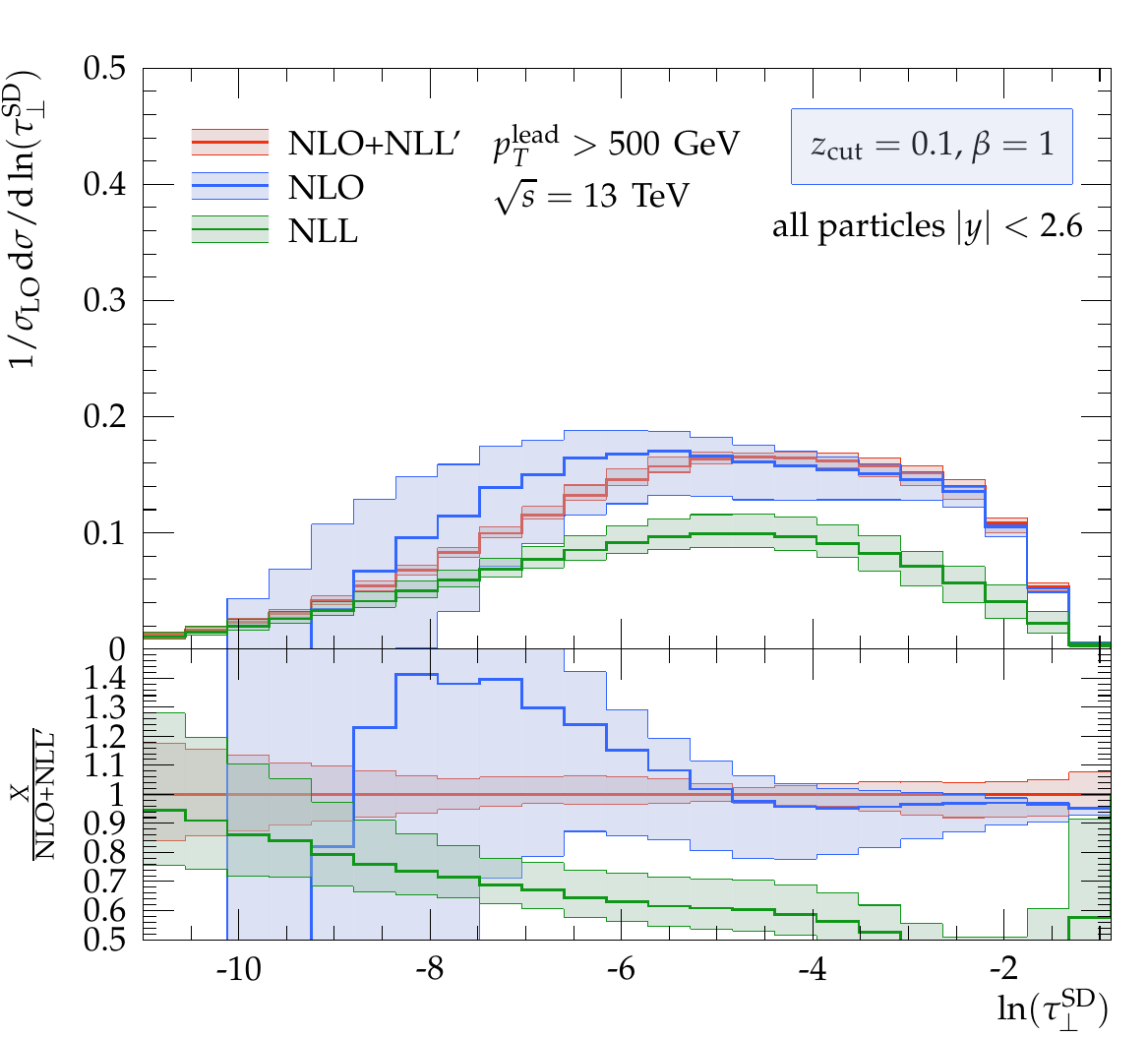}~
		\includegraphics[width=0.32\textwidth]{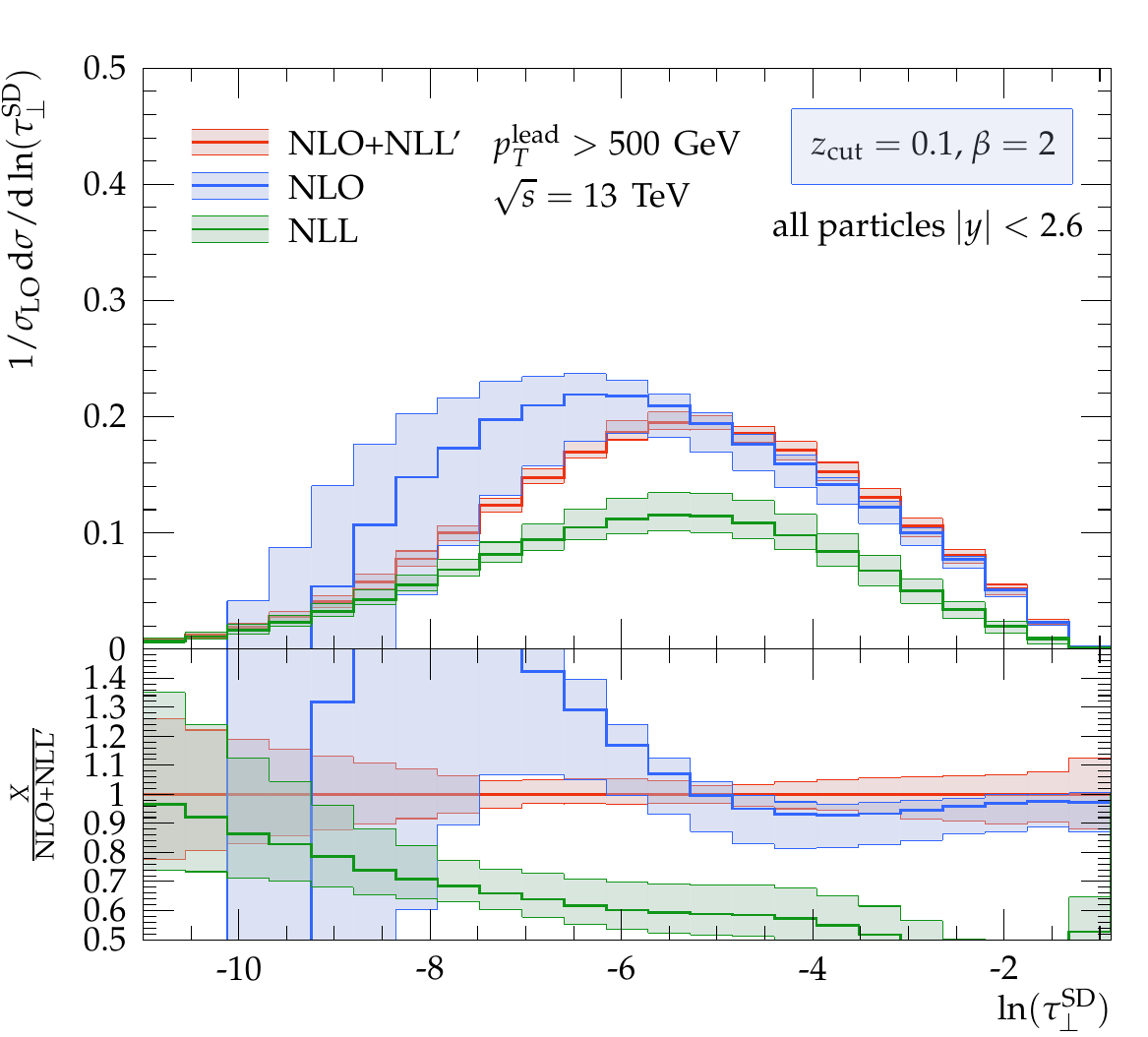}\\
		\includegraphics[width=0.32\textwidth]{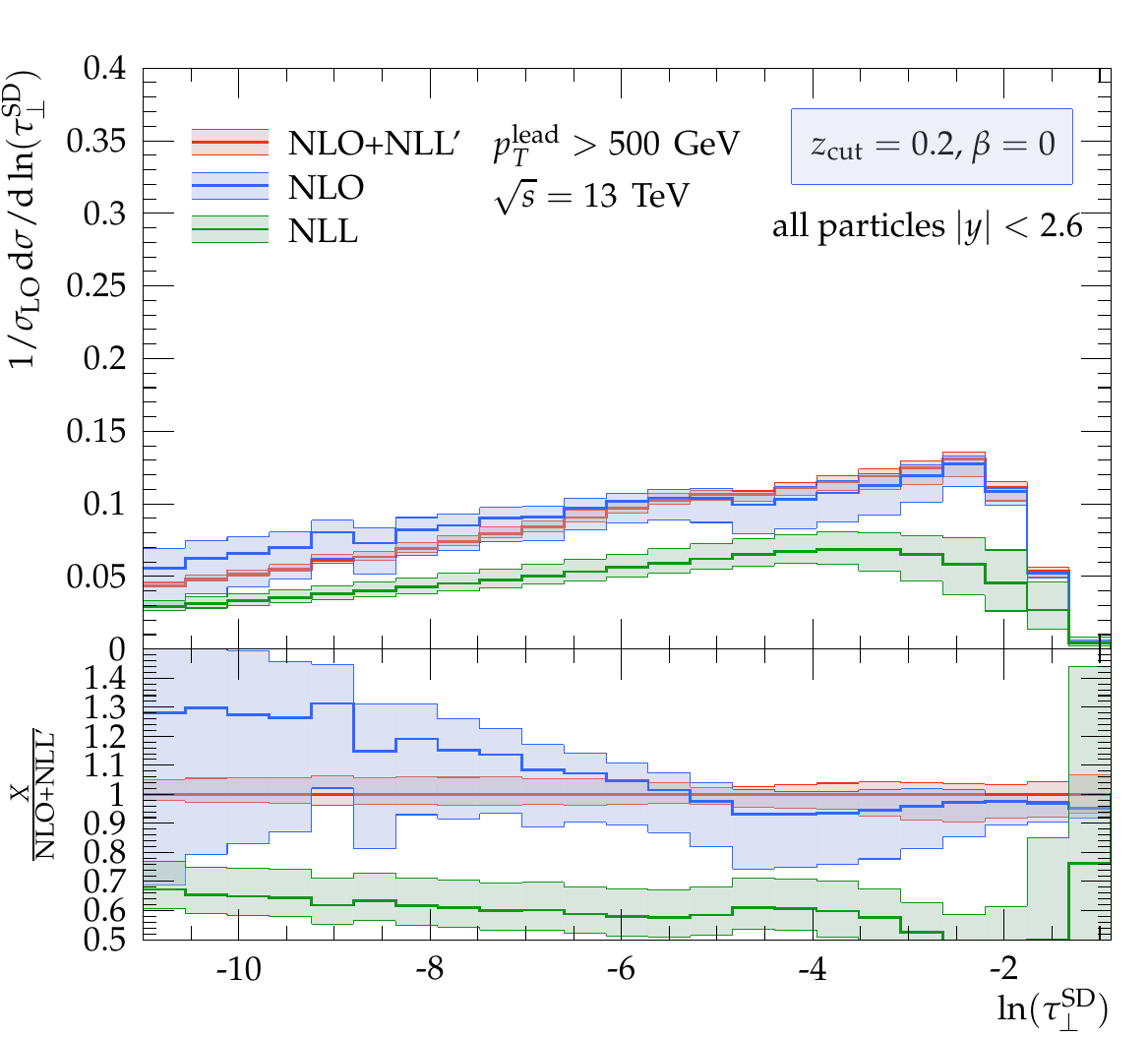}~
		\includegraphics[width=0.32\textwidth]{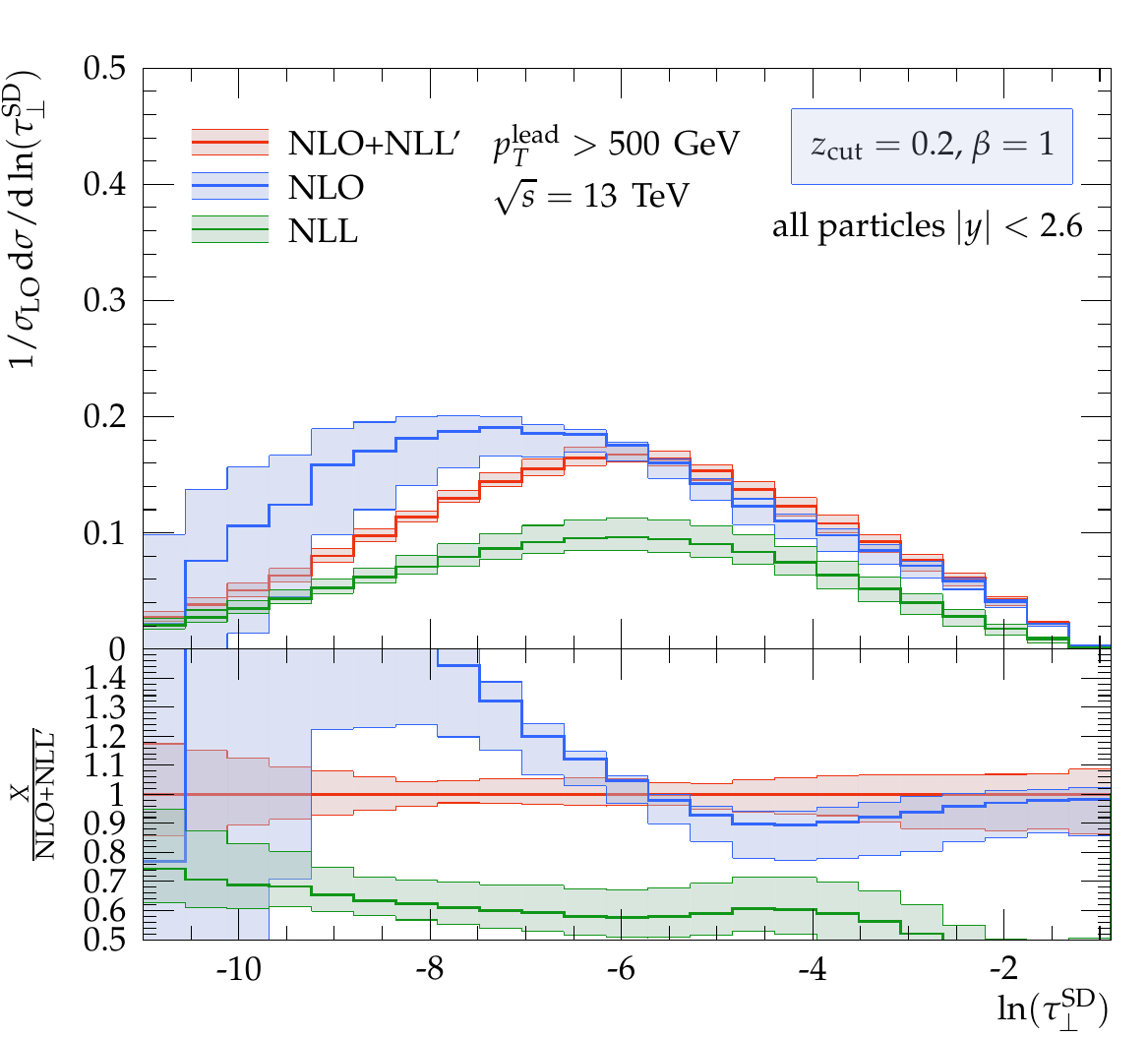}~
		\includegraphics[width=0.32\textwidth]{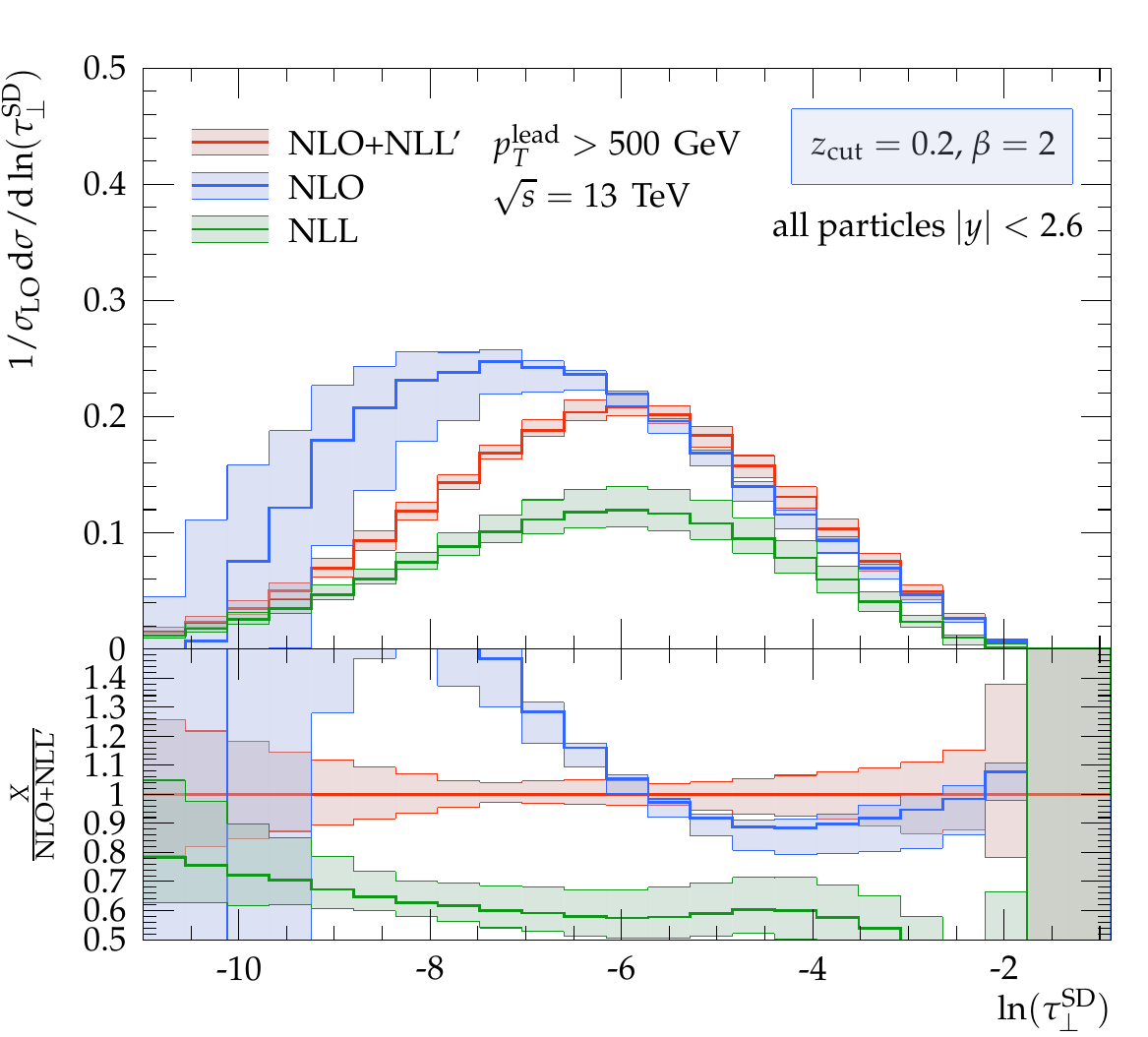}\\
		\includegraphics[width=0.32\textwidth]{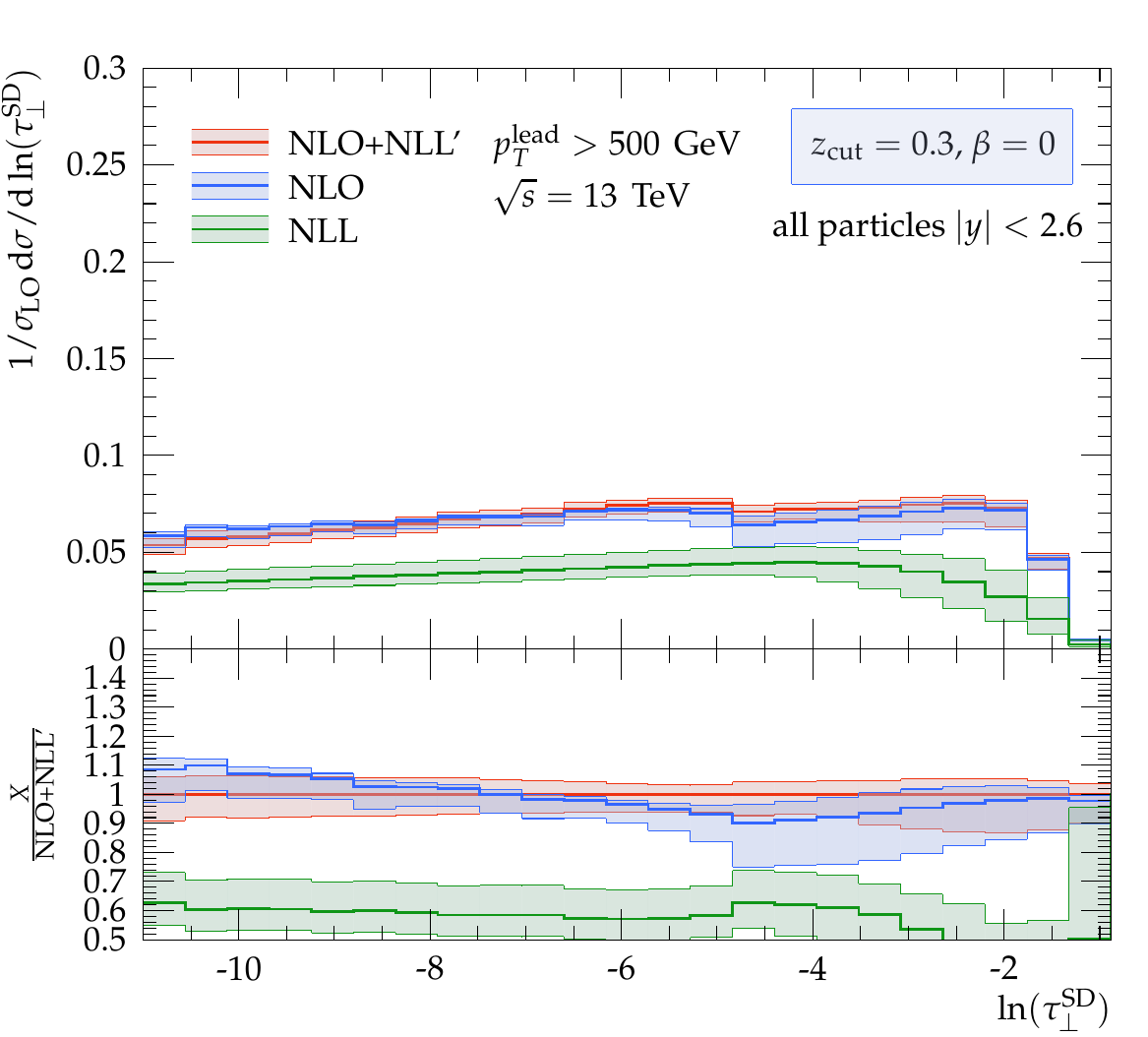}~
		\includegraphics[width=0.32\textwidth]{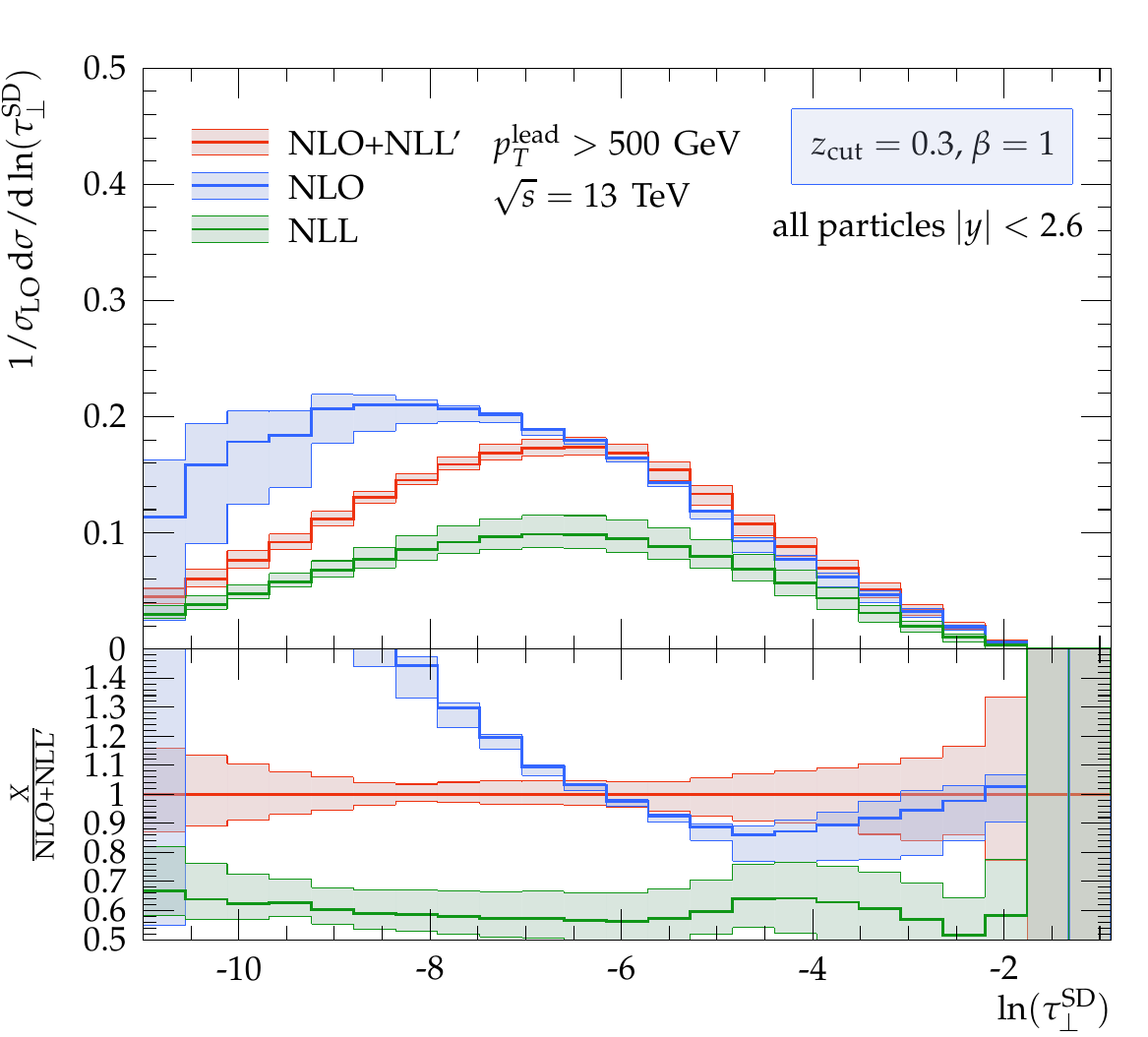}~
		\includegraphics[width=0.32\textwidth]{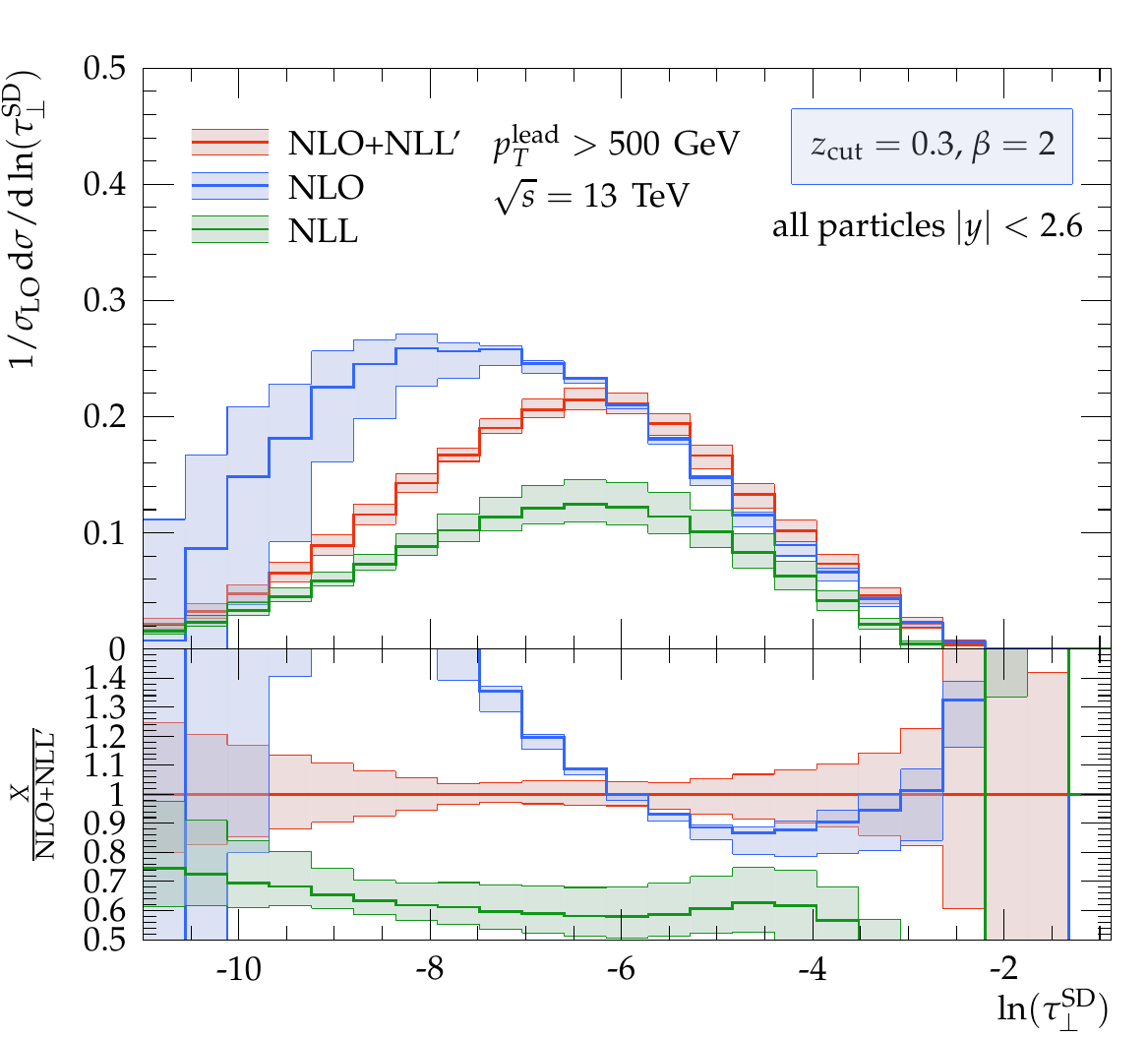}
	\end{center}
	\caption{Same as Fig.~\ref{fig:Resum} but for the $p_{T,\text{min}}=500\;\text{GeV}$ event selection. }
	\label{fig:Resum_500}
\end{figure}

In Fig.~\ref{fig:Resum_500} we present corresponding predictions for the $500\;\text{GeV}$ event
selection. The results are in fact very similar to the $p_{T,\text{min}}=200\;\text{GeV}$ case.
The good agreement of the \NLOpNLLp prediction with the NLO calculation and the pure NLL resummation
in the hard and soft region respectively is confirmed. Furthermore, we recover the reduction of scale-variation
uncertainties for the matched calculation. In Fig.~\ref{fig:Resum_500_aux} in
Appendix~\ref{app:aux_results} we compile further results for smaller values of the
grooming threshold, namely $\zcut\in\{0.01,0.02,0.05\}$, that will be used in Sec.~\ref{sec:ue_mitigation}
 when studying the potential of soft-drop grooming for underlying-event mitigation.

\clearpage

\section{Phenomenological studies of soft-drop groomed thrust}\label{sec:pheno}

In this section we present hadron-level predictions for the soft-drop groomed thrust
distribution in proton--proton collisions at $13$ TeV centre-of-mass energy. However,
we begin by comparing our \NLOpNLLp accurate predictions to multijet merged
parton-level simulations obtained with \Sherpa. We then focus on the potential
of soft-drop grooming to reduce the impact of the underlying event on the event-shape
distribution. To this end we consider particle-level simulations obtained with \Sherpa,
\Herwig and \Pythia.

\subsection{Monte Carlo simulations -- multijet merging, underlying event}\label{sec:MCsims}

To study the event-shape variable at the particle level, and to extract non-perturbative
corrections, we compile Monte Carlo predictions based on the \Sherpa event
generator version 2.2.10~\cite{Bothmann:2019yzt}. We simulate inclusive dijet production using
the \MEPSatNLO formalism~\cite{Hoeche:2012yf,Hoeche:2012fm}, thereby merging the
NLO QCD matrix elements for $2-$ and $3-$ and the LO matrix element for
the $4-$parton final states obtained from \Comix~\cite{Gleisberg:2008fv}, with
virtual corrections obtain from \OpenLoops~\cite{Cascioli:2011va}, dressed by
the \Sherpa dipole parton shower~\cite{Schumann:2007mg}. The merging-scale
parameter we set to $Q_{\text{cut}}=30\,\text{GeV}$.

To estimate the perturbative uncertainty of the Monte Carlo predictions, we
again consider $7-$point variations of the factorisation and renormalisation
scales, both in the hard matrix elements and the parton shower, evaluated using
on-the-fly reweighting~\cite{Bothmann:2016nao}.

The nominal perturbative scales are defined according to the CKKW-style scale setting
prescription~\cite{Catani:2001cc}, \emph{cf.}~\cite{Hoeche:2009rj} for details. In this
procedure the hard-process partons get clustered into a Born-like $2\to 2$ configuration
that defines the $2-$jet \emph{core process} with an associated scale
$\mu_{\text{core}}$, that we set to
\begin{equation}
  \mu_{\text{core}}= \frac12 H_T\,.
\end{equation}
This corresponds to the jet transverse momentum for the reconstructed $2-$jet system
and is also used to define the factorisation scale and the shower-starting scale of the core
process, \emph{i.e.}\
\begin{equation}
  \mu_{\text{F}}=\mu_{\text{Q}}=\mu_{\text{core}}\,.
\end{equation}
The effective renormalisation scale, $\mu_{\text{CKKW}}$, of the $n$-parton hard matrix
elements corresponds to
\begin{equation}
\alphaS^n(\mu^2_{\text{CKKW}})=\alphaS^2(\mu^2_{\text{core}})\prod\limits_{i=1}^{n-2}\alphaS(t_{i})\,,
\end{equation}
with $t_i$ the reconstructed shower-branching scales. 

To account for hadronisation effects we employ \Sherpa's cluster fragmentation
model~\cite{Winter:2003tt}. The underlying event simulation uses the \Sherpa
implementation of the Sj\"ostrand--Zijl multiple-parton interaction
model~\cite{Sjostrand:1987su}. In both models the default set of tuning parameters
is used, see~\cite{Bothmann:2019yzt} for details. To contrast and underpin the
\Sherpa predictions with other theoretical approaches to parton showering and
in particular non-perturbative effects~\cite{Buckley:2011ms}, we also compile full
particle-level results with \Herwig version 7.2.1~\cite{Bellm:2015jjp,Bellm:2019zci}
and \Pythia version 8.240~\cite{Sjostrand:2014zea}. In both generators we simulate
inclusive dijet production at leading order, using the default models and parameters
for the underlying event and hadronisation simulations. We also make use of the codes'
default scale-setting that in both cases relates the parton-shower starting scale to
the $p_T$ of the hard-process jets.

For event selection and analysis we employ the \Rivet analysis package~\cite{Buckley:2010ar}.

\subsection{Parton-level predictions}

In order to further validate our analytic calculations and benchmark the
perturbative inputs to our full particle-level Monte Carlo simulations, we
compare the \Sherpa \MEPSatNLO parton-level (PL) predictions against our
previously presented resummation results. To be more precise, by PL we here mean
terminating the event evolution right after the parton-shower stage, determined
by the dipole-shower cutoff $p^{\text{CSS}}_{T,\text{min}}=1\;\text{GeV}$. We
then apply our usual event selection to this partonic final state, \emph{cf.}\ 
Sec.~\ref{sec:eventselections}, and include all generated partons with $|y|<2.6$
in the observable evaluation. 

\begin{figure}[t!]
  \begin{center}
  	\includegraphics[width=0.32\textwidth]{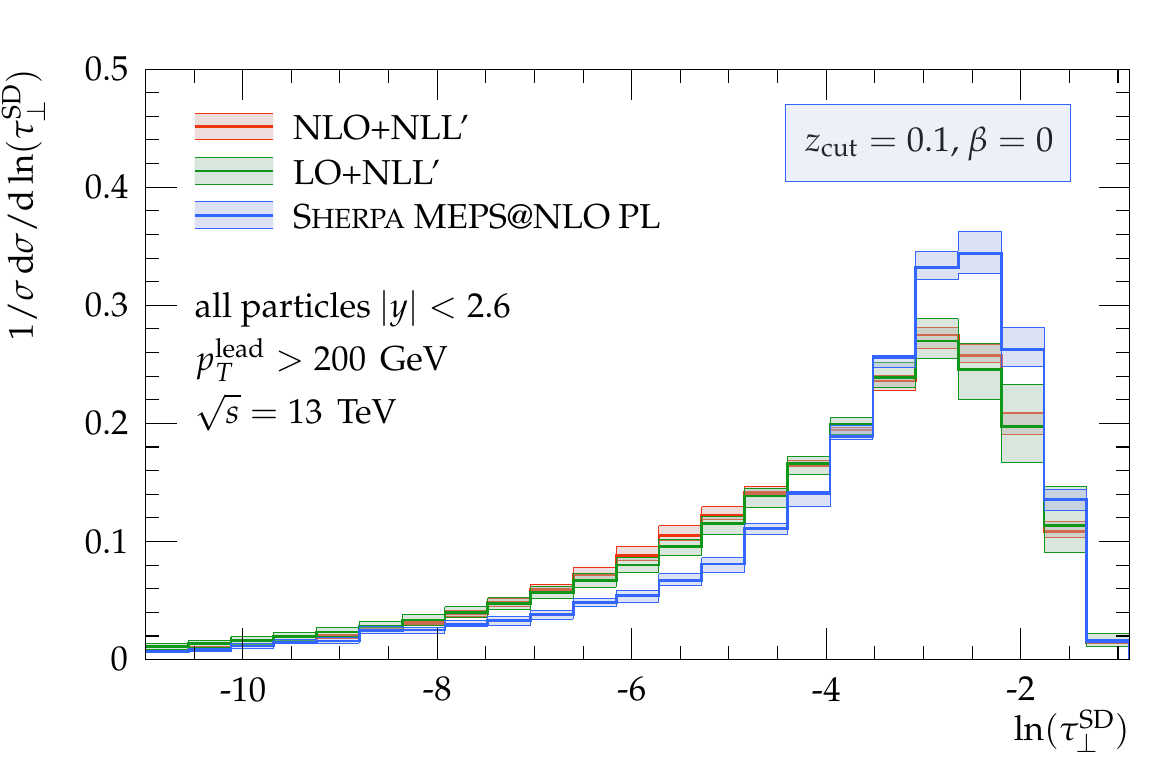}~
  	\includegraphics[width=0.32\textwidth]{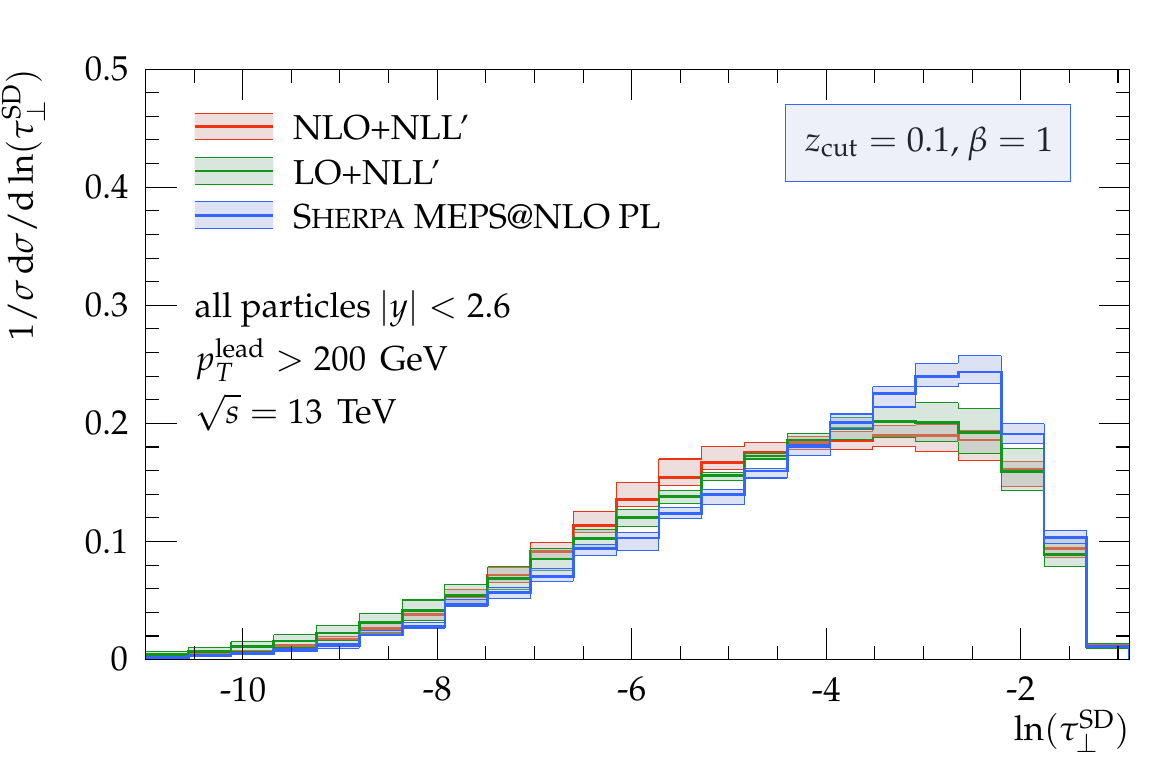}~
  	\includegraphics[width=0.32\textwidth]{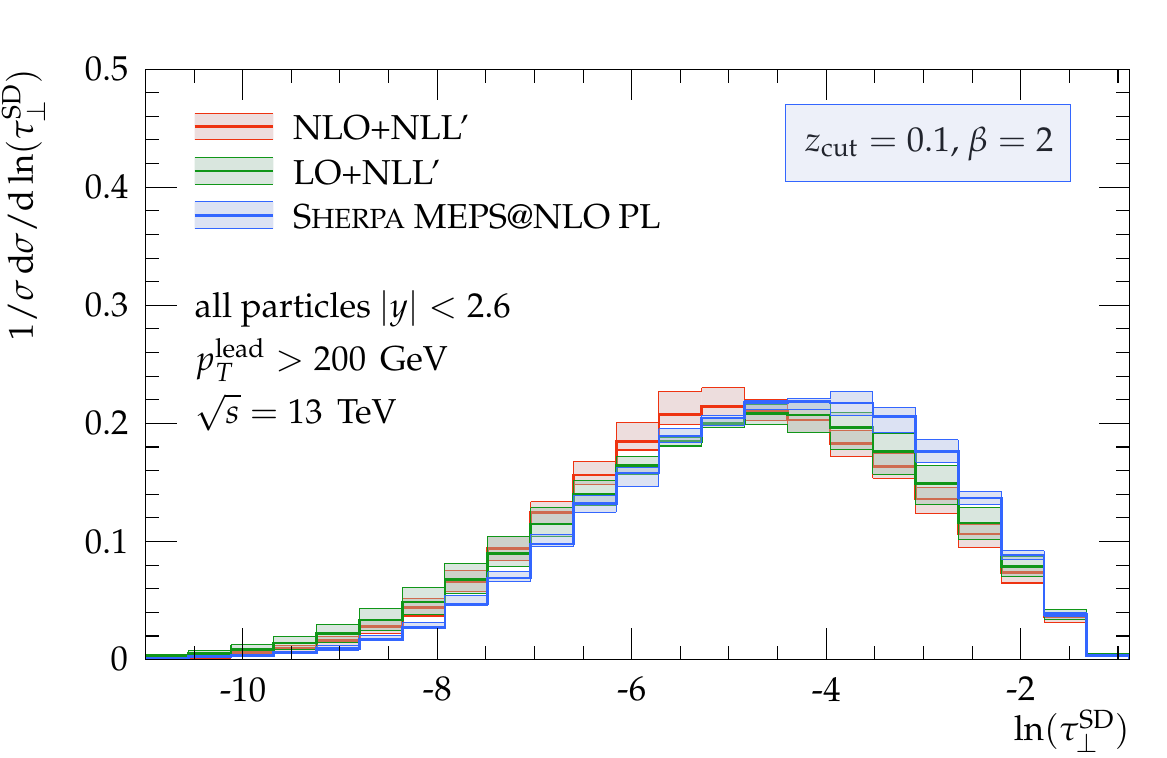}\\
  	\includegraphics[width=0.32\textwidth]{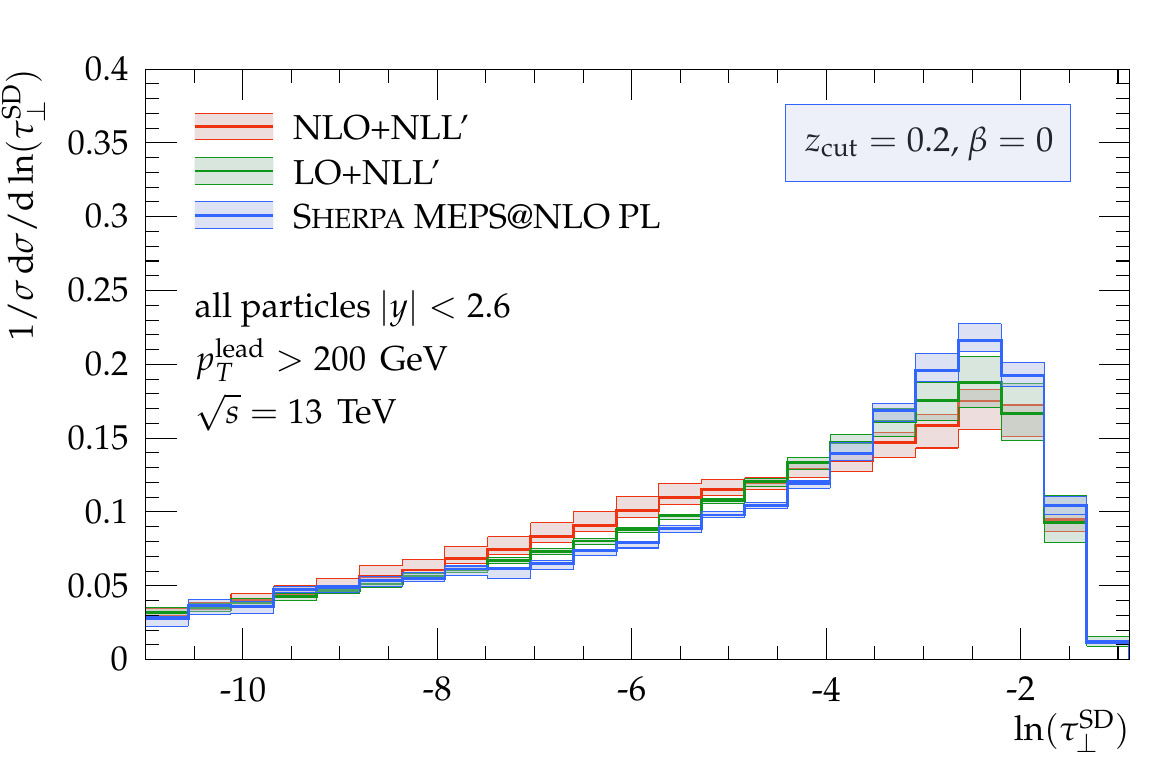}~
  	\includegraphics[width=0.32\textwidth]{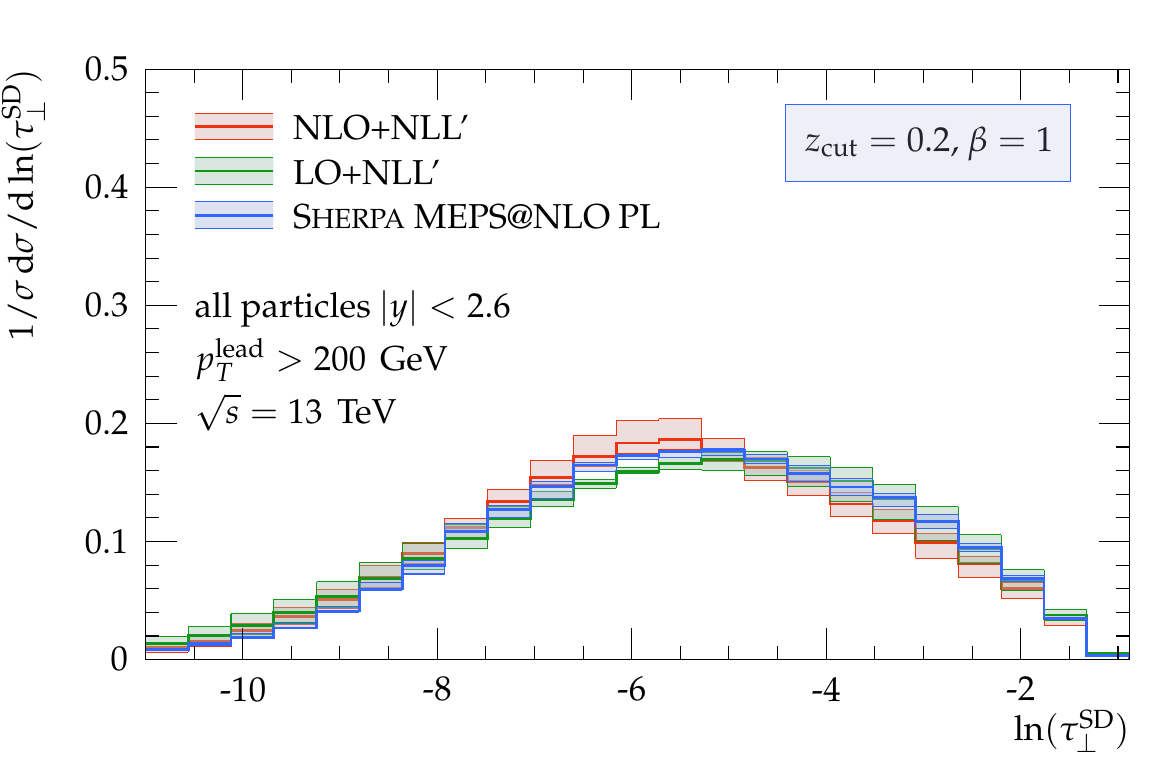}~
  	\includegraphics[width=0.32\textwidth]{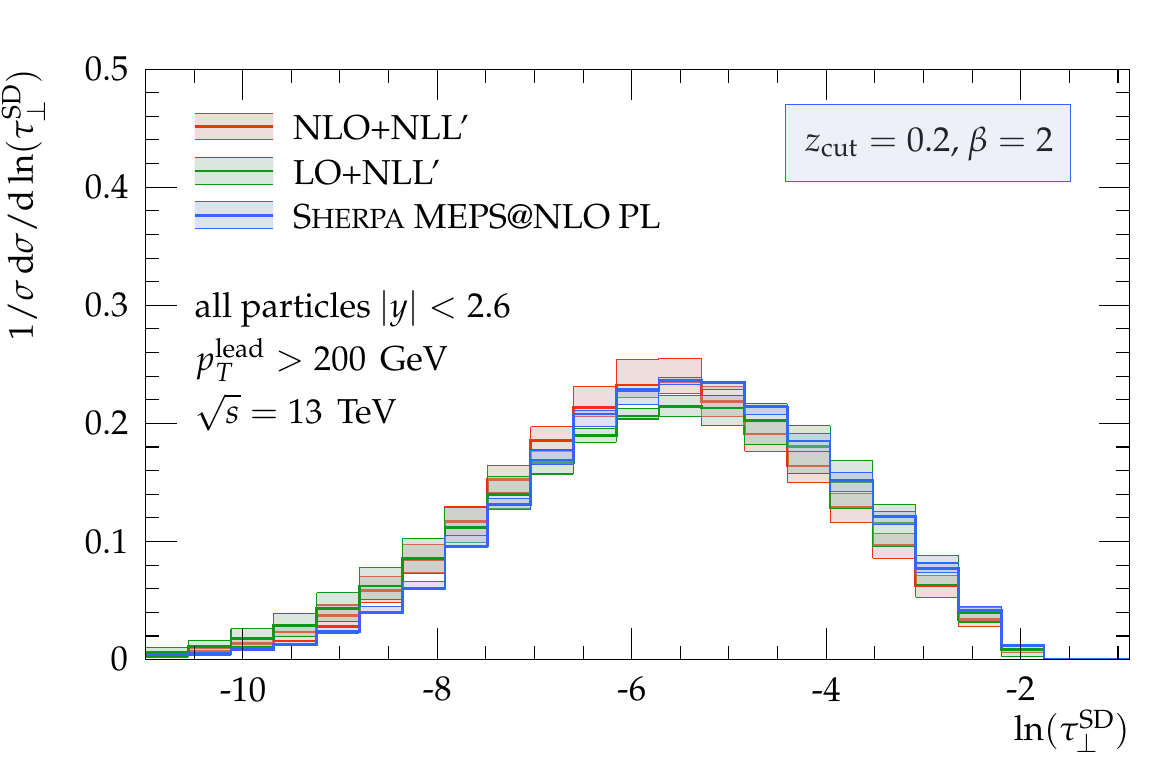}\\
  	\includegraphics[width=0.32\textwidth]{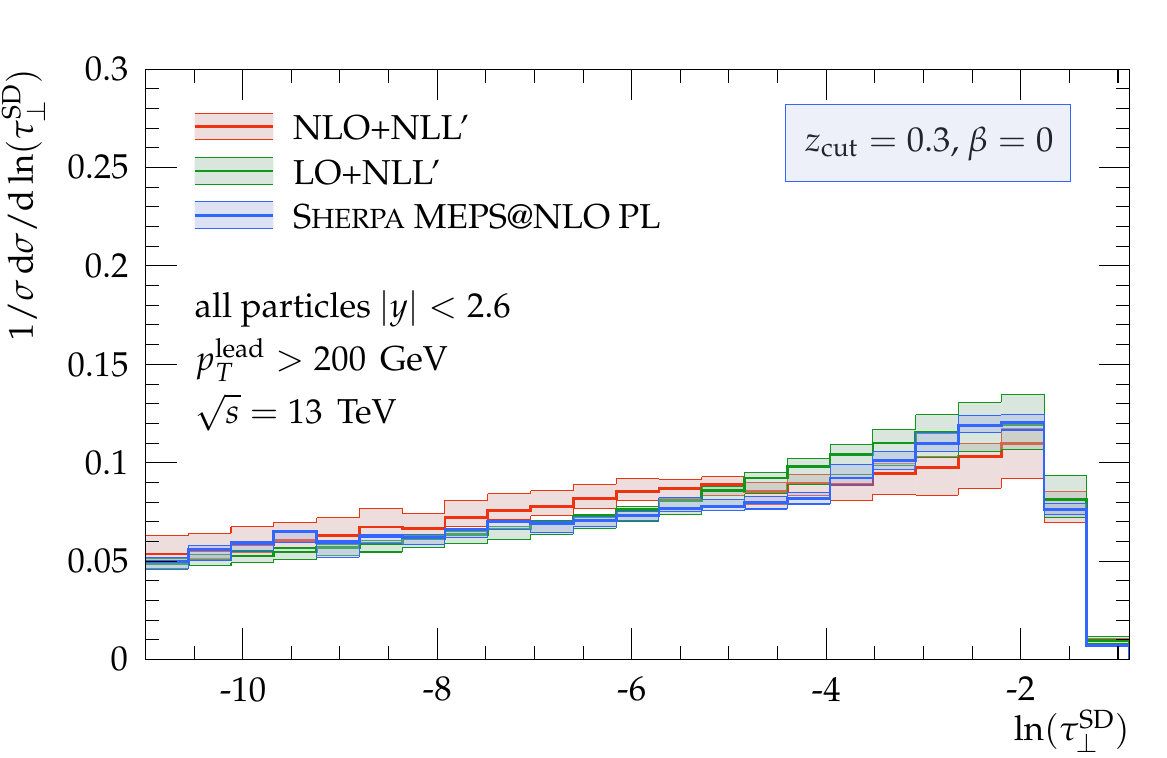}~
  	\includegraphics[width=0.32\textwidth]{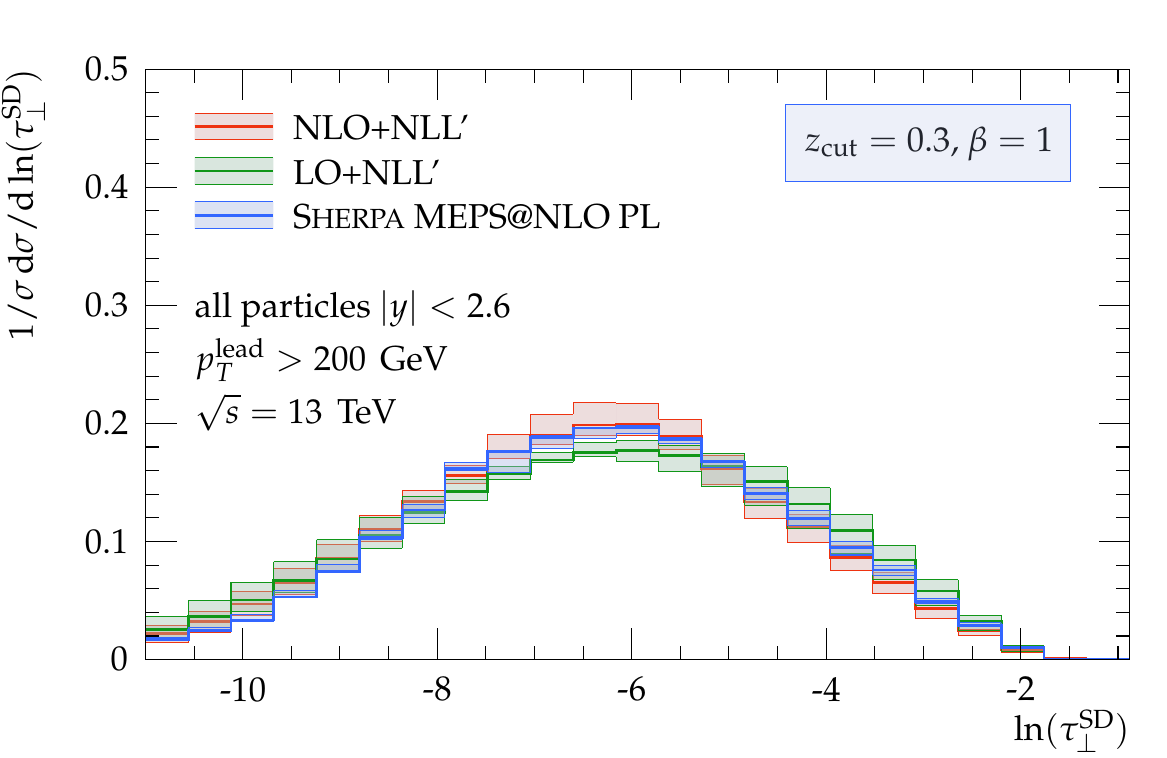}~
  	\includegraphics[width=0.32\textwidth]{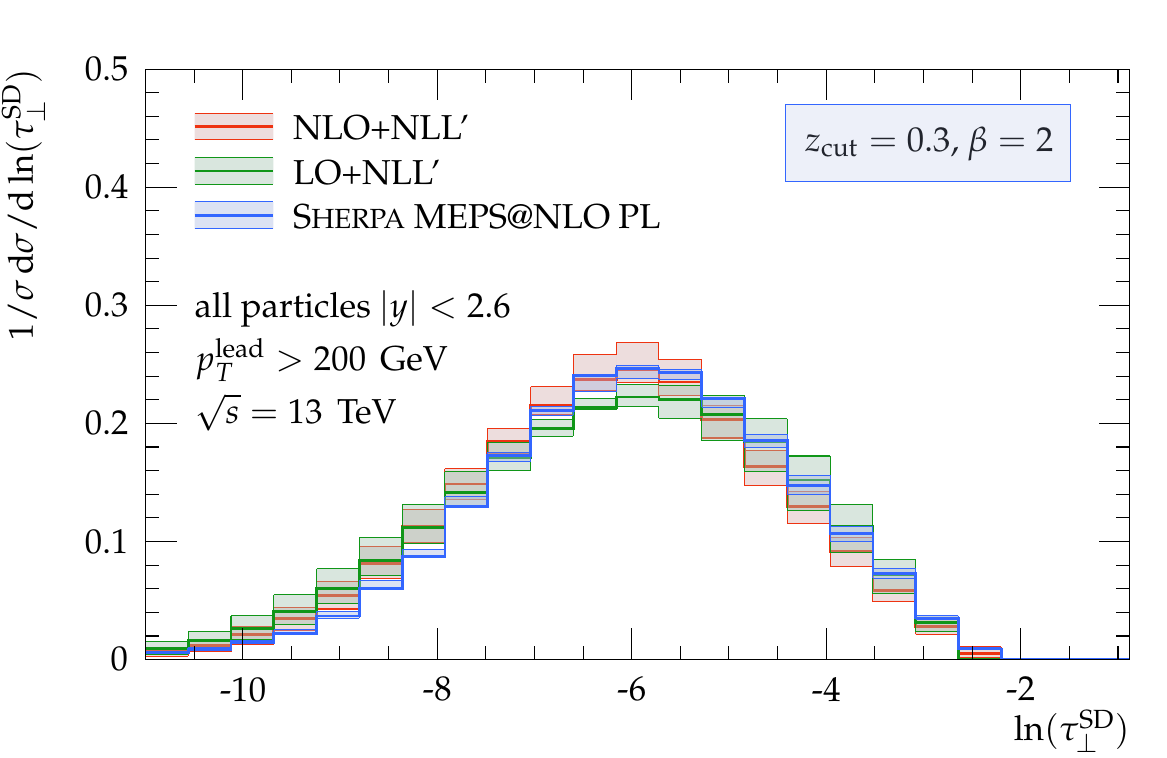}
  \end{center}
  \caption{\NLOpNLLp and \LOpNLLp predictions for groomed transverse thrust for
    $\beta\in\{0,1,2\}$ (columns) and $\zcut\in\{0.1,0.2,0.3\}$ (rows) for the
    $p_{T,\text{min}}=200\;\text{GeV}$ event selection in comparison to parton-level
    \MEPSatNLO results from \Sherpa. }
  \label{fig:ResumVSMC}
\end{figure}

We consider our resummed predictions matched to LO and NLO, \emph{i.e.}\
\LOpNLLp and \NLOpNLLp, respectively. Complementary to the analytic resummation
approach, in parton-shower simulations emissions off the Born process, as well
as subsequent emissions thereof, get stochastically generated, thereby
accounting for momentum conservation as well as finite recoil effects. In this
context, it is important to stress that we also do not attempt to adjust
the scale choices and evolution schemes in shower or resummation to
particularly match each other which, together with the aforementioned
recoil effects, can lead to significant practical differences
\cite{Hoeche:2017jsi}. We here rather aim to compare the resummed results to the
exact perturbative input of our phenomenological studies.  We hence should not
necessarily expect the central values to agree, and rather be prepared to accept 
incompatibilities as insufficiencies in our error estimates. 

\begin{figure}[t!]
  \begin{center}
  	\includegraphics[width=0.32\textwidth]{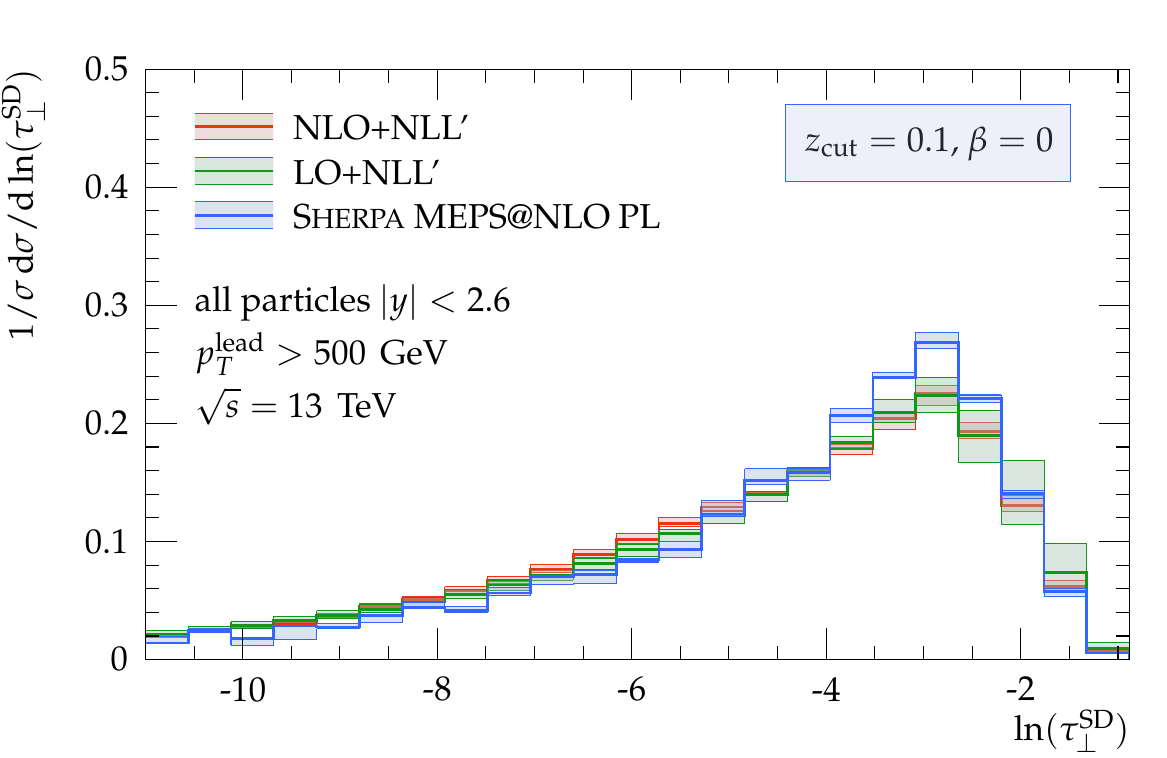}~
  	\includegraphics[width=0.32\textwidth]{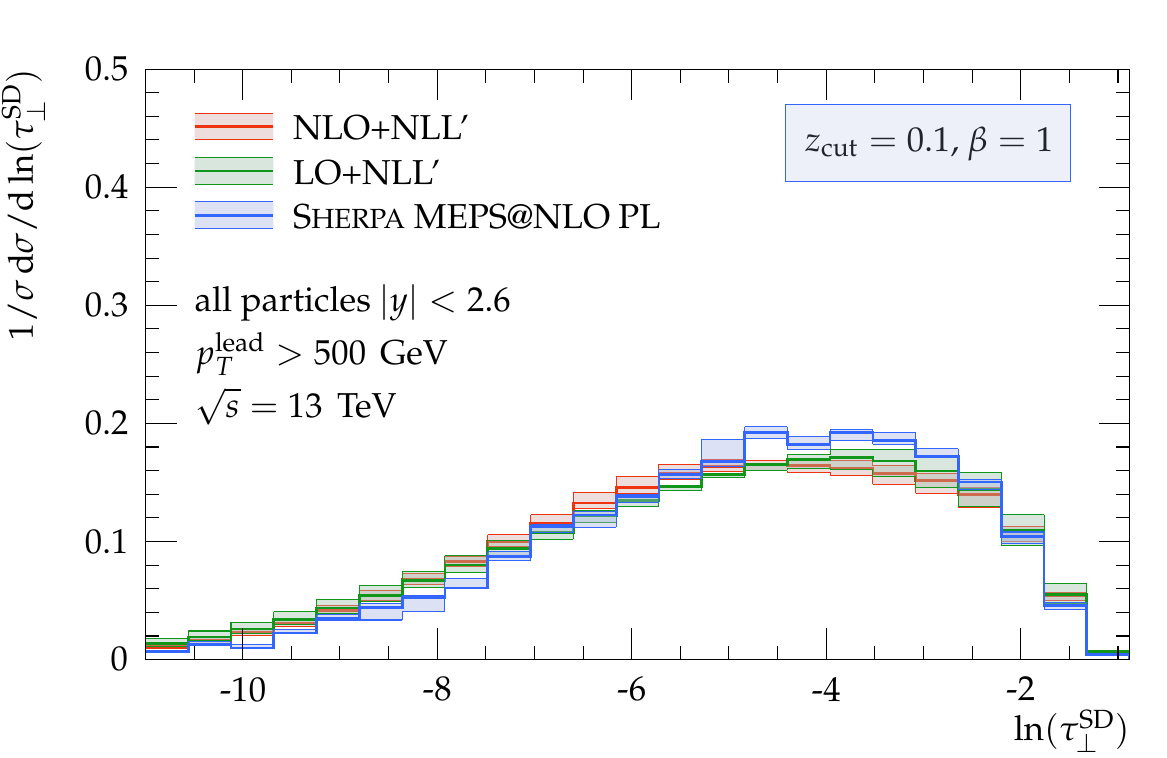}~
  	\includegraphics[width=0.32\textwidth]{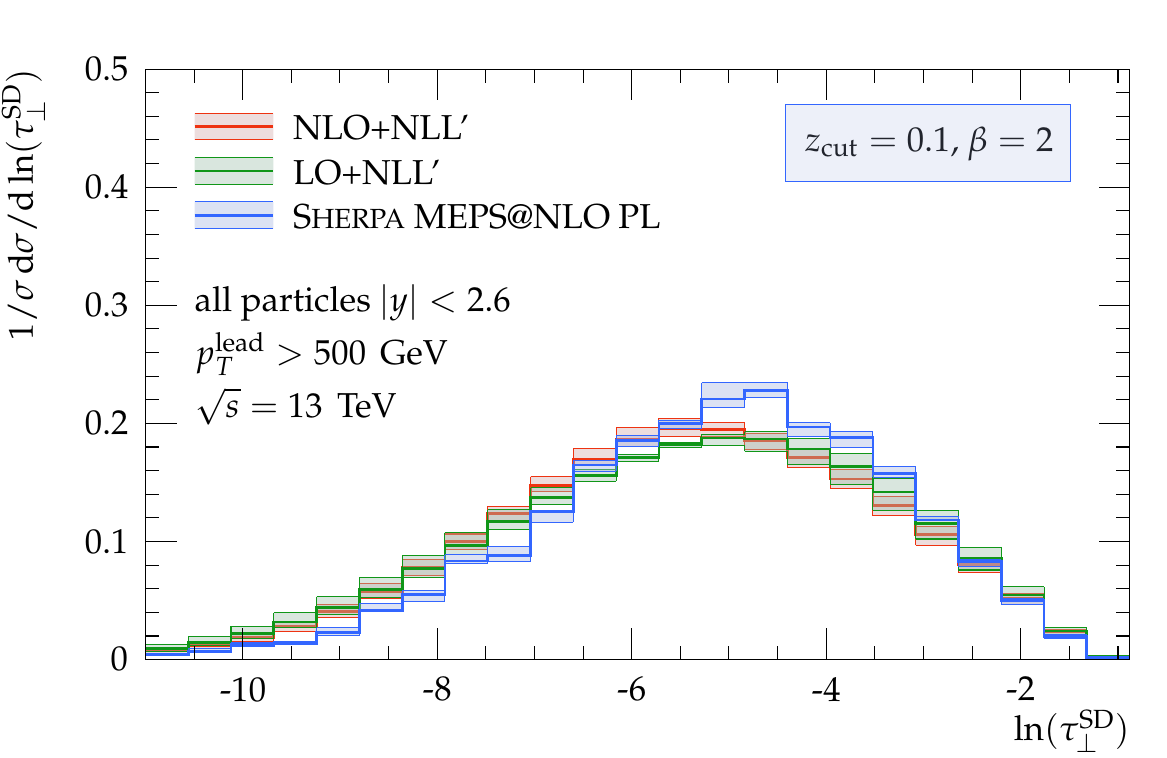}\\
  	\includegraphics[width=0.32\textwidth]{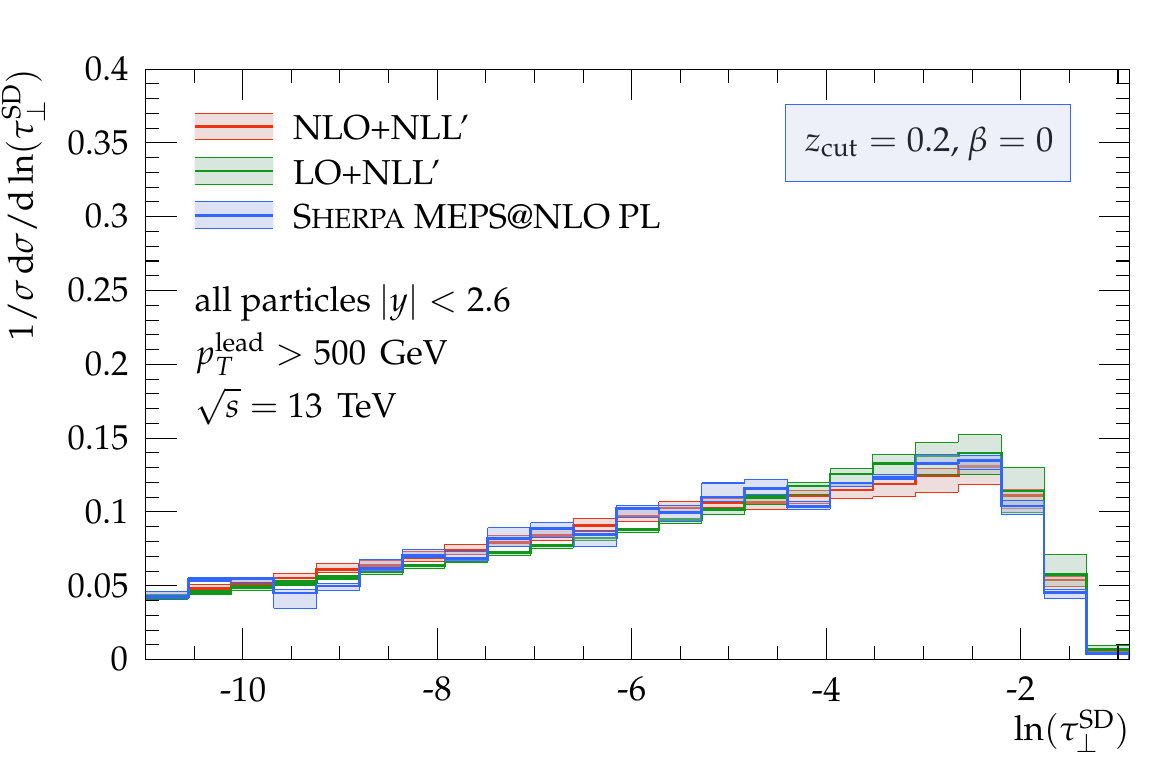}~
  	\includegraphics[width=0.32\textwidth]{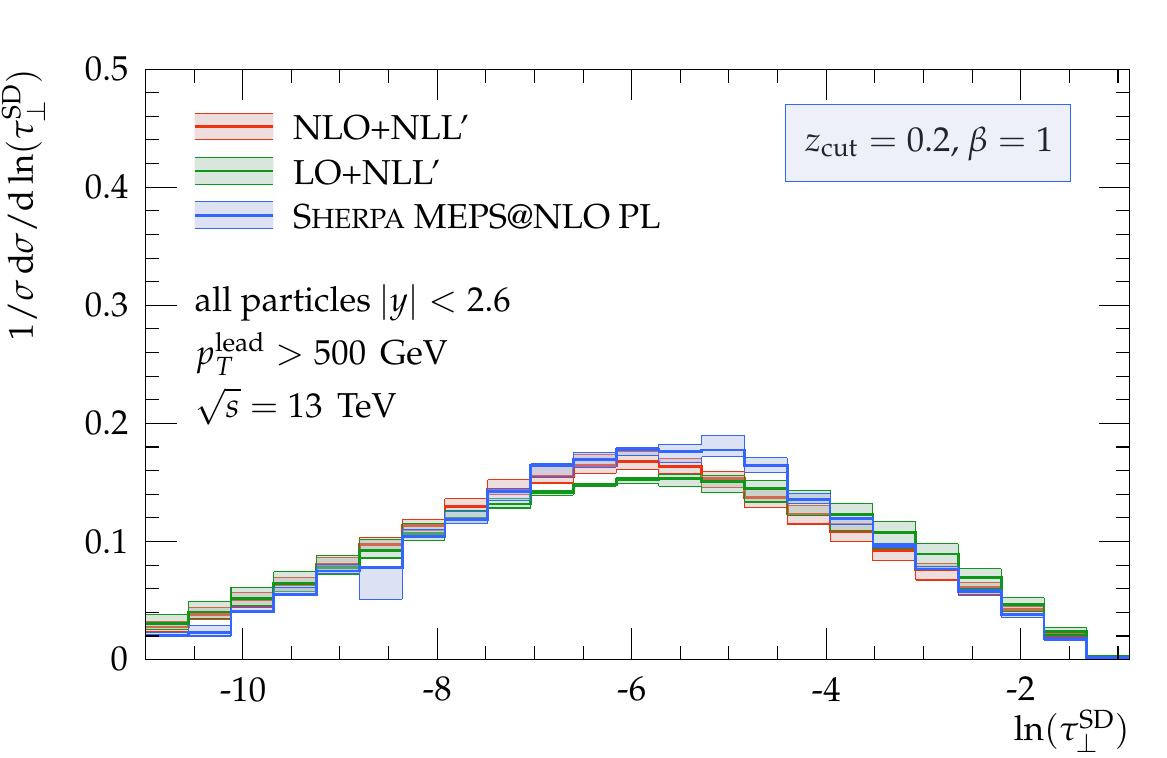}~
  	\includegraphics[width=0.32\textwidth]{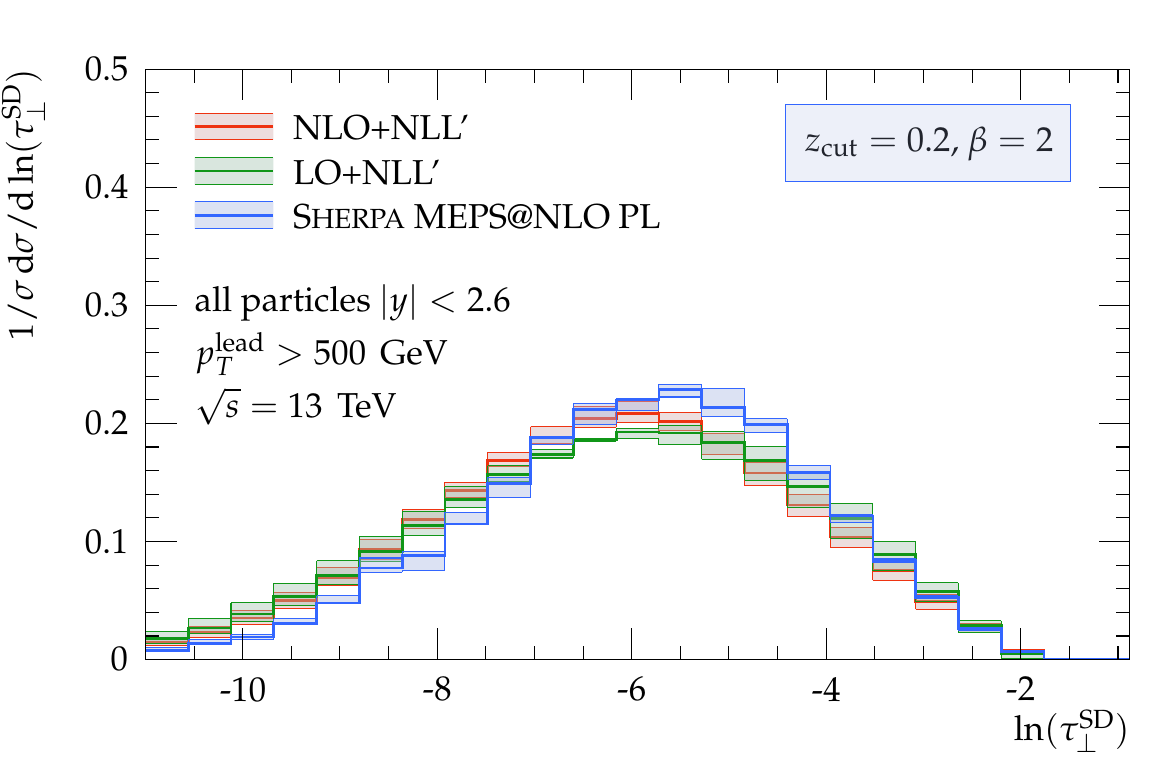}\\
  	\includegraphics[width=0.32\textwidth]{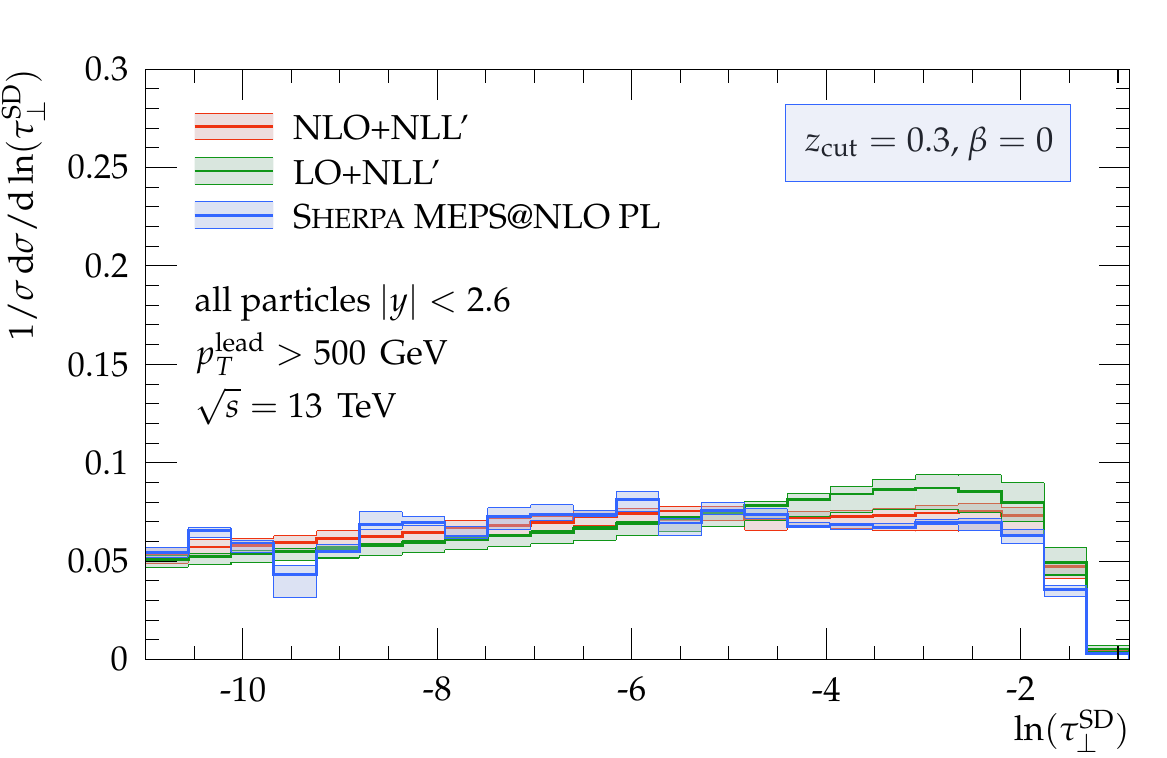}~
  	\includegraphics[width=0.32\textwidth]{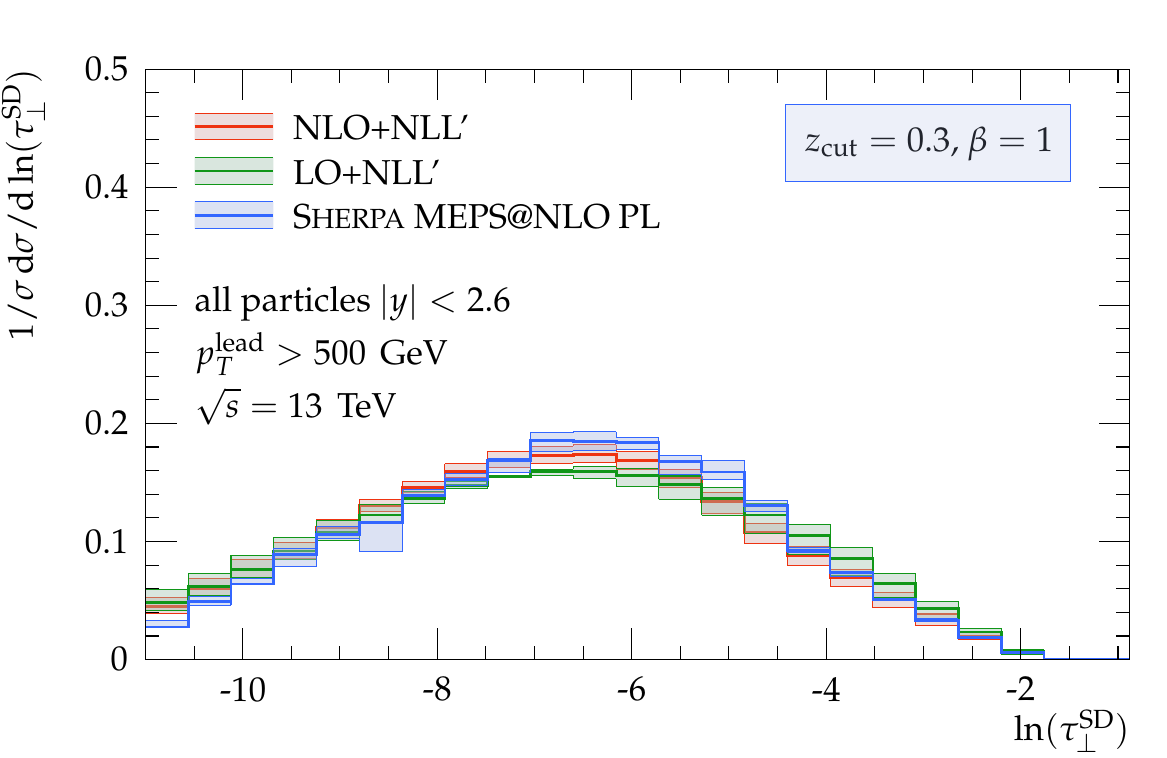}~
  	\includegraphics[width=0.32\textwidth]{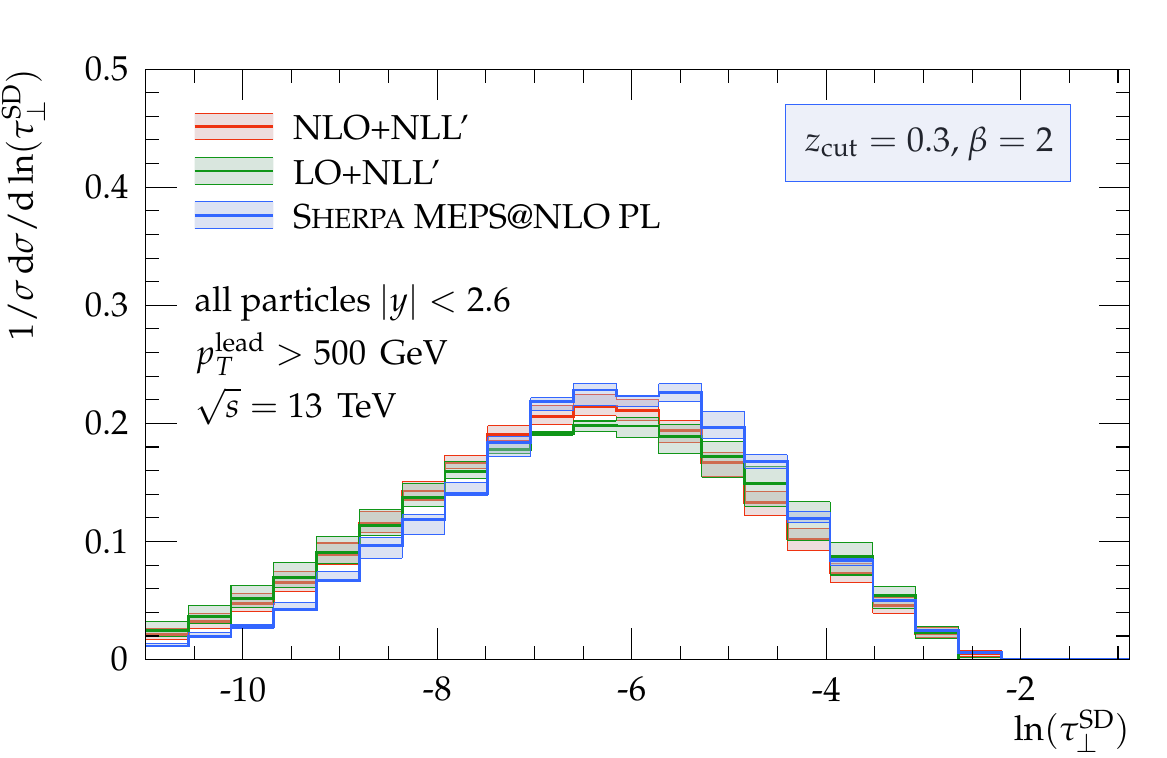}
  \end{center}
  \caption{Same as Fig.~\ref{fig:ResumVSMC} but for the $p_{T,\text{min}}=500\;\text{GeV}$ event selection.}
  \label{fig:ResumVSMC_500}
\end{figure}

In Fig.~\ref{fig:ResumVSMC} we compare the three perturbative calculations,
with the same grooming parameters as previously, $\beta \in \{0,1,2\}$ and
$\zcut\in\{0.1,0.2,0.3\}$, for the $p_{T,\text{min}}=200\;\text{GeV}$ event
selection. The \MEPSatNLO results, including the scale-uncertainty band, are
shown in blue, \NLOpNLLp in red, and \LOpNLLp in green. 
In Fig.~\ref{fig:ResumVSMC_500} we compile the corresponding results for the
$p_{T,\text{min}}=500\;\text{GeV}$ event selection, considering the same set of
grooming parameters. We furthermore present predictions for smaller values of
the grooming threshold, \emph{i.e.}\  $\zcut\in\{0.01,0.02,0.05\}$ in
Fig.~\ref{fig:ResumVSMC_500_aux} in App.~\ref{app:aux_results}.

For both choices of $p_{T,\text{min}}$, the three predictions yield very similar
qualitative results, especially as far as the effect of grooming and the
dependence on the parameters $\zcut$ and $\beta$ goes. This holds for the
hard end of the spectrum, dominated by the fixed-order components, through the
intermediate transition region, as well as for the resummation (multiple
emission) dominated small-$\tauSD$ limit. The  matrix-element improved
parton-shower simulation fully confirms the observations on the impact of
grooming on the distributions shape. By increasing $\zcut$, events are pushed to
lower observable values. For $\beta>0$ this even affects the position of the
peak of the distributions. We take this agreement as further confirmation of
the general effect of our proposed grooming procedure.
 
For the $200\;\text{GeV}$ selection we observe some quantitative differences, particularly so for
$\zcut=0.1$, most significant for the case of $\beta=0$ and $\beta=1$.
The \MEPSatNLO simulation here predicts a somewhat narrower distribution with a more
pronounced peak around the transition point between groomed and ungroomed
regions. For the case of ungroomed transverse thrust a similar level of
deviation has been observed in previous studies~\cite{Banfi:2010xy}. However,
for stronger grooming these differences become smaller and we find in fact a
remarkably good agreement, given the theoretical uncertainty. This behaviour is
consistent with naive expectations. Soft drop, while not necessarily designed
with perturbative ambiguities in mind, after all mainly removes wide-angle
radiation that is less constrained.

For the $500\;\text{GeV}$ selection, the results fully confirm our general
observations for the $200\;\text{GeV}$ case. The \MEPSatNLO and the resummed
predictions  mostly agree as long as the final state is significantly
groomed. Even for the grooming setups where differences were obvious,
\emph{e.g.} $\zcut=0.1$ and $\beta=0,1$, they appear to be reduced in the
$500\;\text{GeV}$ case. Again, \Sherpa predicts a somewhat narrower distribution
than both the \LOpNLLp and \NLOpNLLp for those parameters.

We conclude from this initial parton-level comparison that the effects of soft-drop
grooming are consistently modelled between the \Sherpa PL and our resummed
calculations, at least for sufficient grooming. To fully interpret the
significance of the remaining differences, we should determine and confirm to
what level hadronisation and underlying-event corrections are reduced by
grooming. We will do exactly that in the following section.

\FloatBarrier
\subsection{Underlying event mitigation}\label{sec:ue_mitigation}

After having validated the resummed predictions and compared them to
parton-level matrix-element plus parton-shower simulations, the focus
shall now shift to studying the impact of non-perturbative corrections
on the observable and the potential of soft-drop grooming to reduce their impact.
In Fig.~\ref{fig:NP_ungroomed} we already indicated
the typical size of underlying event and hadronisation corrections as they arise
for plain transverse thrust. In particular the underlying event results
in a significant shift of the distribution towards higher observable
values. While the parton-to-hadron fragmentation has a smaller impact
for the bulk of the events, it is significant in particular in the low-$\tauPerp$
tail. As the event display in Fig.~\ref{fig:eventdisplay} illustrates,
we can expect that soft-drop grooming can be quite efficient in removing
the largely uniform underlying-event activity from the final
state. This can open the possibility for a more direct reproducibility of
experimental measurements by perturbative predictions.

\begin{figure}[ht!]
	\begin{center}
	  \includegraphics[width=0.32\textwidth]{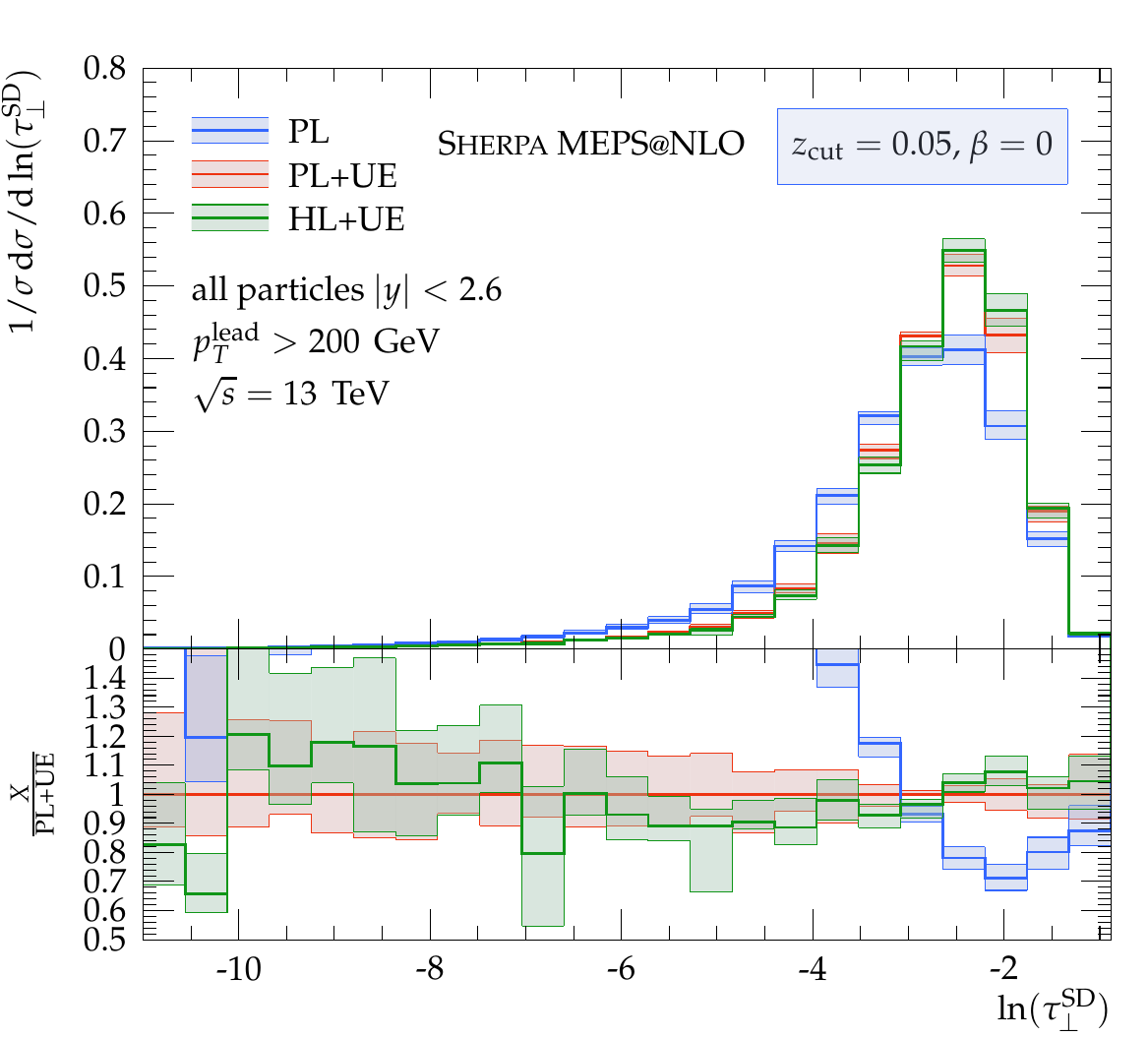}~
	  \includegraphics[width=0.32\textwidth]{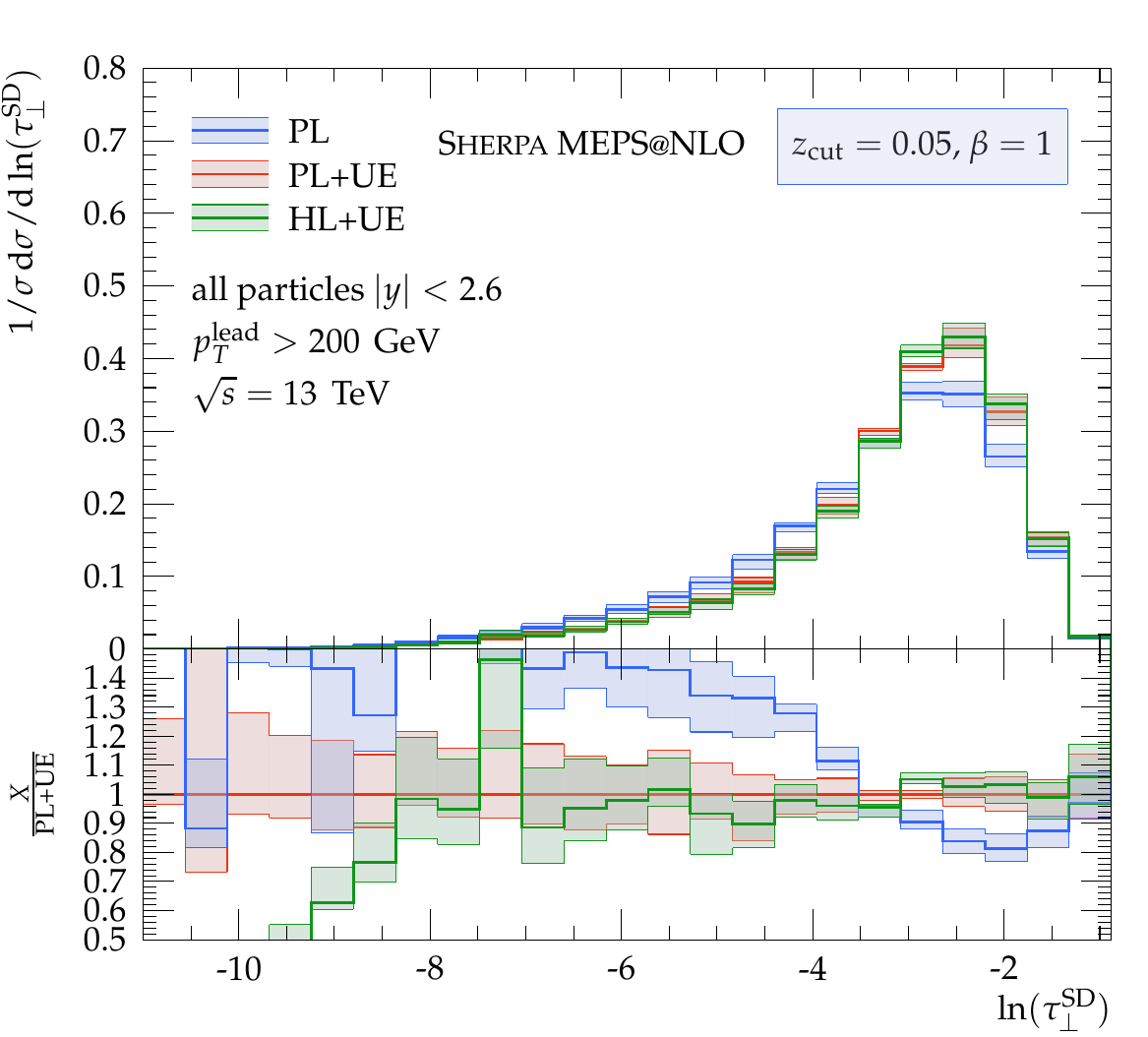}~
	  \includegraphics[width=0.32\textwidth]{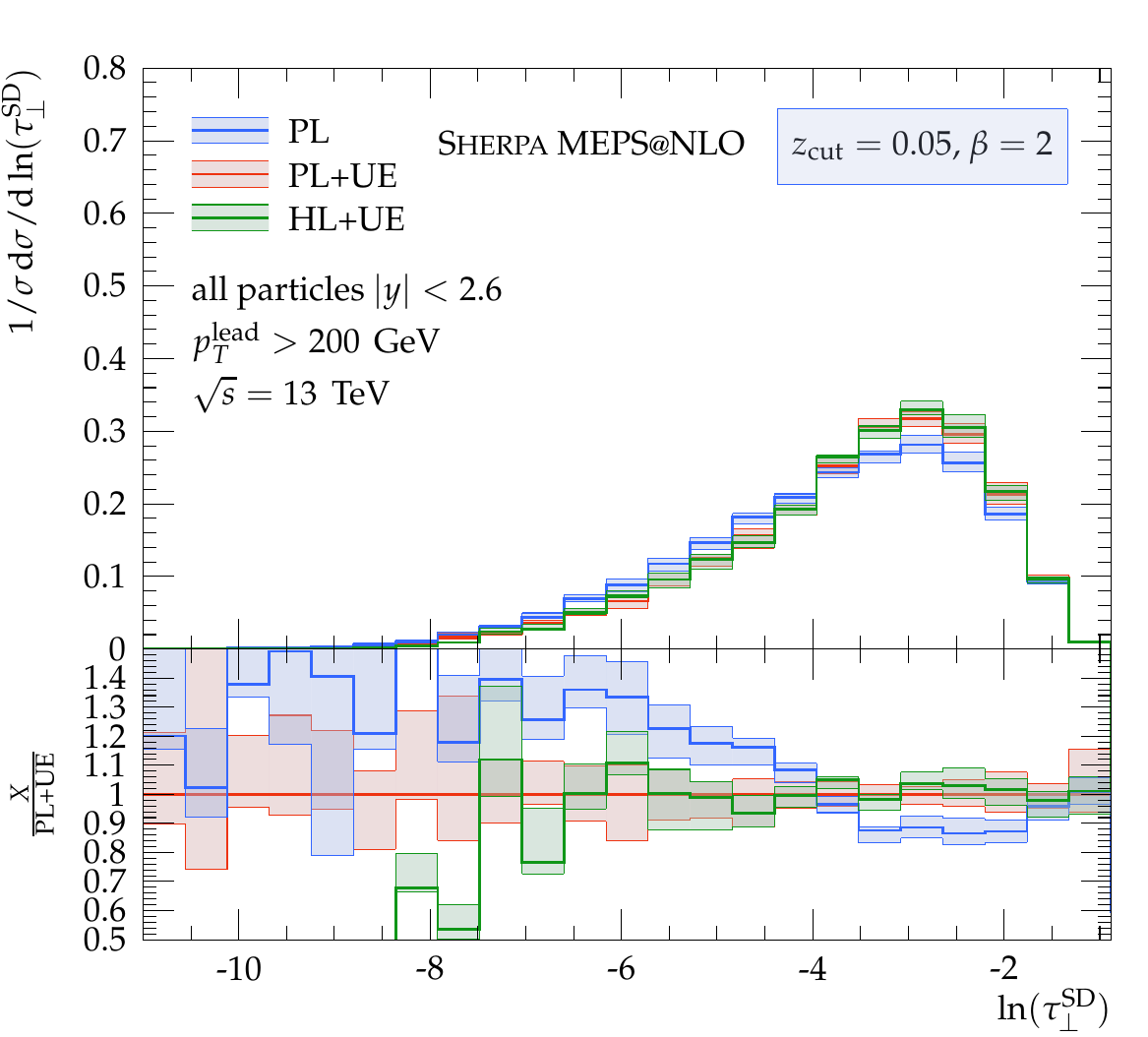}\\
	  \includegraphics[width=0.32\textwidth]{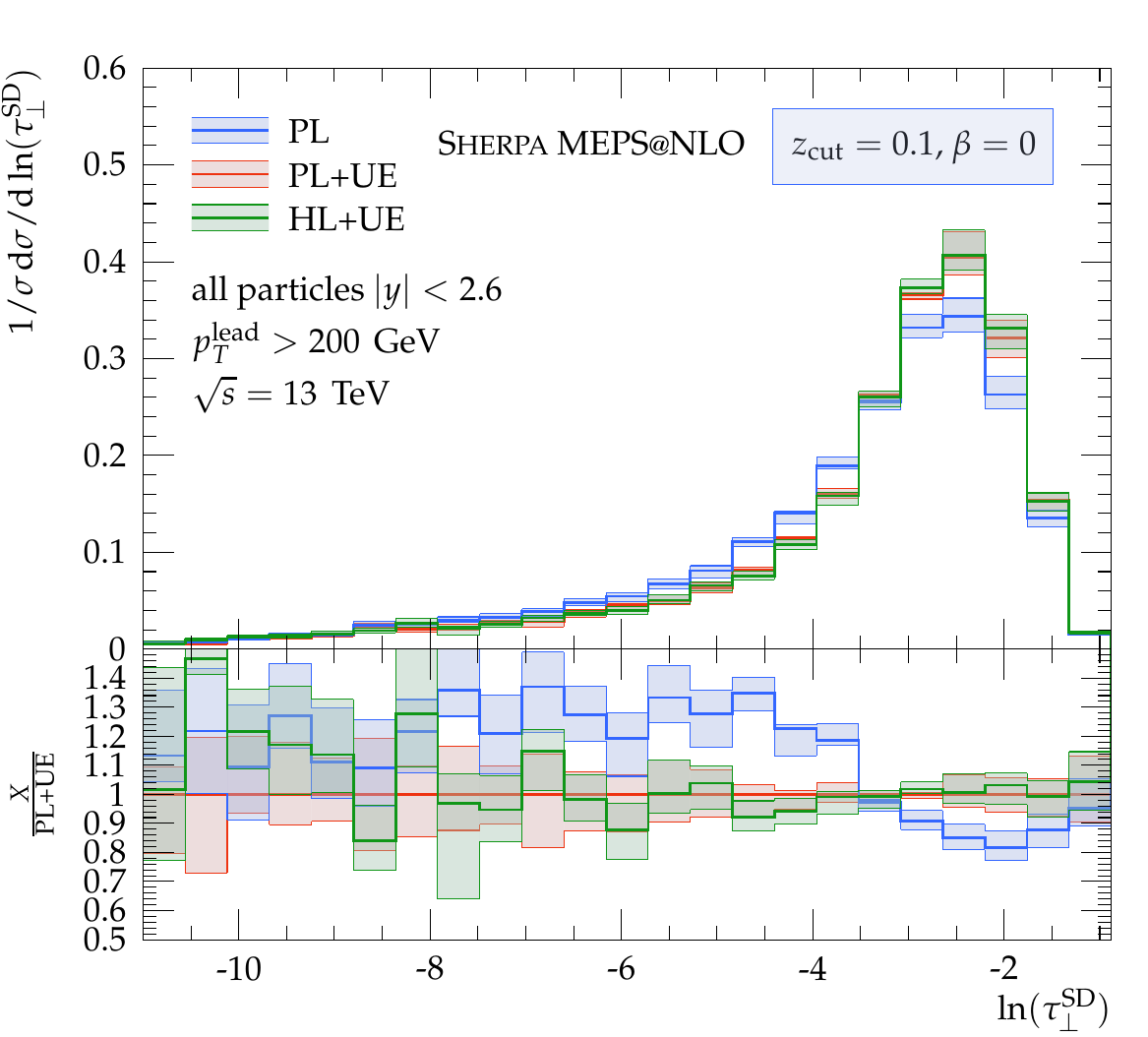}~
	  \includegraphics[width=0.32\textwidth]{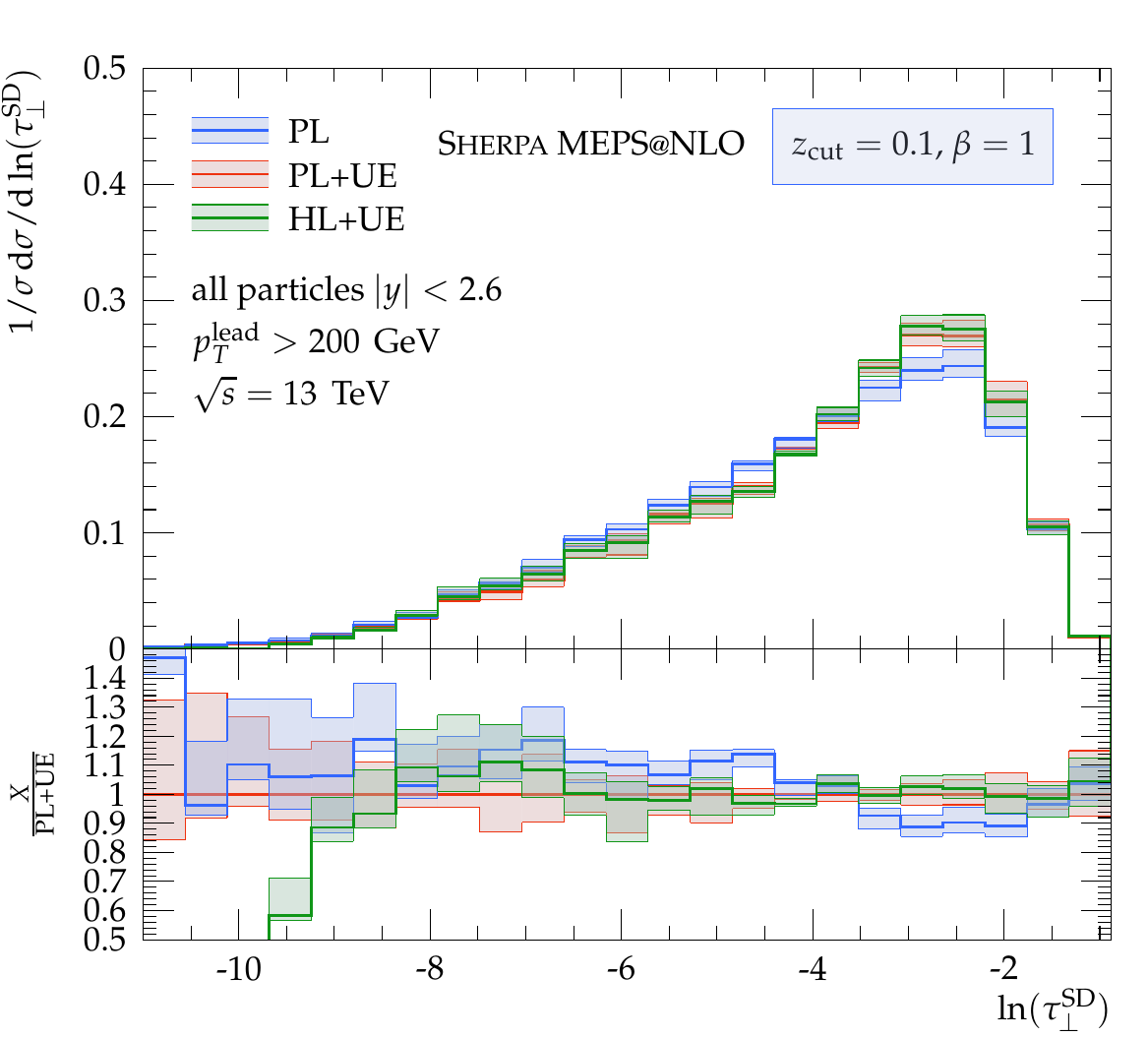}~
	  \includegraphics[width=0.32\textwidth]{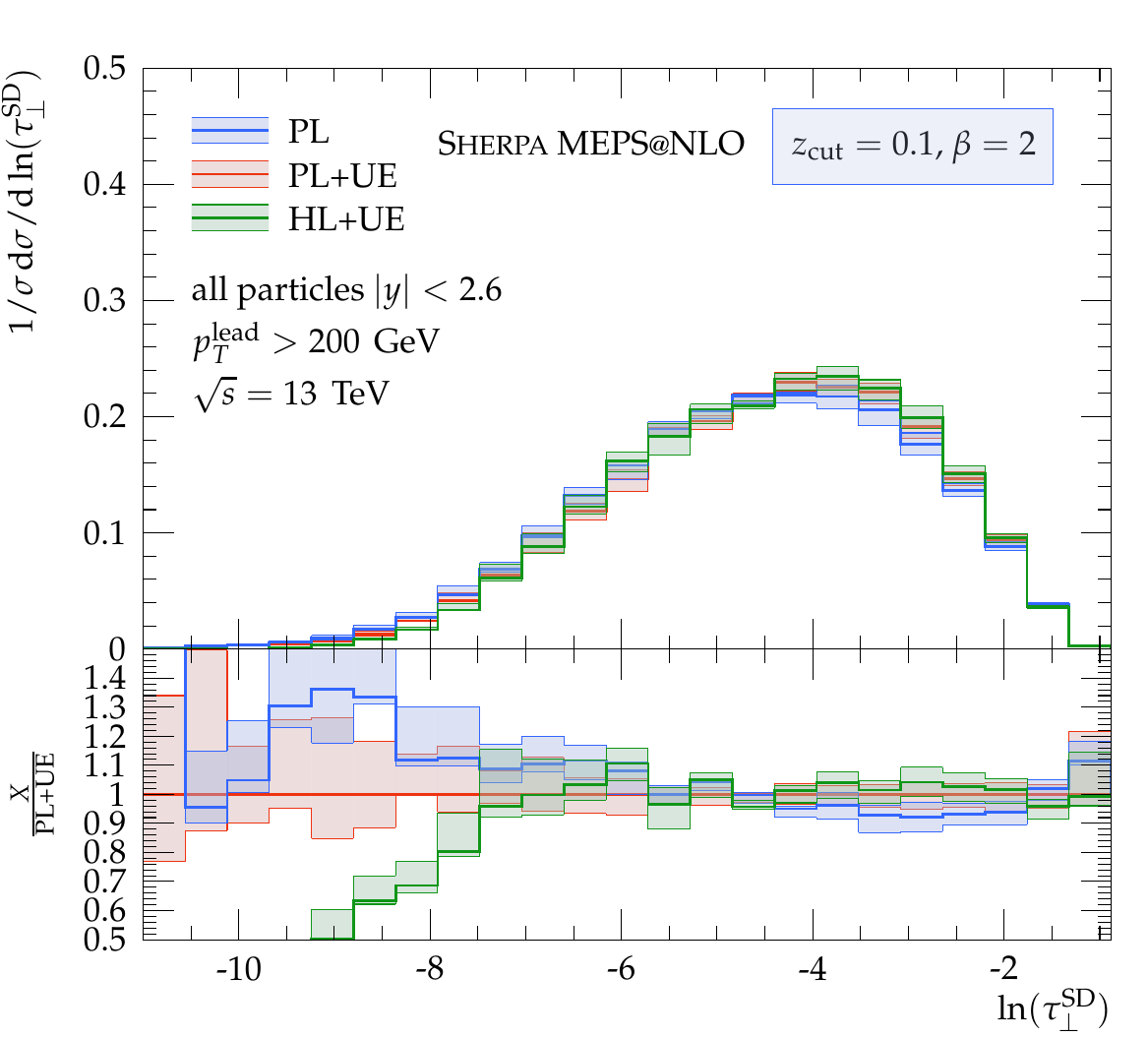}\\
	  \includegraphics[width=0.32\textwidth]{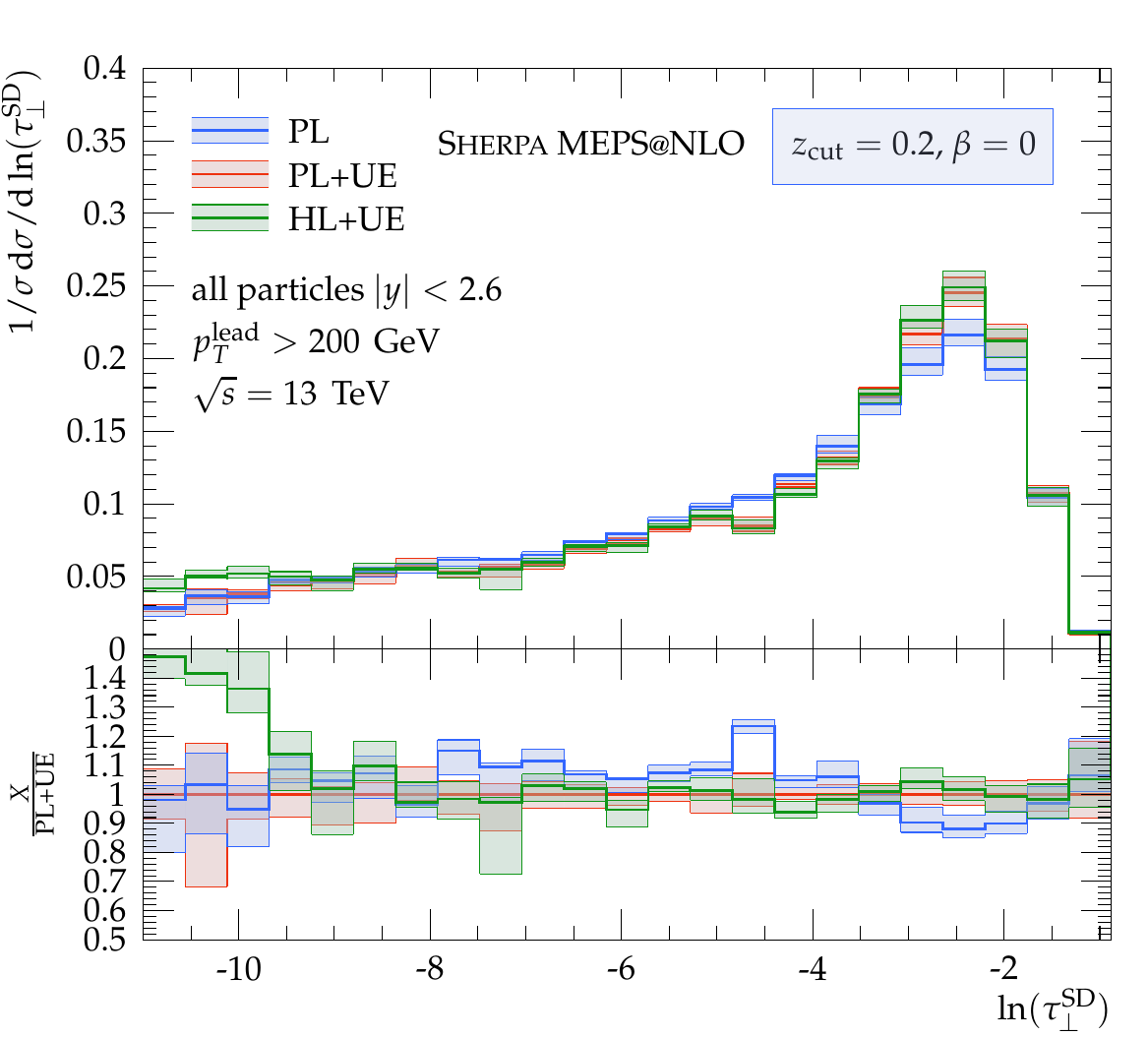}~
	  \includegraphics[width=0.32\textwidth]{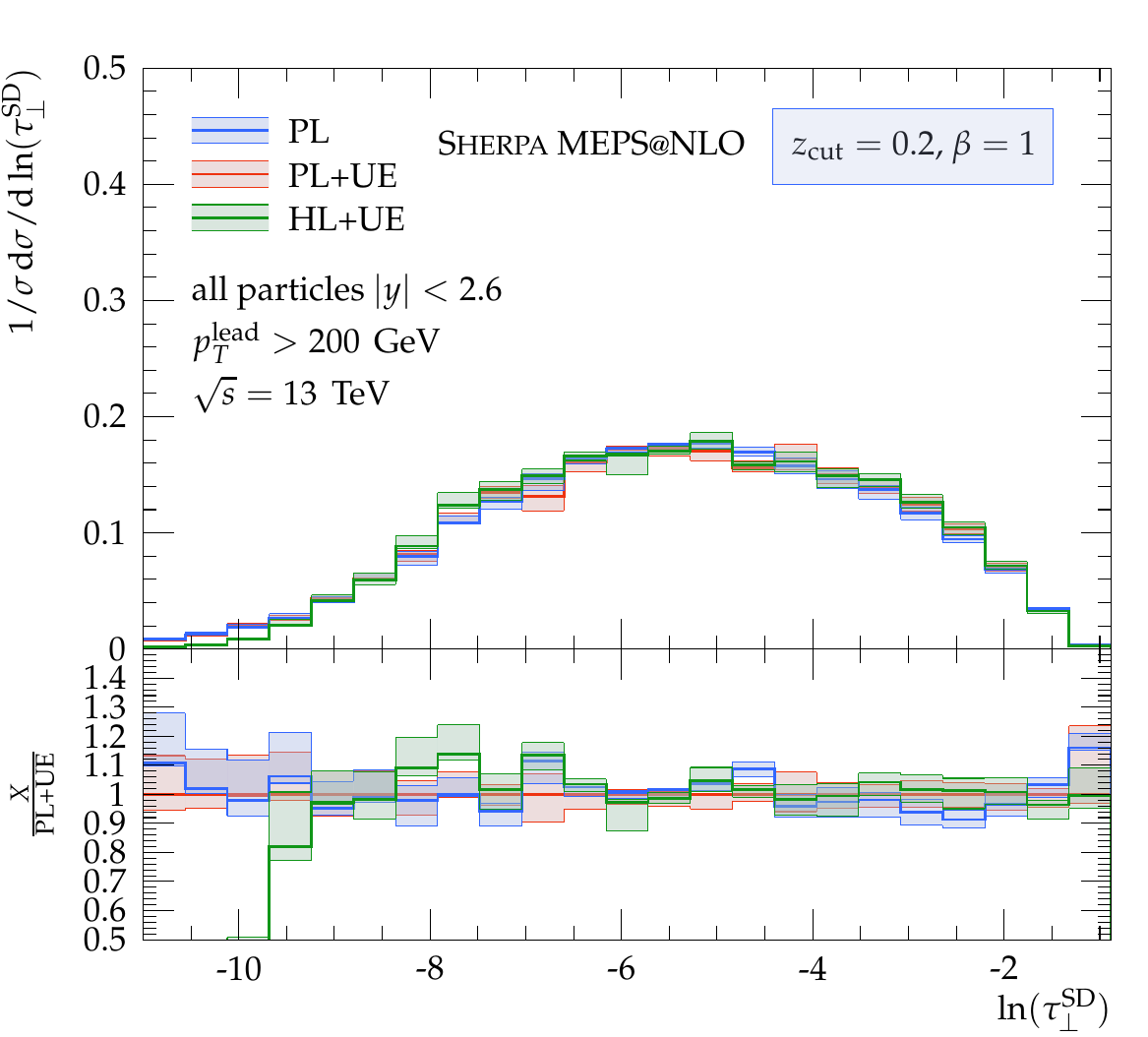}~
	  \includegraphics[width=0.32\textwidth]{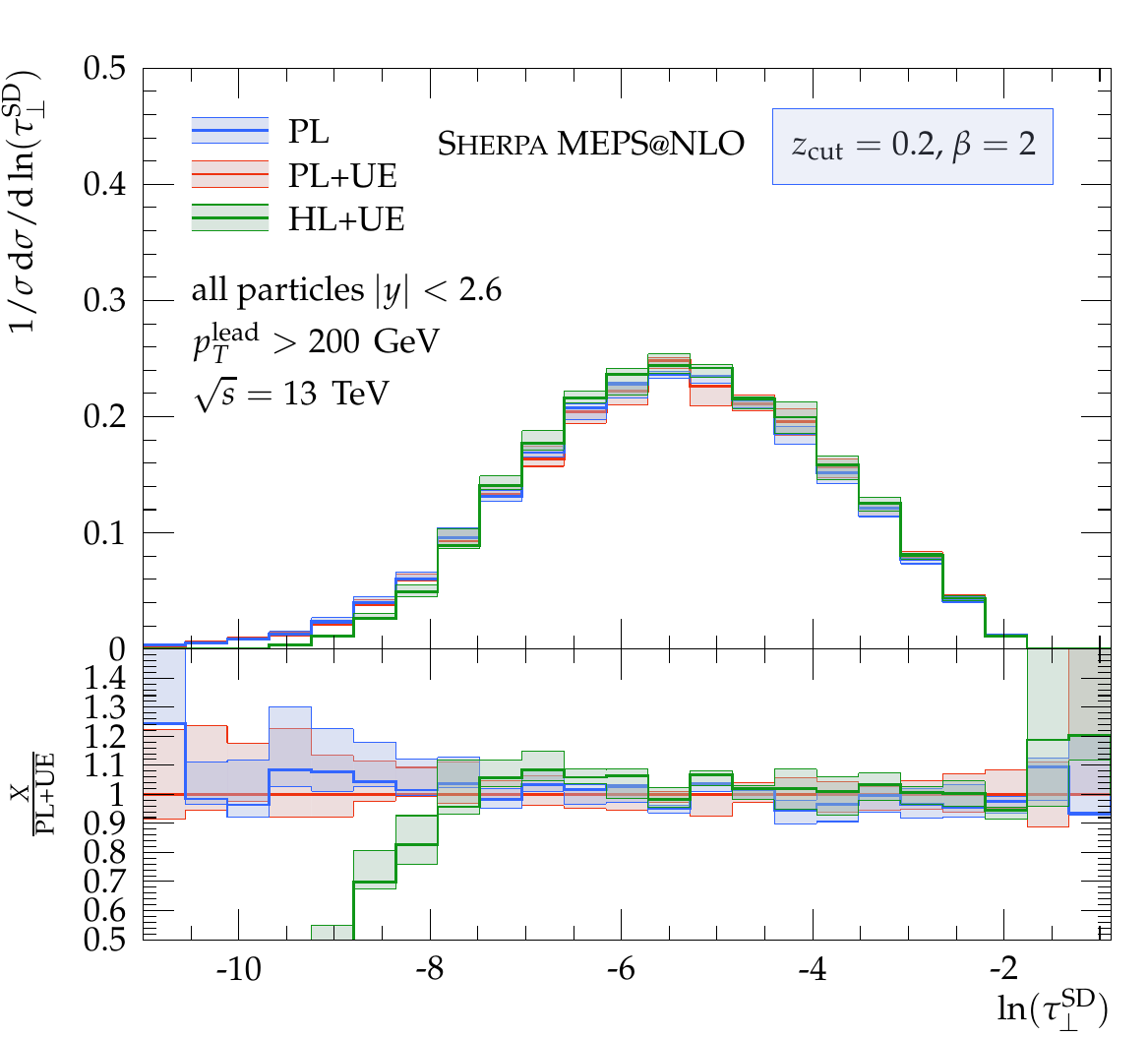}\\
	  \includegraphics[width=0.32\textwidth]{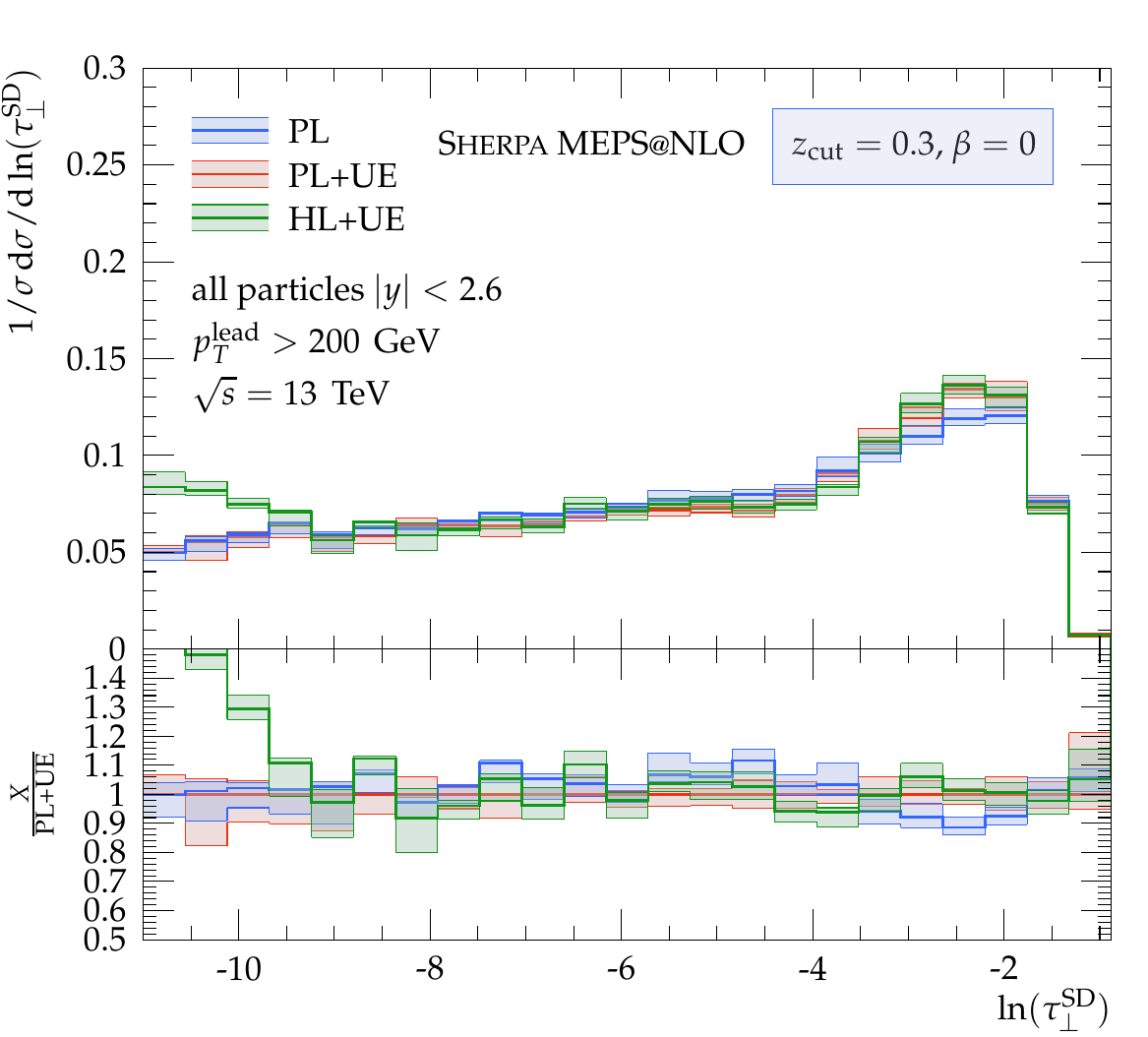}~
	  \includegraphics[width=0.32\textwidth]{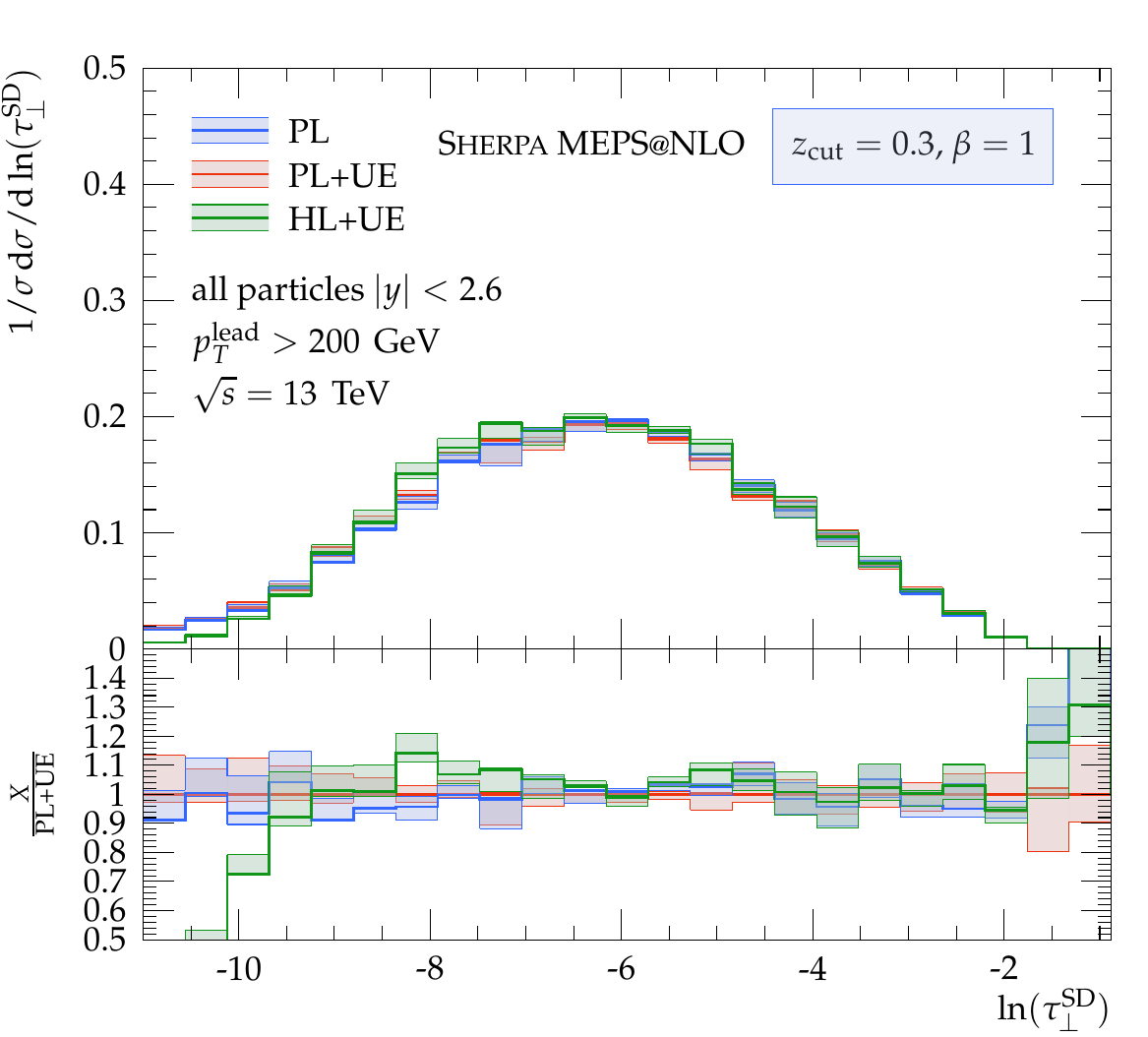}~
	  \includegraphics[width=0.32\textwidth]{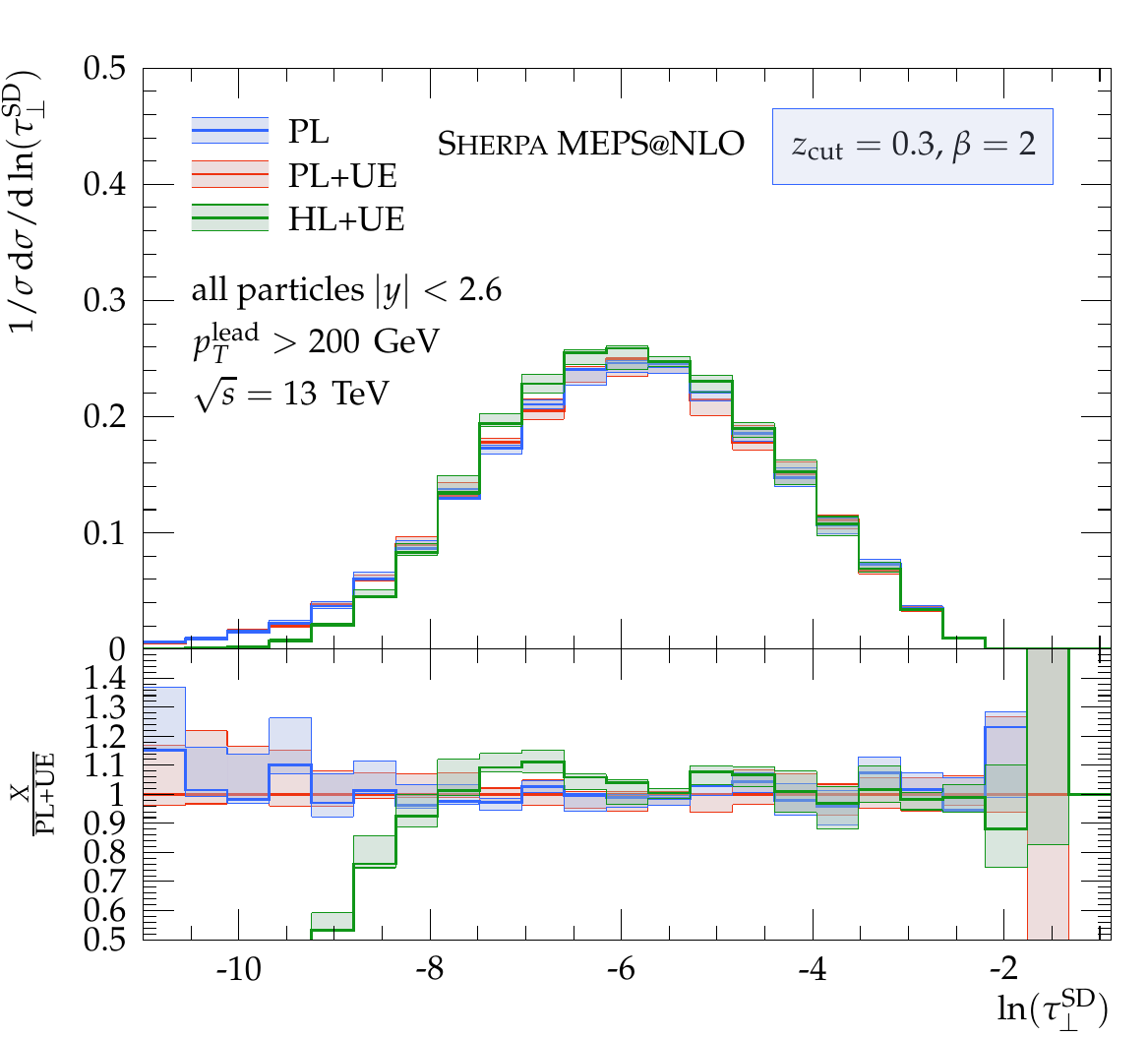}\\
	\end{center}
	\caption{The groomed thrust distributions for $\beta\in\{0,1,2\}$ (columns) and $\zcut\in\{0.05,0.1,0.2,0.3\}$ (rows)
          for the $p_{T,\text{min}}=200\;\text{GeV}$ event selection. The lower panels show the ratios with respect to the
          parton-level simulation including the underlying event (PL+UE) prediction. }
	\label{fig:NP}
\end{figure}

In Fig.~\ref{fig:NP} we present predictions for groomed transverse thrust
for the $p_{T,\text{min}}=200\;\text{GeV}$ event selection obtained with \Sherpa
taking into account \emph{all} particles with $|y|\leq y_{\text{max}}$. Results are
given after showering the hard process (PL), after including the underlying
event (PL+UE), and at full hadron level (HL+UE). For the grooming we again use
$\beta \in \{0,1,2\}$, however in order to better illustrate the 
approach of the ungroomed case, we additionally include a smaller $\zcut$
value and here consider $\zcut\in\{0.05,0.1,0.2,0.3\}$.

For mild grooming with $\zcut=0.05$ and $\beta=0$ underlying-event corrections
are indeed of similar size as for the ungroomed case shown in Fig.~\ref{fig:NP_ungroomed}.  
As we groom harder by enlarging $\zcut$ we observe a more significant reduction
in the impact of the underlying event. This is most evident for the case of
$\beta=0$ (first column in Fig.~\ref{fig:NP}). For $\zcut=0.1$ the peak is still
shifted when including multiple parton interactions, although much less dramatic
than for the ungroomed case. Finally, for $\zcut=0.3$ the impact stays below
$10\%$. When increasing $\beta$ the effect of the underlying event is further
reduced, to the level of at most $20\%$ even for $\zcut=0.1$ for $\ln(\tauSD)>-8$.
For $\zcut\geq 0.2$ the underlying-event correction stays well below $10\%$,
with the exception of a few seemingly statistical exceptions.

When comparing to the ungroomed case in Fig.~\ref{fig:NP_ungroomed}, hadronisation
effects, similar to the $e^+e^-$ case~\cite{Baron:2018nfz}, are pushed to lower
observable values through grooming. This affects predominantly the low-$\tauSD$
region, where very collimated parton-level jets get spread out. Notably, in the
region $\ln(\tauSD)<-8$ the hadronisation corrections change sign when going
from $\beta=0$ to $\beta=1,2$. For the latter choices grooming is suppressed
for particles with $\Delta R < R_{\text{SD}}$ from the hard jets, accordingly,
as in the ungroomed case, hadronisation shifts events towards larger $\tauSD$.
However, even for $\beta=2$ there is a sizeable observable range for which
also hadronisation effects are small. Qualitatively, the observed dependence on
$\beta$ and $\zcut$ for the transition to the hadronisation dominated regime
is well reflected by the estimate derived in Eq.~\eqref{eq:vNP_groomed} for
$a_l=b_l=1$, also illustrated in Fig.~\ref{fig:vNP}. Note, in the Monte-Carlo
simulations the parton-to-hadron transition sets in at the parton-shower cut-off
scale ${\cal{O}}(1\,\text{GeV})$ already, rather than the Landau pole $\Lambda_{\text{QCD}}$.
Furthermore, we are sensitive to the hadronisation of partons originating from
the underlying event. 

\begin{figure}[ht!]
	\begin{center}
	  \includegraphics[width=0.32\textwidth]{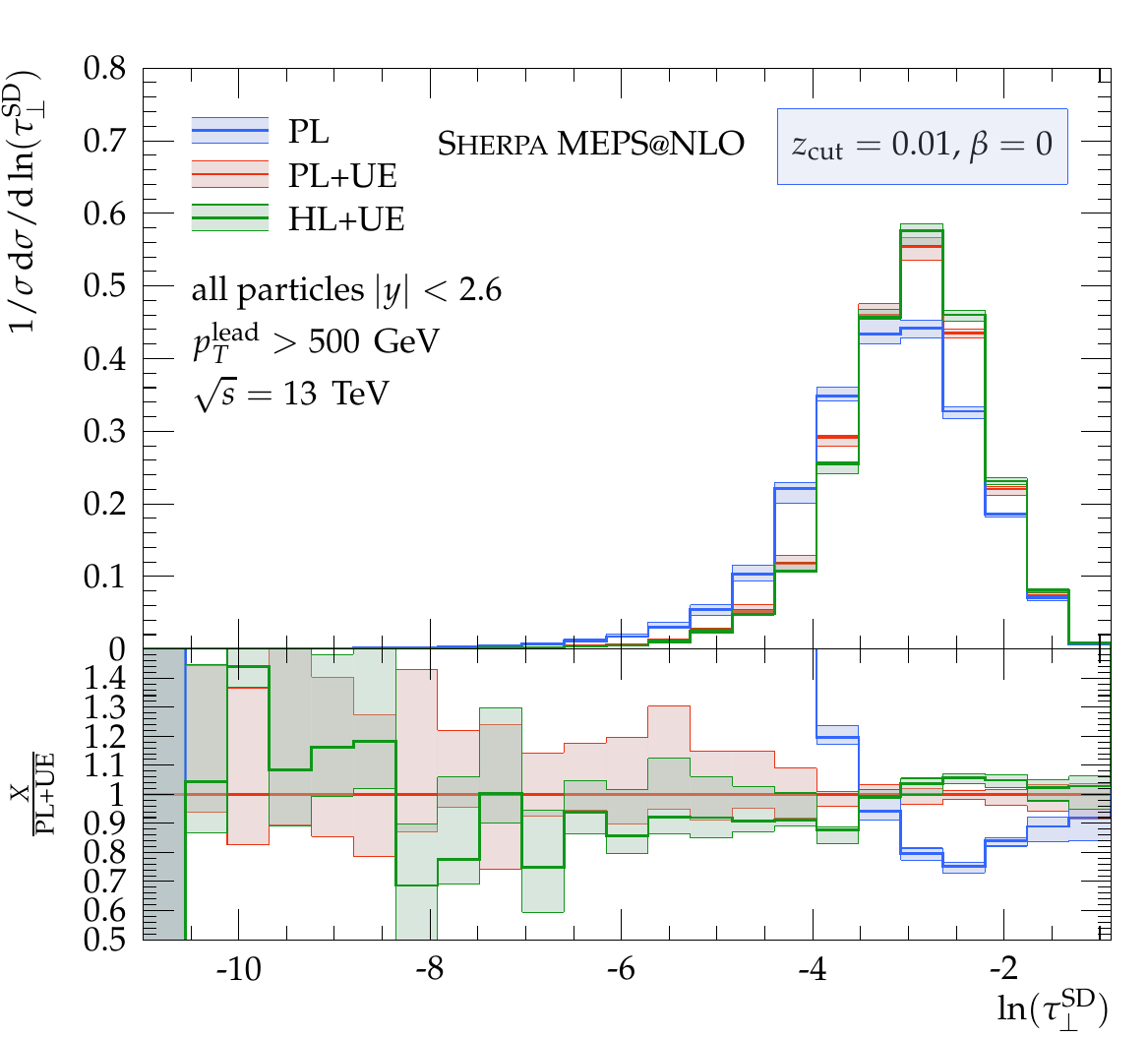}~
	  \includegraphics[width=0.32\textwidth]{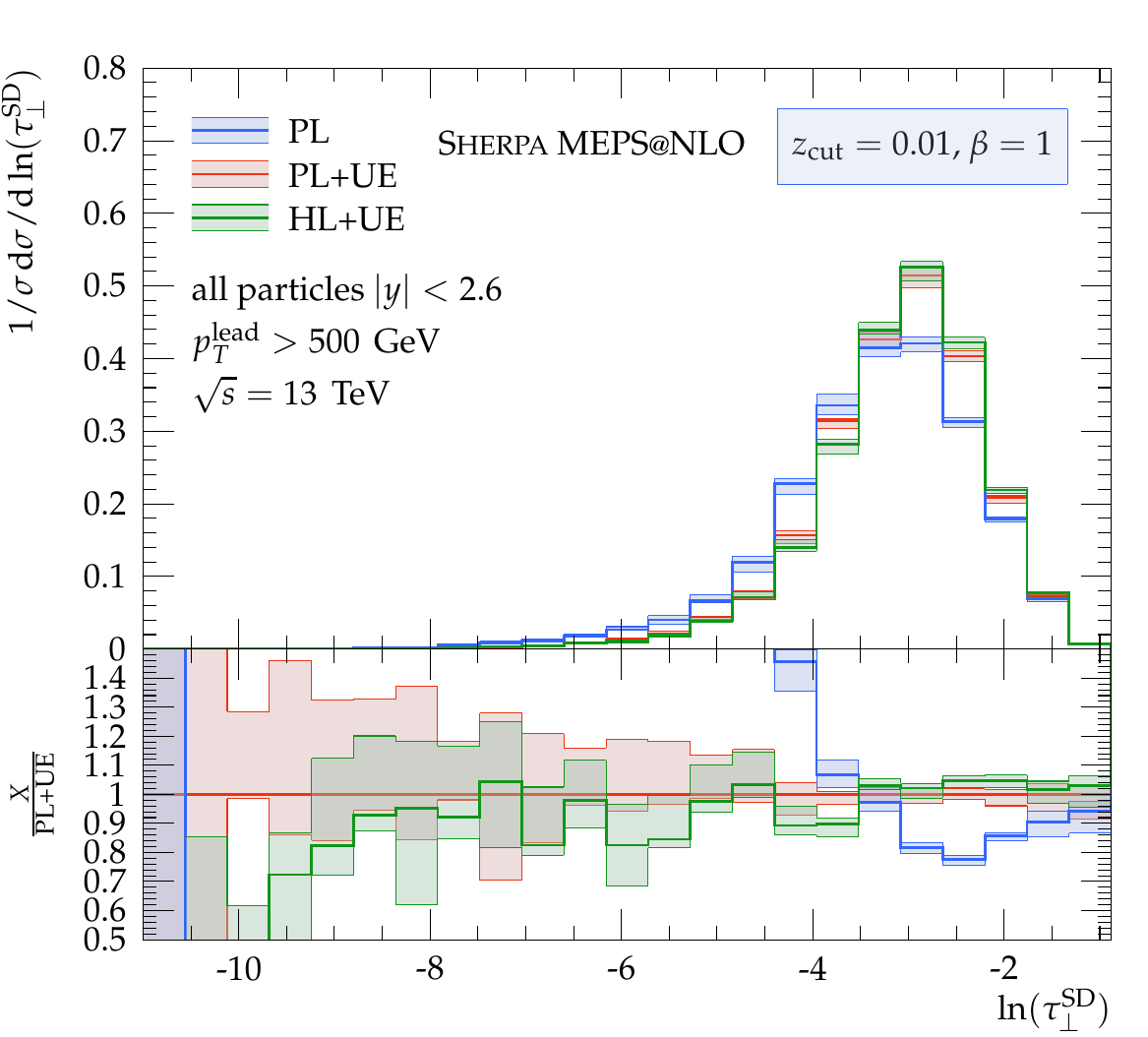}~
	  \includegraphics[width=0.32\textwidth]{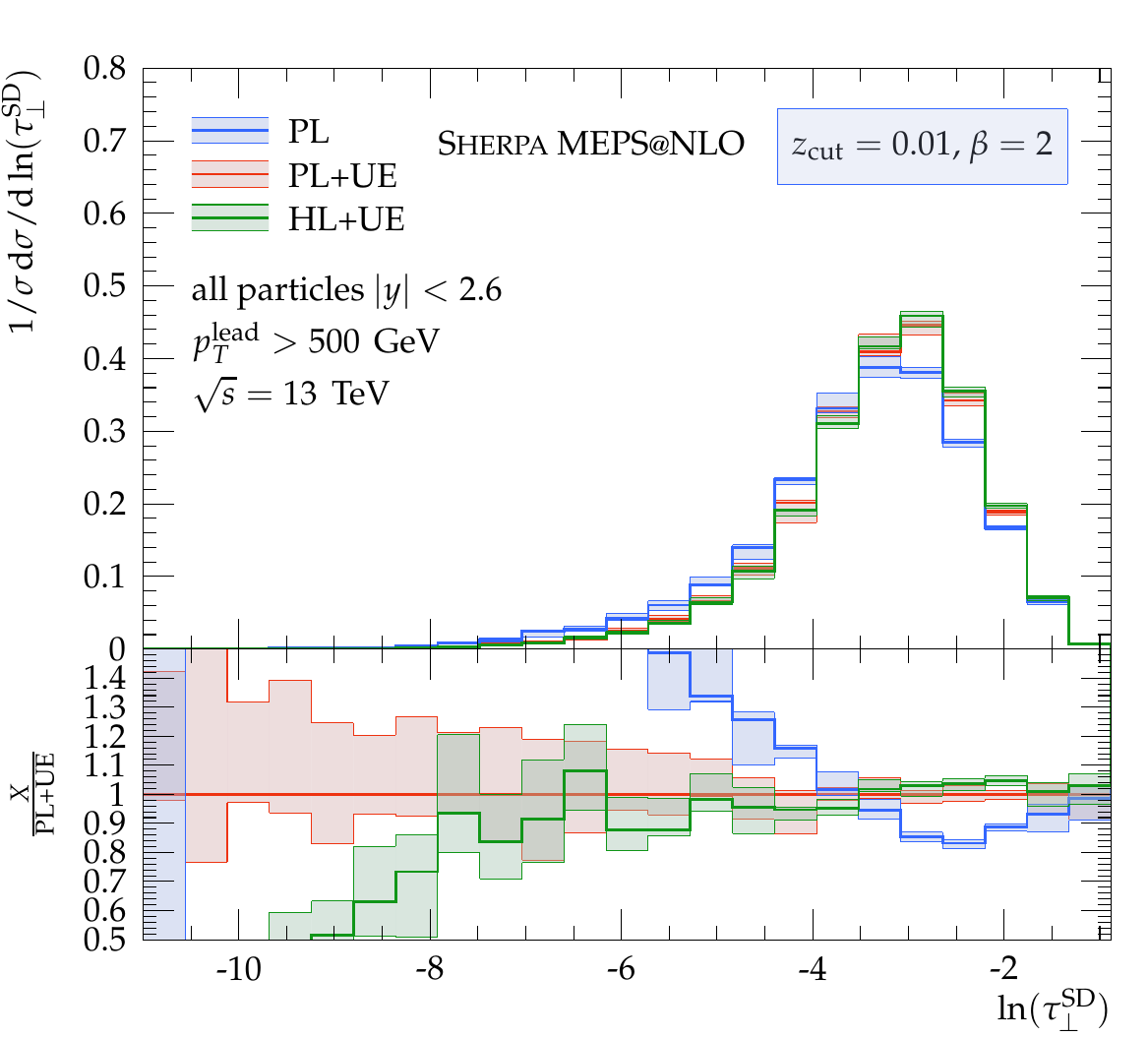}\\
	  \includegraphics[width=0.32\textwidth]{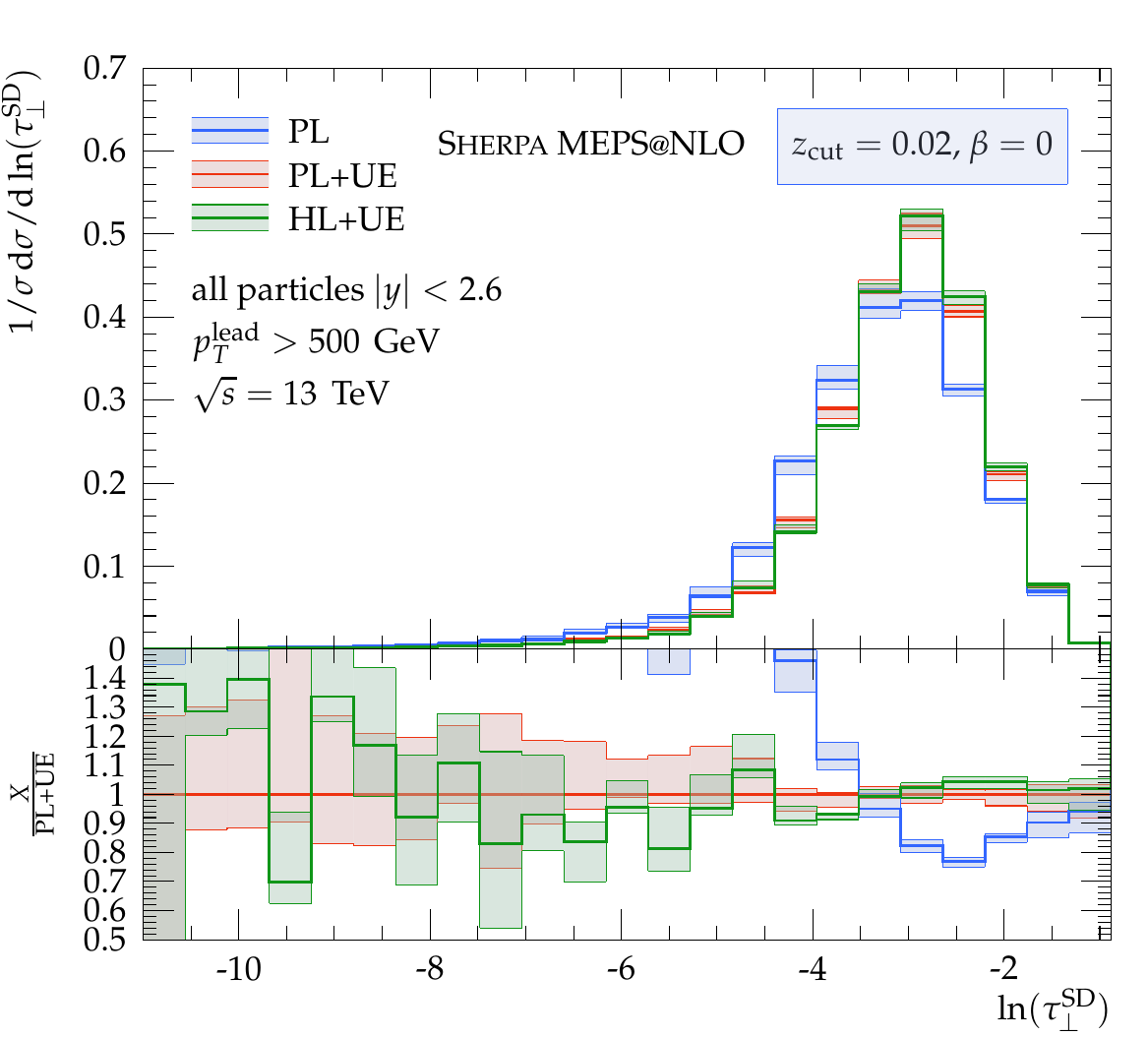}~
	  \includegraphics[width=0.32\textwidth]{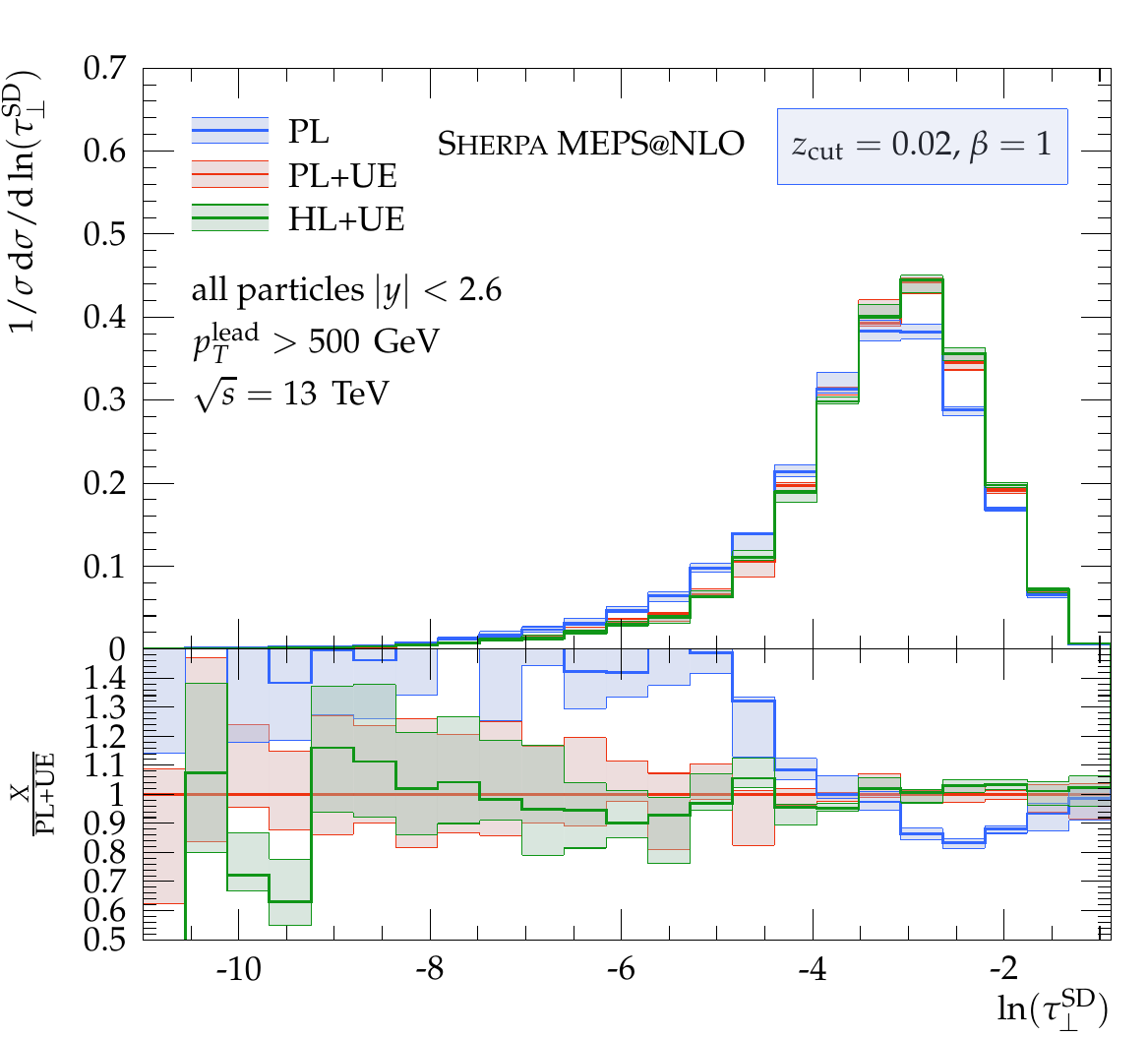}~
	  \includegraphics[width=0.32\textwidth]{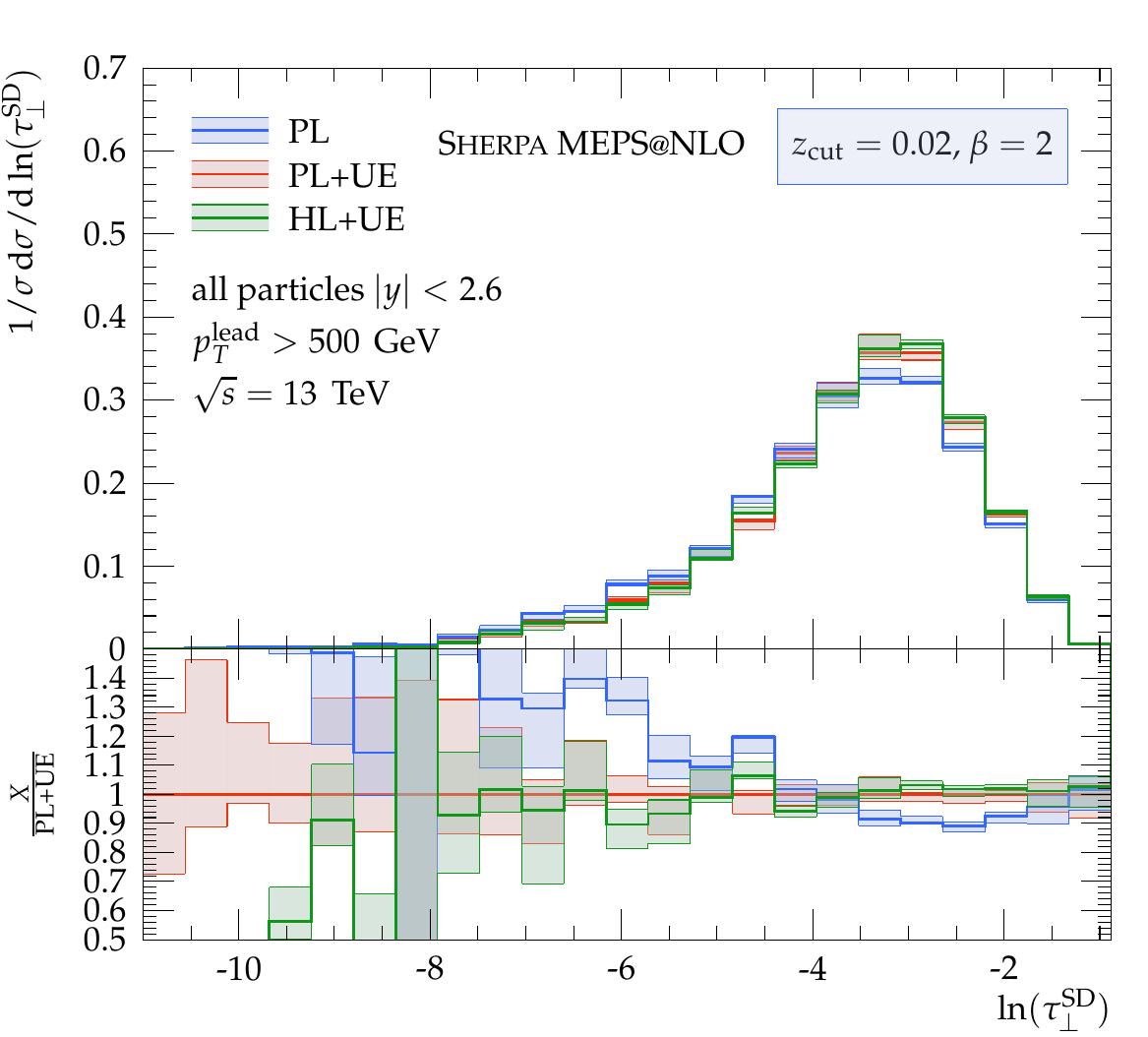}\\
	  \includegraphics[width=0.32\textwidth]{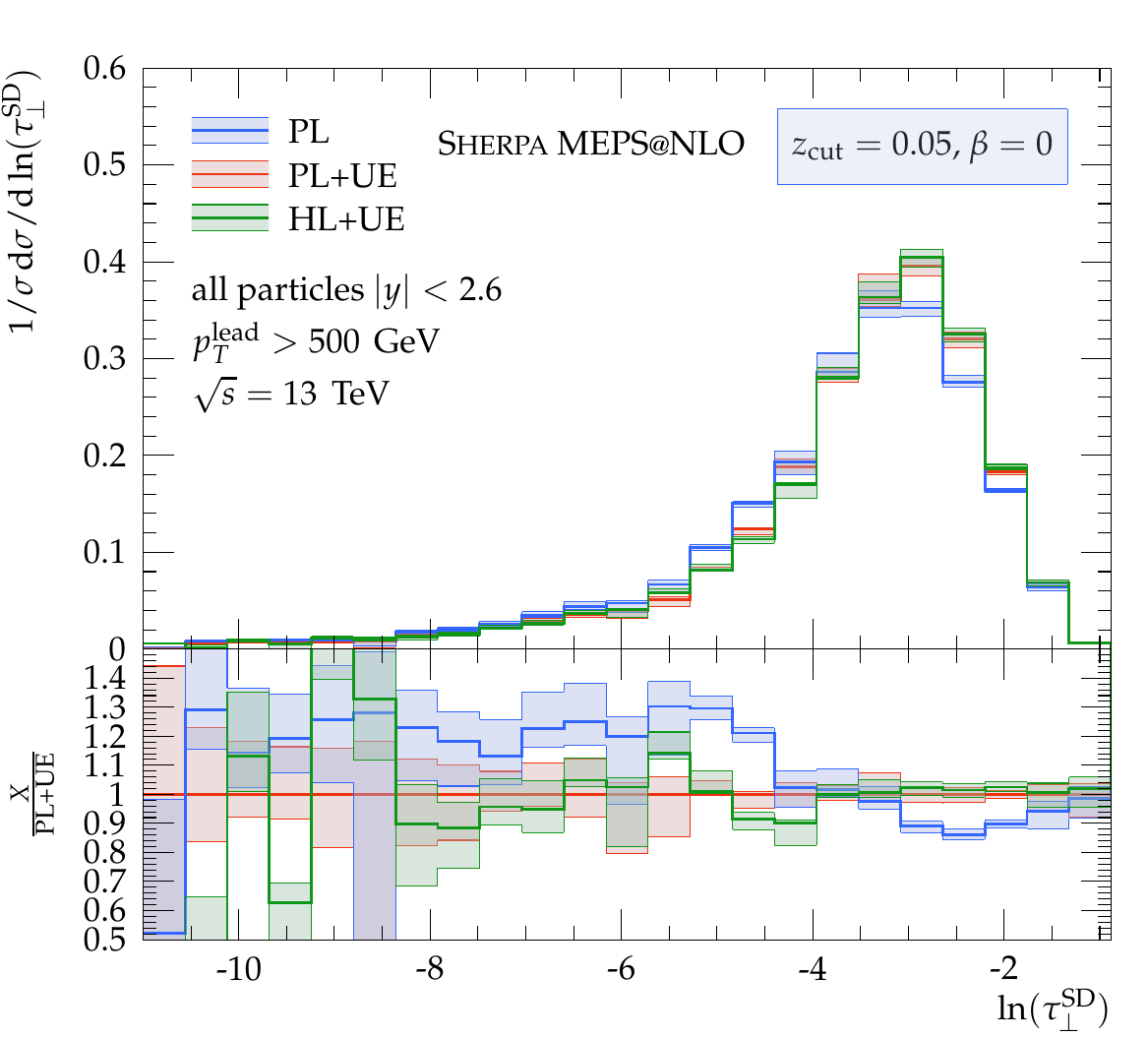}~
	  \includegraphics[width=0.32\textwidth]{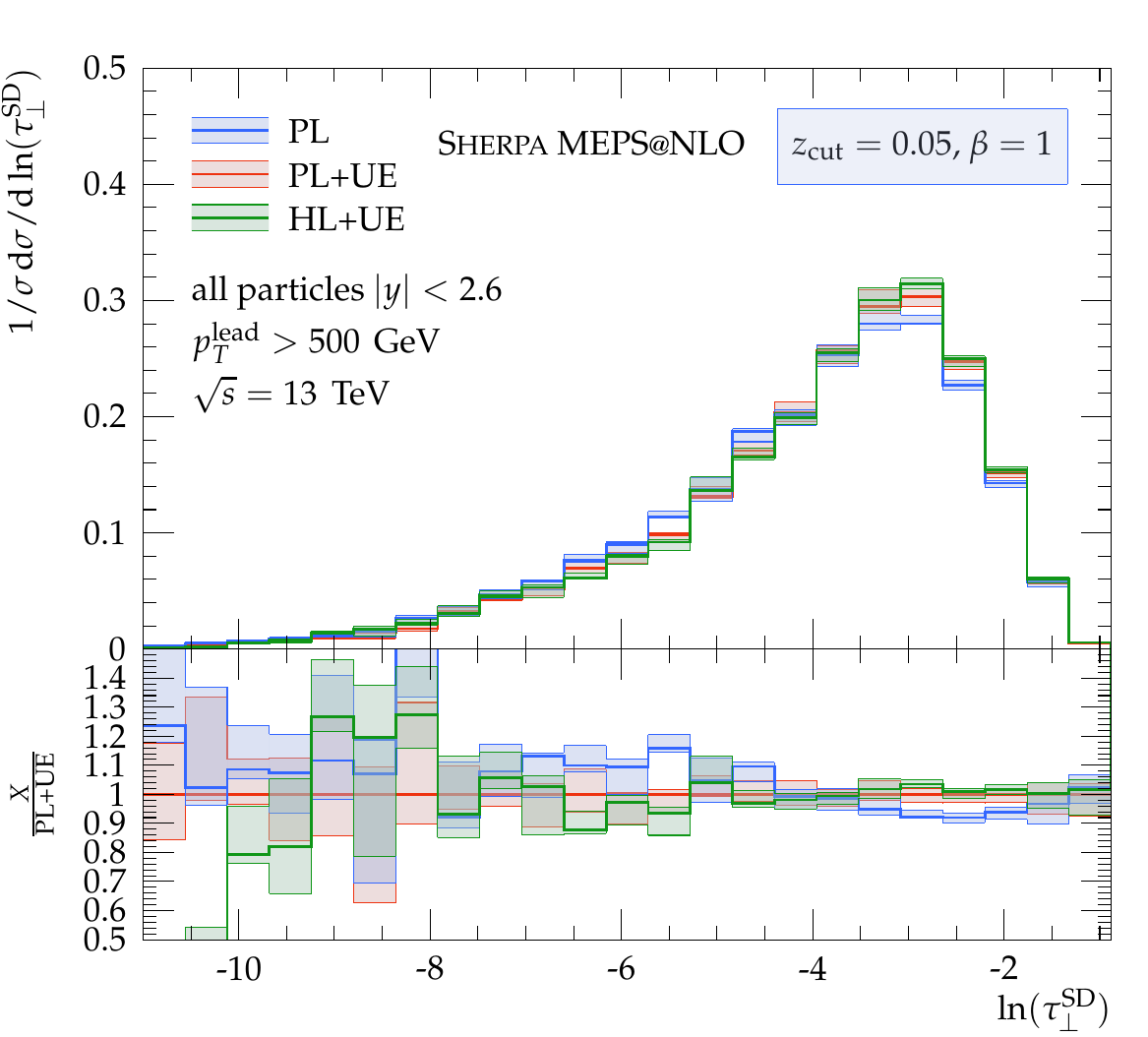}~
	  \includegraphics[width=0.32\textwidth]{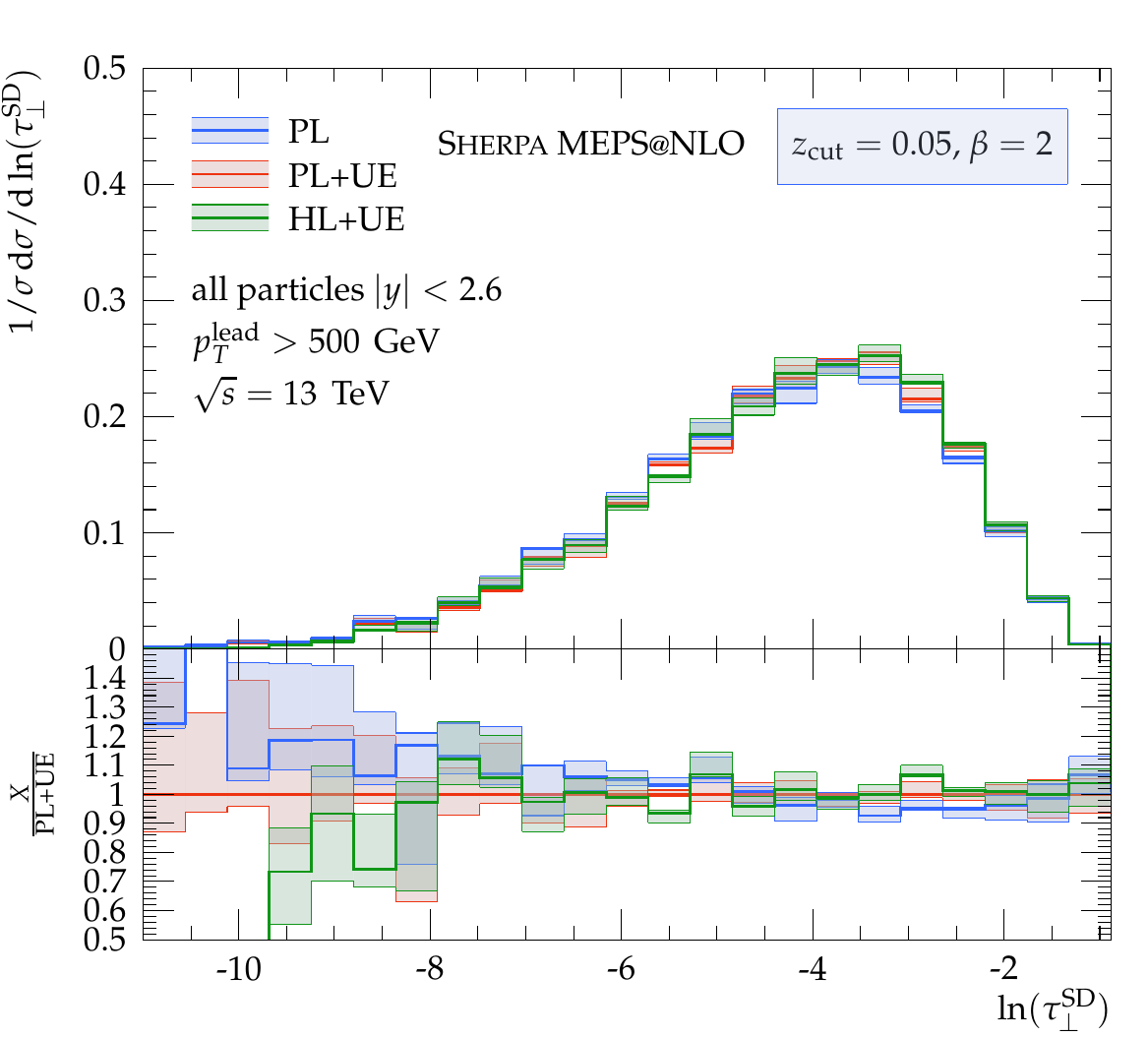}\\
	  \includegraphics[width=0.32\textwidth]{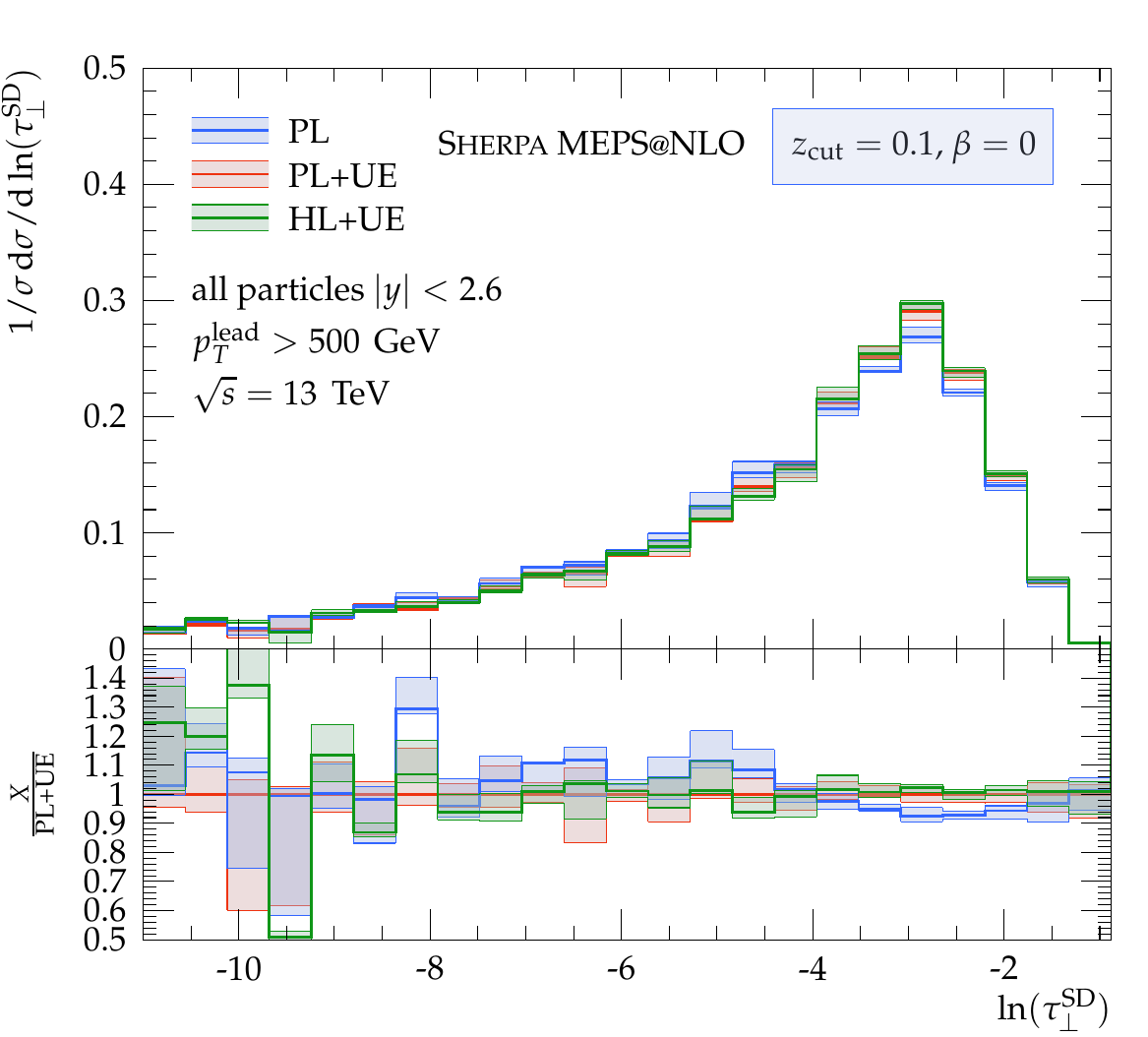}~
	  \includegraphics[width=0.32\textwidth]{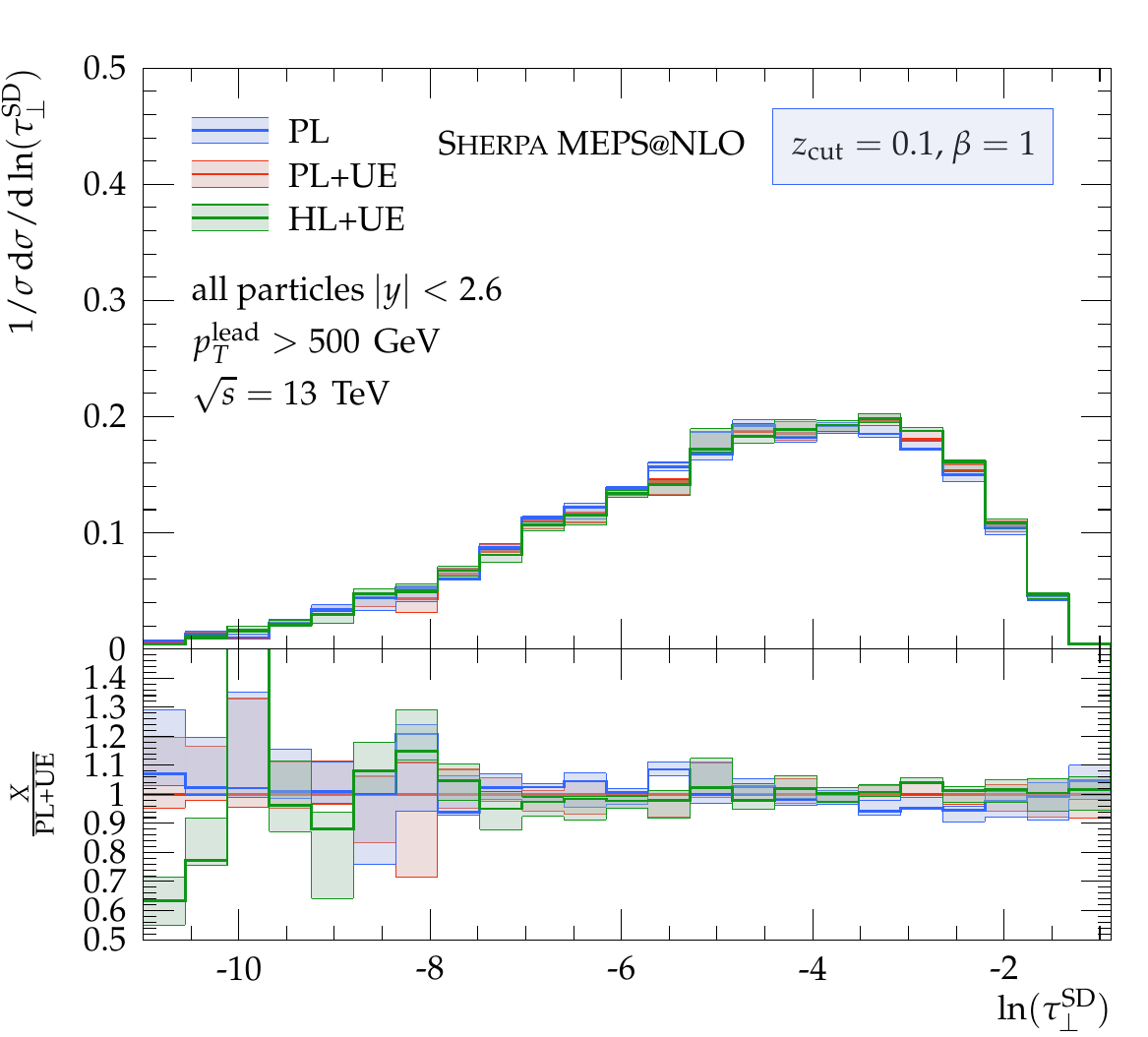}~
	  \includegraphics[width=0.32\textwidth]{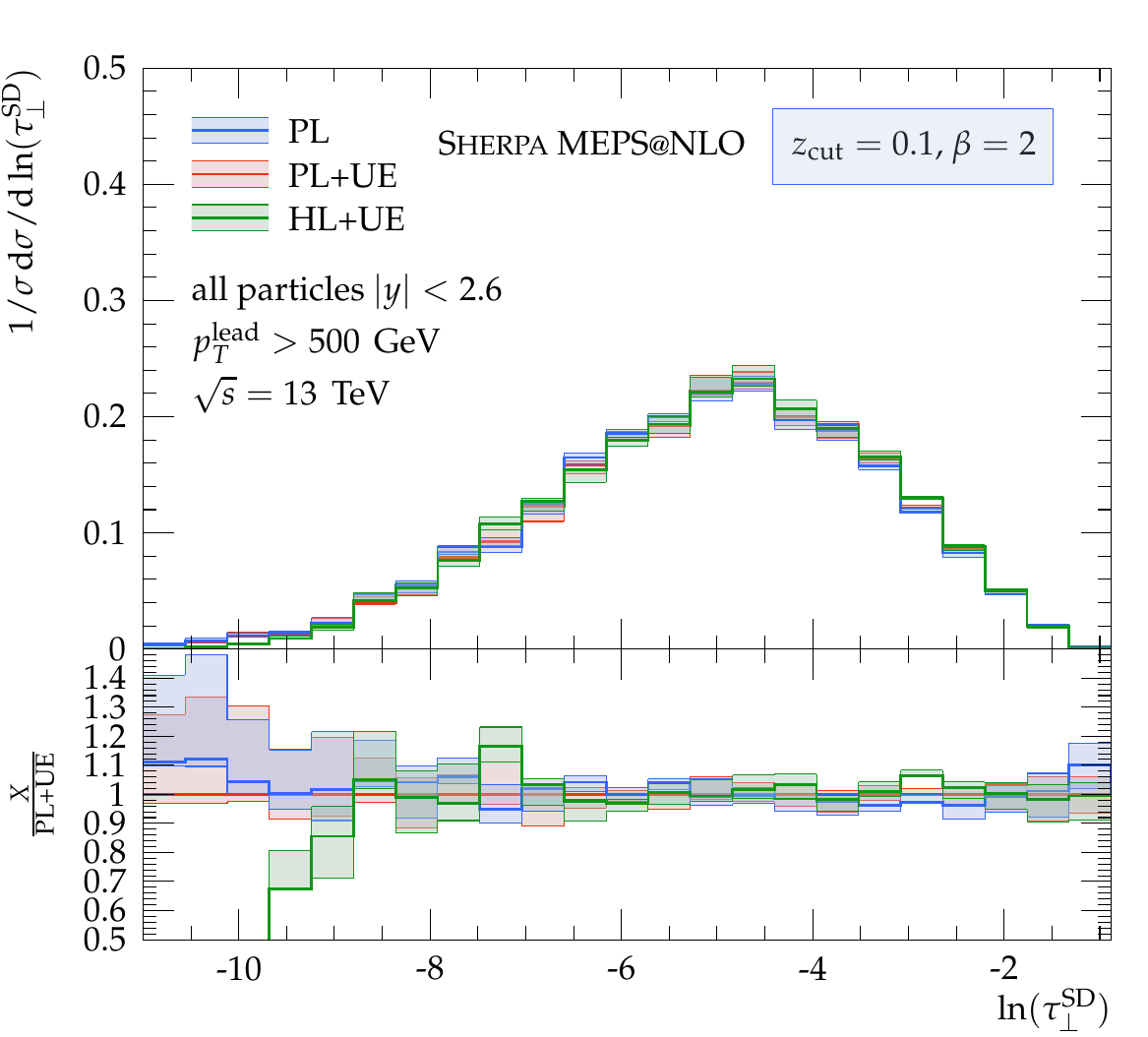}\\
	\end{center}
	\caption{The groomed thrust distributions for $\beta\in\{0,1,2\}$ (columns) and $\zcut\in\{0.01, 0.02, 0.05, 0.1\}$
          (rows) for the $p_{T,\text{min}}=500\;\text{GeV}$ event selection. The lower panels show the ratios with respect
          to the parton-level simulation including the underlying event (PL+UE) prediction.}
	\label{fig:NP_500}
\end{figure}

When increasing the scale of the hard dijet-production process by raising
$p_{T,\text{min}}$ to $500\;\text{GeV}$, the scale separation of the hard
process and the underlying event increases. Accordingly, lower values of $\zcut$ 
are sufficient to achieve a sizeable reduction of the underlying-event
corrections. If the underlying event would be associated with the exact same
scales irrespective of the hard-jet $p_T$ requirement, we would expect $\zcut$
values scaled by a factor $200/500 = 0.4$ to yield results in the
$p_{T,\text{min}}=500\;\text{GeV}$ case that are comparable to
$p_{T,\text{min}}=200\;\text{GeV}$, \emph{cf.} Eq.~\eqref{eq:SD}. We therefore
consider $\zcut \in  \{0.01, 0.02, 0.05, 0.1\}$ here and present
corresponding results in Fig.~\ref{fig:NP_500}. For $\zcut=0.01$, independent of $\beta$,
grooming does not yet significantly reduce the underlying-event
contribution. The results for $\zcut=0.02$ are very similar to the findings for
$\zcut=0.05$ for $p_{T,\text{min}}=200\;\text{GeV}$ and the reduction observed for
the $500\;\text{GeV}$ selection with $\zcut=0.05$ are quite close to those for
$\zcut=0.1$ with the lower $p_{T,\text{min}}$. When finally
increasing $\zcut$ to $0.1$ we observe agreement between the parton-level
predictions with and without the underlying event to better than
$10\%$, again with the exception of a few fluctuations due to the limited
statistics we could afford for these very demanding simulations. Notably,
overall the impact of hadronisation corrections is reduced due to the
increased $p_{T,\text{min}}$, and correspondingly $\muR$, consistent again with the
findings in Sec.~\ref{sec:np_scales}. For $\ln(\tauSD)>-8$ they are in fact rather
mild for all values of $\zcut$ and $\beta$ considered here. 

\subsubsection*{Track-based observable evaluations}

Our current observable definition, based on \emph{all} final-state particles within the rapidity range
$|y|\leq y_{\text{max}}$, requires the use of sophisticated experimental techniques such as
\emph{particle flow}~\cite{Sirunyan:2017ulk}. As a simpler
alternative we want to consider the charged final-state particles only, accessible by conventional
tracking detectors, for tracks with sufficient transverse momentum in order to be reliably detectable.
While these measurement constraints can be imposed on a fully exclusive Monte Carlo simulation, they obscure
the comparability with purely perturbative calculations and in addition invoke systematic uncertainties related to the modelling
of low-energetic particles in the generators' hadronisation models. Here we want to study the effect of these
constraints on our generator predictions and in particular explore to what extent grooming ameliorates the
correspondence with perturbative predictions. To gain statistical power for quantifying the impact
of the track-level final-state selections, we produced a high-statistics simplified \Sherpa sample
setup based on the pure parton shower, \emph{i.e.}\ without the inclusion of higher-order matrix
elements, denoted \Sherpa PS in the following.

\begin{figure}[ht!]
\begin{center}
	\includegraphics[width=0.32\textwidth]{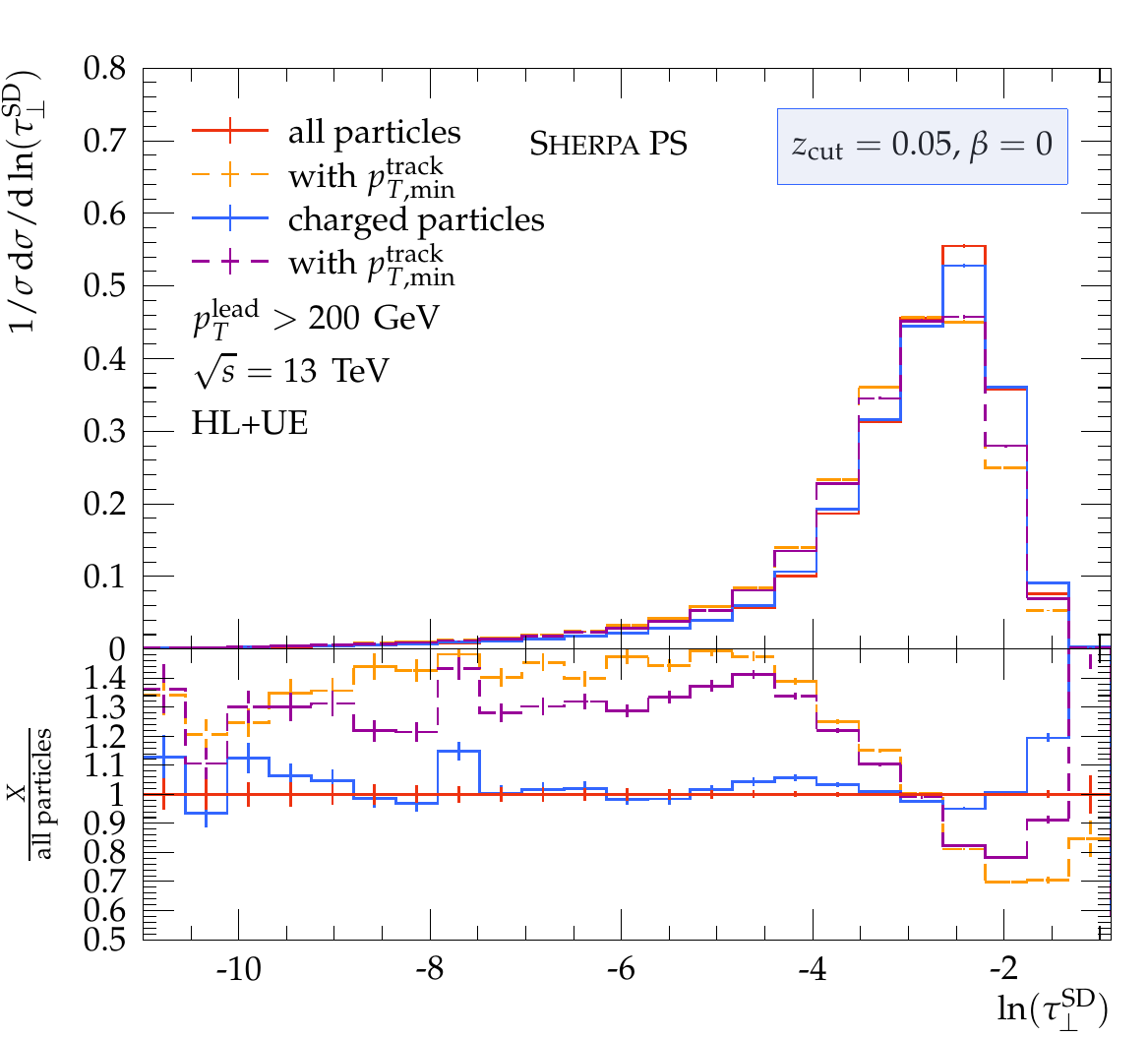}~
	\includegraphics[width=0.32\textwidth]{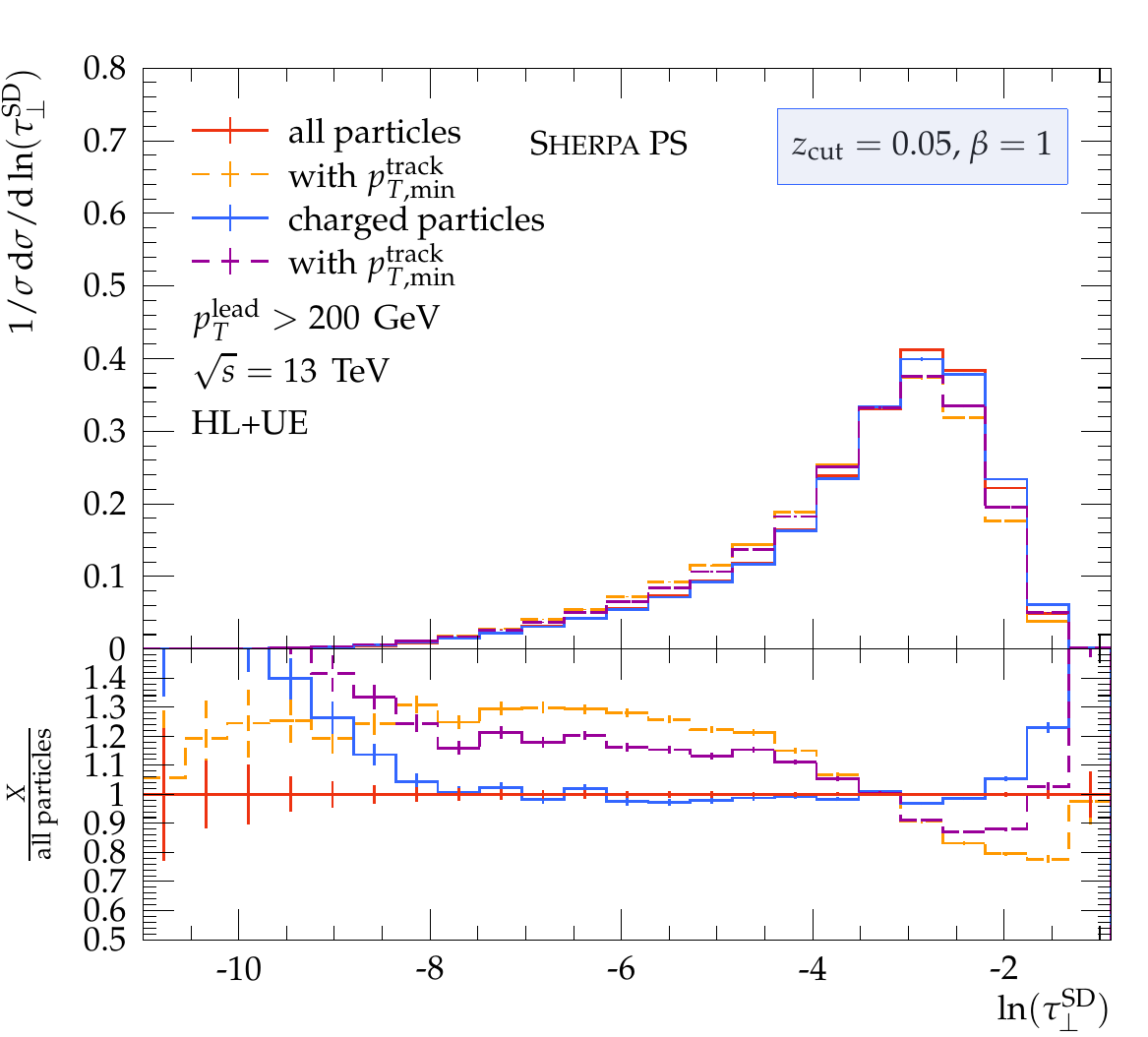}~
	\includegraphics[width=0.32\textwidth]{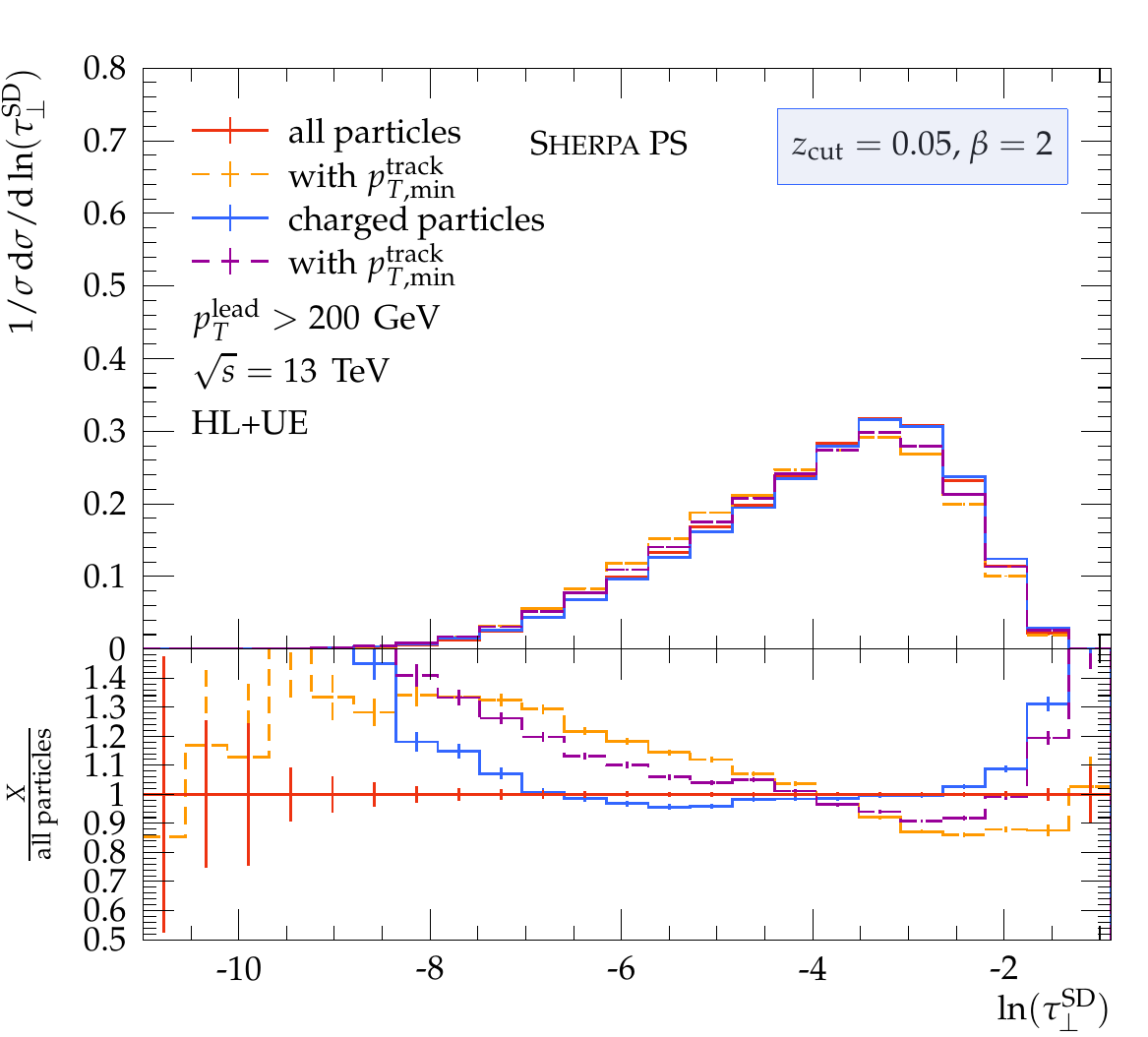}\\
	\includegraphics[width=0.32\textwidth]{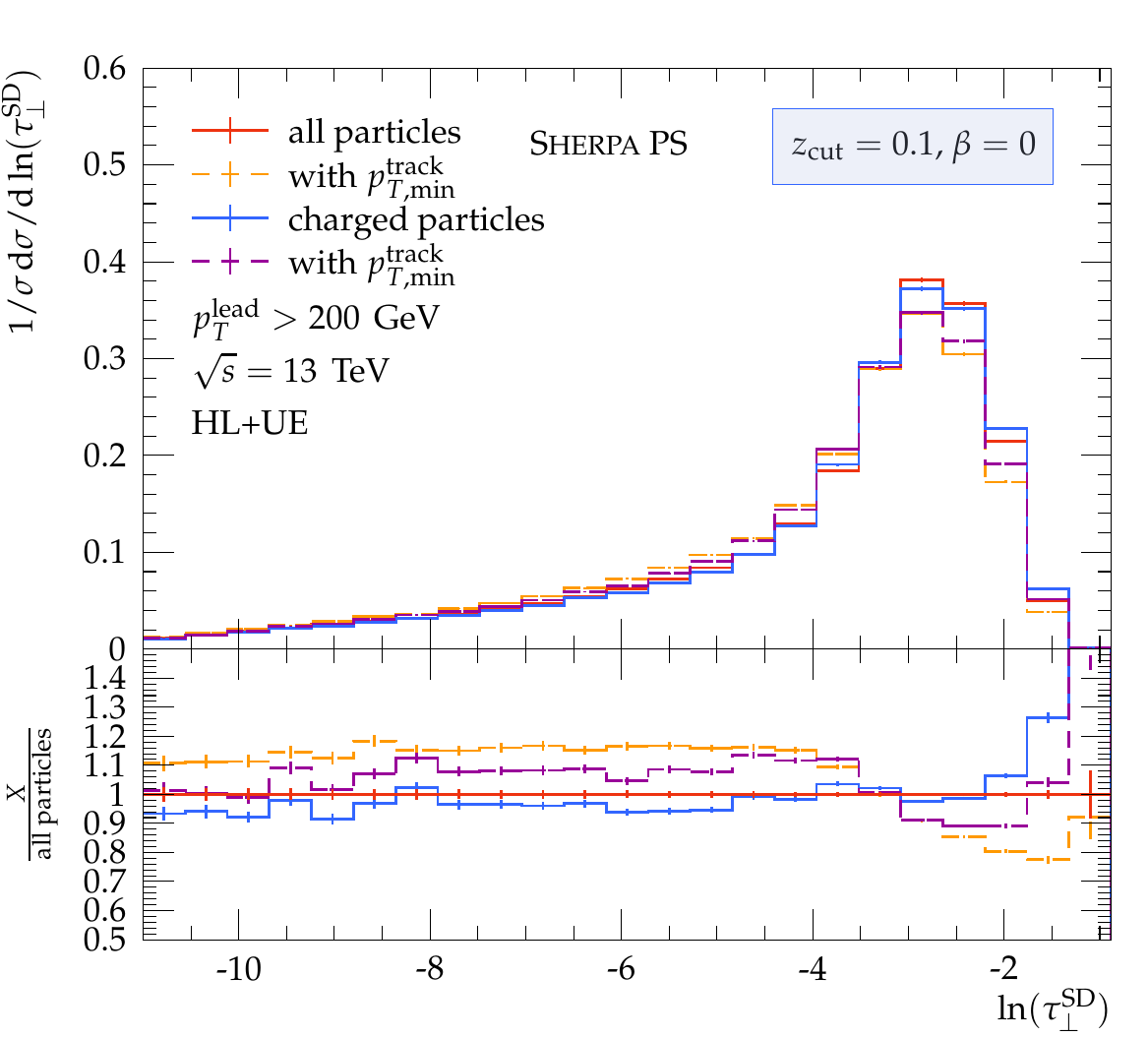}~
	\includegraphics[width=0.32\textwidth]{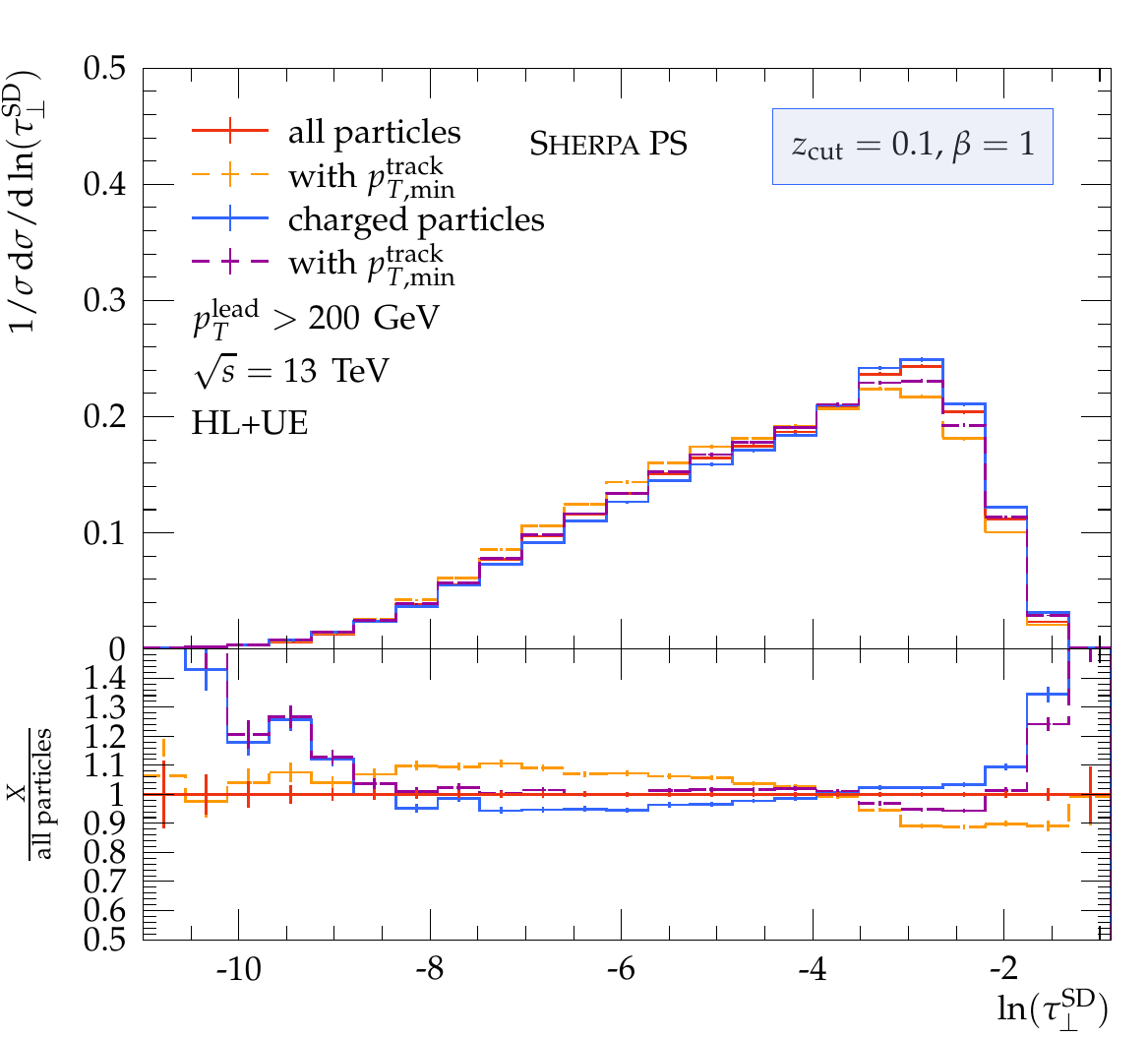}~
	\includegraphics[width=0.32\textwidth]{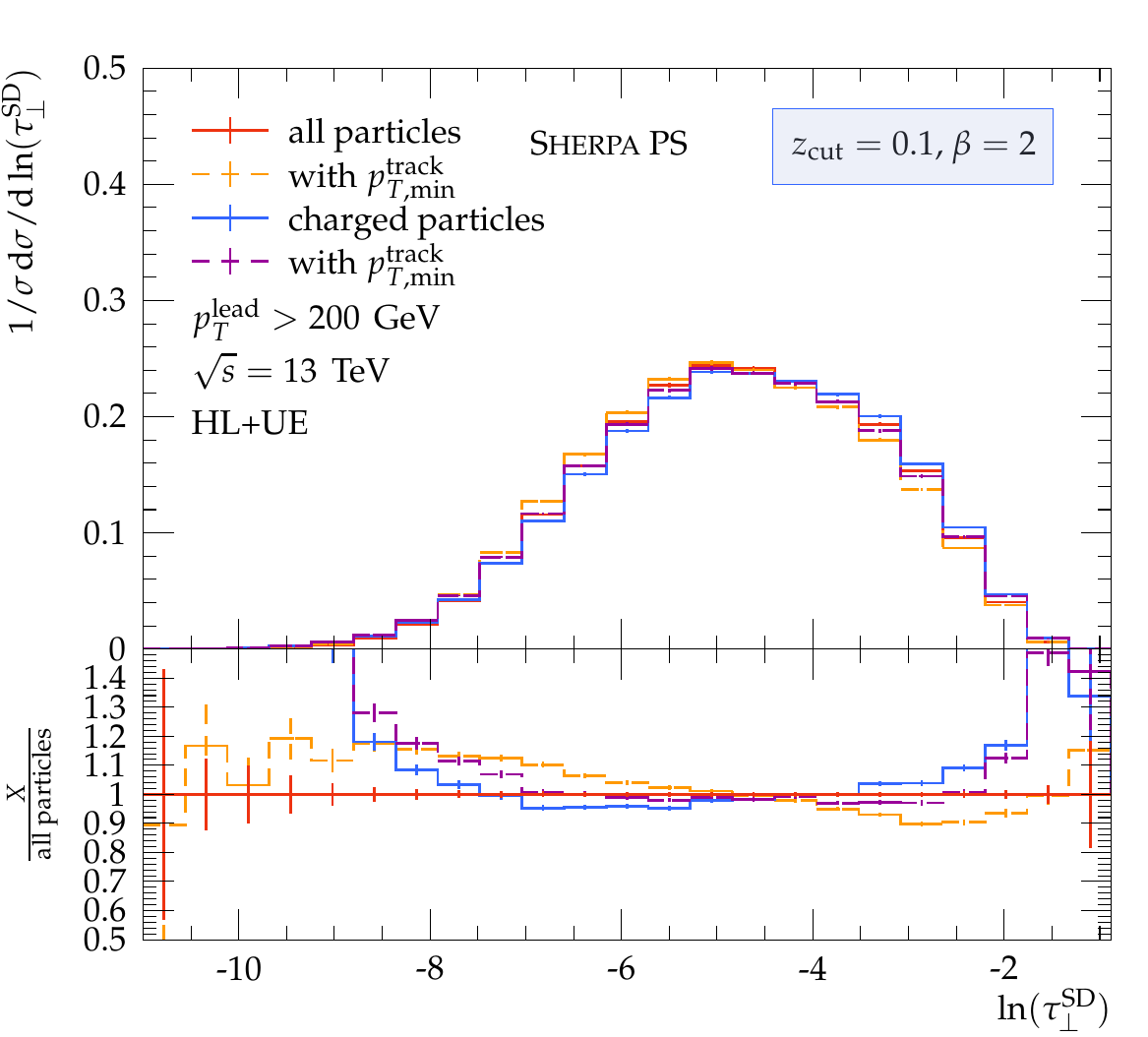}\\
	\includegraphics[width=0.32\textwidth]{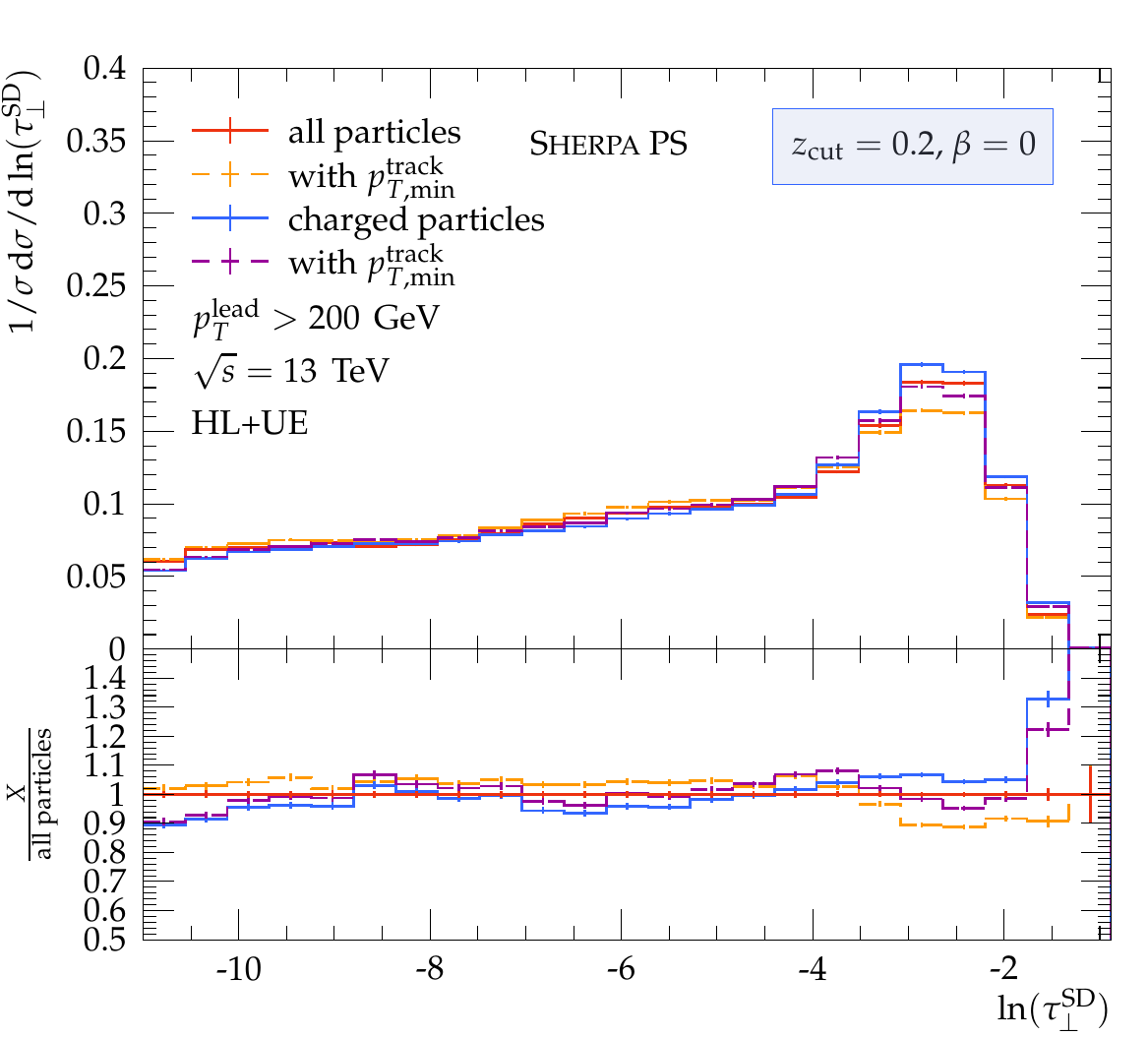}~
	\includegraphics[width=0.32\textwidth]{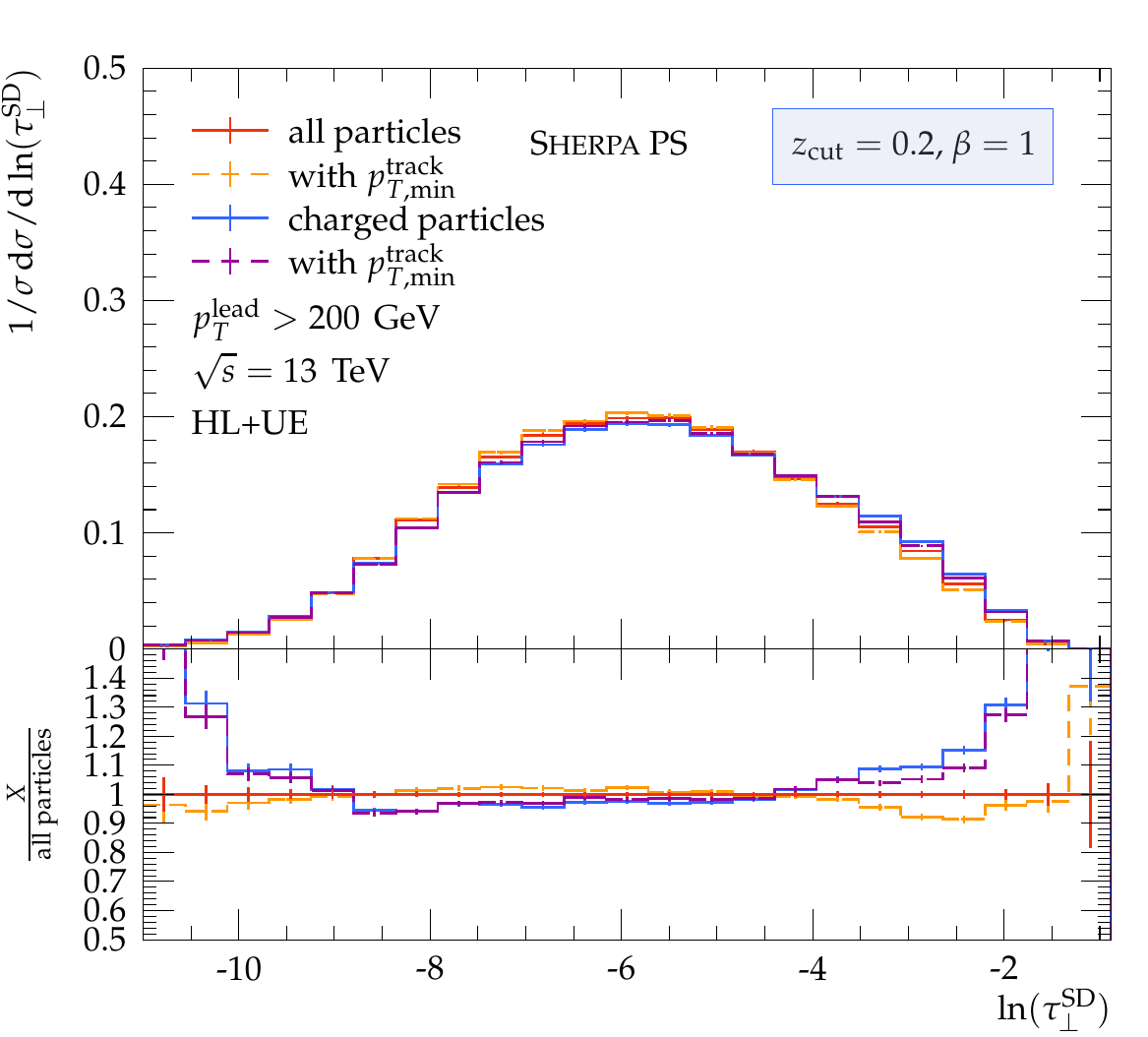}~
	\includegraphics[width=0.32\textwidth]{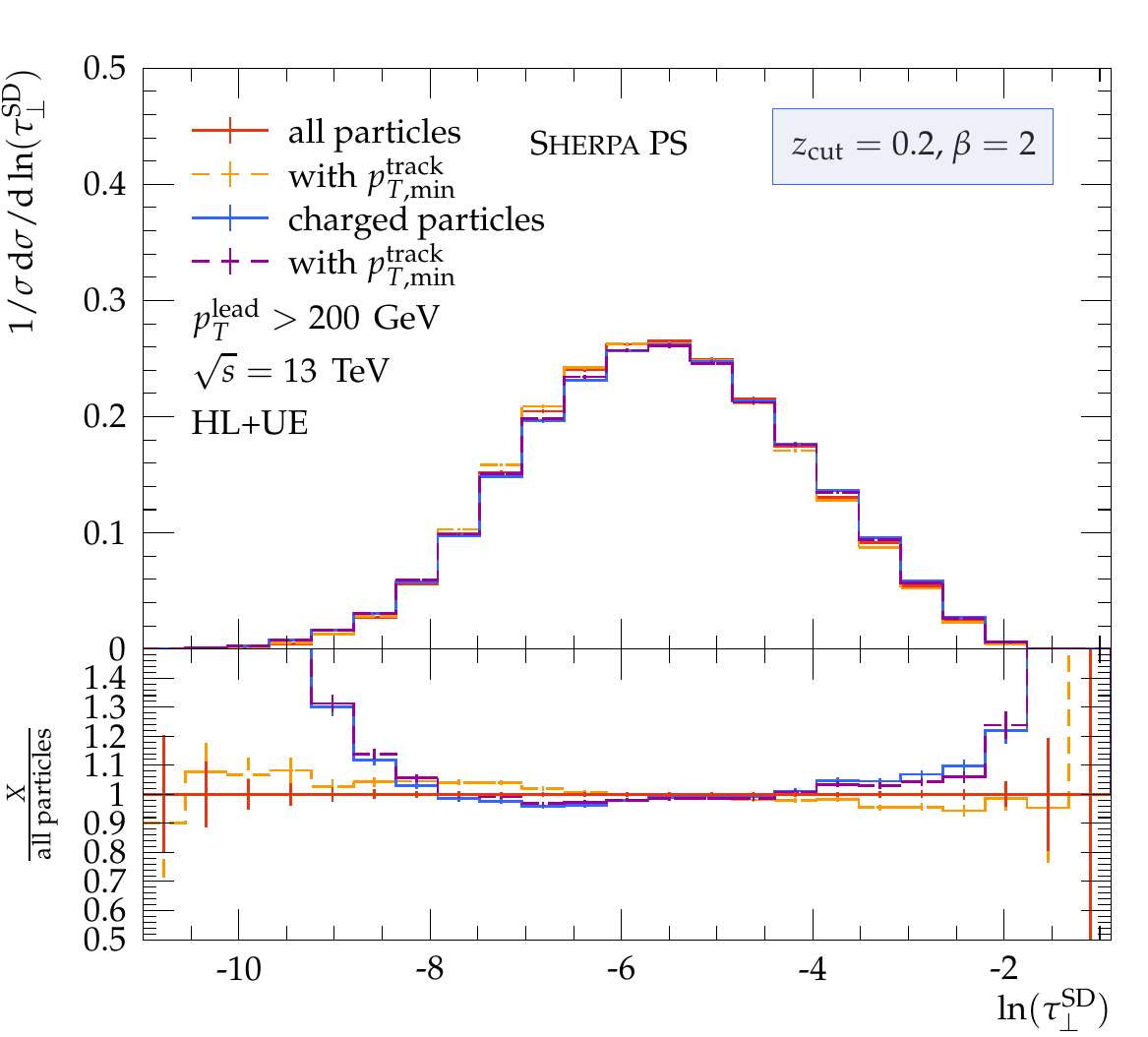}\\
	\includegraphics[width=0.32\textwidth]{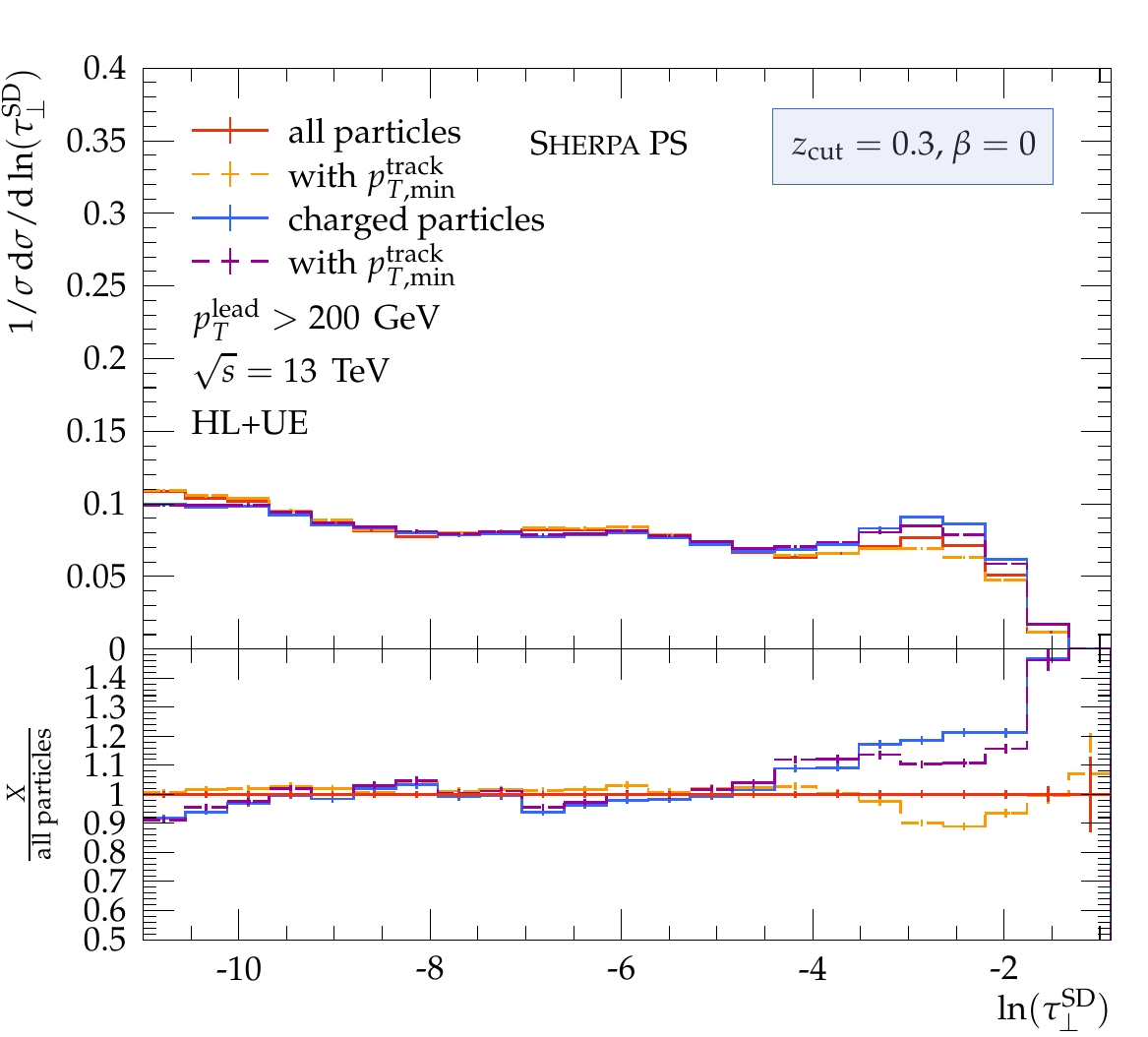}~
	\includegraphics[width=0.32\textwidth]{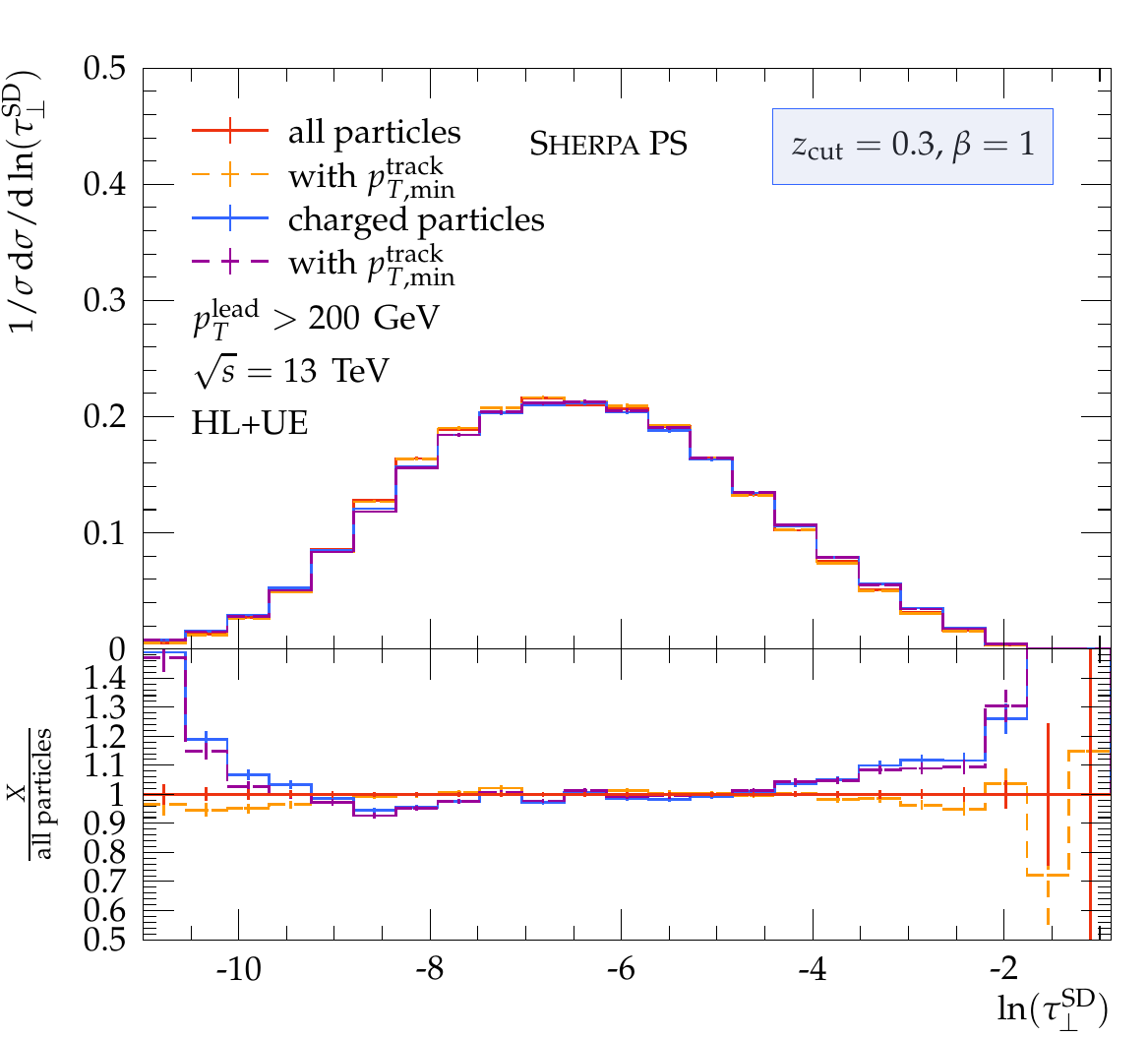}~
	\includegraphics[width=0.32\textwidth]{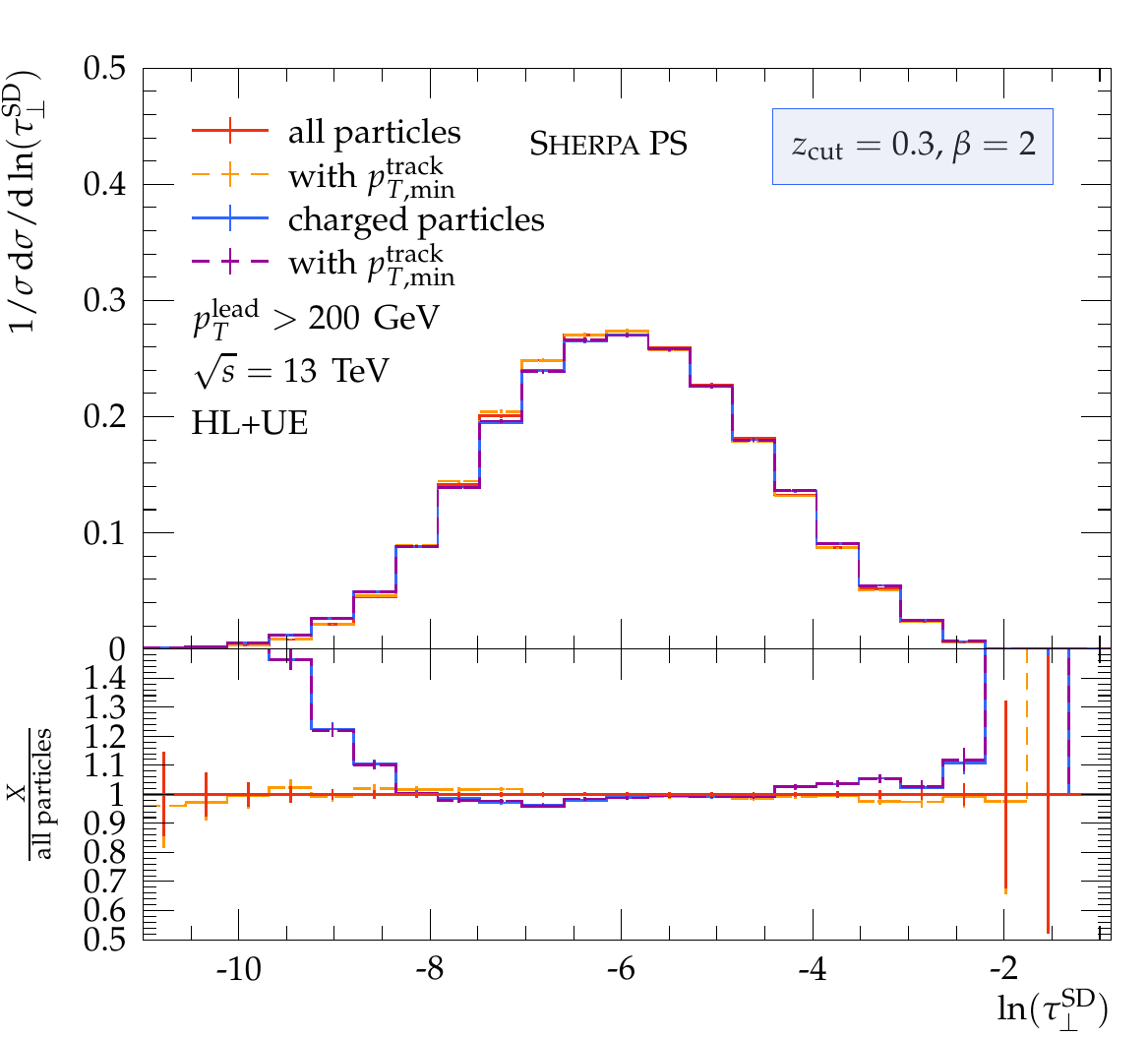}\\
\end{center}
\caption{Hadron-level results for groomed thrust for $\beta\in\{0,1,2\}$ (columns)
  and $\zcut\in\{0.05, 0.1, 0.2, 0.3\}$ (rows) for the $p_{T,\text{min}}=200\;\text{GeV}$ event selection.
  Shown are predictions from a parton-shower based \Sherpa simulation, where the observable gets
  determined from \emph{all} or \emph{charged} final-state particles only. In addition, the effect of
  a minimal particle transverse-momentum cut of $p^{\text{track}}_{T,\text{min}}= 500\;\text{MeV}$ is studied.
  The lower panels show ratios with respect to the \emph{all} particles and \emph{no} track-$p_T$ cut prediction.}
\label{fig:charge-pt}
\end{figure}

In Fig.~\ref{fig:charge-pt} we compile hadron-level predictions based on \Sherpa PS for the
$p_{T,\text{min}}=200\;\text{GeV}$ selection for four different inputs to the observable evaluation.
We consider the previous default, \emph{i.e.}\ \emph{all} final-state particles without any
particle $p_T$ requirement, impose an additional track $p_T$ cut of $p^{\text{track}}_{T,\text{min}}= 500\;\text{MeV}$,
and limit to \emph{charged} particles, with and without the $p^{\text{track}}_{T}$ threshold. As before we
consider $\beta\in \{0,1,2\}$ with $\zcut\in\{0.05,0.1,0.2,0.3\}$. We can observe that the
impact of calculating the observable on all \emph{vs.} charged particles only is rather mild, with the
exception of regions where the cross section is rather tiny, \emph{i.e.}\ towards the kinematic
end-point and for very low $\tauSD$. For the latter region this effect is somewhat more pronounced
for $\beta>0$ where grooming is suppressed for objects $\Delta R < R_{\text{SD}}$ away from the hard
jets and hadronisation corrections are sizeable. In contrast, the track-quality
cut $p_{T,\text{min}}^{\text{track}}$ has a much stronger impact, in particular for $\zcut=0.05$. Here it
significantly shifts the distribution, for $\beta=0$ it results in corrections in the peak region
of up to $30\%$. Increasing $\beta$ improves the agreement for the bulk of the events. However,
when raising $\zcut$ to $0.2$ or $0.3$ all observable definitions agree with the \emph{all}
particles \emph{no} $p^{\text{track}}_{T,\text{min}}$ staying below $10\%$ for a wide range of
$\tauSD$ values. For events with $p_{T,\text{min}}=500\;\text{GeV}$, corresponding plots are shown in
App.~\ref{app:aux_results}, Fig.~\ref{fig:charge-pt_500}. The conclusions on the impact of the
restriction to charged tracks only and the track-$p_T$ cut are in fact very similar. However,
as for the underlying-event suppression, the $\zcut$ values needed to reduce the impact of
the $p^{\text{track}}_{T,\text{min}}$ criterion scale inversely with the hardness of the hard process
and are thus significantly lowered.

In summary, our findings indicate that in particular for heavier grooming resummed predictions
could quite directly be compared to experimental data, without significant non-perturbative or
track-level corrections.

\FloatBarrier
\clearpage

\subsubsection*{Generator-level predictions}

Based on the \Sherpa \MEPSatNLO simulations we have demonstrated the potential of
soft-drop grooming to reduce the impact of the underlying event on the transverse-thrust
distribution. Furthermore, we could show that the effect of constraining the observable
evaluation to charged-particle tracks above a minimal $p^{\text{track}}_{T,\text{min}}$ threshold
is significantly reduced by soft-drop grooming. This, in fact, allows for a
more direct comparison of experimental measurements with perturbative predictions.
Furthermore, the tunable sensitivity to the underlying event for different
grooming parameters provides means to constrain Monte Carlo generator models
for non-perturbative phenomena. Given corresponding measurements, this can
be employed in the validation and the tuning of the models and their parameters.

To confirm the viability and robustness of our conclusions and to establish the
two use cases, it remains to be studied to what extent our findings are possibly generator
specific. To this end, we contrast our hadron-level track-based \MEPSatNLO
predictions from \Sherpa, based on the $2-$ and $3-$jet NLO matrix elements,
with track-level results from the \Herwig and \Pythia generators. For the latter
we simulated leading-order dijet production, dressed with parton showers,
employing the generators' default underlying-event and hadronisation models
and parameter settings.

\begin{figure}[t]
\begin{center}
	\includegraphics[width=0.32\textwidth]{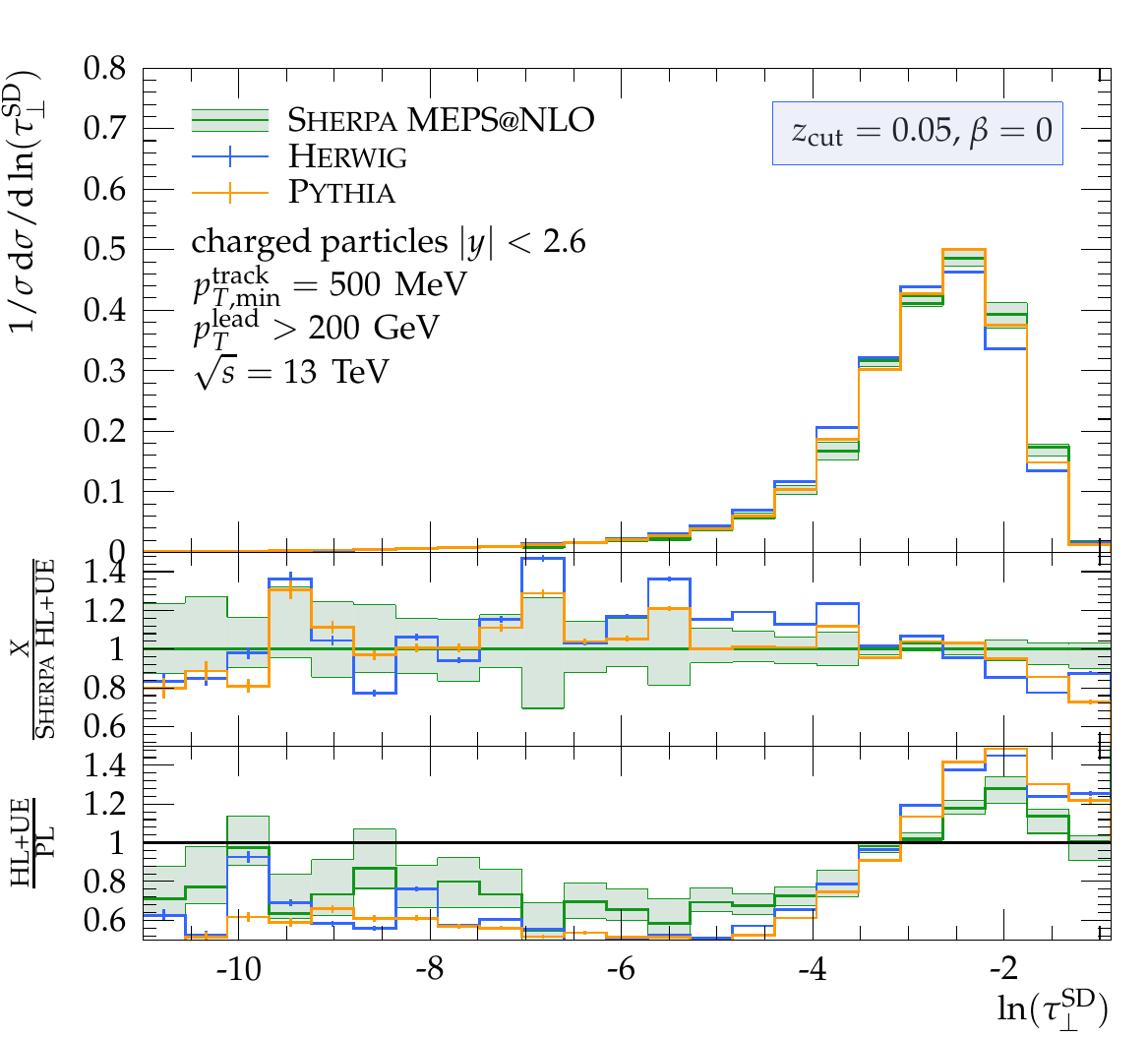}~
	\includegraphics[width=0.32\textwidth]{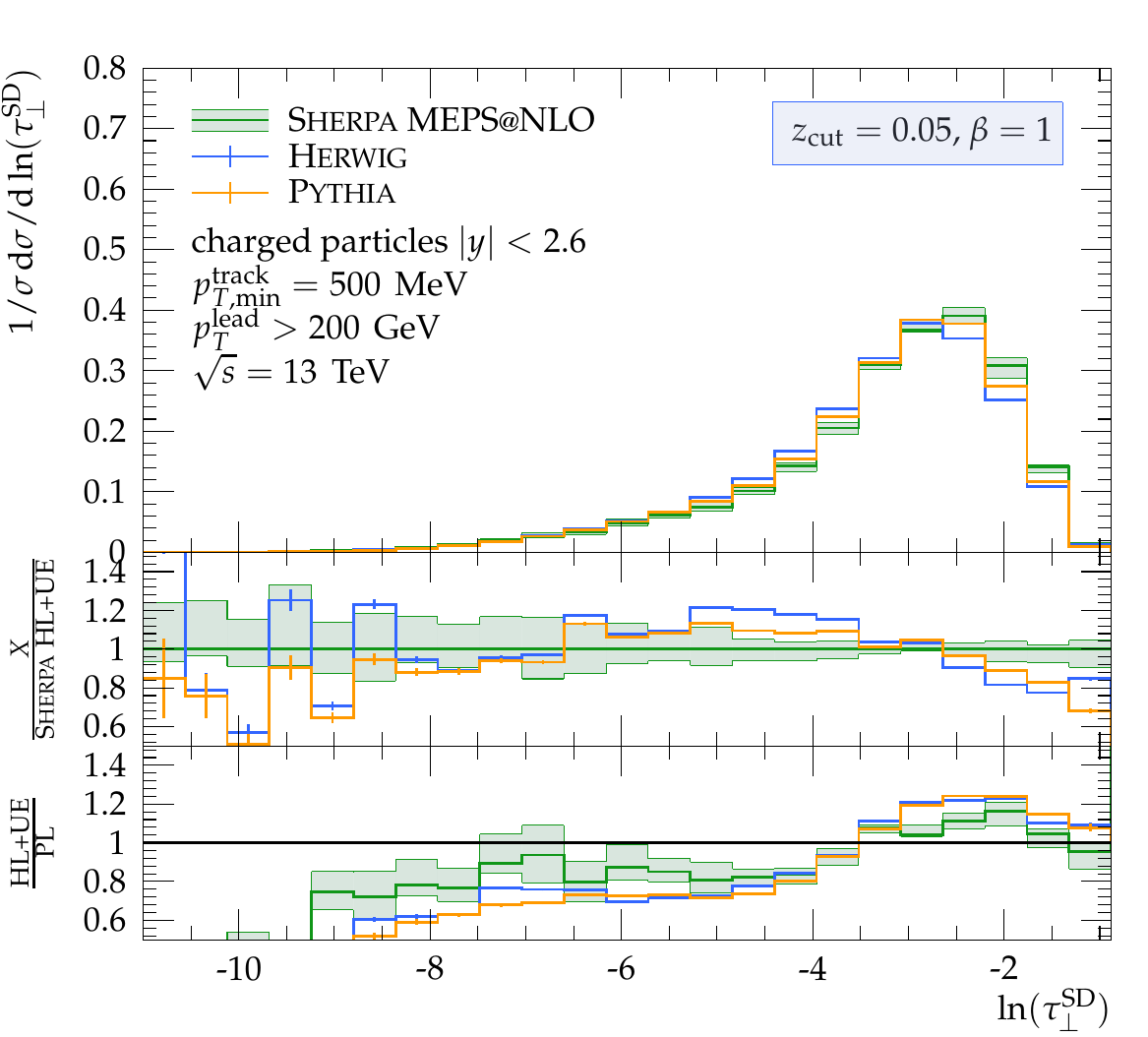}~
	\includegraphics[width=0.32\textwidth]{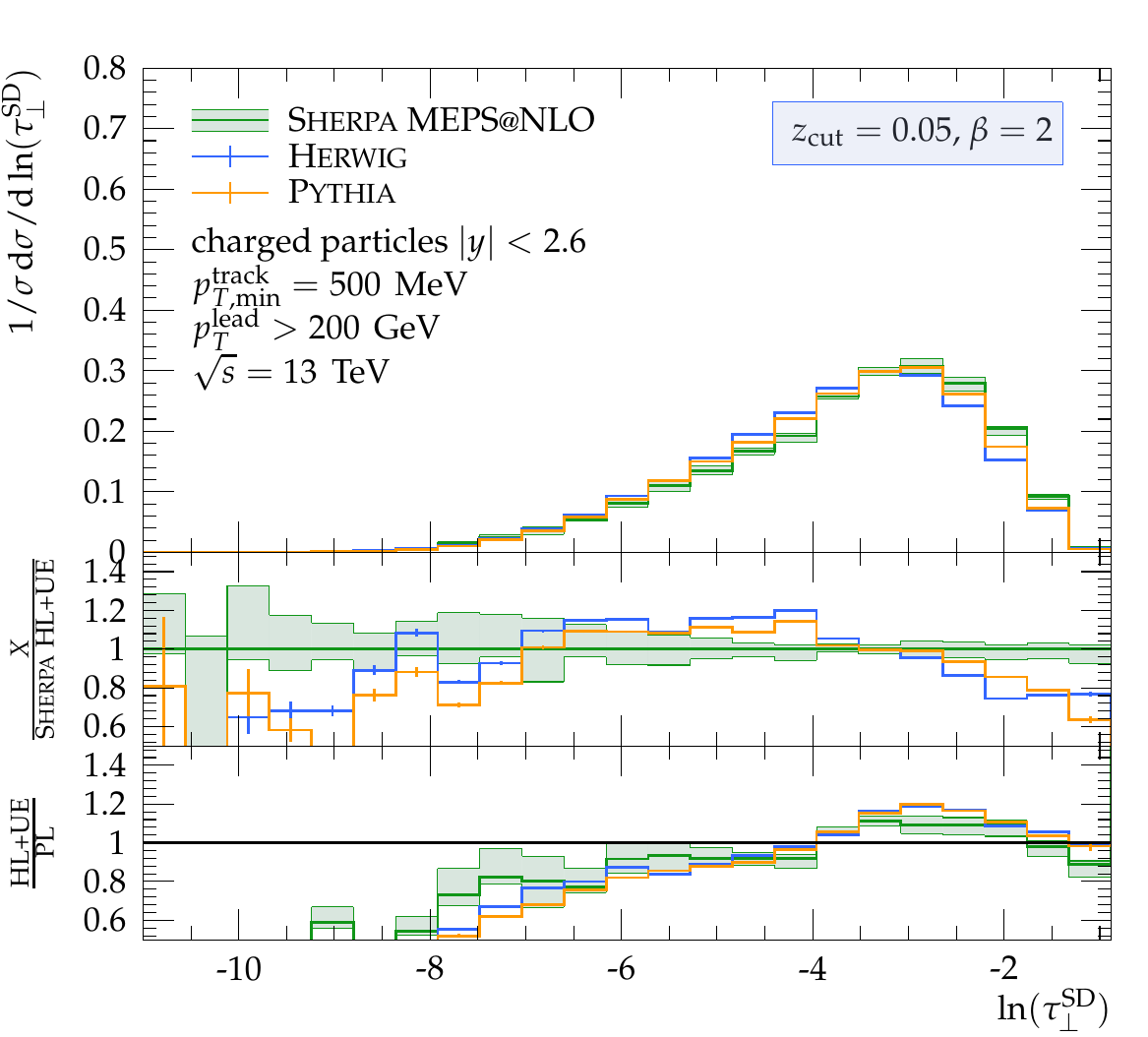}\\
	\includegraphics[width=0.32\textwidth]{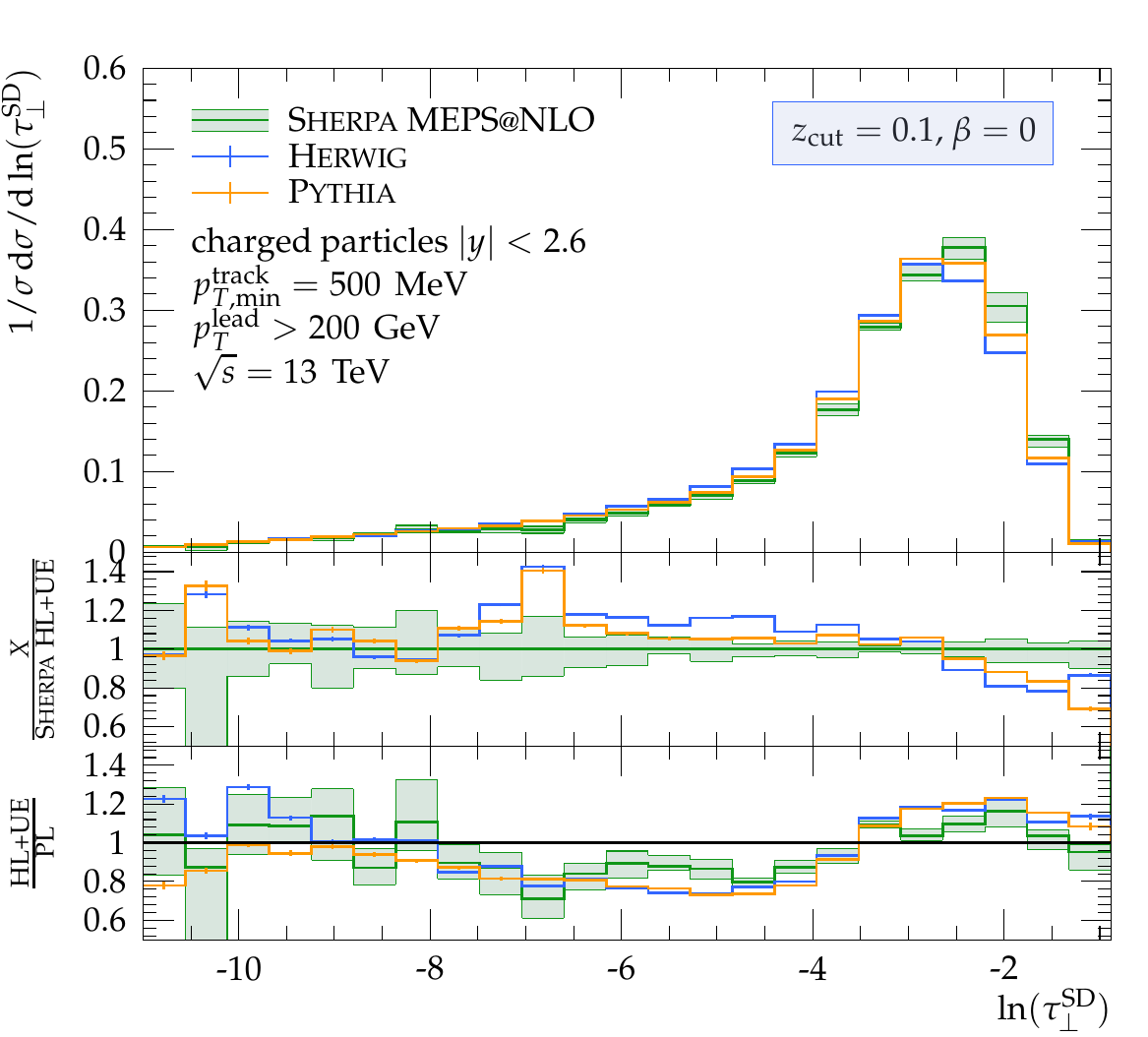}~
	\includegraphics[width=0.32\textwidth]{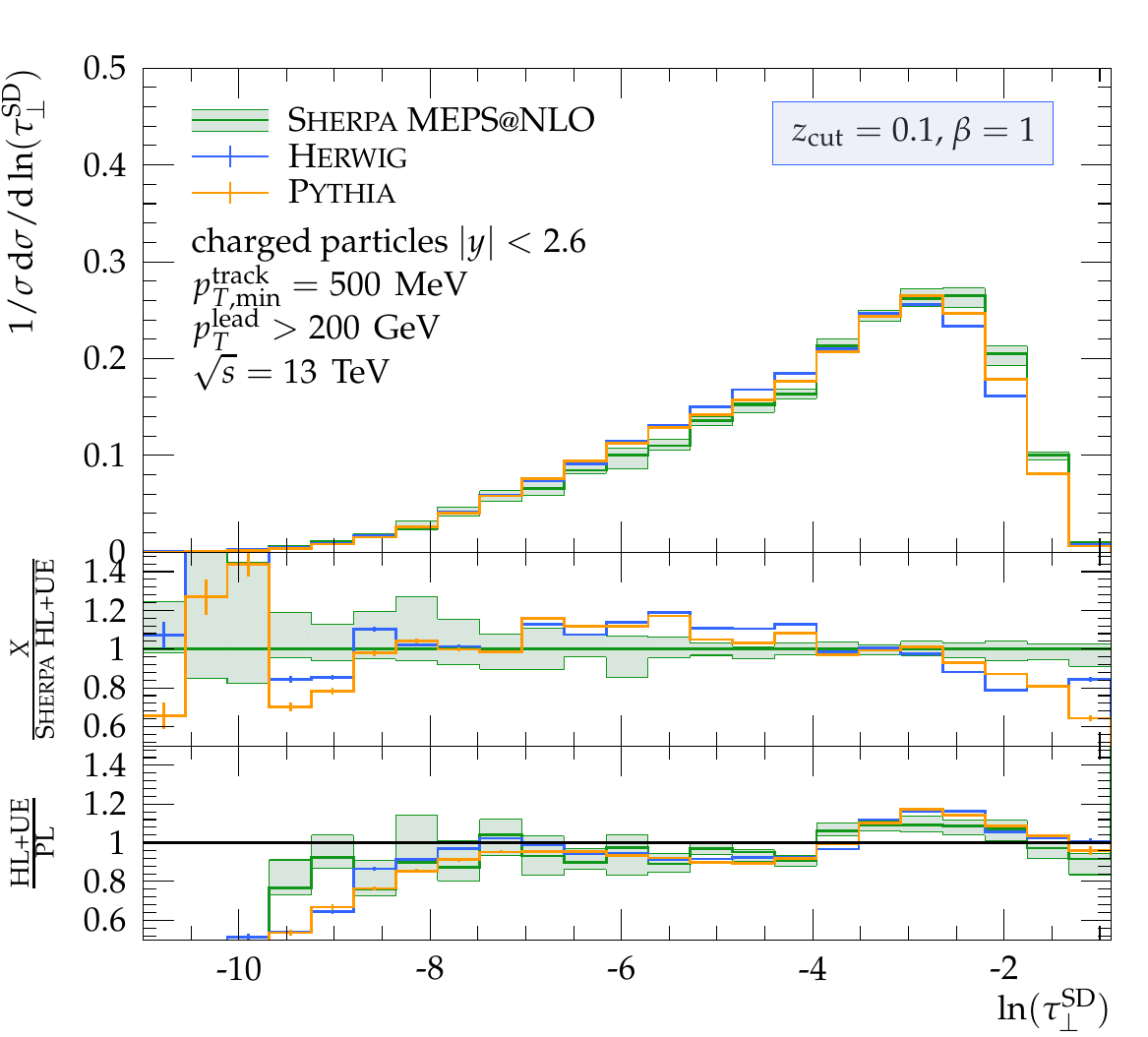}~
	\includegraphics[width=0.32\textwidth]{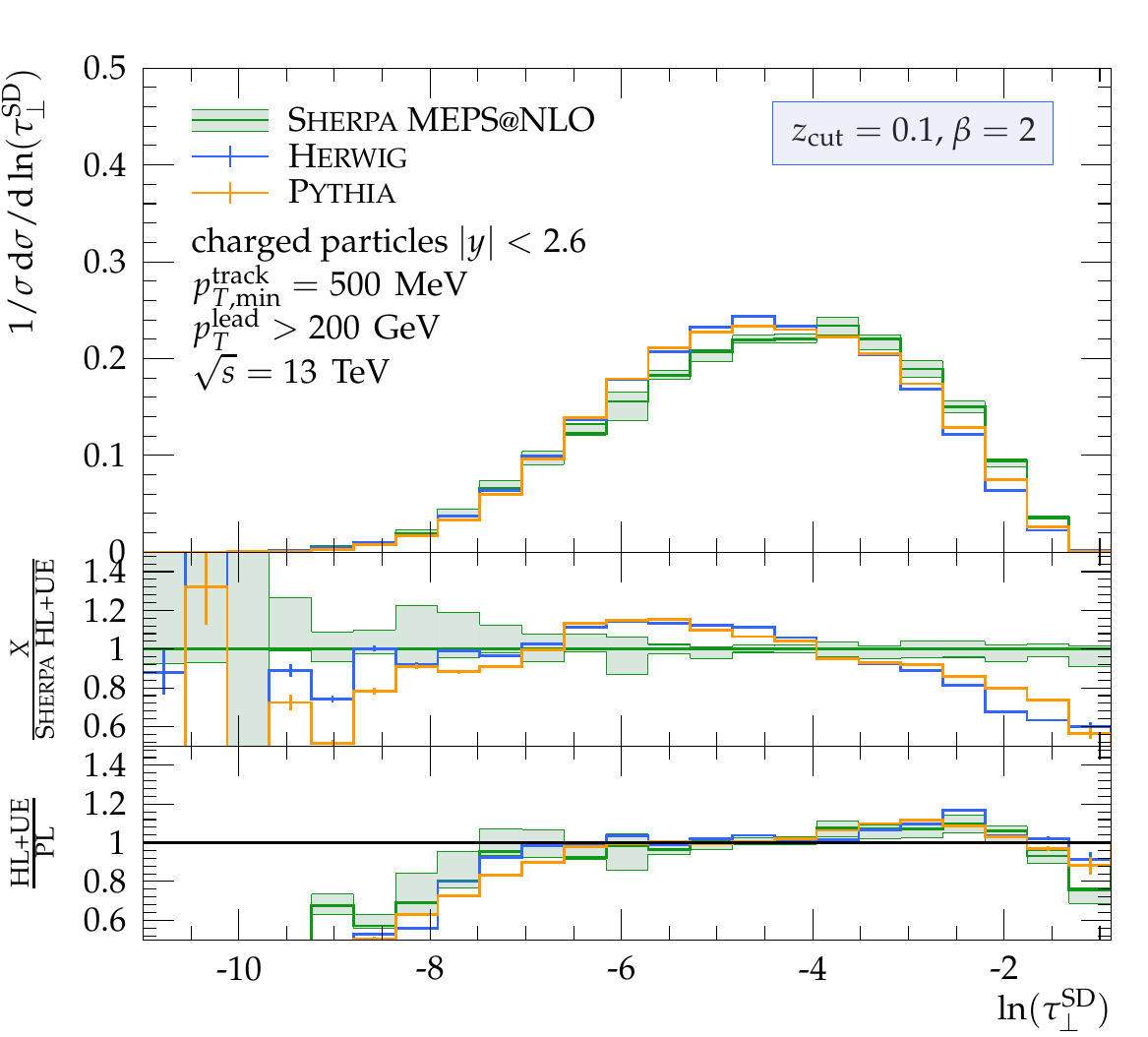}\\
	\includegraphics[width=0.32\textwidth]{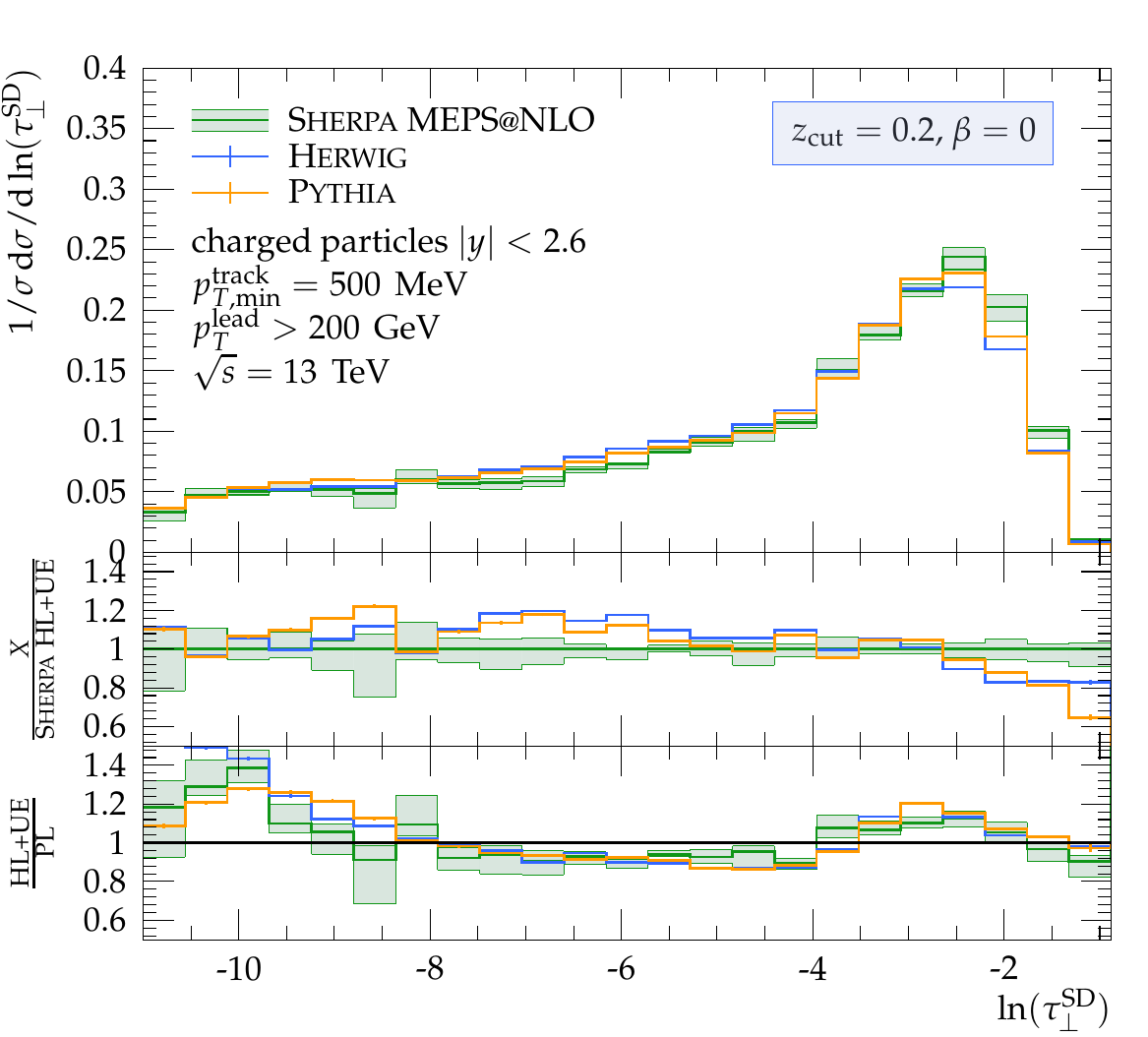}~
	\includegraphics[width=0.32\textwidth]{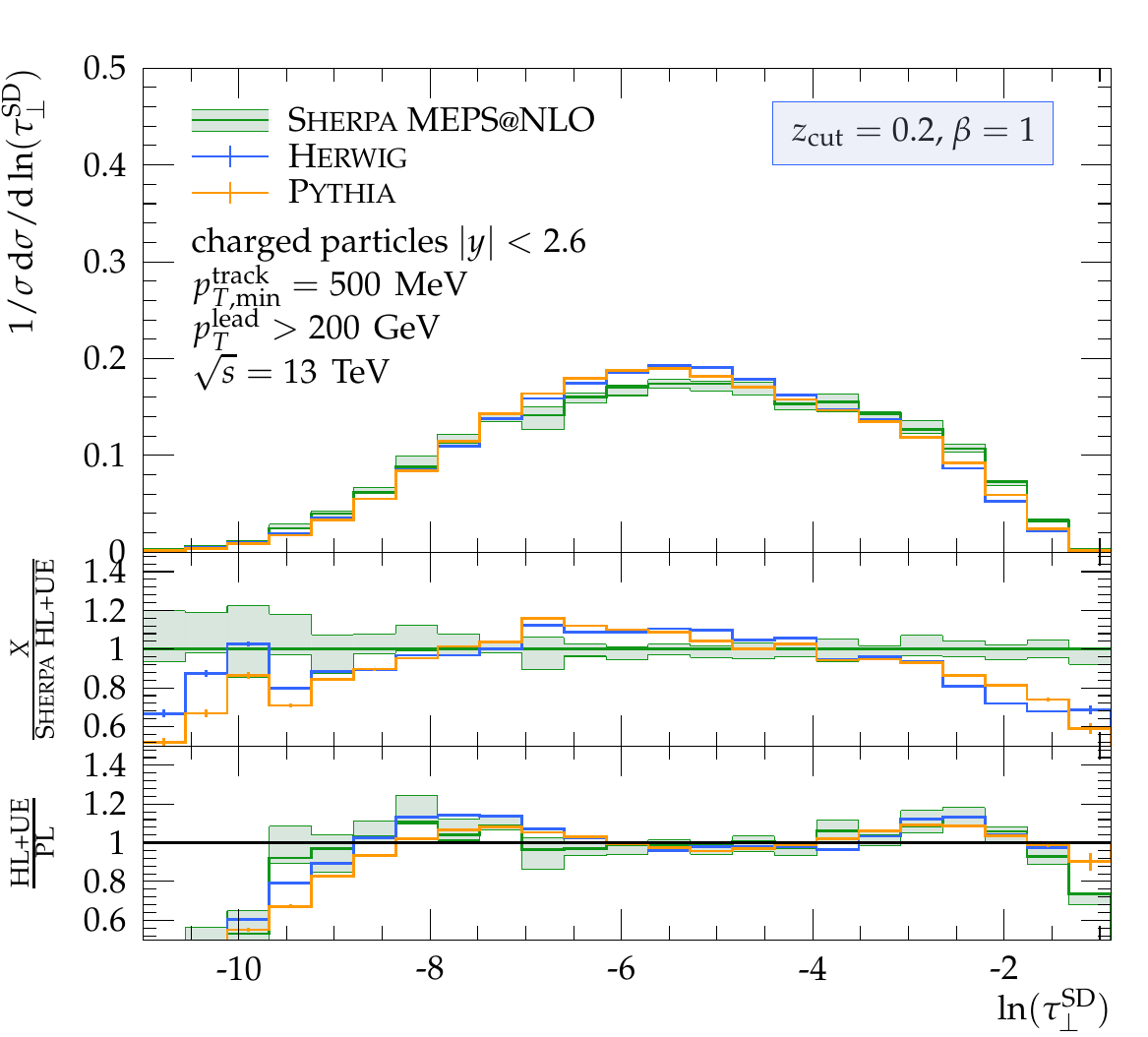}~
	\includegraphics[width=0.32\textwidth]{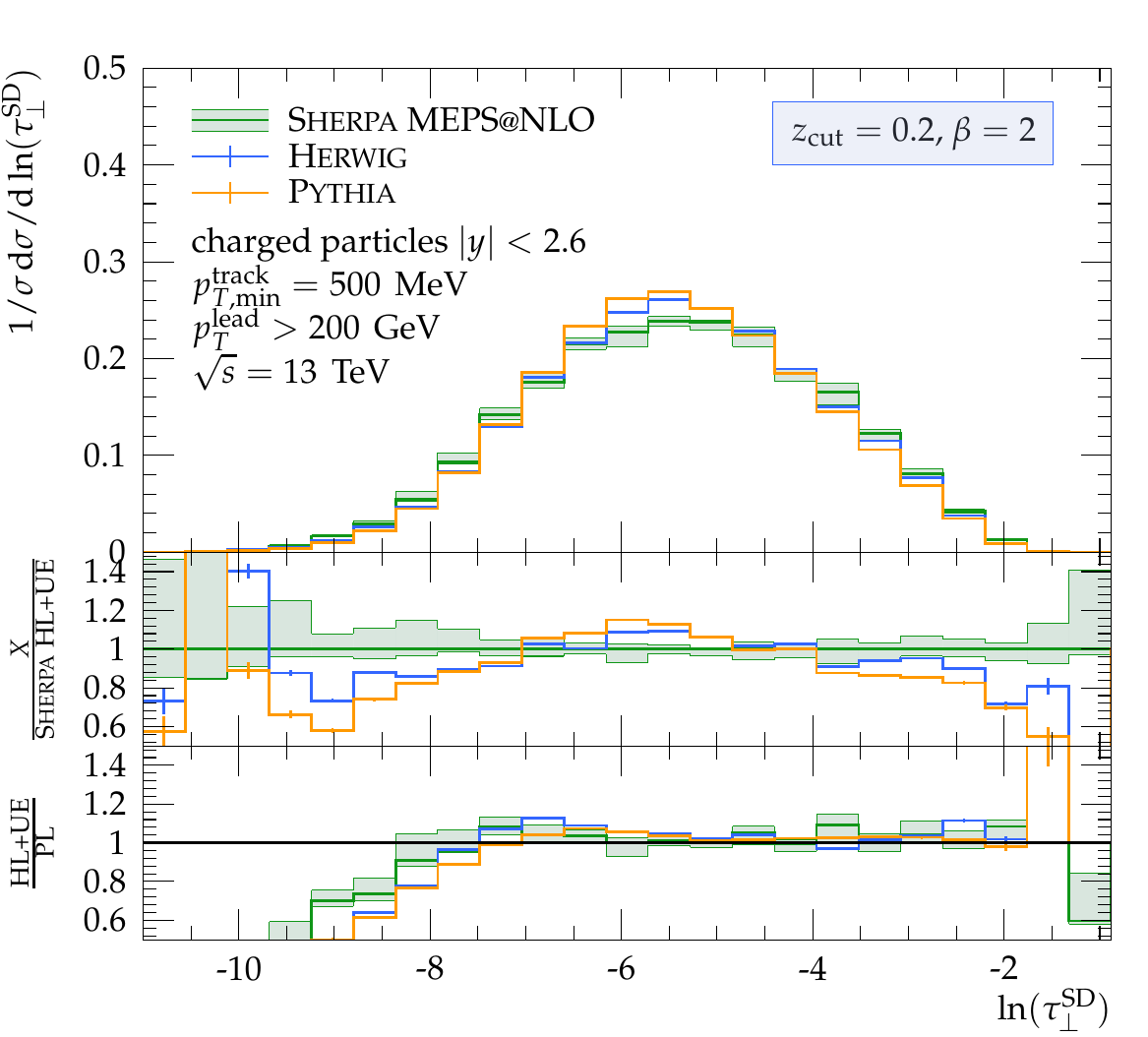}\\
	\includegraphics[width=0.32\textwidth]{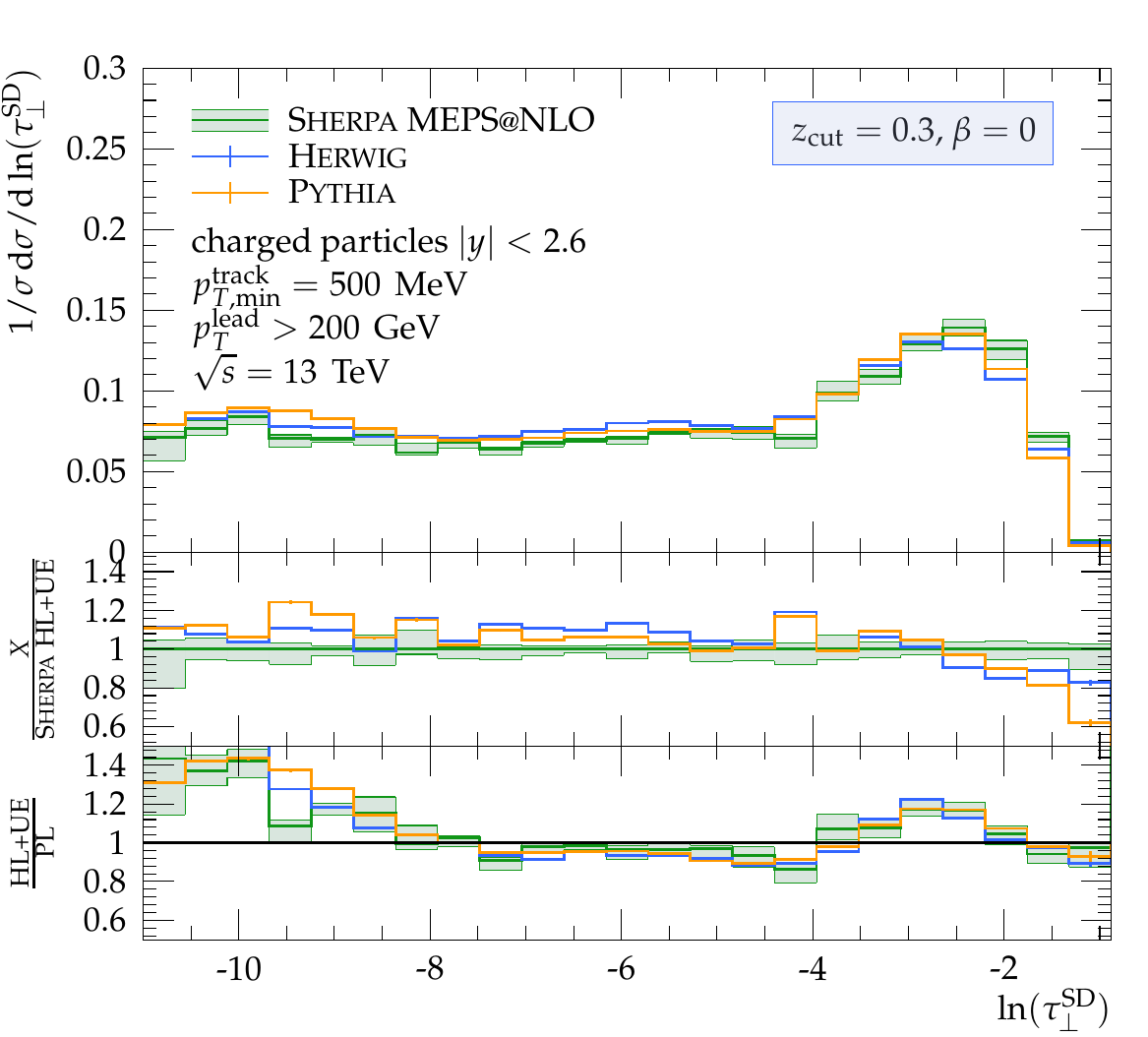}~
	\includegraphics[width=0.32\textwidth]{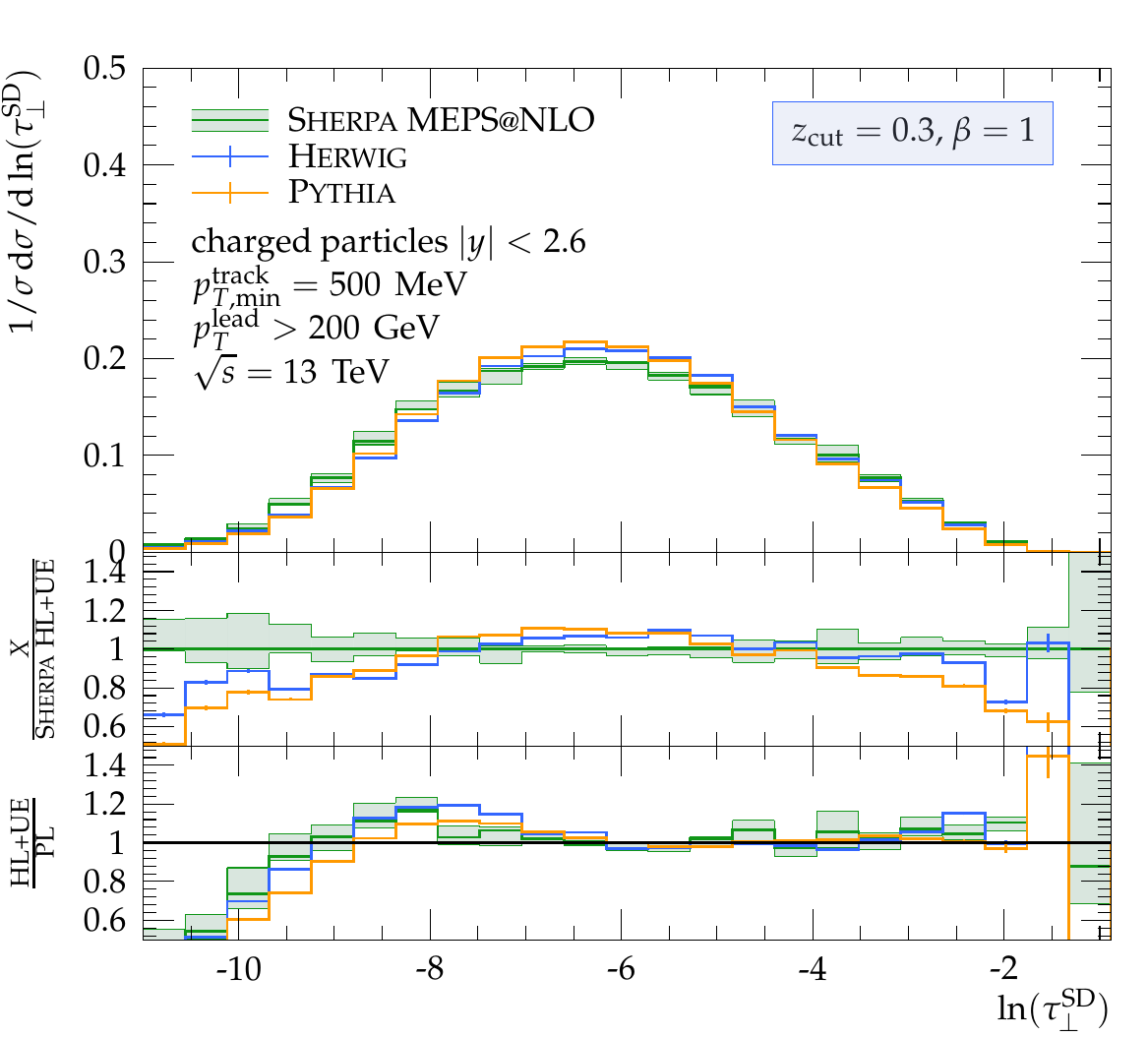}~
	\includegraphics[width=0.32\textwidth]{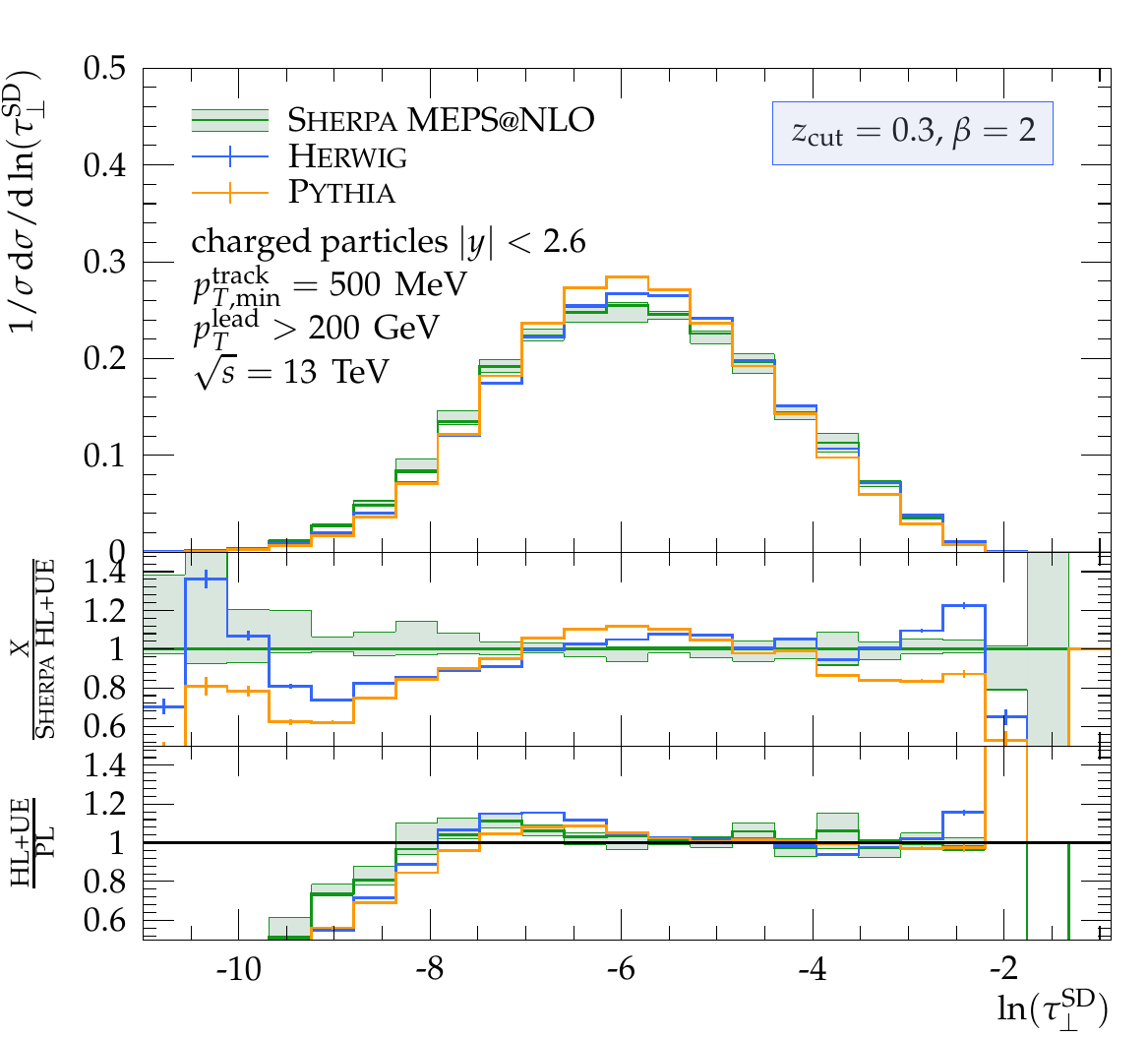}\\
\end{center}
\caption{Hadron-level results based on charged-particle tracks with $p^{\text{track}}_{T,\text{min}}>500\;\text{MeV}$
  for groomed thrust with $\beta\in\{0,1,2\}$ (columns) and $\zcut\in\{0.05, 0.1, 0.2, 0.3\}$ (rows) for the
  $p_{T,\text{min}}=200\;\text{GeV}$ event selection. Shown are predictions based on the leading-order dijet
  production from \Pythia and \Herwig, as well as the \MEPSatNLO result from \Sherpa. 
  The two lower panels present the ratios with respect to the \Sherpa hadron-level
  prediction and the generators' parton-level prediction, respectively.}
\label{fig:MC}
\end{figure}

\begin{figure}[t]
\begin{center}
	\includegraphics[width=0.32\textwidth]{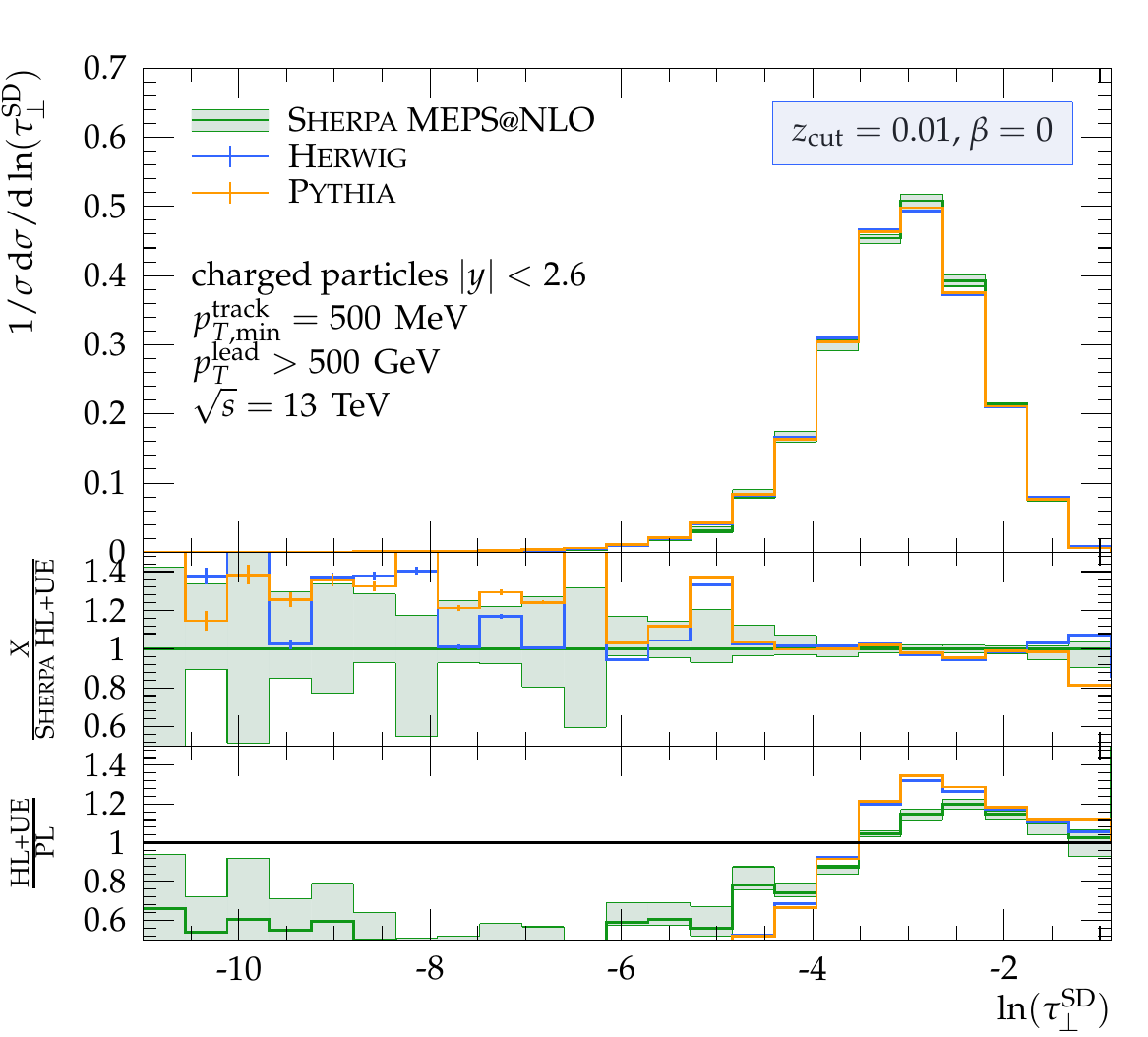}~
	\includegraphics[width=0.32\textwidth]{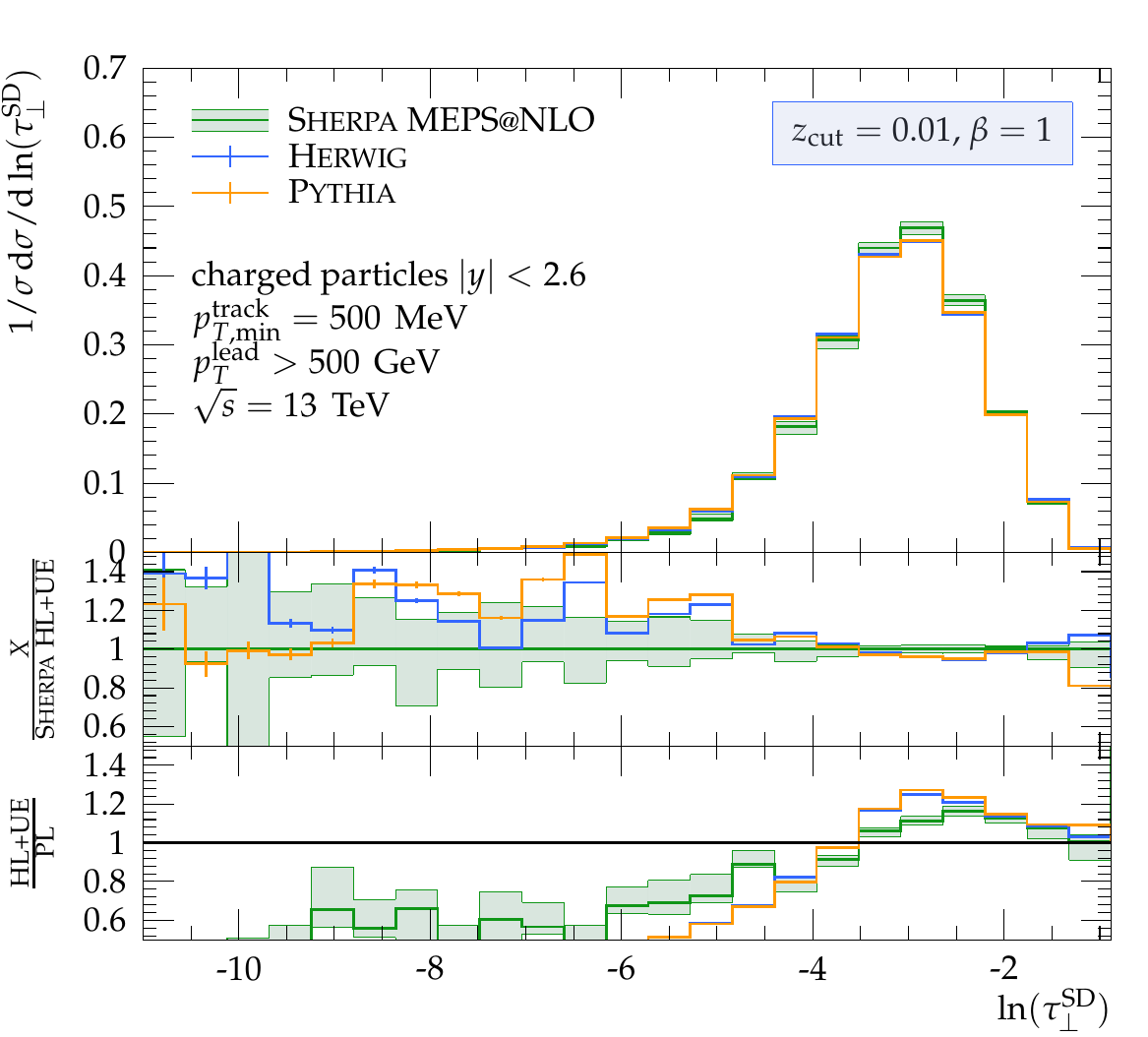}~
	\includegraphics[width=0.32\textwidth]{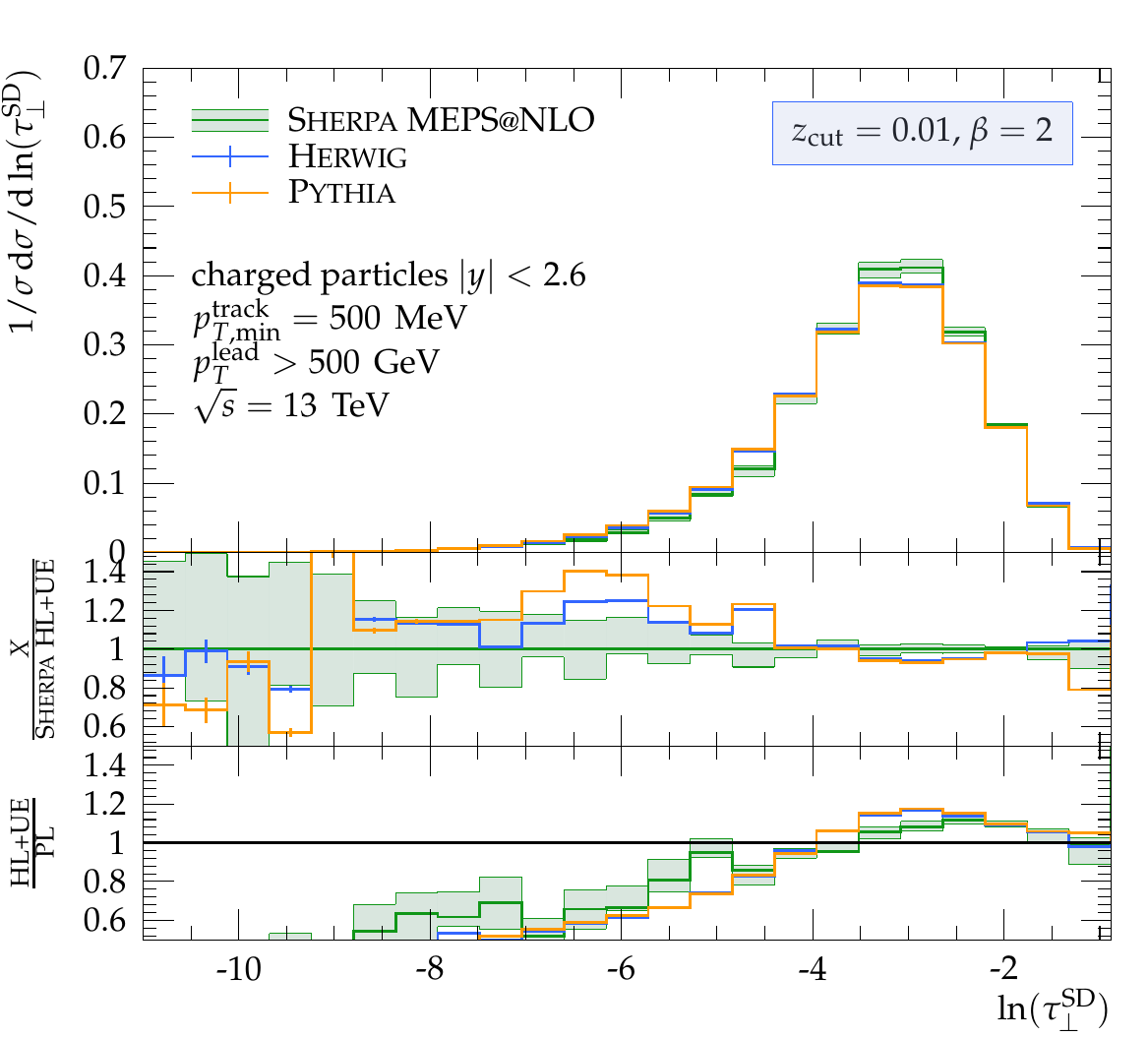}\\
	\includegraphics[width=0.32\textwidth]{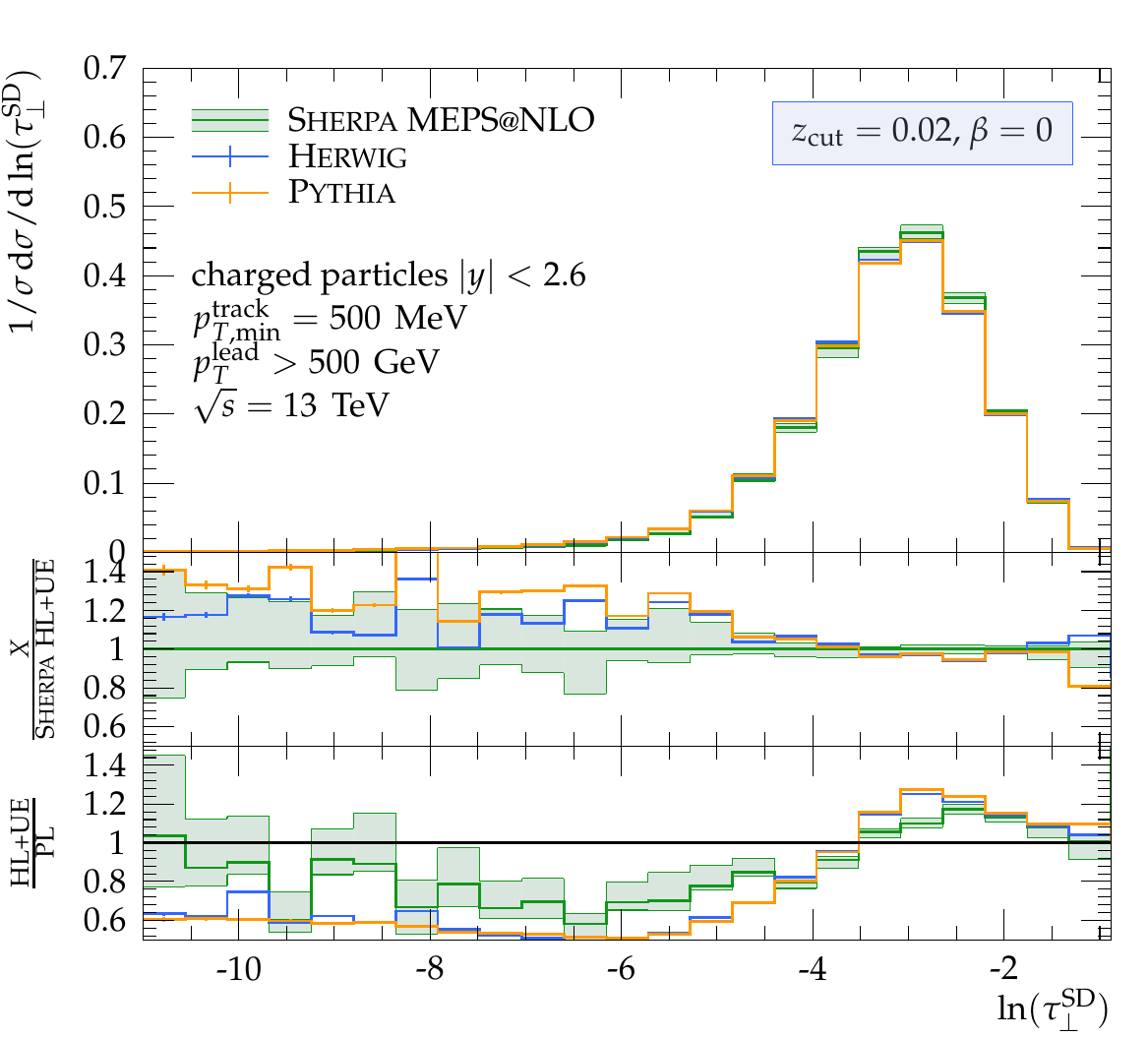}~
	\includegraphics[width=0.32\textwidth]{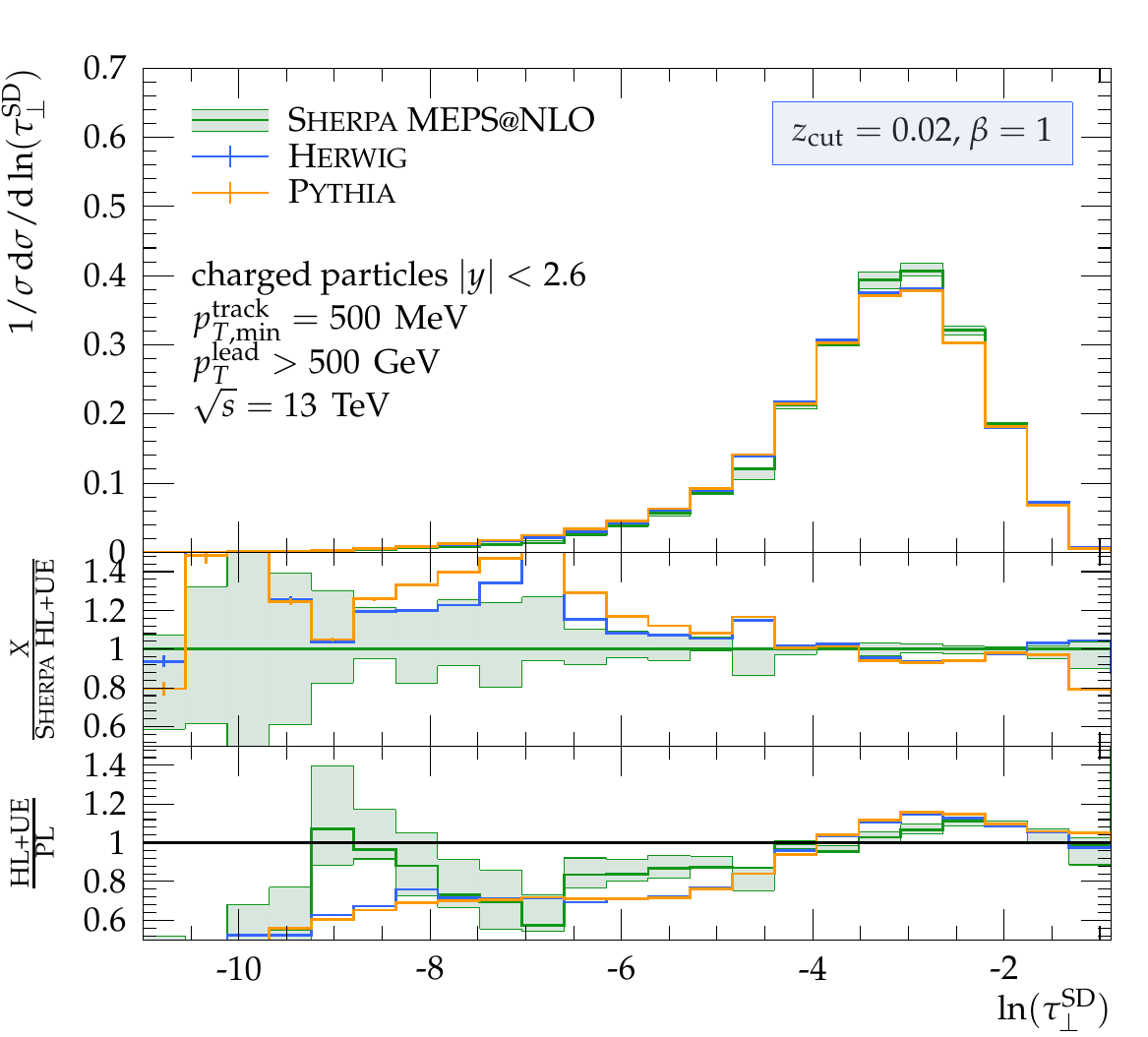}~
	\includegraphics[width=0.32\textwidth]{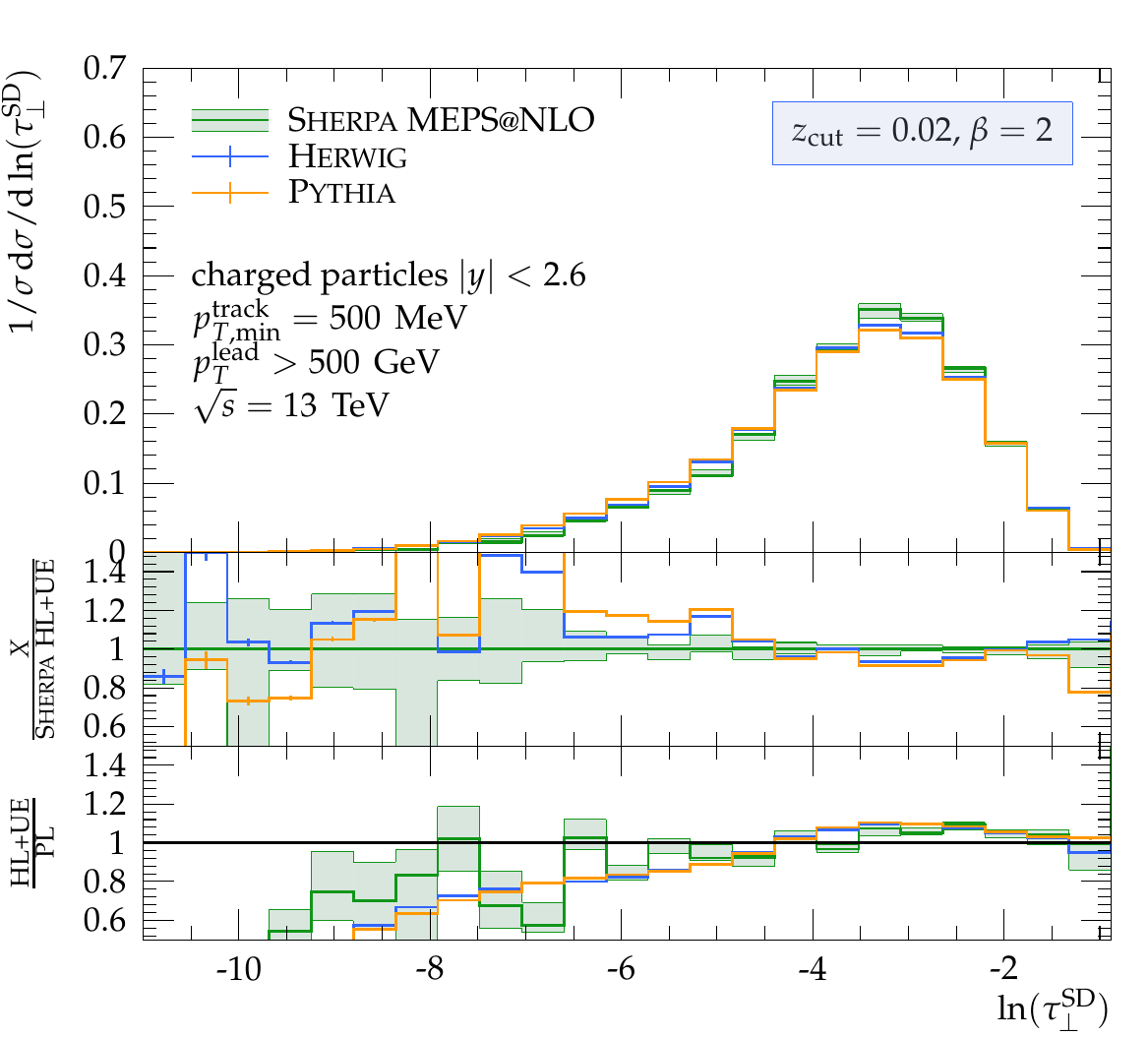}\\
	\includegraphics[width=0.32\textwidth]{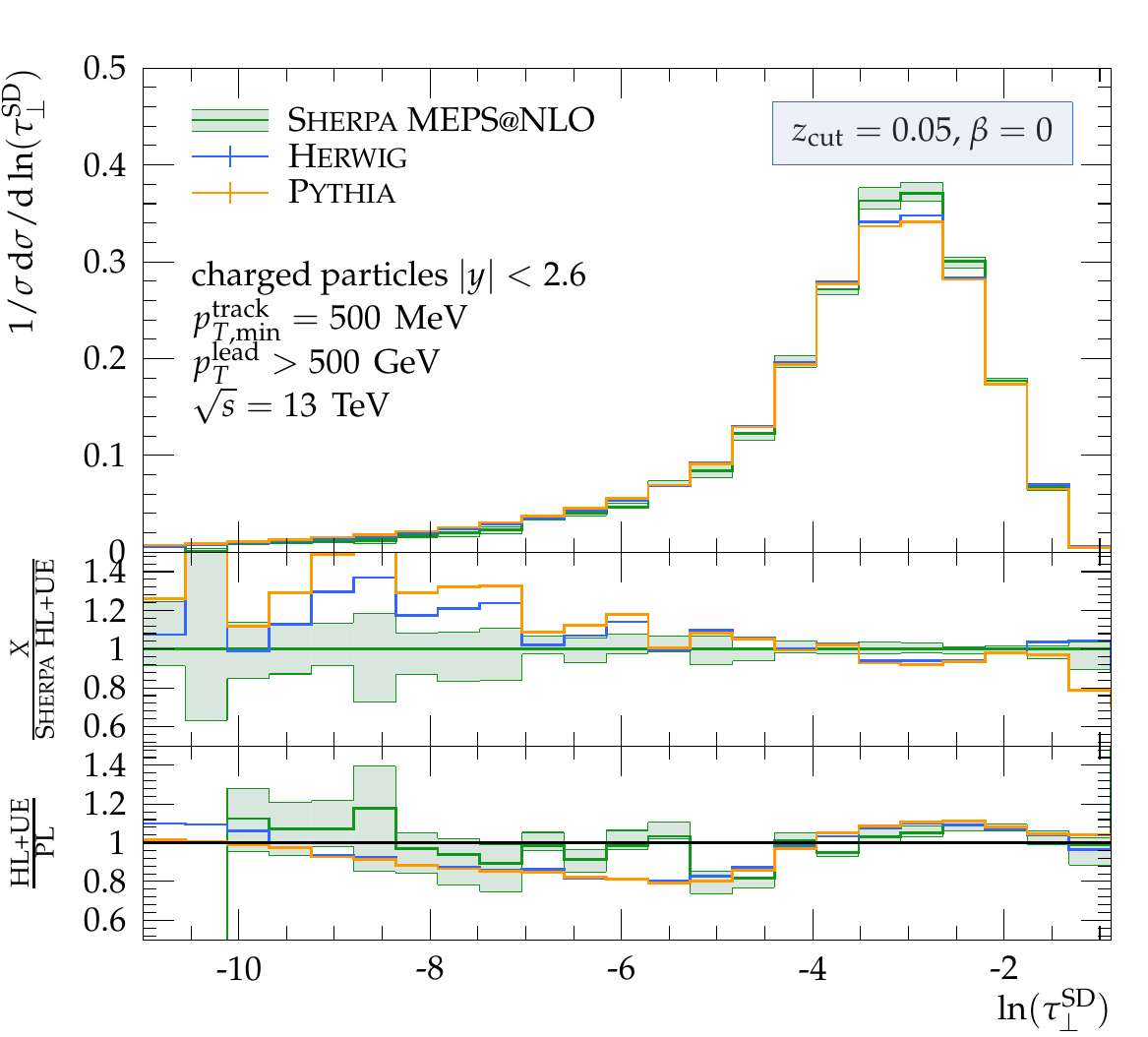}~
	\includegraphics[width=0.32\textwidth]{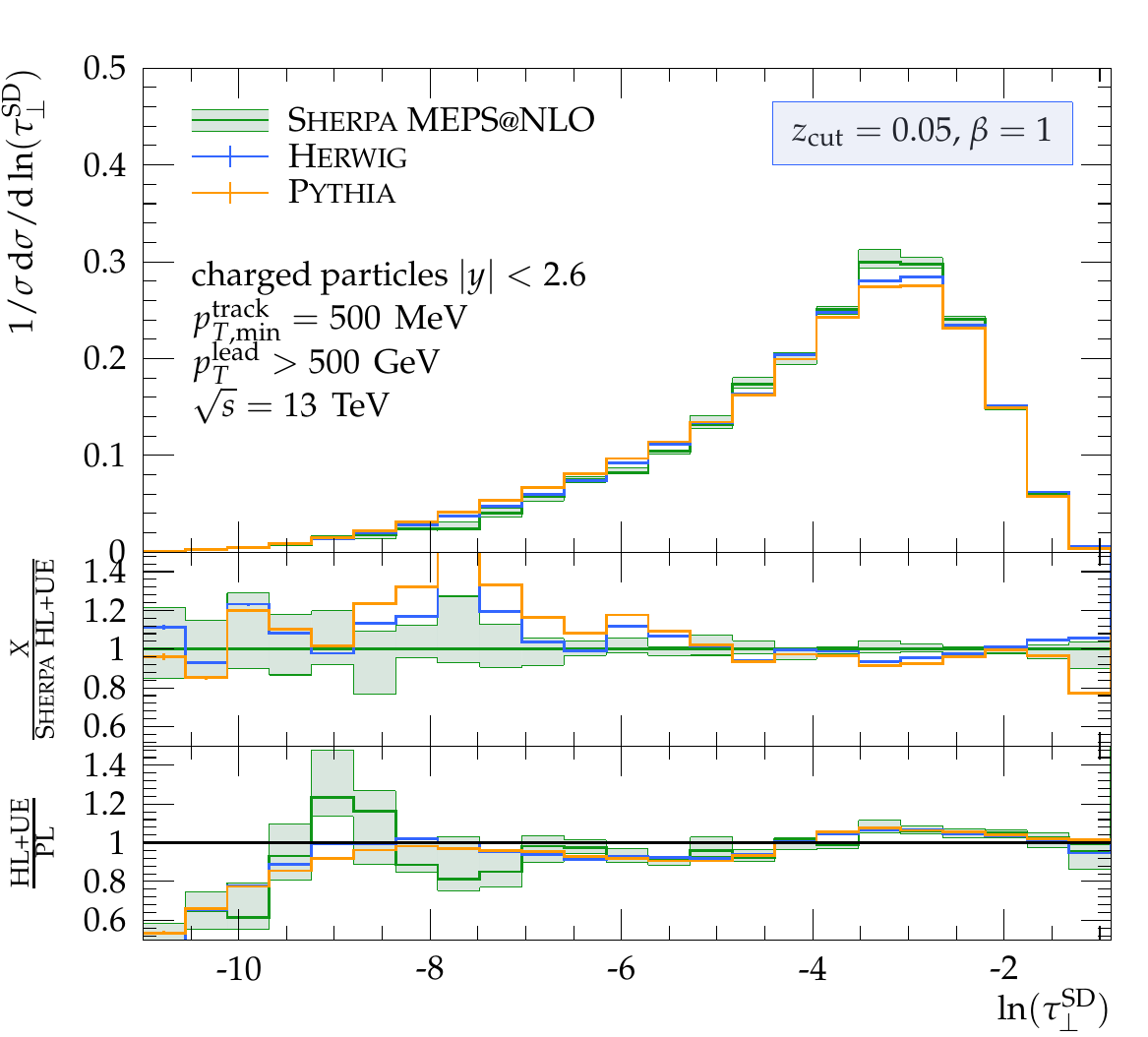}~
	\includegraphics[width=0.32\textwidth]{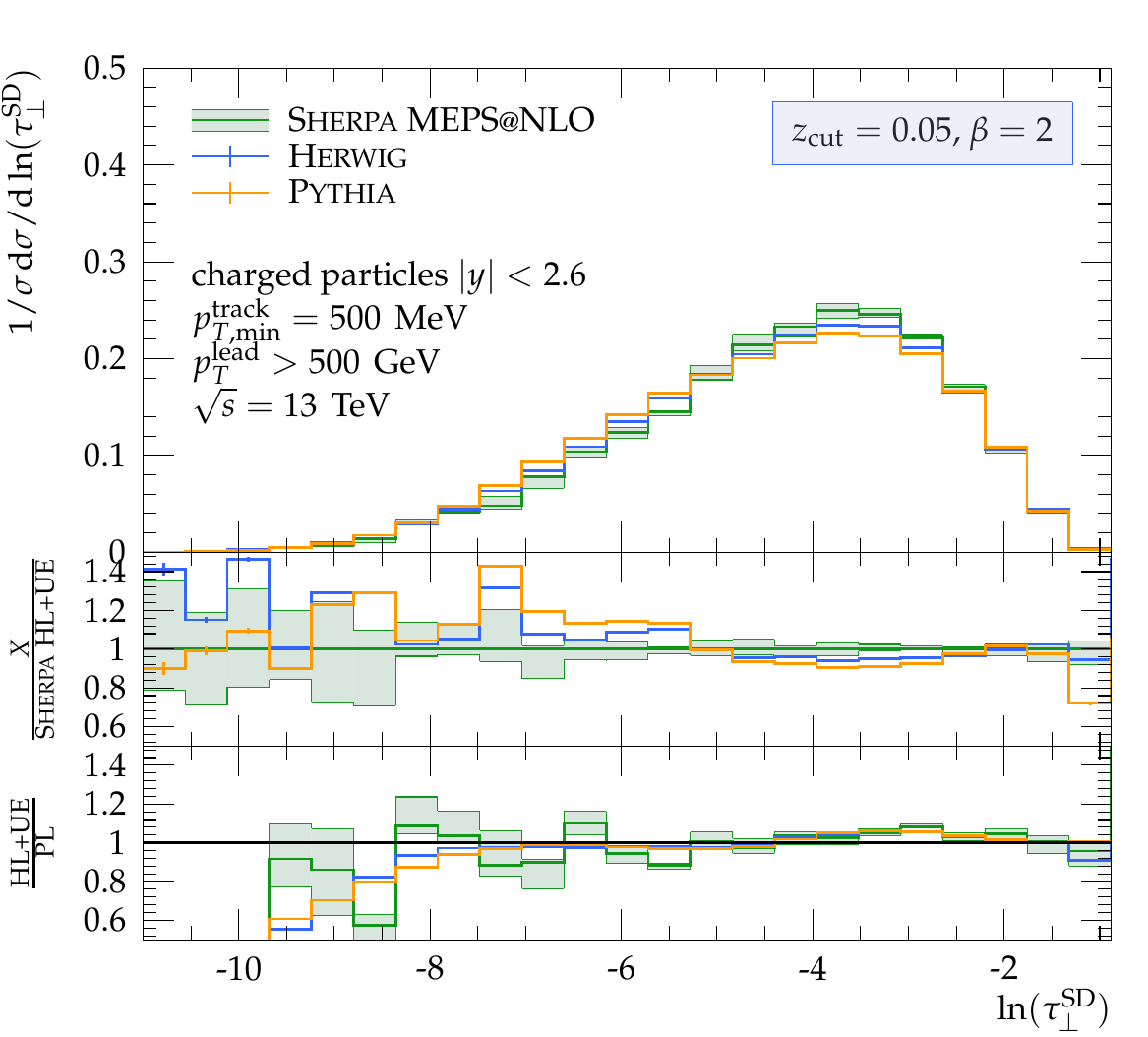}\\
	\includegraphics[width=0.32\textwidth]{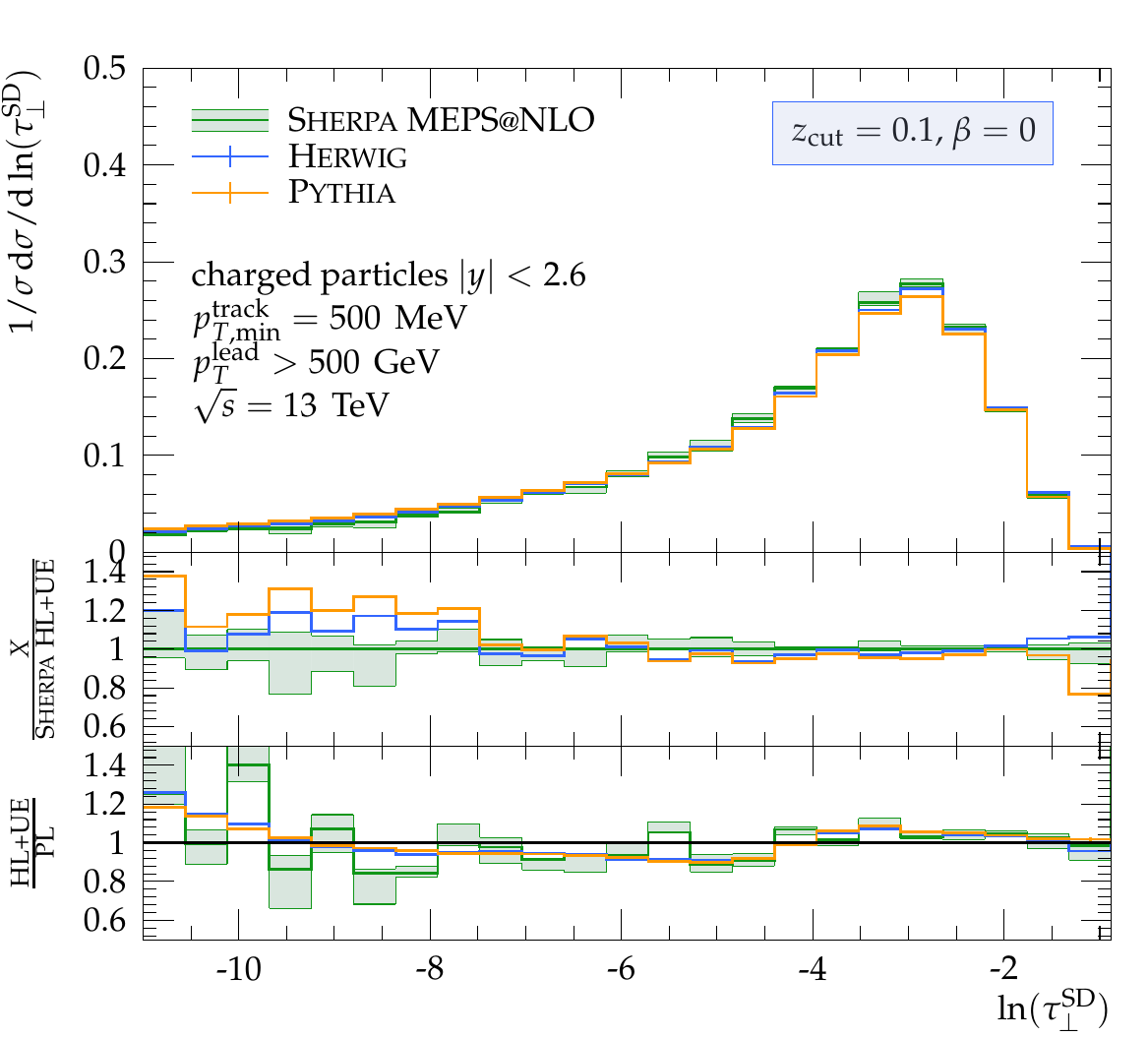}~
	\includegraphics[width=0.32\textwidth]{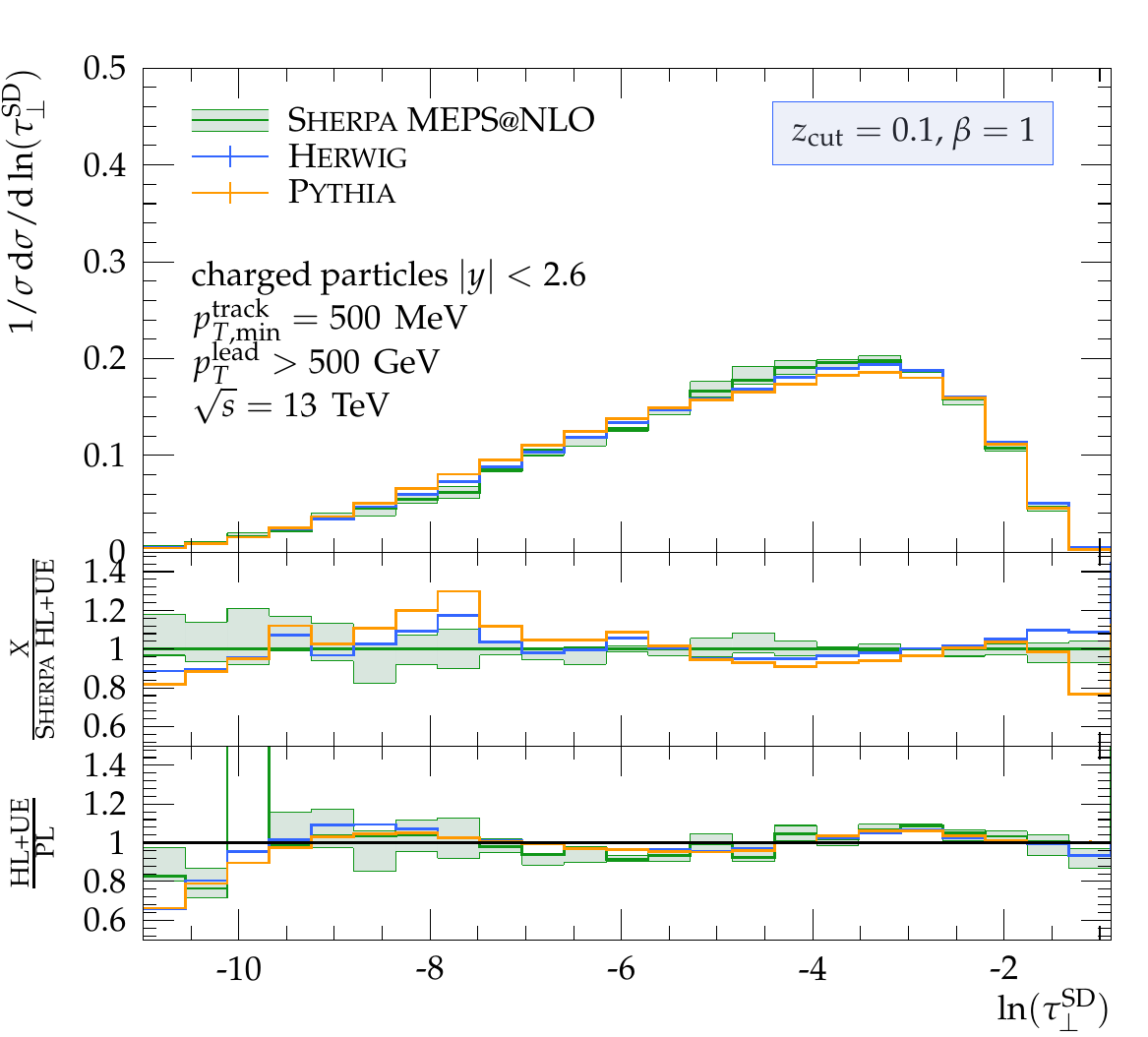}~
	\includegraphics[width=0.32\textwidth]{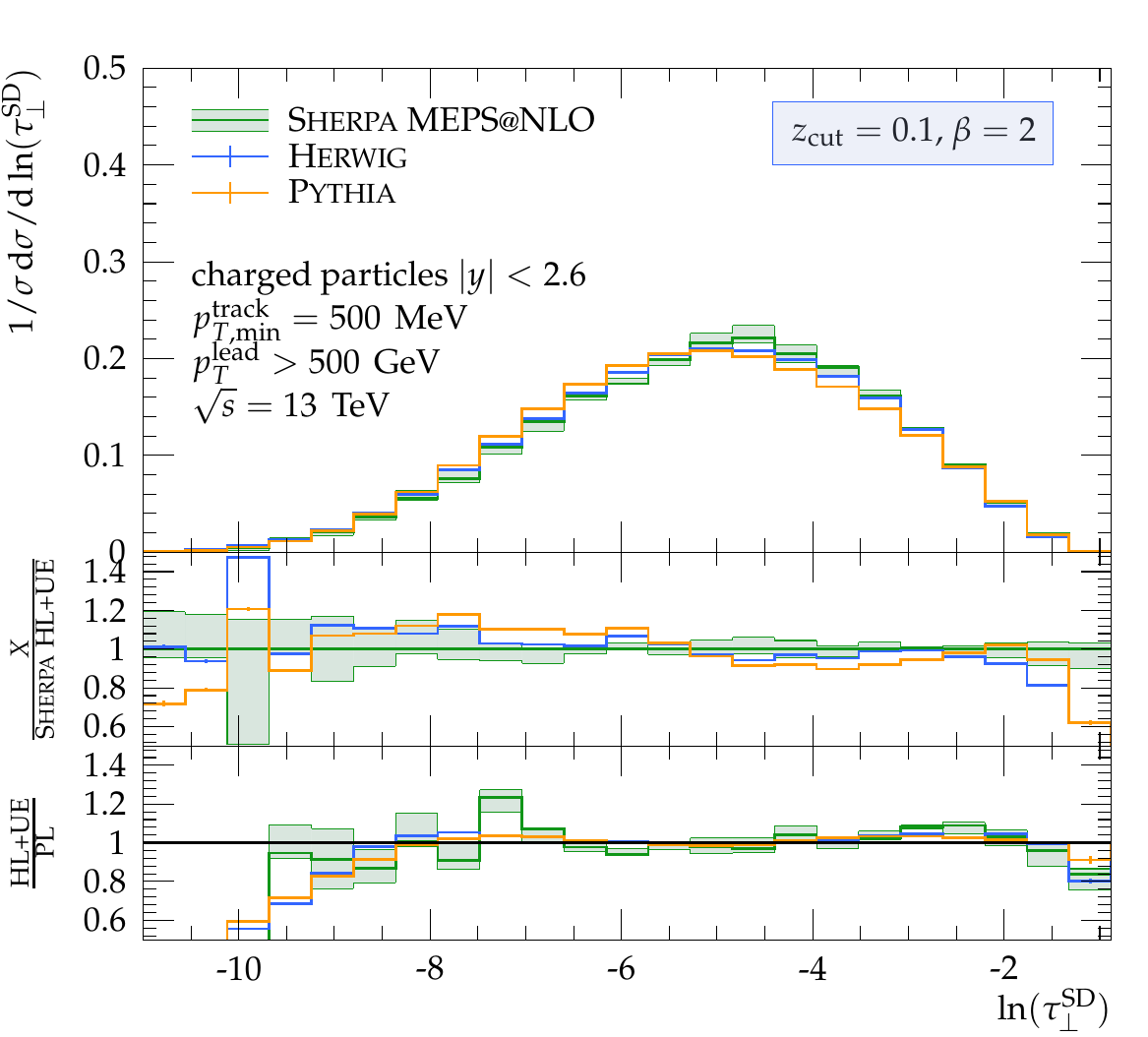}\\
\end{center}
\caption{Same as Fig.~\ref{fig:MC} but for the $p_{T,\text{min}}=500\;\text{GeV}$ event selection
  and $\zcut\in\{0.01, 0.02, 0.05, 0.1\}$. }
\label{fig:MC_500}
\end{figure}

In Figs.~\ref{fig:MC} and \ref{fig:MC_500} we compile the
corresponding predictions for the $p_{T,\text{min}}=200\;\text{GeV}$ and
$500\;\text{GeV}$ event selections, respectively. We consider the cases
$\beta\in\{0,1,2\}$ and $\zcut\in \{0.05,0.1,0.2,0.3\}$
($p_{T,\text{min}}=200\;\text{GeV}$) and $\zcut\in\{0.01,0.02,0.05,0.1\}$
($p_{T,\text{min}}=500\;\text{GeV}$). For each set of grooming
parameters $\zcut$ and $\beta$ we provide the ratios with respect to the
\Sherpa hadron-level prediction (first ratio panels), as well as to the
corresponding parton-level result of the respective generator, \emph{i.e.}\
$\frac{\text{HL+UE}}{\text{PL}}$ for each Monte Carlo (lower ratio
panels).

Overall we observe that the three generator predictions agree well
for the shape of the track-level thrust distribution for all considered
combinations of grooming parameters and both $p_{T,\text{min}}$ selections.
In particular in the peak region the differences observed for the hadron-level
predictions rarely exceed $10\%$. The results from \Herwig and \Pythia are
very similar and certainly consistent within LO uncertainties. However,
the \Sherpa \MEPSatNLO simulation, using exact NLO matrix elements, produces
somewhat harder emission spectra, resulting in more events with larger values
of $\tauSD$. Also in the low-$\tauSD$ tails the differences can get more
significant, in particular for $\beta>0$.

In the lowest panel of the plots we compare for each simulation the fully
hadronised prediction with the underlying event included against the corresponding
parton-level result. This allows us to extract the non-perturbative
corrections for each simulation and provides an estimate for the
generator-model dependence. The corrections in fact are very similar for
the three simulations. As observed before for the \Sherpa simulation,
non-perturbative corrections get significantly reduced through grooming
also for \Herwig and \Pythia. For $\zcut \geq 0.2$ for the $200\;\text{GeV}$
selection and $\zcut\geq 0.05$ for $p_{T,\text{min}}=500\;\text{GeV}$ they
stay around or below $10\%$ when we consider the region $-7<\ln(\tauSD)<-2$
for all three generators.  

The observations in these generator comparisons support the earlier observations.
They robustly confirm the finding that soft-drop grooming is very efficient
in removing contributions of non-perturbative phenomena from the events'
final state. We further conclude that for the event selections considered
here the remaining differences between the various generator predictions
are rather mild and mainly affect the tails of the distribution
where, however, the event rates are rather small. 

This highlights the potential of groomed transverse thrust for precision
QCD studies, using fixed- or all-orders predictions. For sufficient grooming
the remaining non-perturbative corrections are rather small and seemingly under
good control. By lowering the grooming parameters more underlying-event activity
can be blended in, providing valuable input to the validation and tuning of
phenomenological models in Monte Carlo event generators. 

\clearpage

\clearpage

\FloatBarrier
\clearpage

\section{Conclusions}\label{sec:conclusions}

In this article we considered soft-drop grooming final states of hadronic
collisions prior to the evaluation of QCD event-shape variables. This offers
great potential to largely remove final-state contributions originating from
the underlying event, enabling more direct comparisons of accurate theoretical
predictions with experimental data. 

To compile first accurate perturbative results, we have extended the
well-known \Caesar formalism for the resummation of NLL soft-gluon corrections
to groomed event-shape observables in hadronic collisions. We correspondingly
generalised the \Caesar implementation in the \Sherpa event-generator framework.
To take into account exact fixed-order corrections we worked out the matching
of the resummation up to NLO accurately split up into flavour channels. With this
at hand we provide the means to make \NLOpNLLp accurate predictions for a wide
range of hadronic event shapes, with and without grooming.

We focused the first application on groomed transverse thrust in inclusive
dijet production at the LHC. This generalisation of the transverse-thrust
variable is based on the division of the event final state into two
hemispheres that separately get soft-drop groomed. The remaining set of
final-state particles, dependent on the chosen set of soft-drop
parameters $\beta$ and $\zcut$, then enter the adapted observable evaluation.

We compared our \NLOpNLLp results against state-of-the-art parton-level
predictions based on NLO matrix-element improved parton-shower simulations
in the \MEPSatNLO formalism from the \Sherpa event generator. Both methods
confirm the feature that grooming shifts the thrust variable to lower values,
with larger grooming thresholds $\zcut$ resulting in larger shifts. While for
mild grooming we observe notable differences between the two theoretical
approaches, the more aggressive the grooming, the better agreement is found.

With the consistency of the two complementary approaches to
describe the observable established, we focused on the phenomenology of the
groomed thrust variable. In particular we explored the potential of soft-drop
grooming to reduce the impact of non-perturbative corrections on the thrust
event shape. By comparing Monte-Carlo predictions at parton level with and
without the underlying event included, as well as full hadron-level simulations,
we found enormous potential of soft-drop grooming to unmask the hard-process
final state from non-perturbative contributions. In fact, through sufficient
grooming the underlying event can almost entirely be removed, resulting in
non-perturbative corrections to the parton-level predictions of less than
$10\%$ for a wide range of the observable.  

Our initial theoretical considerations have been based on the analysis of all
final-state particles, \emph{i.e.}\ partons or charged \emph{and} neutral hadrons,
with rapidity $|y|<y_{\text{max}}$. While this can, after grooming,
directly be related to resummed predictions, a corresponding experimental
analysis would have to employ sophisticated particle-flow techniques.
Alternatively, the observable can experimentally be defined on charged-particle
tracks above a certain transverse-momentum threshold $p^{\text{track}}_{T,\text{min}}$.
Accordingly, we studied the impact of restricting the observable input to charged-tracks with
$p^{\text{track}}_{T,\text{min}}=500\;\text{MeV}$. We found that for groomed
thrust there remains a close correspondence between the track-based observable
and its all-particles variant. We furthermore validated our predictions by
comparing to results from \Herwig and \Pythia, that confirmed our findings with
regards to the reduction of non-perturbative corrections for groomed transverse
thrust.

\begin{figure}[ht!]
	\begin{center}
		\includegraphics[width=0.45\textwidth]{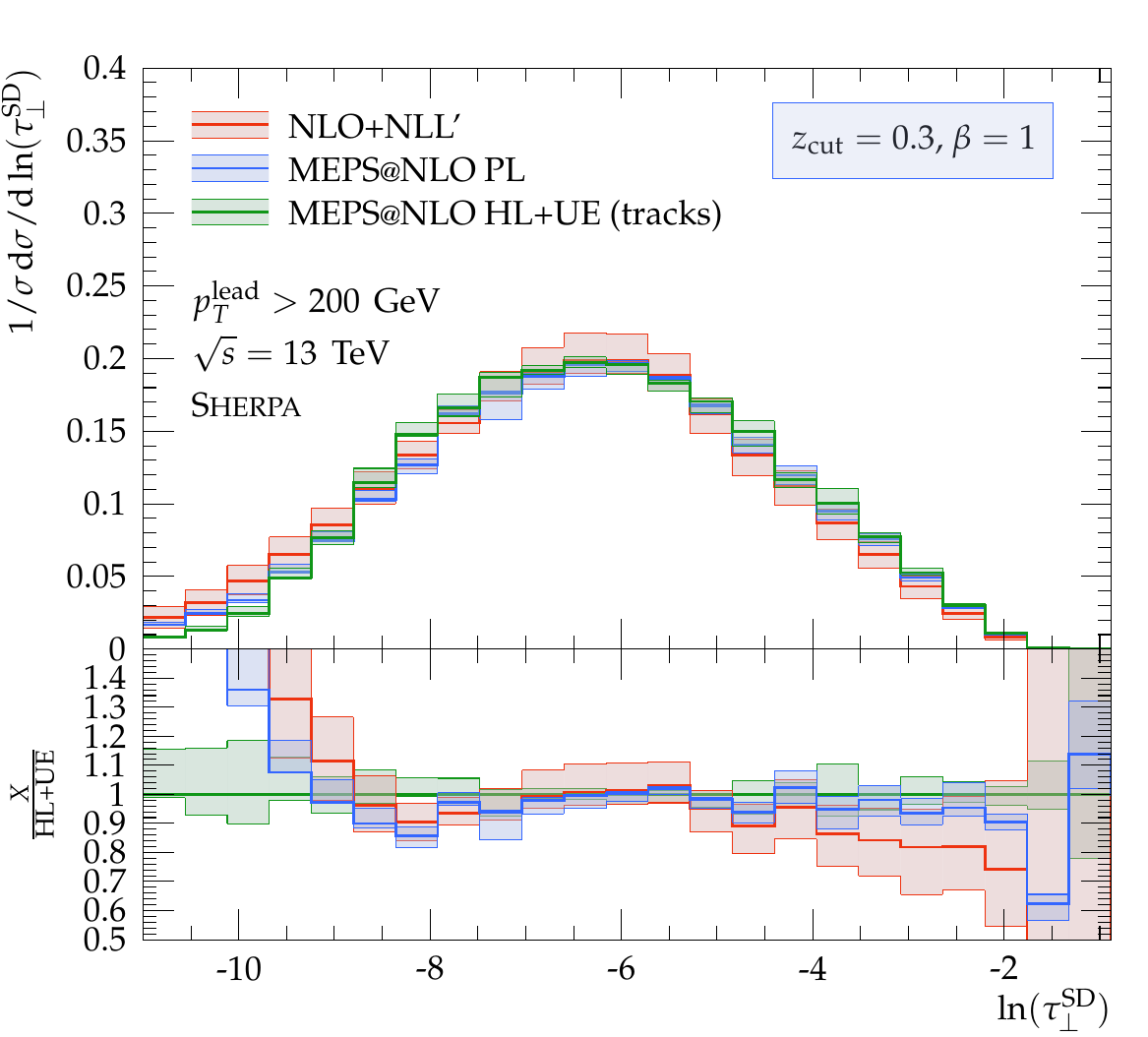}\qquad
		\includegraphics[width=0.45\textwidth]{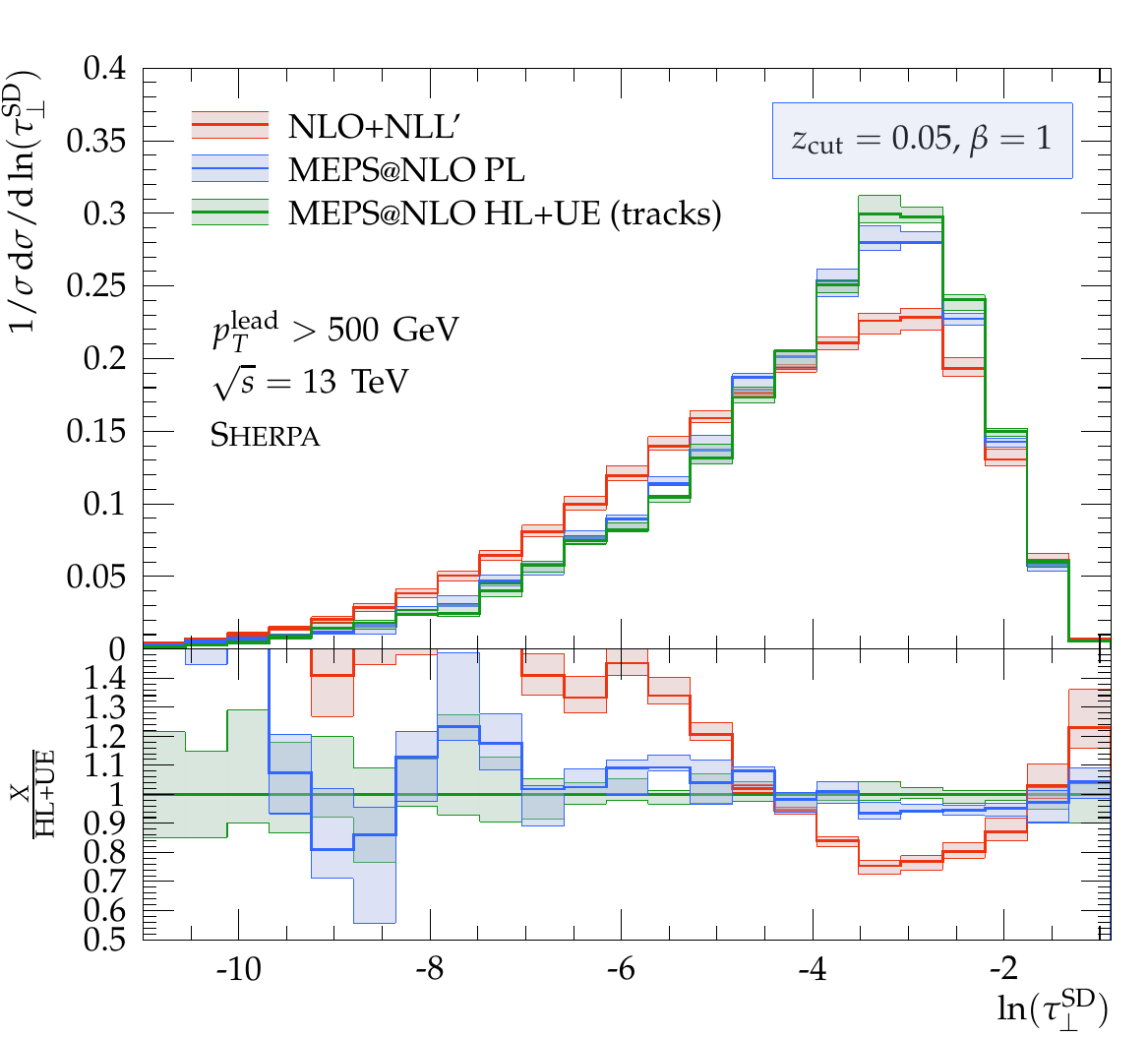}
	\end{center}
	\caption{The groomed transverse-thrust distributions for events with a leading-jet transverse
          momentum $p^{\text{lead}}_T > 200\;\text{GeV}$, with $\zcut=0.3$ and $\beta=1$ (left) and
          $p^{\text{lead}}_T > 500\;\text{GeV}$, with $\zcut=0.05$ and $\beta=1$ (right).
          Shown are results at \NLOpNLLp accuracy as well as \MEPSatNLO\ predictions obtained with \Sherpa at parton level (PL),
          full hadron level (HL+UE), the latter based on charged tracks with $p_{T}^{\text{track}}>500\;\text{MeV}$.
          In all cases we only include final-state particles with $|y|<2.6$ in the observable evaluation. 
          The lower panels show the ratios with respect to the HL+UE prediction.}
	\label{fig:final_groomed}
\end{figure}

In Fig.~\ref{fig:final_groomed} we summarise our main results by comparing
predictions at \NLOpNLLp with \MEPSatNLO simulations at parton level, without
the underlying event included, and at full hadron level. While the parton-level
simulation takes into account \emph{all} partons, the hadron-level prediction
is based on charged-particle tracks with $p^{\text{track}}_{T}>500\;\text{MeV}$.
We show representative results for our two dijet event selections.
For $p_{T,\text{min}}=200\;\text{GeV}$ we consider the case of rather hard
grooming using $\zcut=0.3$ and $\beta=1$. We find that indeed the
remaining non-perturbative corrections are significantly reduced over a wide
range of $\tauSD$. Furthermore, the \NLOpNLLp result nicely agrees with the
generator predictions. In the $p_{T,\text{min}}=500\;\text{GeV}$ example we
instead consider rather mild grooming with $\zcut=0.05$ and $\beta=1$. Even
for such low $\zcut$ the non-perturbative corrections are largely reduced,
in particular when comparing with the ungroomed case, see
Fig.~\ref{fig:NP_ungroomed}. A measurement could hence be sensitive to the
differences between perturbative predictions.

Our results indicate that potential future experimental measurements of
groomed event shapes offer a great opportunity for precision studies on perturbative
QCD, and possibly can differentiate and constrain theoretical approaches based on
fixed-order perturbation theory, all-orders resummation, or parton-shower
simulations. By varying the hardness of the selected dijet events or the grooming
parameters $\zcut$ and $\beta$ the observable offers detailed insights into the
transition away from the strict dijet limit. Complementary to this, such variations
offer means to tune and constrain the non-perturbative components of Monte Carlo
event generators.

Beyond the transverse-thrust shape considered here, grooming the event
final state before evaluating the actual observable can easily be applied
to other variables as well. Our derived generalisation of the \Caesar formalism
and its implementation in the resummation plugin to the \Sherpa framework
provides means for largely automated NLL resummation for a wide range of
hadronic event shapes. We note that this even applies to observables that
measure the deviation from more complex final states than the dijet
configuration, with simple modifications such as suitably generalising the
notion of splitting of the event into two hemispheres.

\clearpage
\section*{Acknowledgements}
JB would like to thank Simone Marzani for his exemplary supervising during his PhD thesis
and the MCnet collaboration for the opportunity of spending a short-term studentship at G\"ottingen University.
This work has received funding from the European Union's Horizon 2020 research and innovation
programme as part of the Marie Sk\l{}odowska-Curie Innovative Training Network MCnetITN3
(grant agreement no. 722104). SS acknowledges support through the Fulbright-Cottrell Award
and from BMBF (contract 05H18MGCA1). DR acknowledges support
during early stages of this work from the German-American Fulbright Commission,
allowing him to stay at Fermi National Accelerator Laboratory (Fermilab), a
U.S. Department of Energy, Office of Science, HEP User Facility. Fermilab is
managed by Fermi Research Alliance, LLC (FRA), acting under Contract No.\
DE--AC02--07CH11359.

\appendix

\section{\protect{Logarithmic contributions of \boldmath{$\zcut$}}}\label{sec:logzc}

Though we do not perform the resummation of logarithms of $\zcut$ we can outline a method
to treat them. First we will elaborate on initial-state radiation, where we follow the
same approach as for final-state radiation discussed in Sec.~\ref{sec:SD-NLL}.
In addition, the treatment for soft contributions and PDF ratios will be discussed, neither of which
contribute at LL and therefore offer some freedom in their treatment at \NLL accuracy.

\subsubsection*{Initial-State Radiators}

\begin{figure}
	\begin{center}
		\includegraphics[width=0.45\textwidth]{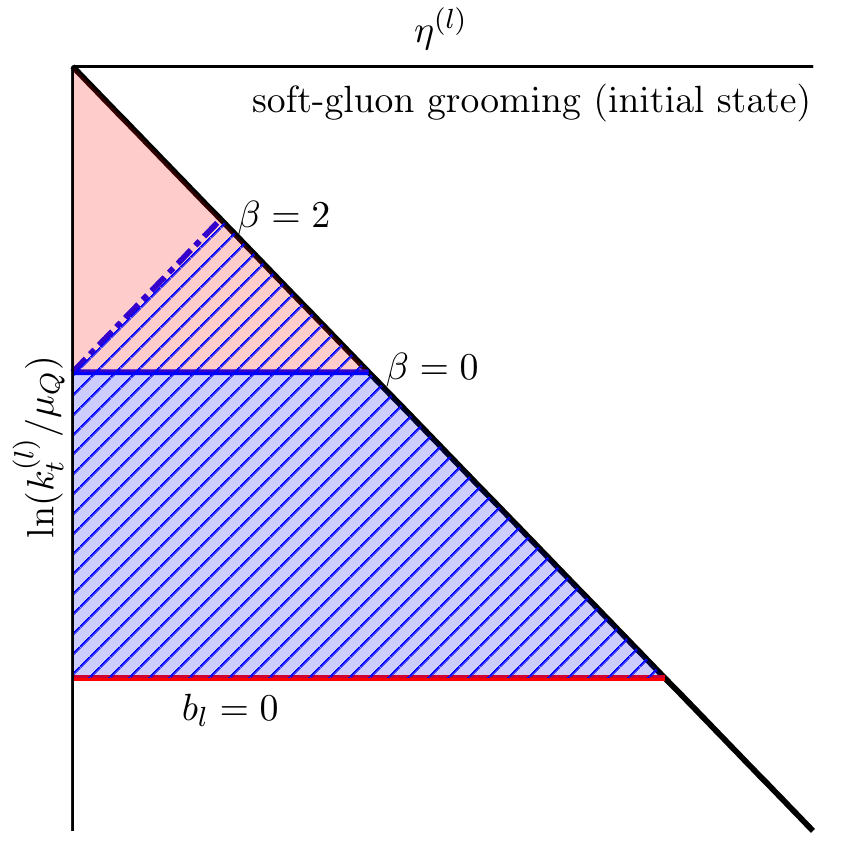}
	\end{center}
	\caption{The Lund diagram showing the kinematic constraints for soft drop applied to initial-state
          emissions. As examples $a_l=1$ with $b_l=0$ and $\beta=0$ (solid line and area) and $\beta=2$
          (dashed line, hatched area) are shown.}
	\label{fig:Lund-In}
\end{figure}

For initial-state emissions the same soft-drop condition is used, however
the criterion is here taken with respect to the nearest final-state
leg in the transverse plane. Accordingly, one can derive 
\begin{eqnarray}
p_{T}^{\left(i\right)}  =k_{t}^{(l)}\,,&& p_{T}^{\left(j\right)} =E_{f}\sin\theta\,,\\
\left|\Delta y\right| =\left|\eta^{\left(l\right)}- y_f^{\left(l\right)}\right|\,,&&\left|\Delta\phi\right| =\phi\,,
\end{eqnarray}
where $E_f$ and $ y_f^{\left(l\right)}$ denote energy and rapidity of the closest
final-state leg to which the emission is clustered. This results in the soft-drop condition:

\begin{equation}
\frac{k_{t}^{(l)}}{E_{f}\sin\theta}>\zcut\left(\frac{\left(\eta^{\left(l\right)}- y_f^{\left(l\right)}\right)^{2}+\phi^{2}}{R_{\text{SD}}^{2}}\right)^{\beta/2}\,,
\end{equation}
which in the limit of large $\eta^{\left(l\right)}$ results in
\begin{equation}
\frac{k_{t}^{(l)}}{E_{f}\sin\theta}>\zcut\left(\frac{\eta^{\left(l\right)}}{R_{\text{SD}}}\right)^{\beta}\,.
\end{equation}
This provides a relation that directly depends on $\eta^{\left(l\right)}$, whereas all other
conditions depend on $e^{\eta^{\left(l\right)}}$ only. Accordingly, we are faced
with a set of conditions that will not result in simple integration boundaries
in $\ln k_t^{(l)}$ and $\eta^{(l)}$ simultaneously. We do not calculate the
integral associated with it, but note again that from Fig.~\ref{fig:Lund-In} it
is clear that it will not result in any logarithms of $v_\text{SD}$ for small enough $v_\text{SD}$,
hence does not contribute at our target accuracy.

In order to get a better idea of the structure for these contributions an alternative distance measure can be used, namely
$\Delta R_{ij}^{2}=2\cosh\left(y_{i}-y_{j}\right)-2\cos\left(\phi_{i}-\phi_{j}\right)$. This definition corresponds to the canonically
used relation in the limit of small angular difference. The soft-drop condition then reads
\begin{equation}
\frac{k_{t}^{(l)}}{E_{f}\sin\theta}>\zcut\left(\frac{2\cosh\left(\eta^{\left(l\right)}- y_f^{\left(l\right)}\right)-2\cos\phi}{R_{\text{SD}}^{2}}\right)^{\beta/2}\,,
\end{equation}
which in the large-$\eta^{\left(l\right)}$ limit instead reduces to 
\begin{equation}
\frac{k_{t}^{(l)}}{E_{f}\sin\theta}>\zcut\left(\frac{e^{\eta^{\left(l\right)}}e^{- y_f^{\left(l\right)}}}{R_{\text{SD}}^{2}}\right)^{\beta/2}=\frac{2}{\sin\theta}\bar{z}_{\text{cut}}e^{\beta\eta^{\left(l\right)}/2}\,,
\end{equation}
where we define $\bar{z}_{\text{cut}}$ such that it absorbs the dependence on $\theta$
and $R_{\text{SD}}$. This condition scales with $\eta^{\left(l\right)}$
in the opposite manner as the final-state contribution, \emph{cf.}\ condition
(iv) in Sec.~\ref{sec:SD-NLL}, and results in an increase in grooming when
$\beta$ is increased, unless $\eta^{\left(l\right)}\approx
y_f^{\left(l\right)}$. The result makes sense, since for 
$\eta^{\left(l\right)}\approx y_f^{\left(l\right)}$ the emission is collinear to the
final-state leg, but otherwise the value of $\Delta R$ becomes large and results
in an enhancement of grooming due to an increase in $\beta$. This can also be
seen in the Lund diagram for this condition depicted in Fig.~\ref{fig:Lund-In}.

The transition point does not work the same as for final-state legs,
in fact in general there exist two transition points. Below the first
transition there are only logarithms of $\zcut$, in between the two there
is a mix of logarithms of $v$ and $\zcut$ and finally above the
second transition point there are only logarithms of the observable.
Here the two transition points are given by $\bar{z}_{\text{cut}}^{a_{l}},\quad\bar{z}_{\text{cut}}^{\frac{2\left(a_{l}+b_{l}\right)}{2+\beta}}$. 
Whichever of the two has smaller value defines the first transition
point, the other consequently the second one. This does mean that for
$2b_{l}=a_{l}\beta$ these two transition points are identical. For transverse
thrust the \Caesar parametrisation for initial-state emissions yields $b_{l}=0$,
and accordingly, there is only a single transition point when $\beta = 0$. 

This can be illustrated by the Lund diagram in Fig.~\ref{fig:Lund-In}.
As the observable line moves up and the value of the observable increases
it will eventually cross the soft-drop line leading to the transition.
When $2b_{l}=a_{l}\beta$ the observable and soft-drop bounds are parallel
and there will be only a single transition. However, for the $\beta=2$
line we can see that when the observable crosses the left-hand side of
the soft-drop bound both conditions contribute to the integral. Once
it passes the right-hand side of the soft-drop boundary the observable
is the only limit that matters. 

In a similar way as was explained for final-state legs in Sec.~\ref{sec:matching_nllp}
we can shift the transition point by factors and introduce a compensating term,
while maintaining \NLL accuracy. This means we can choose to shift the
$\bar{z}_{\text{cut}}^{a_{l}}$ transition point to be equal to the final-state-leg
transition point. However, the second type of transition point cannot be made
equal as it depends on a different power of $\zcut$.

\subsubsection*{Soft Emissions and PDF Contributions}

The argument of logarithms from soft wide-angle emissions is directly related to
the $k_t/\mu_Q$ scaling of the lower left-hand corner of a Lund diagram.
For both initial- and final-state emissions, in the limit $v\ll\zcut\ll1$, this
is $\zcut$ up to overall factors, which are beyond \NLL accuracy. When the distribution
transitions for larger values of the observable the scaling changes to the ungroomed
case $v^{\frac{1}{a_{l}}}$. This allows us to shift the argument of the logarithms and
the transition point for these types of emissions at \NLL accuracy to $\zcut^{a_{l}}$.
When extending this argument beyond \NLL accuracy a separate treatment for initial- and
final-state emissions will be needed for the soft function ${\cal{S}}$.

The scale for the ratio of the pdfs also first enters at \NLL accuracy.
These contributions are linked to the initial-state collinear scaling.
This $k_t$ scale can be found at the right-hand corner of the Lund diagram,
which results in a scaling $\zcut^{\frac{2}{2+\beta}}$. As for the soft function,
for larger values of the observable this contribution transitions back to the
ungroomed case, which scales as $v^{\frac{1}{a_{l}+b_{l}}}$. This would place the
transition point at $\zcut^{\frac{2\left(a_l+b_l\right)}{2+\beta}}$ similar to the second
transition for initial-state emissions.

\FloatBarrier

\section{Auxiliary results}\label{app:aux_results}

\begin{figure}[h]
	\begin{center}
		\includegraphics[width=0.32\textwidth]{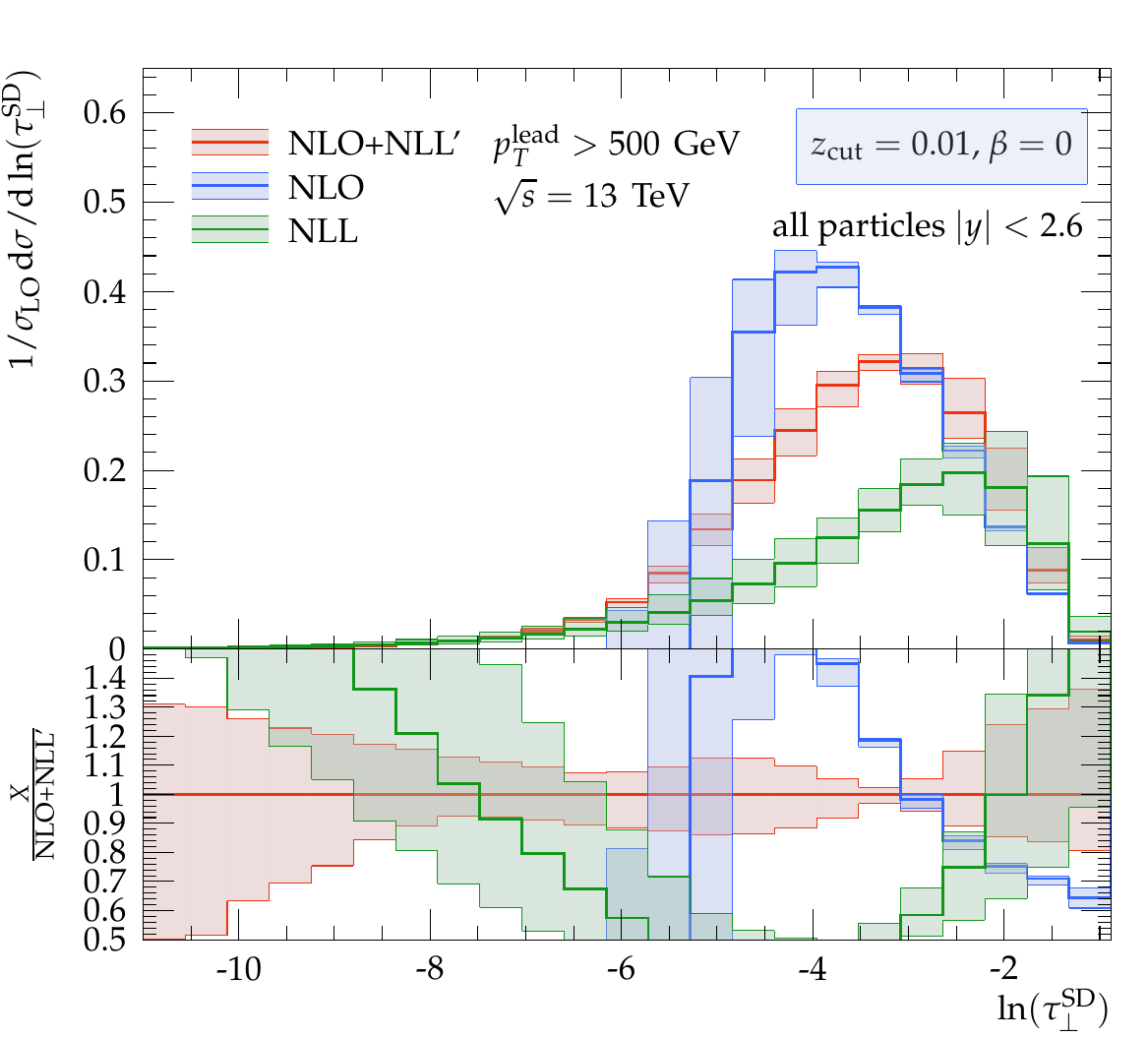}~
		\includegraphics[width=0.32\textwidth]{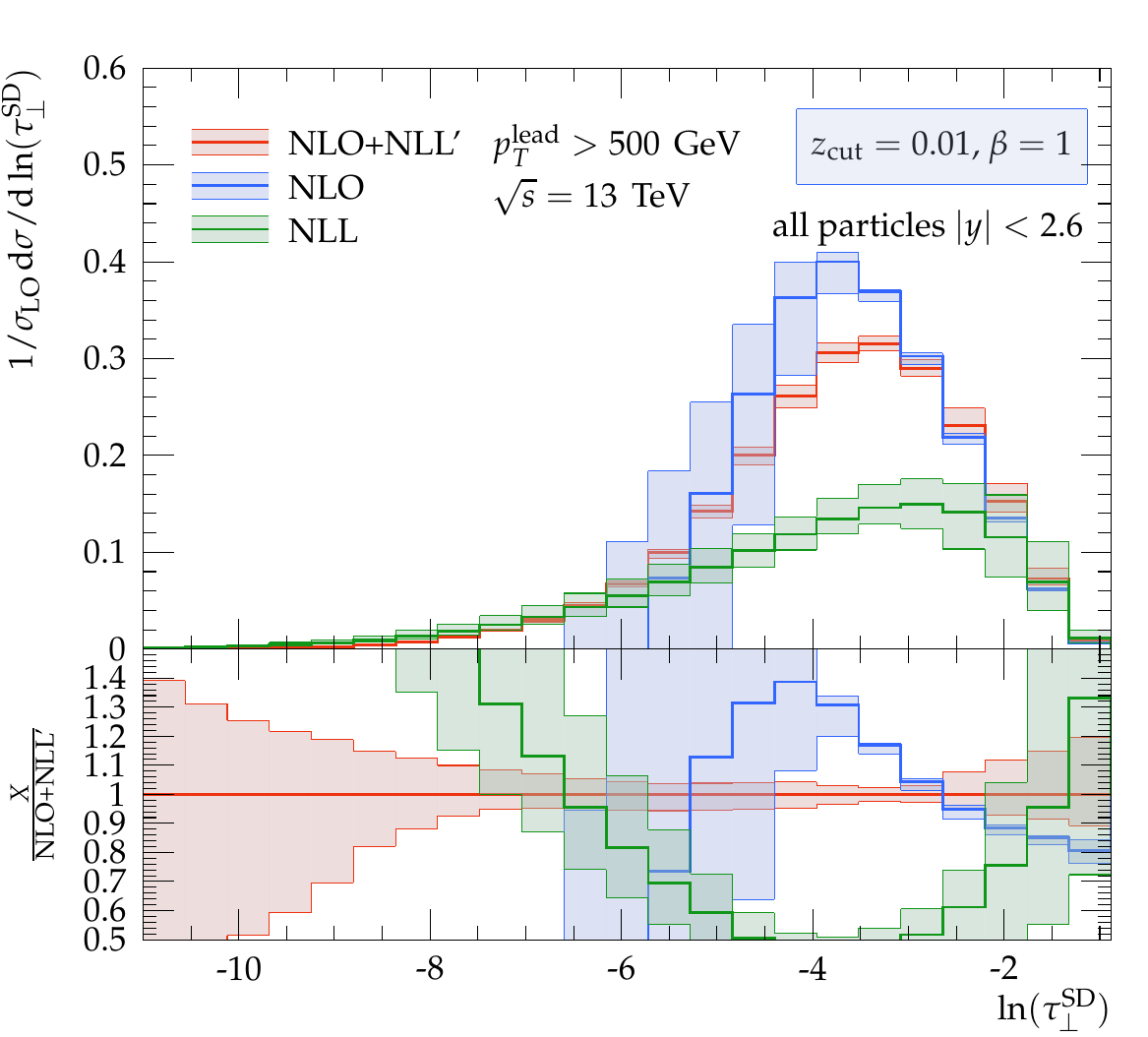}~
		\includegraphics[width=0.32\textwidth]{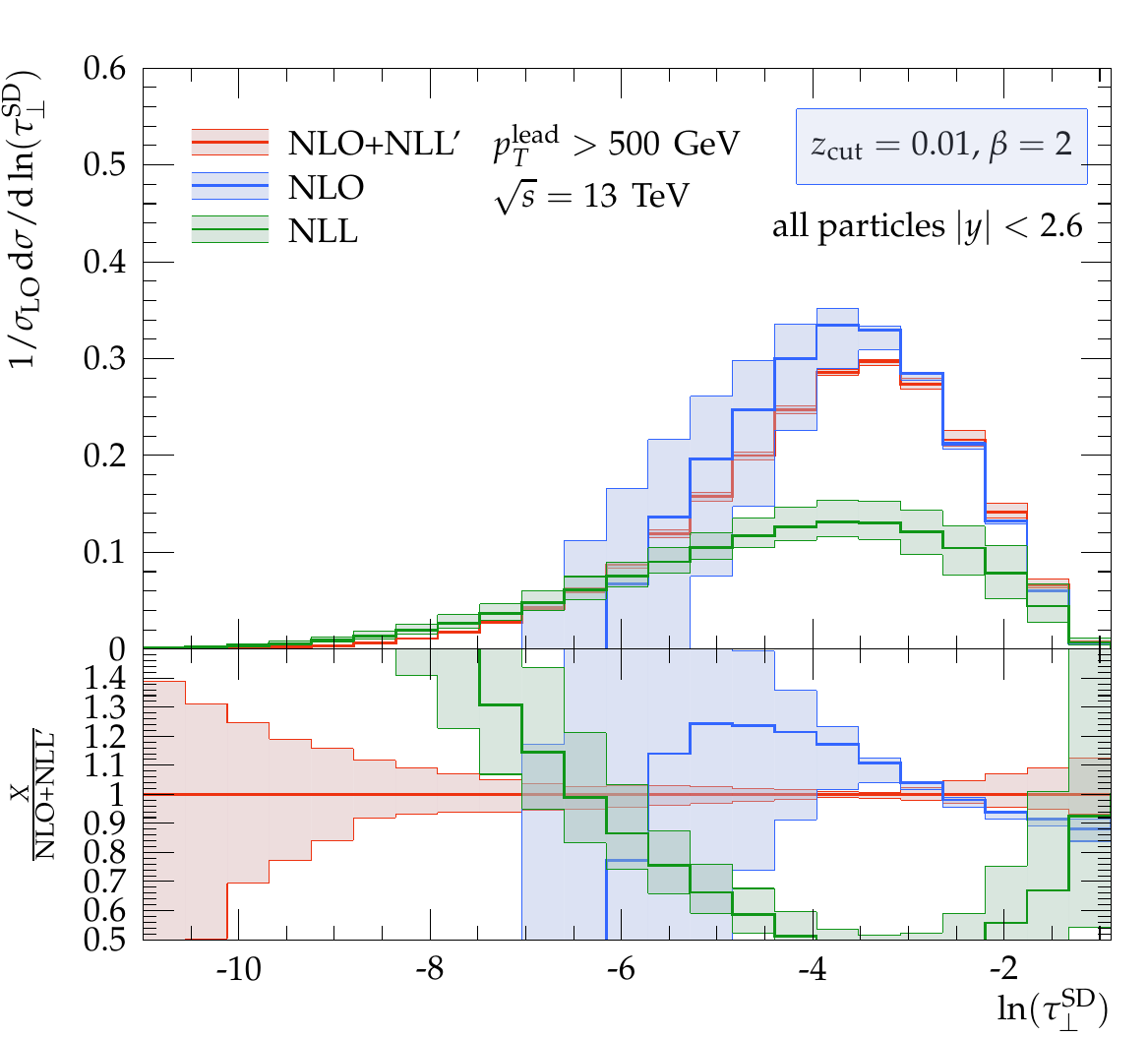}\\
		\includegraphics[width=0.32\textwidth]{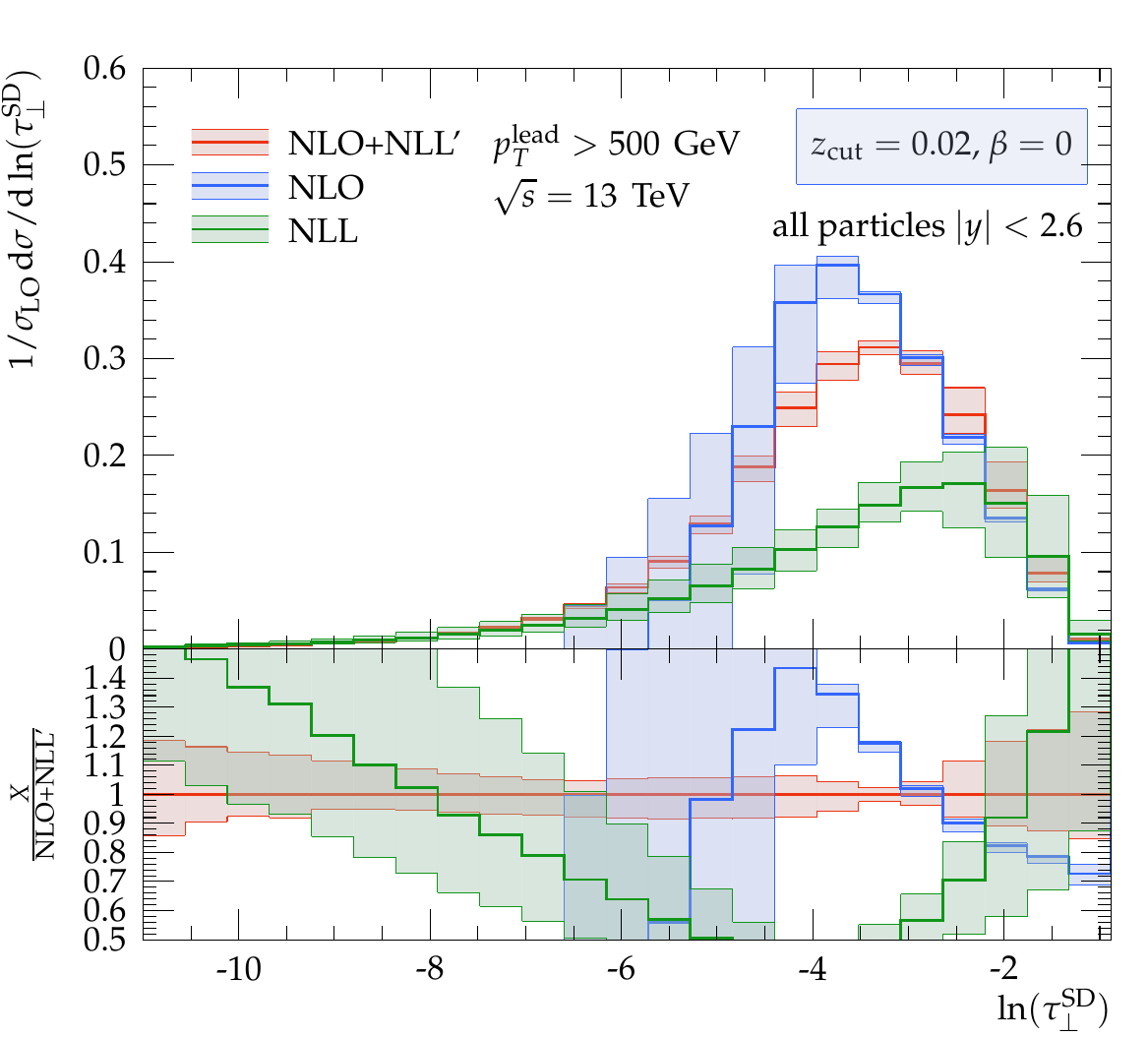}~
		\includegraphics[width=0.32\textwidth]{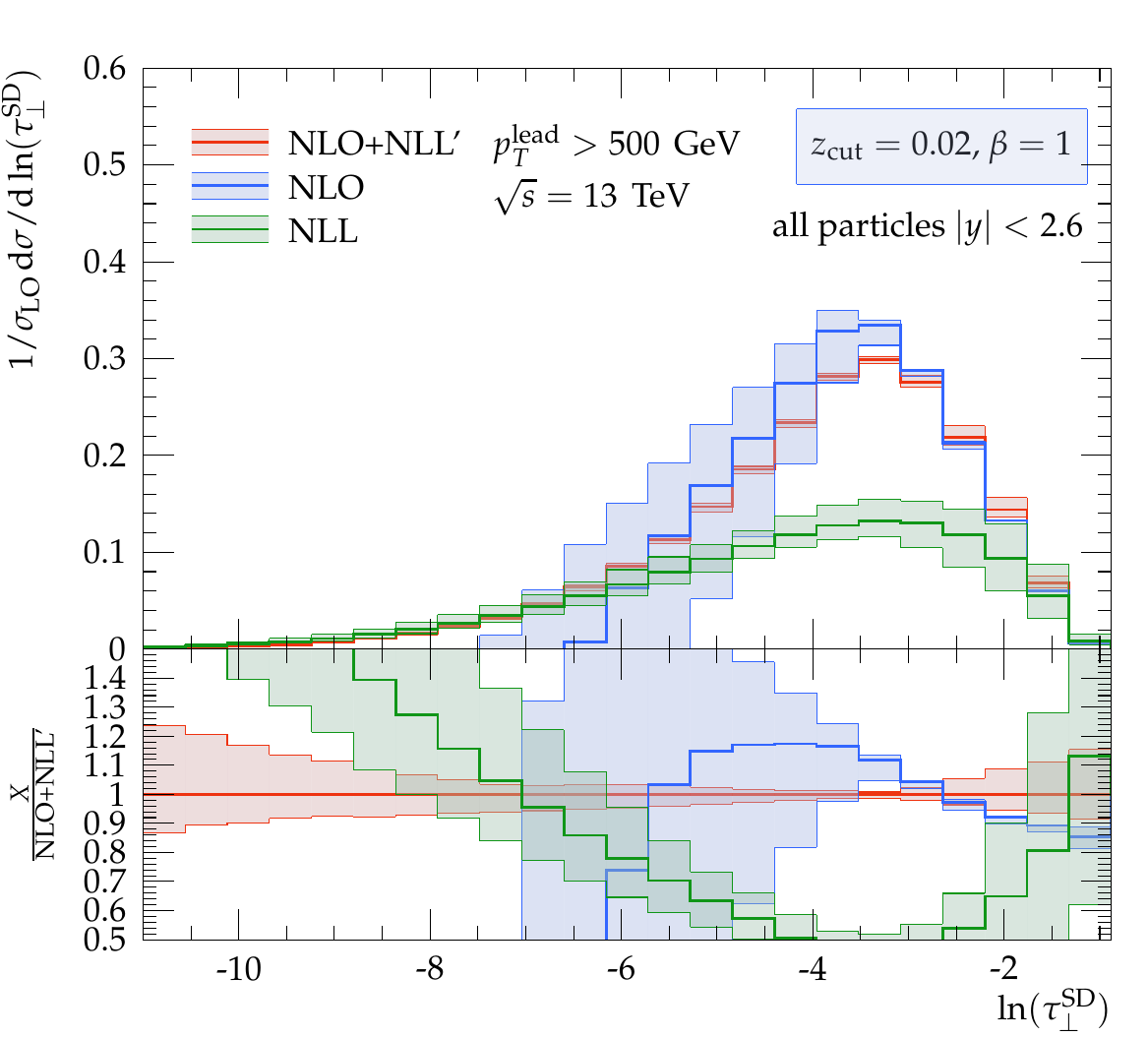}~
		\includegraphics[width=0.32\textwidth]{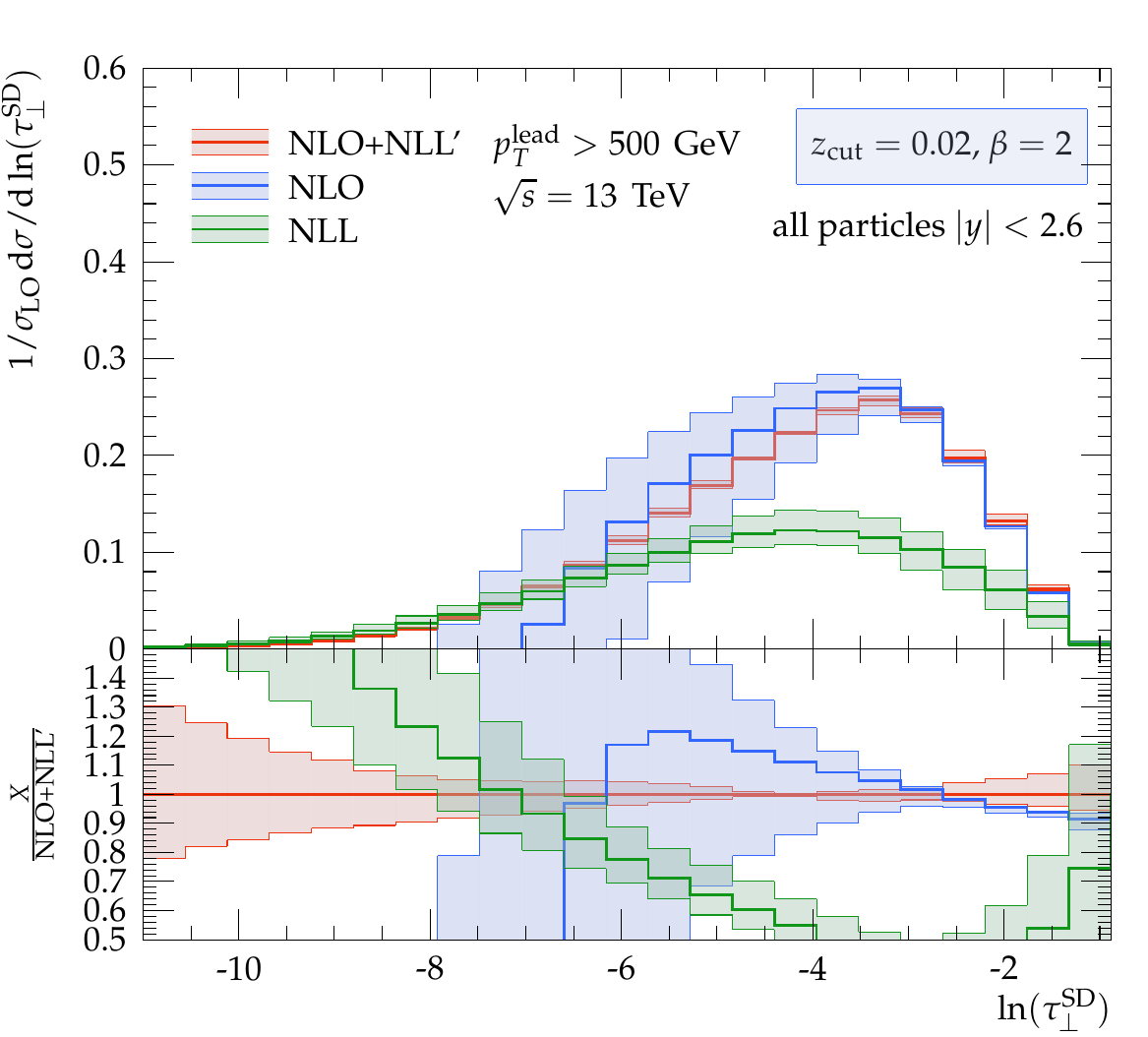}\\
		\includegraphics[width=0.32\textwidth]{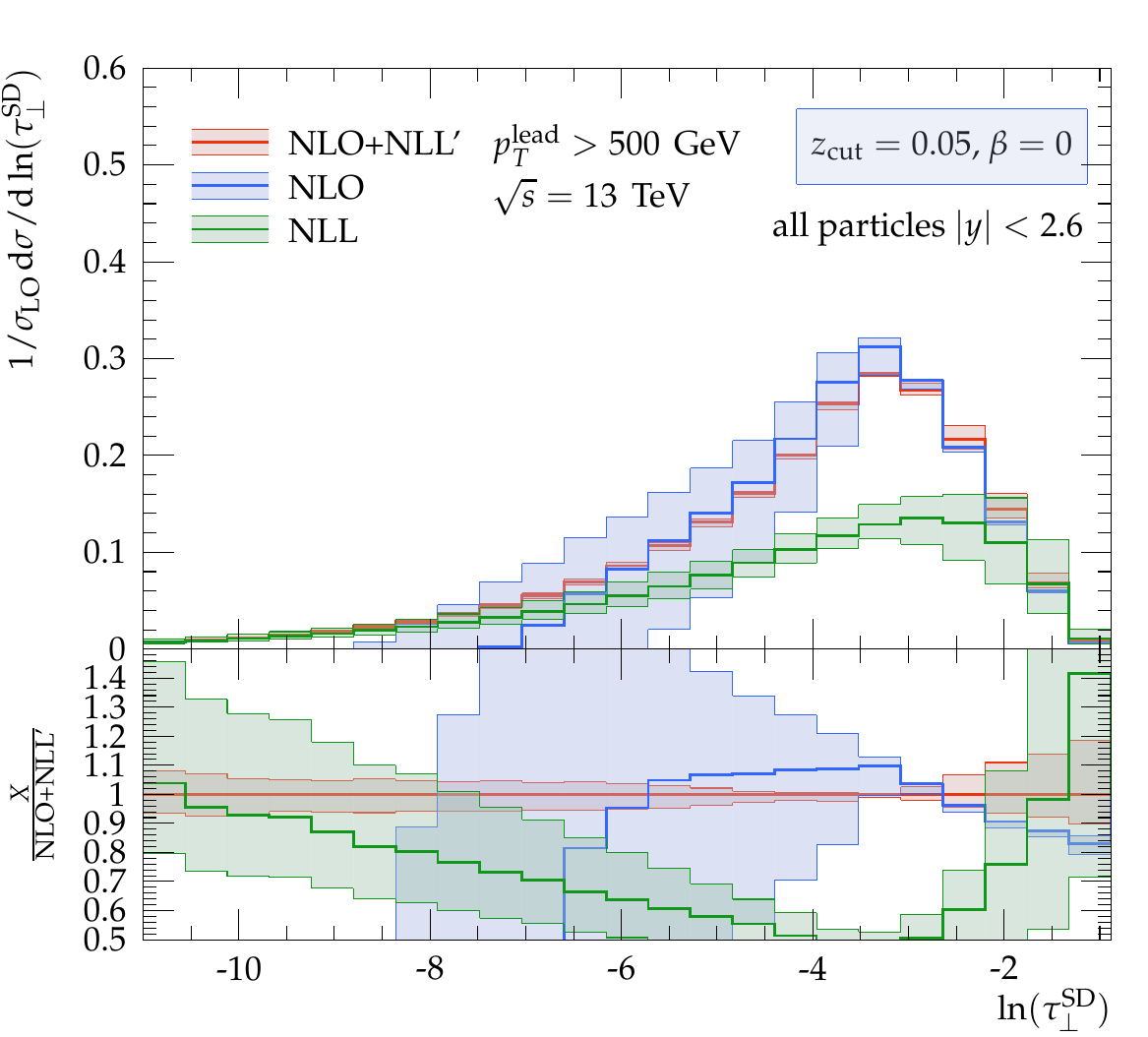}~
		\includegraphics[width=0.32\textwidth]{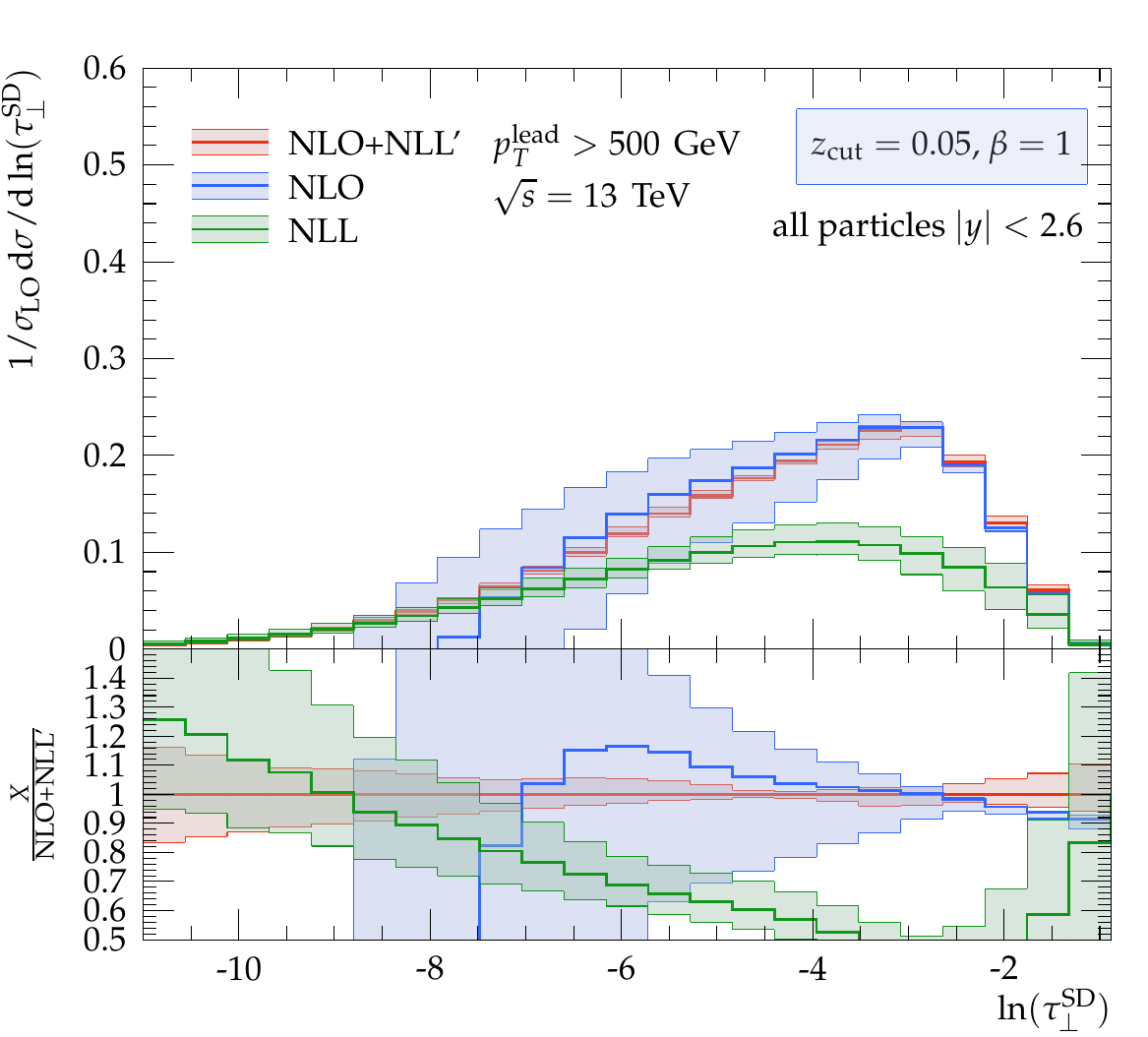}~
		\includegraphics[width=0.32\textwidth]{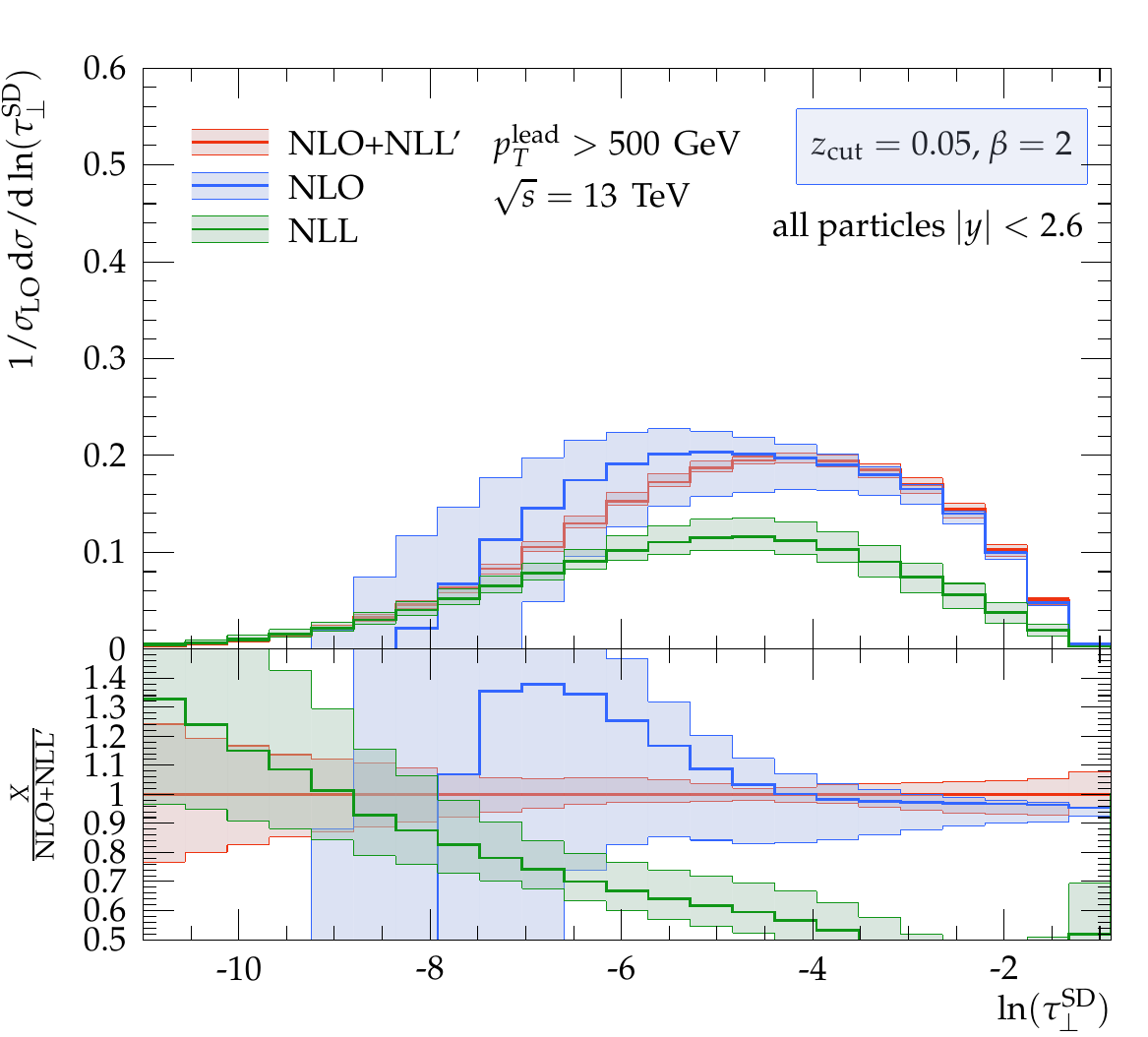}
	\end{center}
	\caption{NLL resummed predictions matched to LO and \NLO for groomed transverse thrust for
          $\beta\in\{0,1,2\}$ (columns) and $\zcut\in\{0.01, 0.02, 0.05\}$ (rows) for the
          $p_{T,\text{min}}=500\;\text{GeV}$ event selection, \emph{cf.}\ Fig.~\ref{fig:Resum} for
          details.}
	\label{fig:Resum_500_aux}
\end{figure}

\begin{figure}[hb!]
	\begin{center}
		\includegraphics[width=0.32\textwidth]{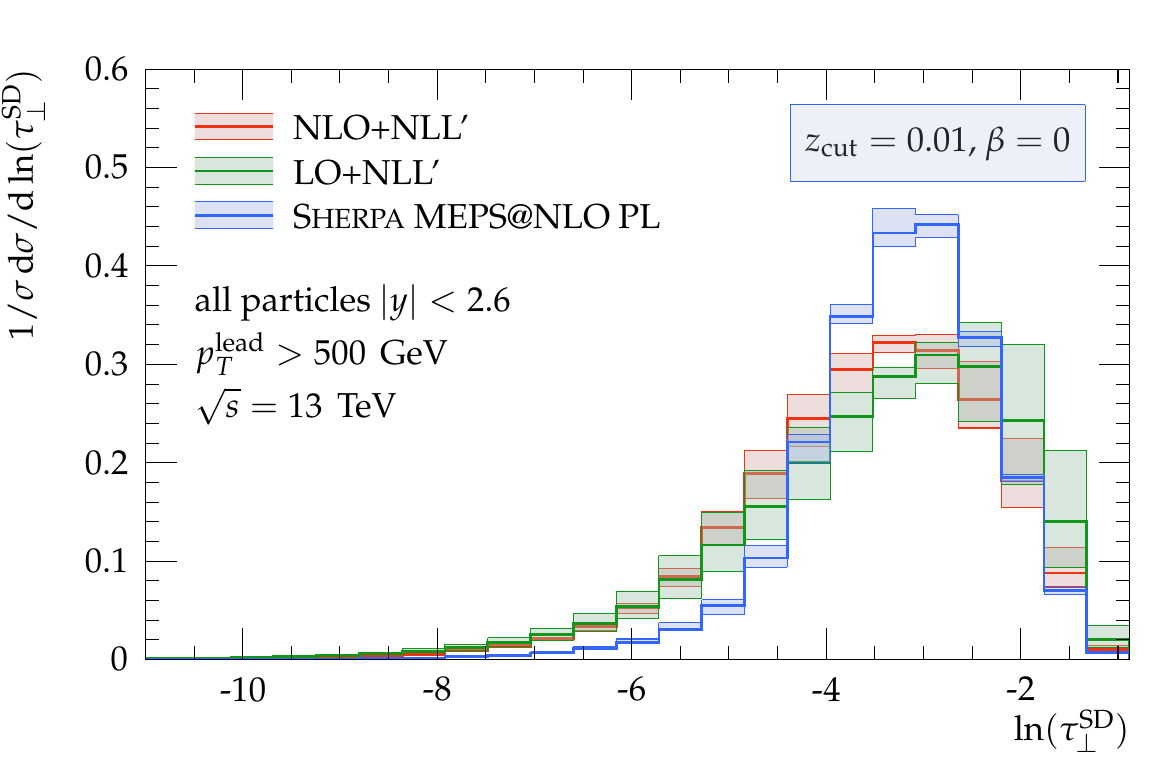}~
		\includegraphics[width=0.32\textwidth]{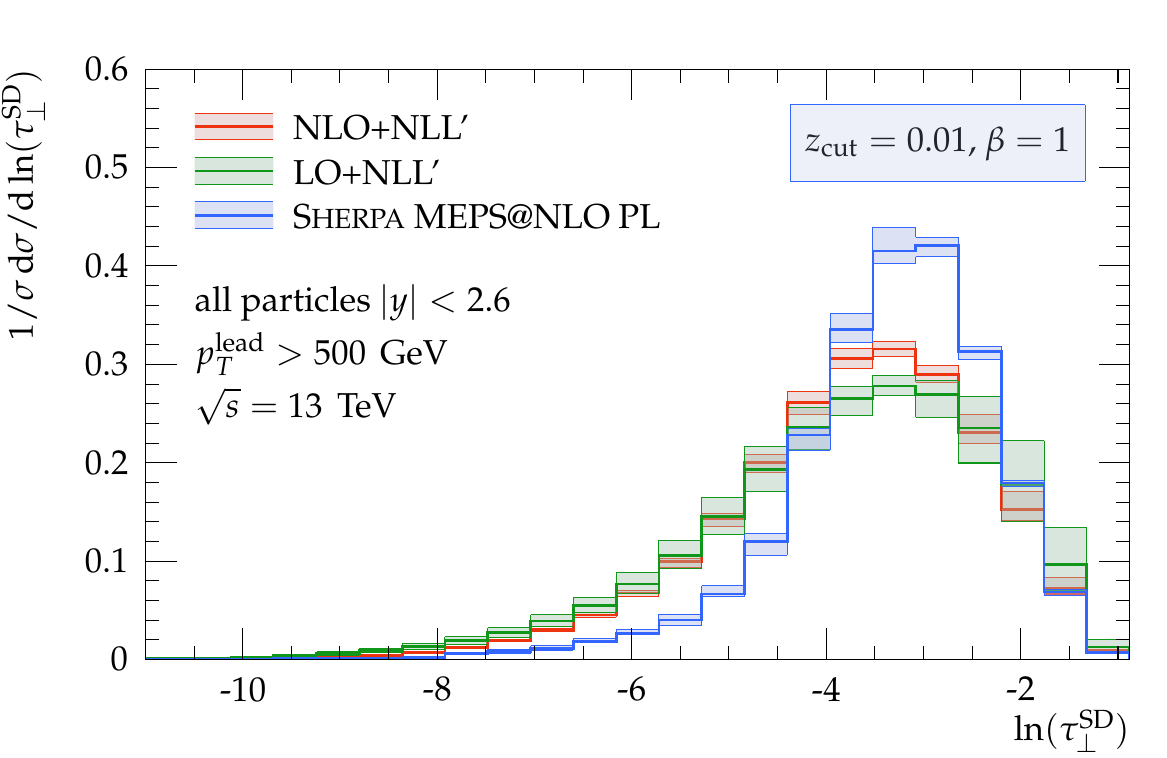}~
		\includegraphics[width=0.32\textwidth]{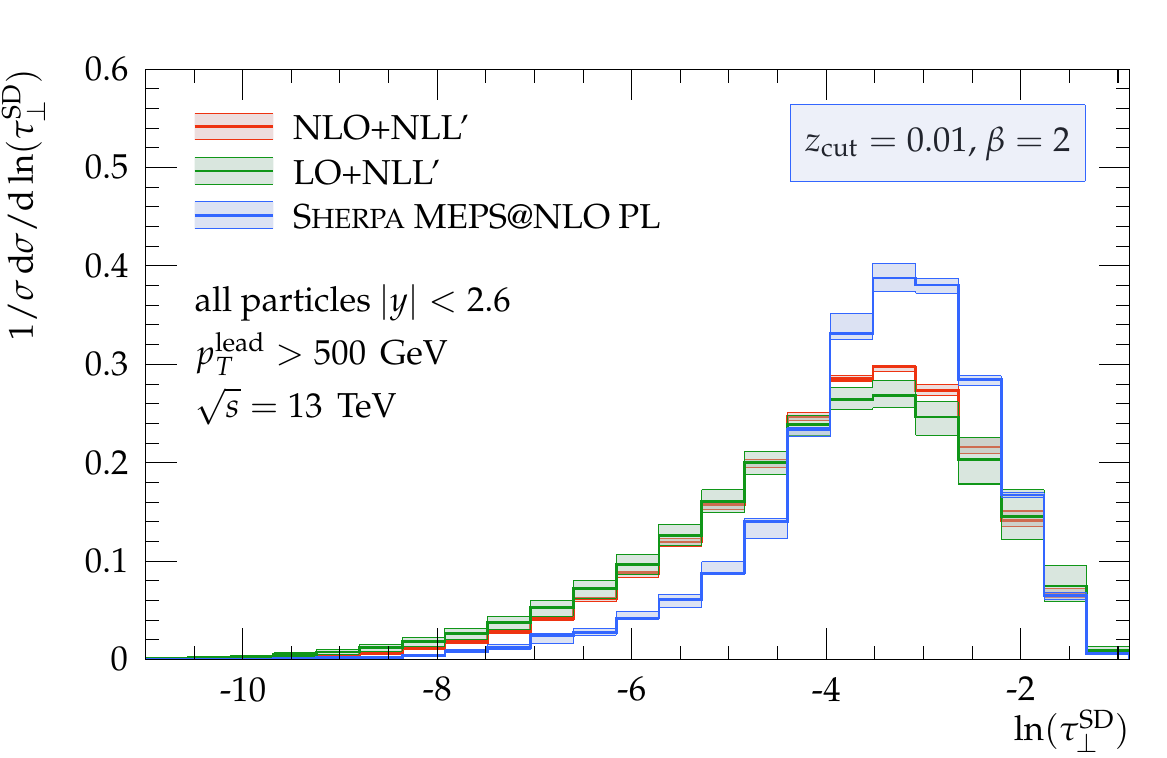}\\
		\includegraphics[width=0.32\textwidth]{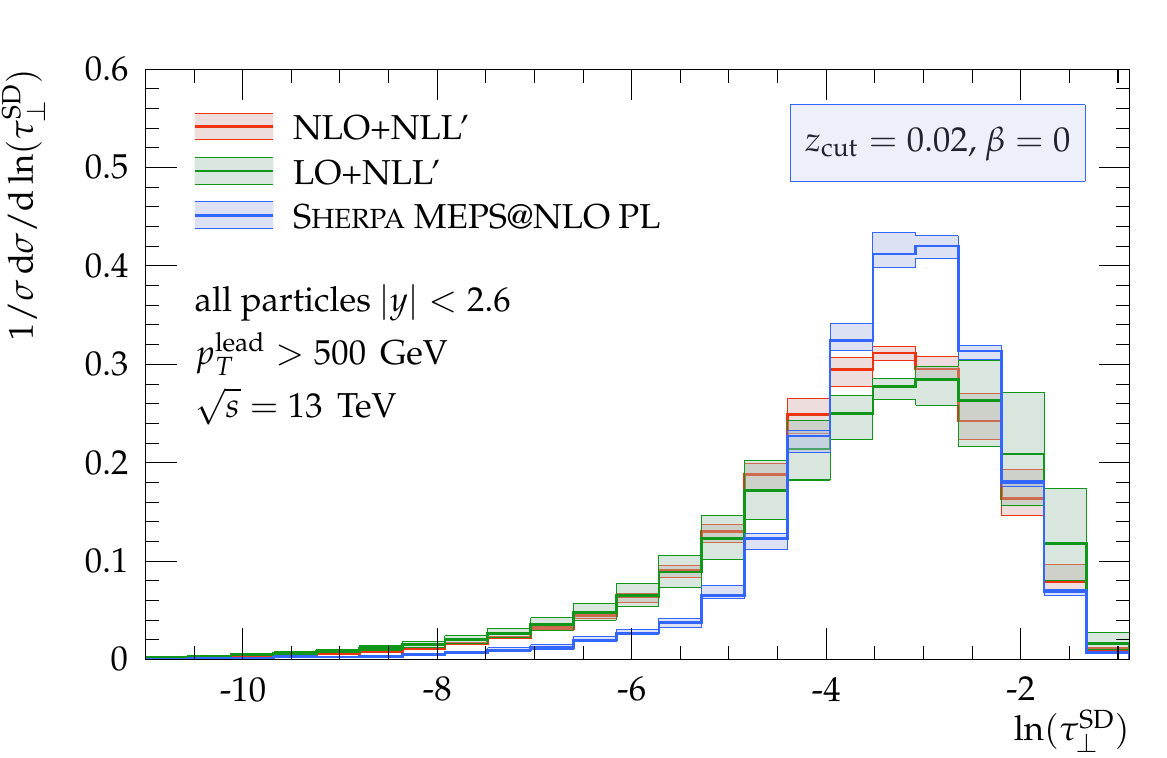}~
		\includegraphics[width=0.32\textwidth]{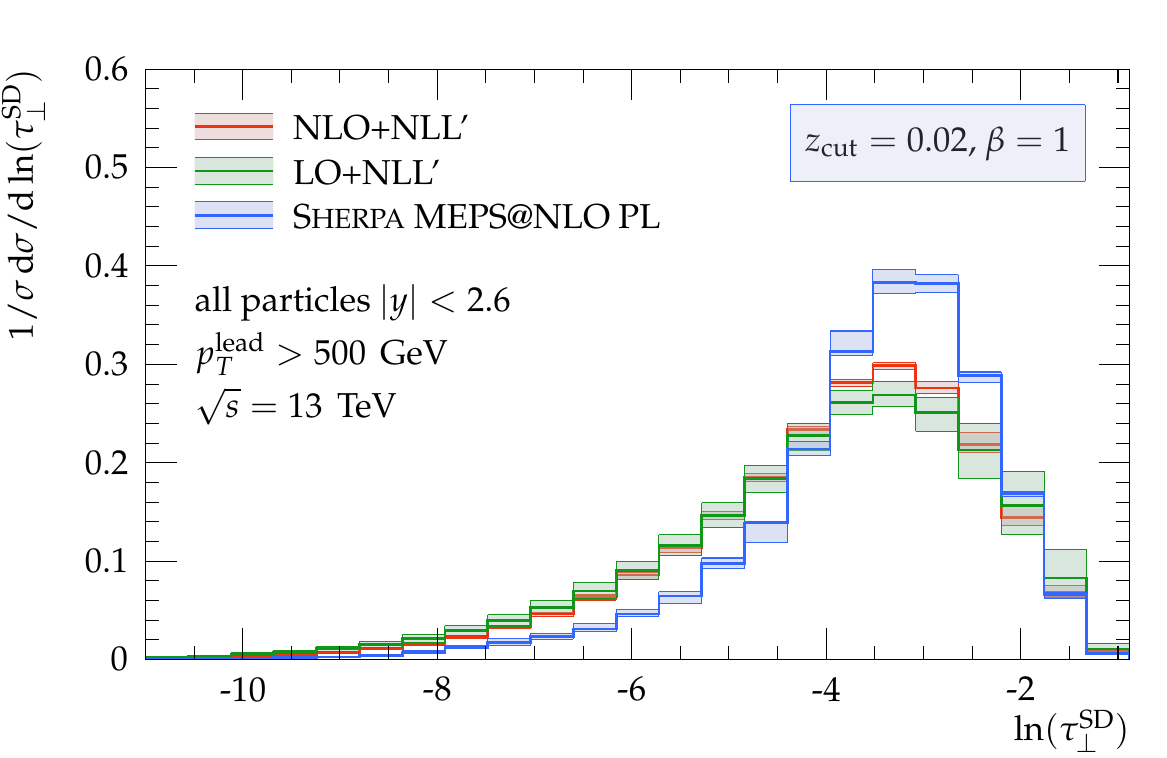}~
		\includegraphics[width=0.32\textwidth]{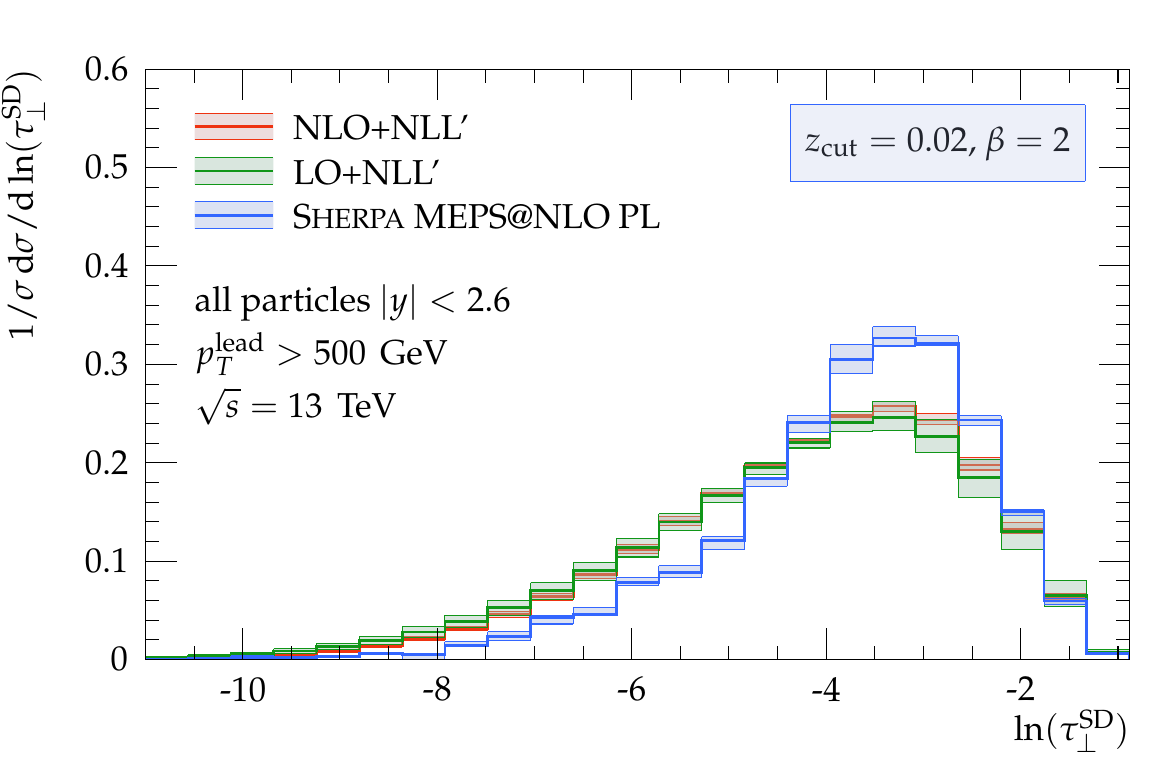}\\
		\includegraphics[width=0.32\textwidth]{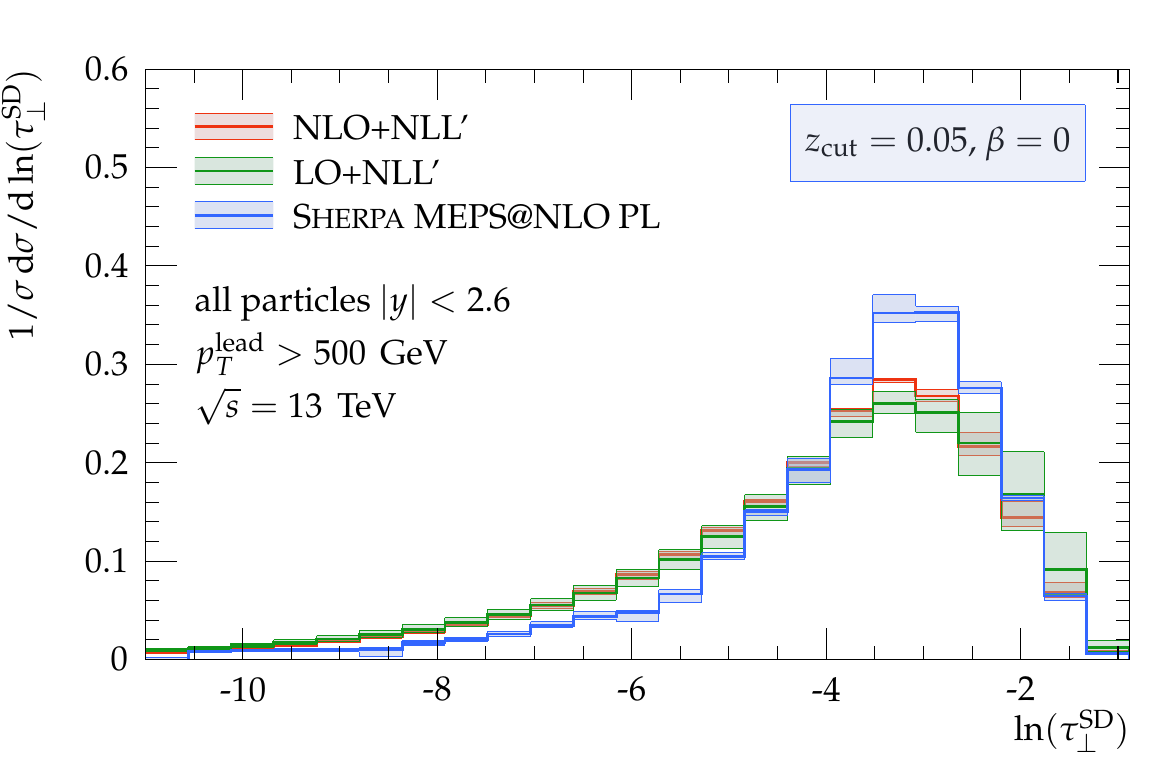}~
		\includegraphics[width=0.32\textwidth]{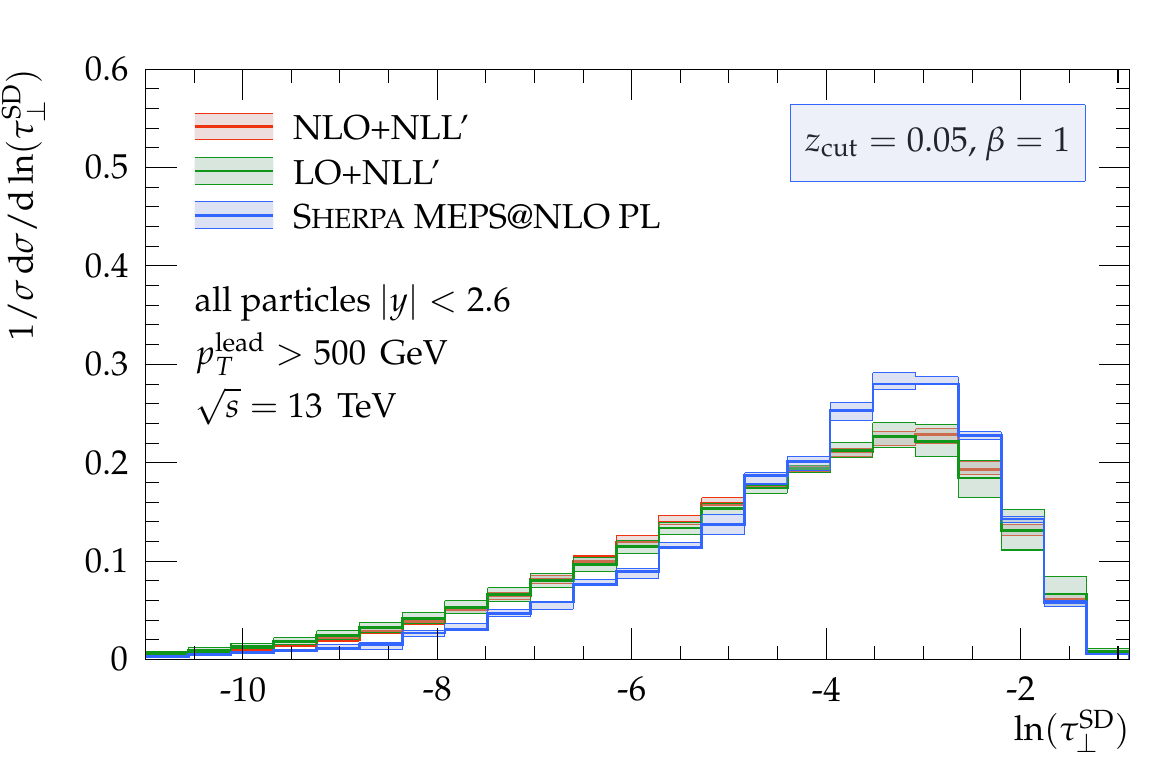}~
		\includegraphics[width=0.32\textwidth]{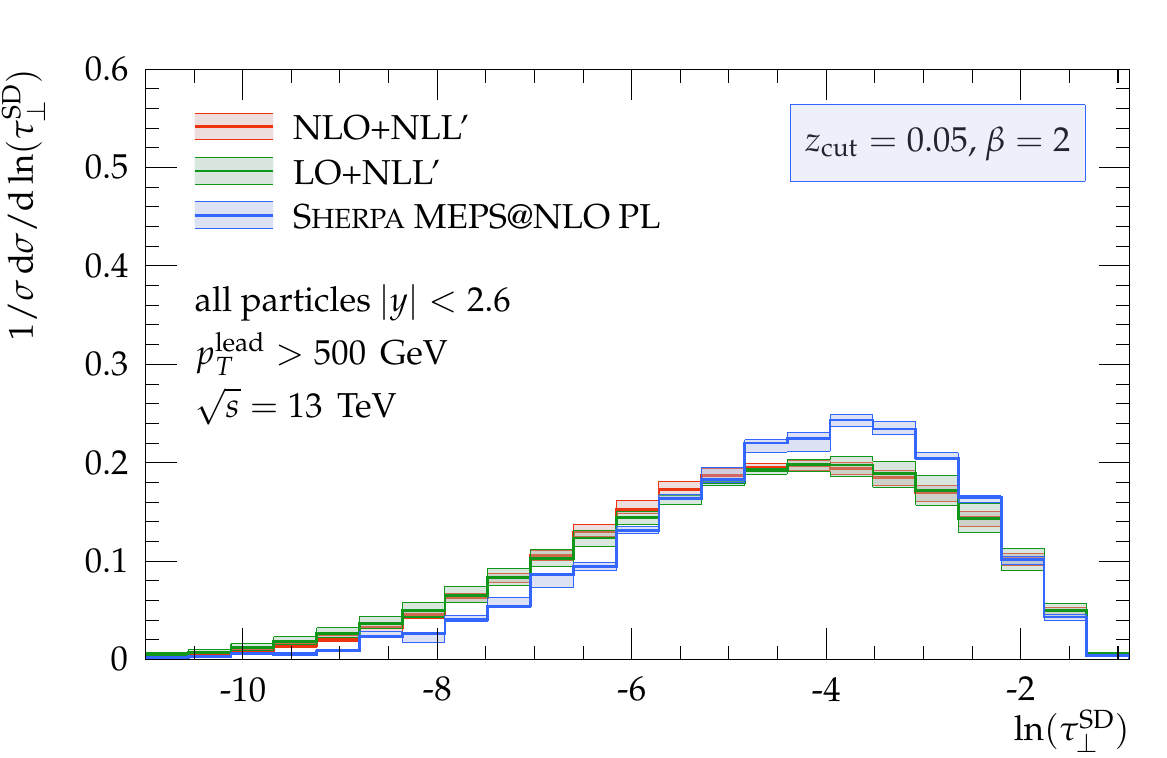}
	\end{center}
	\caption{Comparison of \NLL resummed results matched to LO and \NLO and \MEPSatNLO
          parton-level predictions from \Sherpa for groomed transverse thrust for
          $\beta\in\{0,1,2\}$ (columns) and $\zcut=0.01, 0.02, 0.05$ (rows) for the
          $p_{T,\text{min}}=500\;\text{GeV}$ event selection.}
	\label{fig:ResumVSMC_500_aux}
\end{figure}

\begin{figure}[hb!]
\begin{center}
	\includegraphics[width=0.32\textwidth]{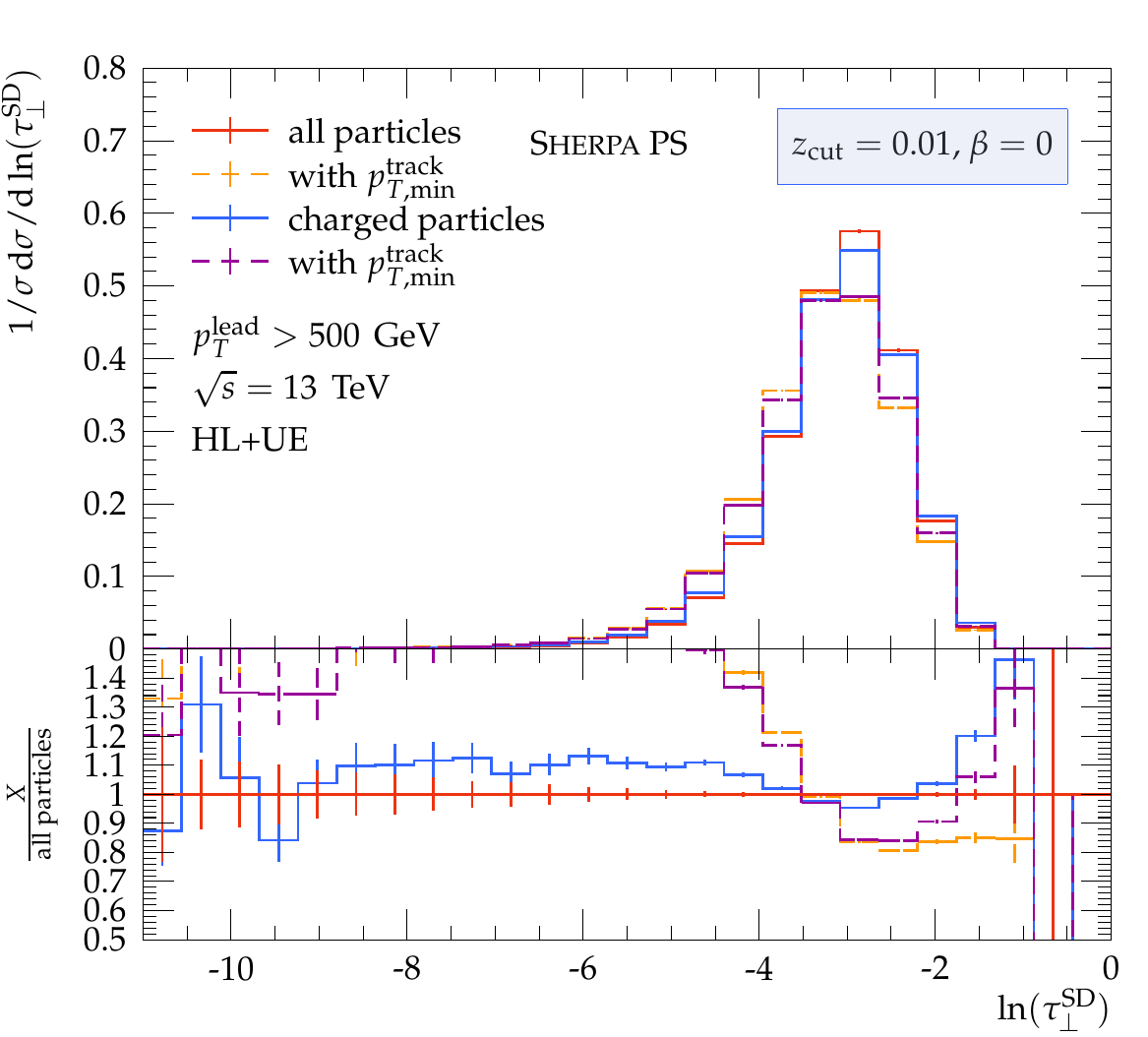}~
	\includegraphics[width=0.32\textwidth]{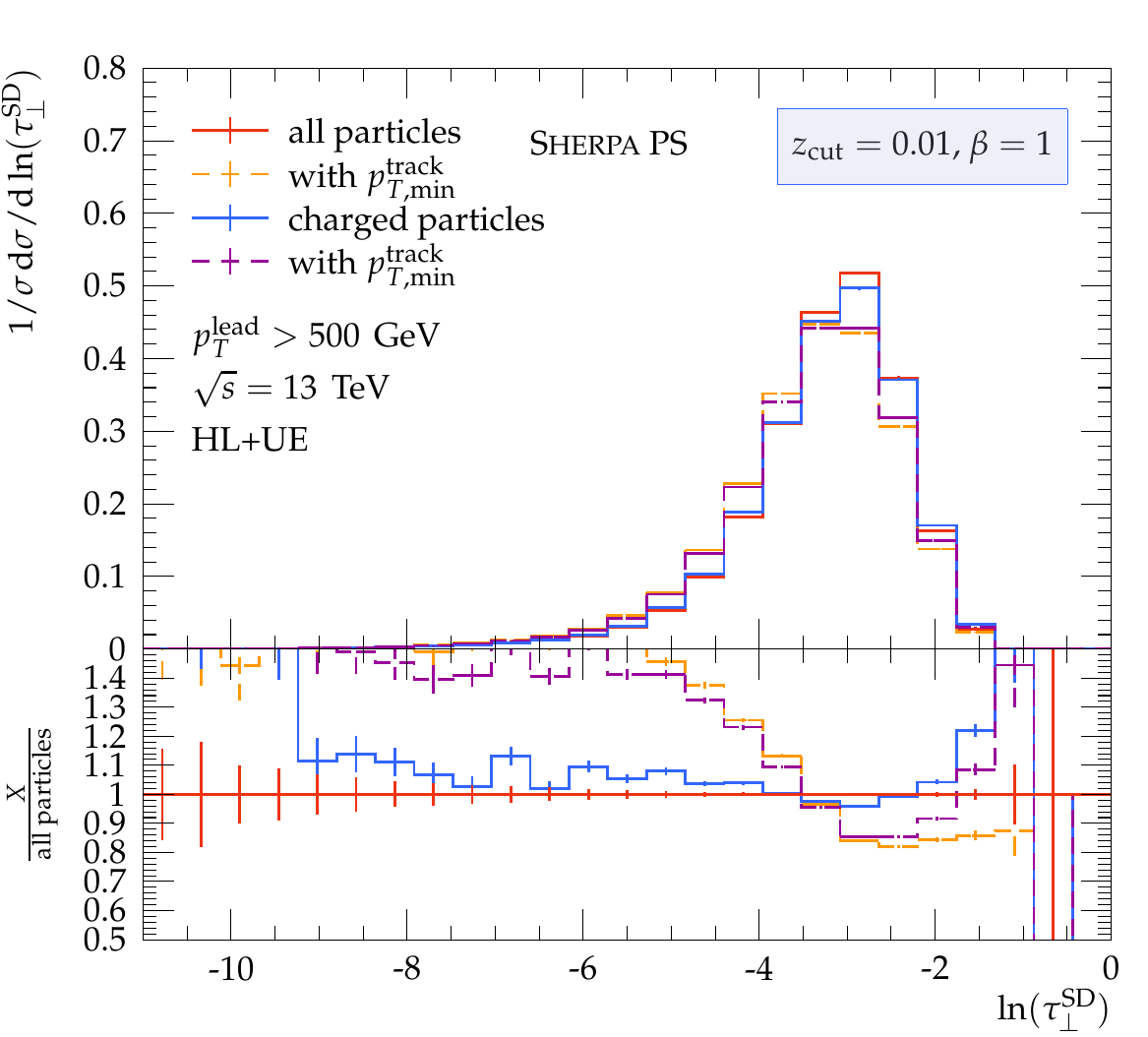}~
	\includegraphics[width=0.32\textwidth]{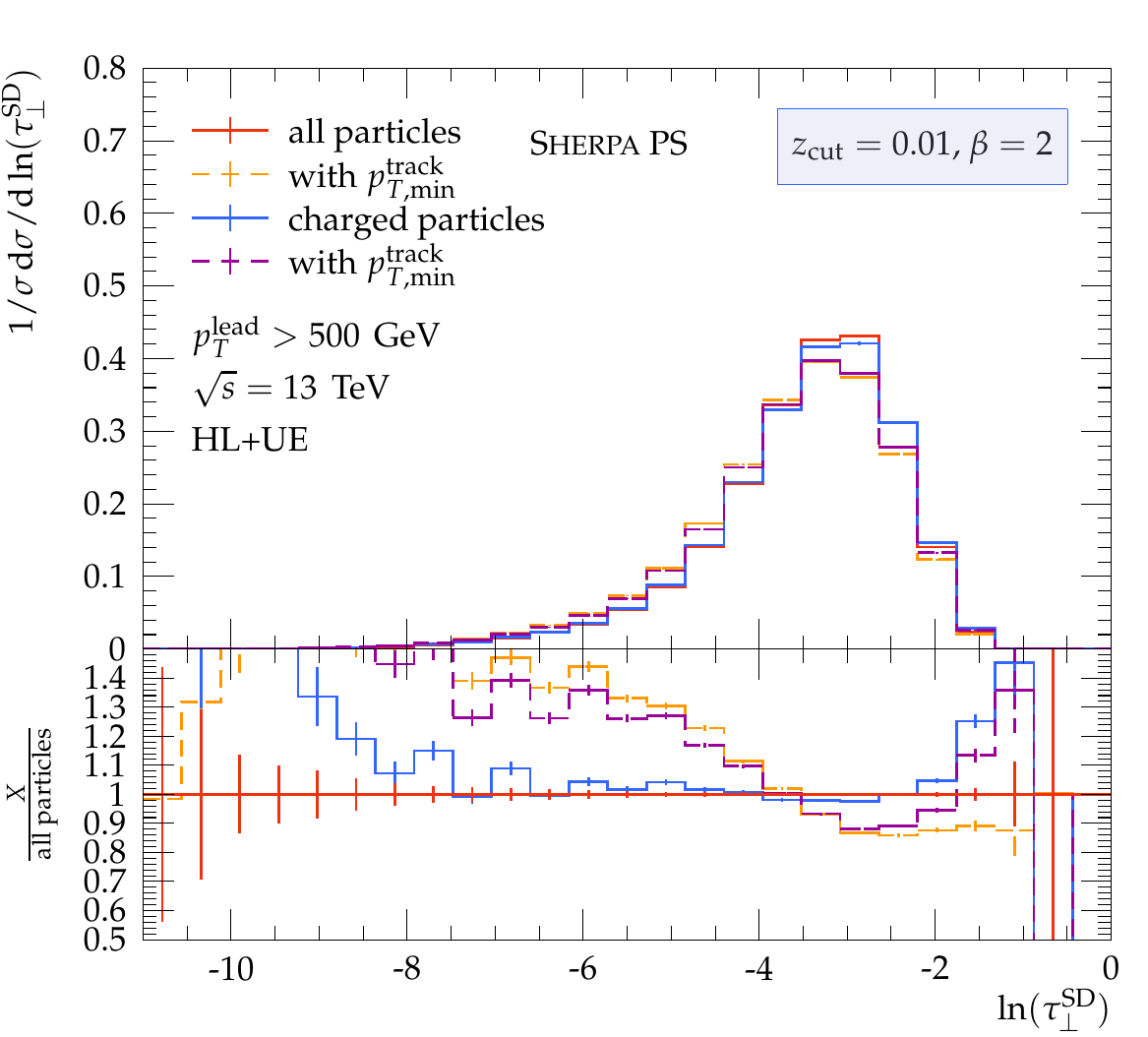}\\
	\includegraphics[width=0.32\textwidth]{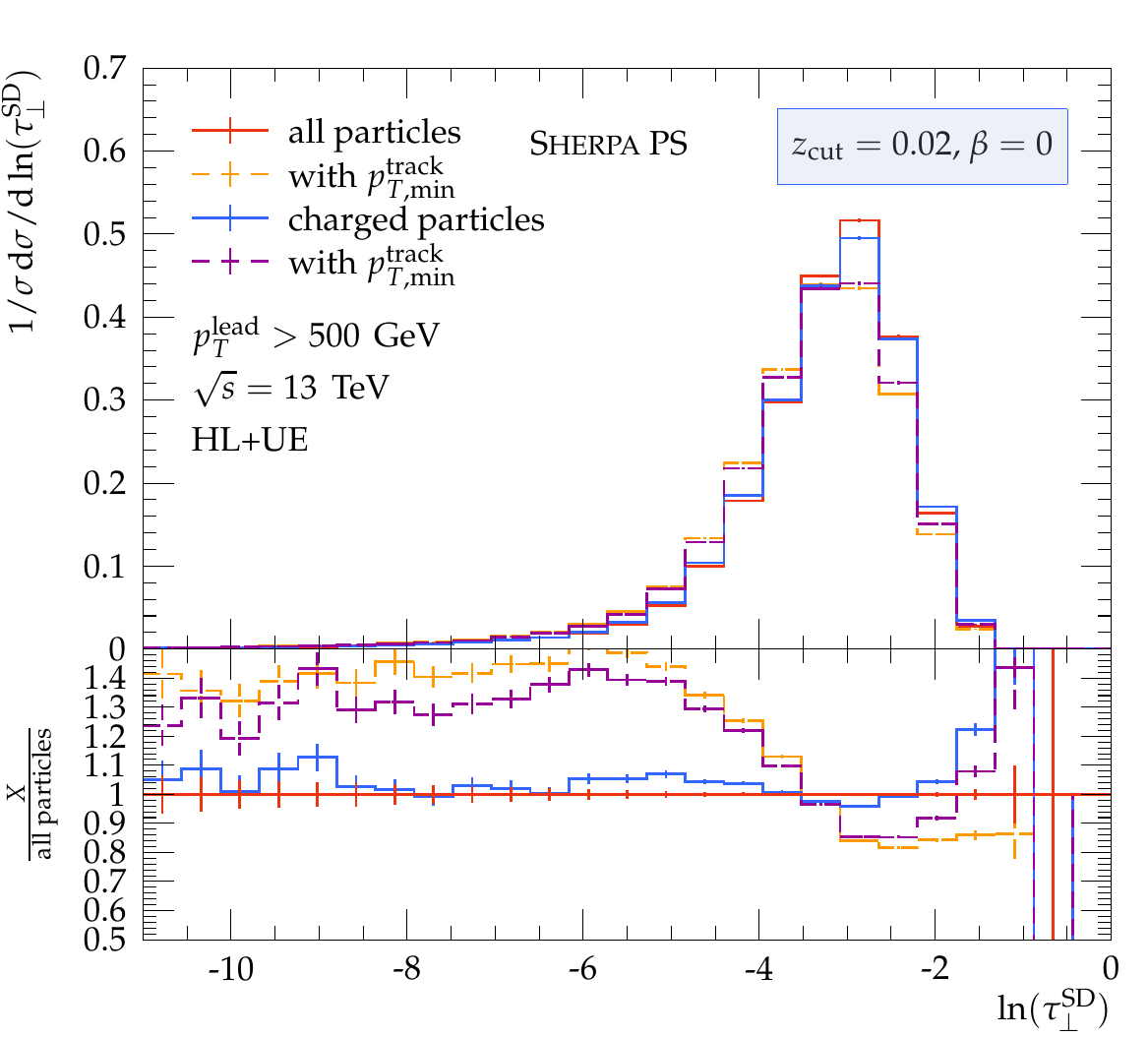}~
	\includegraphics[width=0.32\textwidth]{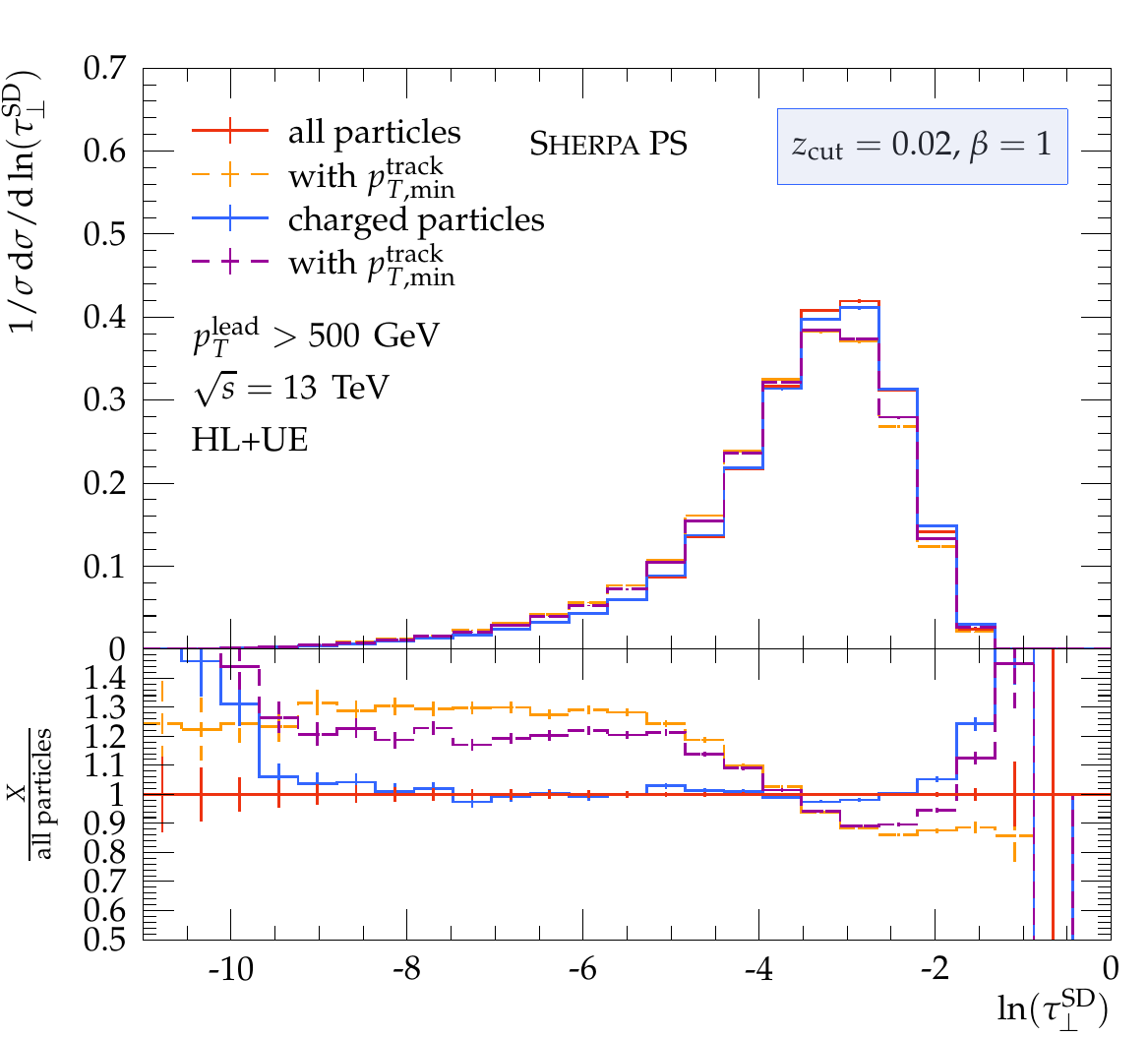}~
	\includegraphics[width=0.32\textwidth]{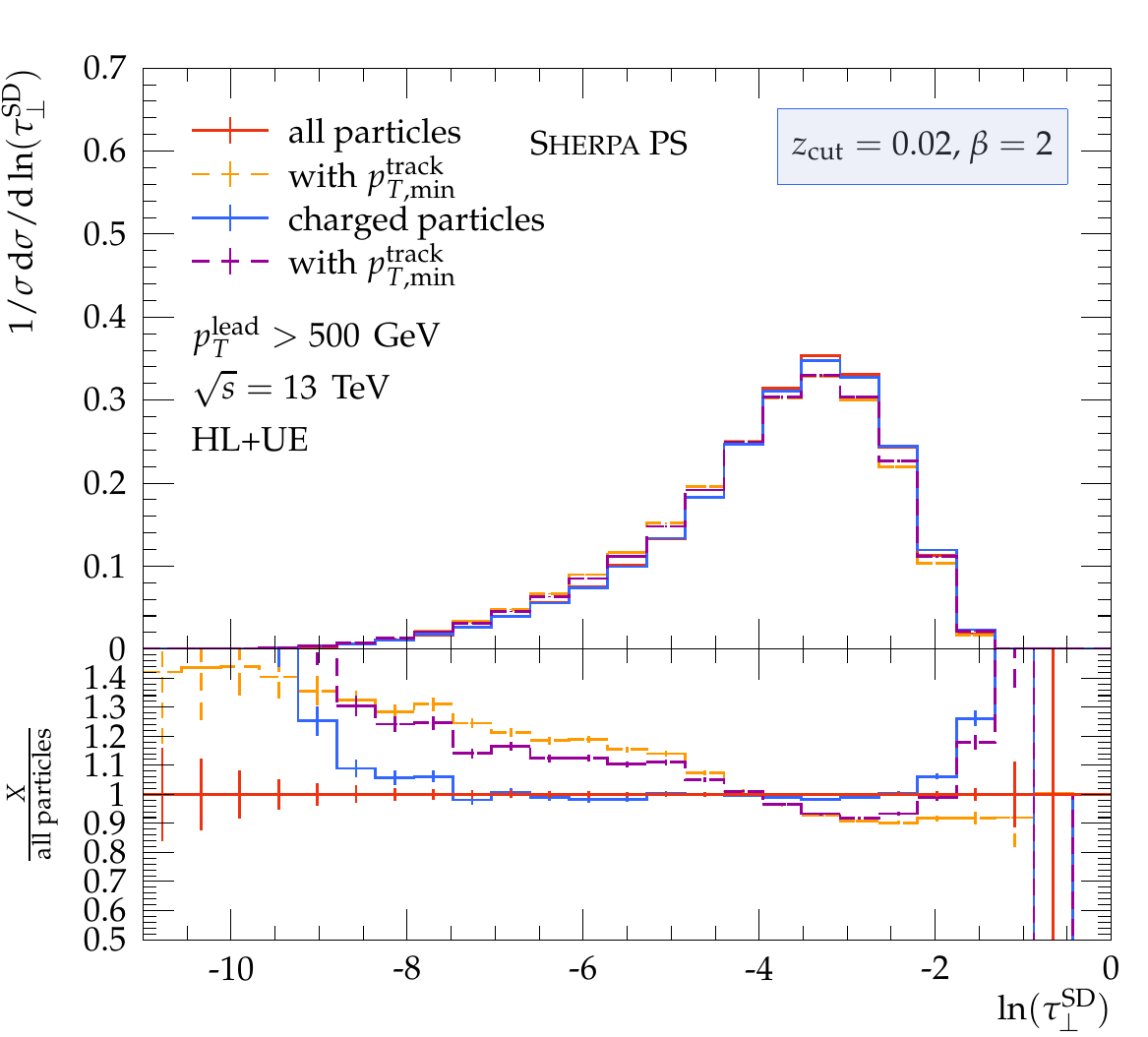}\\
	\includegraphics[width=0.32\textwidth]{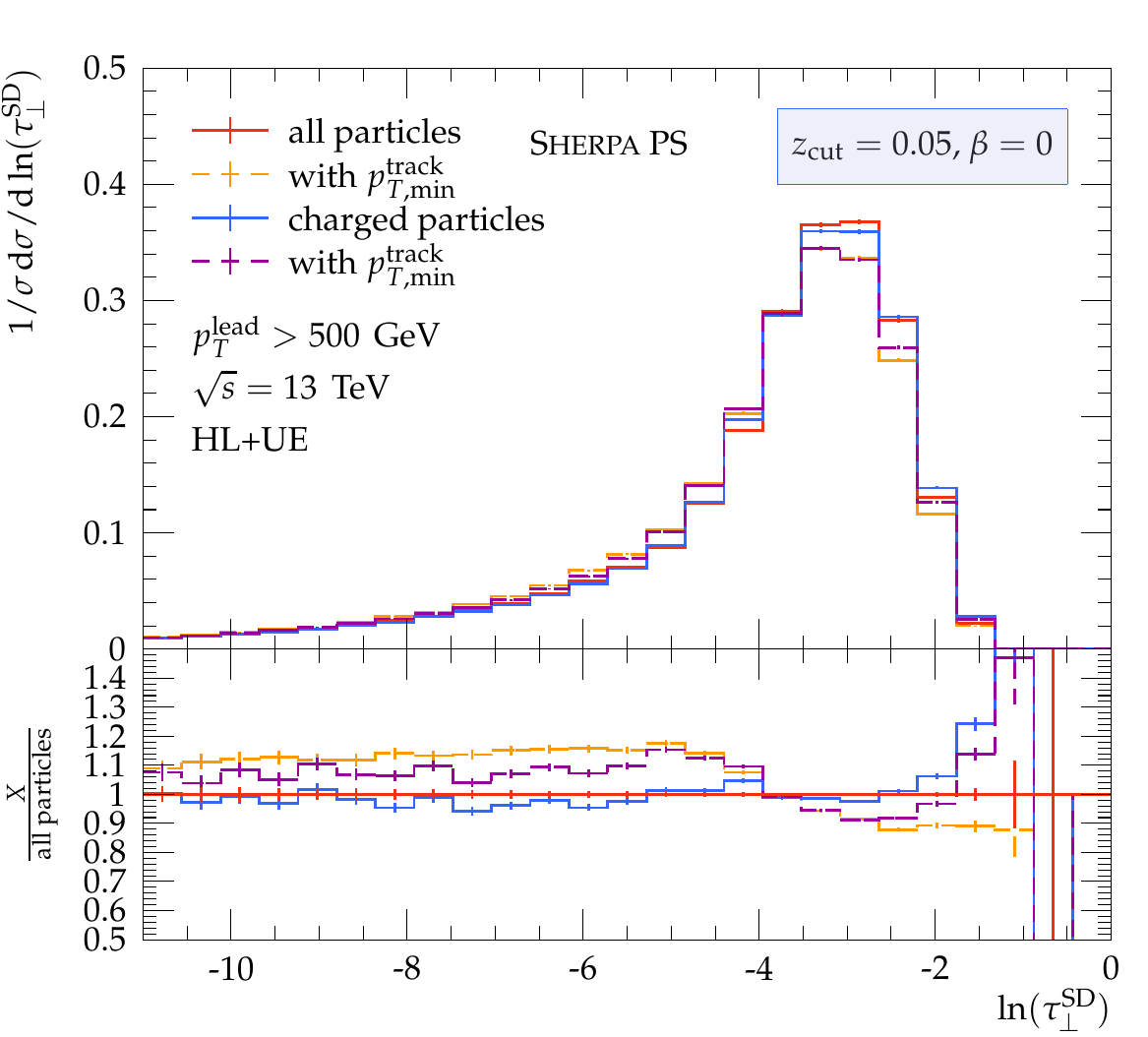}~
	\includegraphics[width=0.32\textwidth]{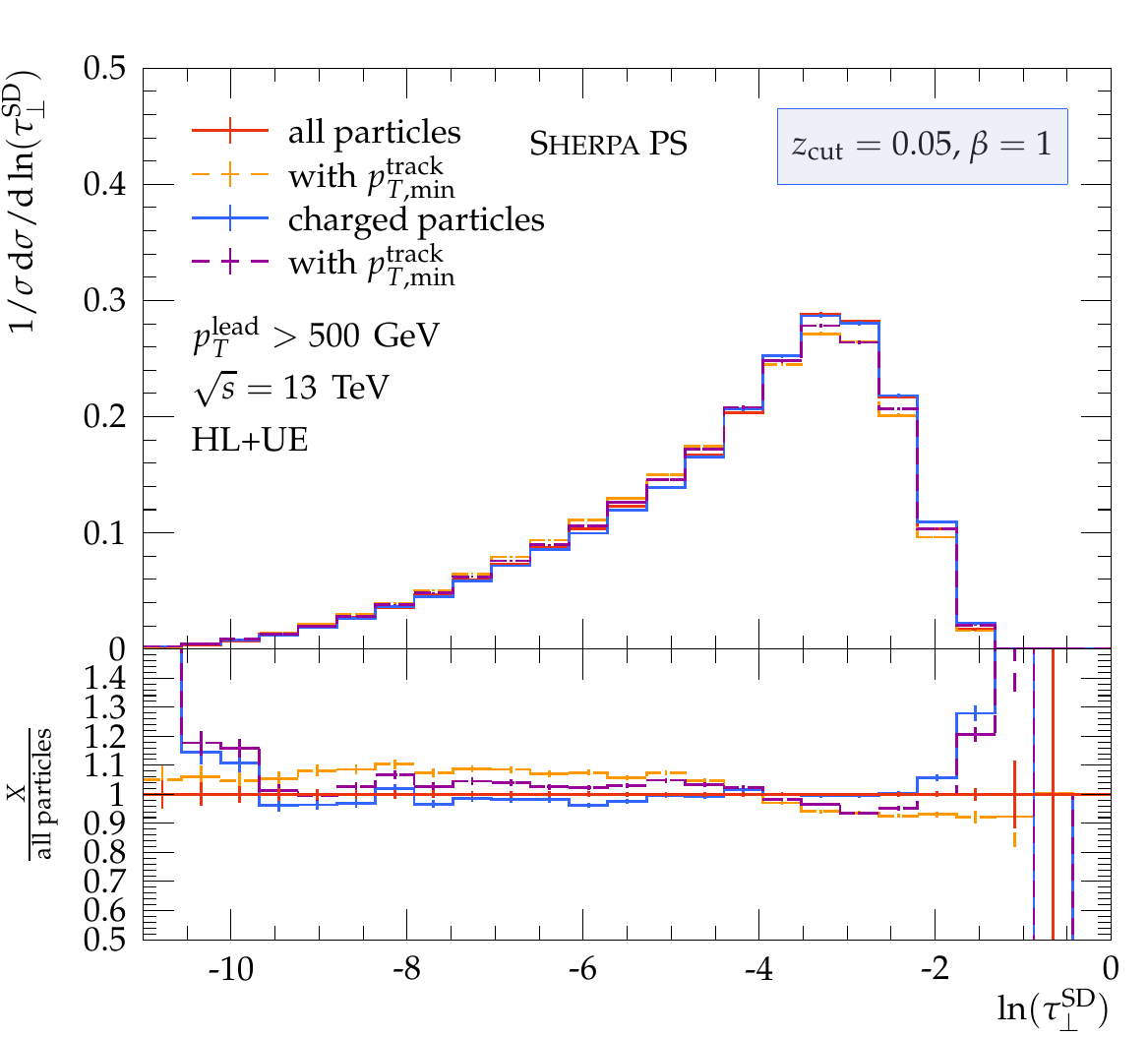}~
	\includegraphics[width=0.32\textwidth]{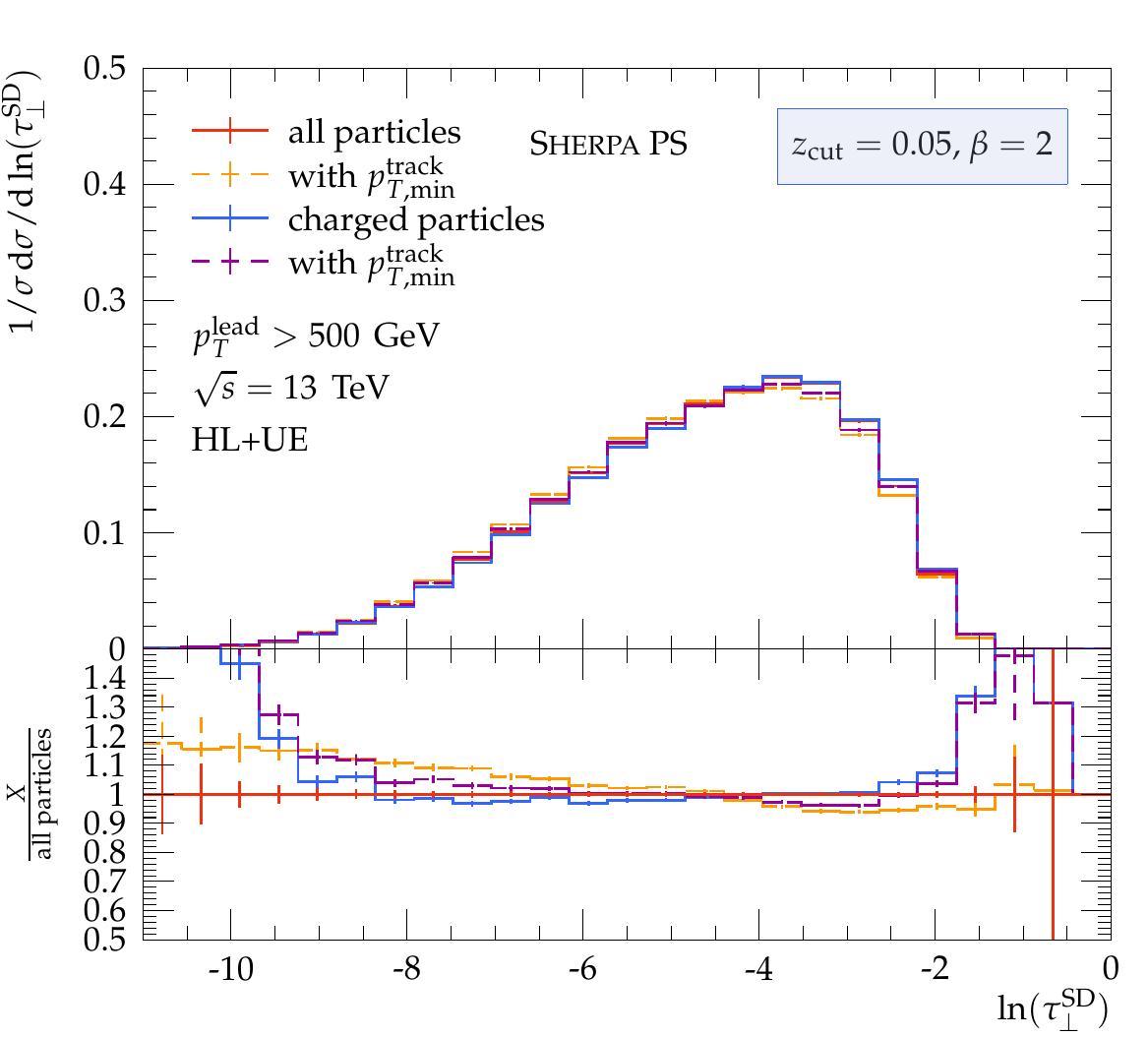}\\
	\includegraphics[width=0.32\textwidth]{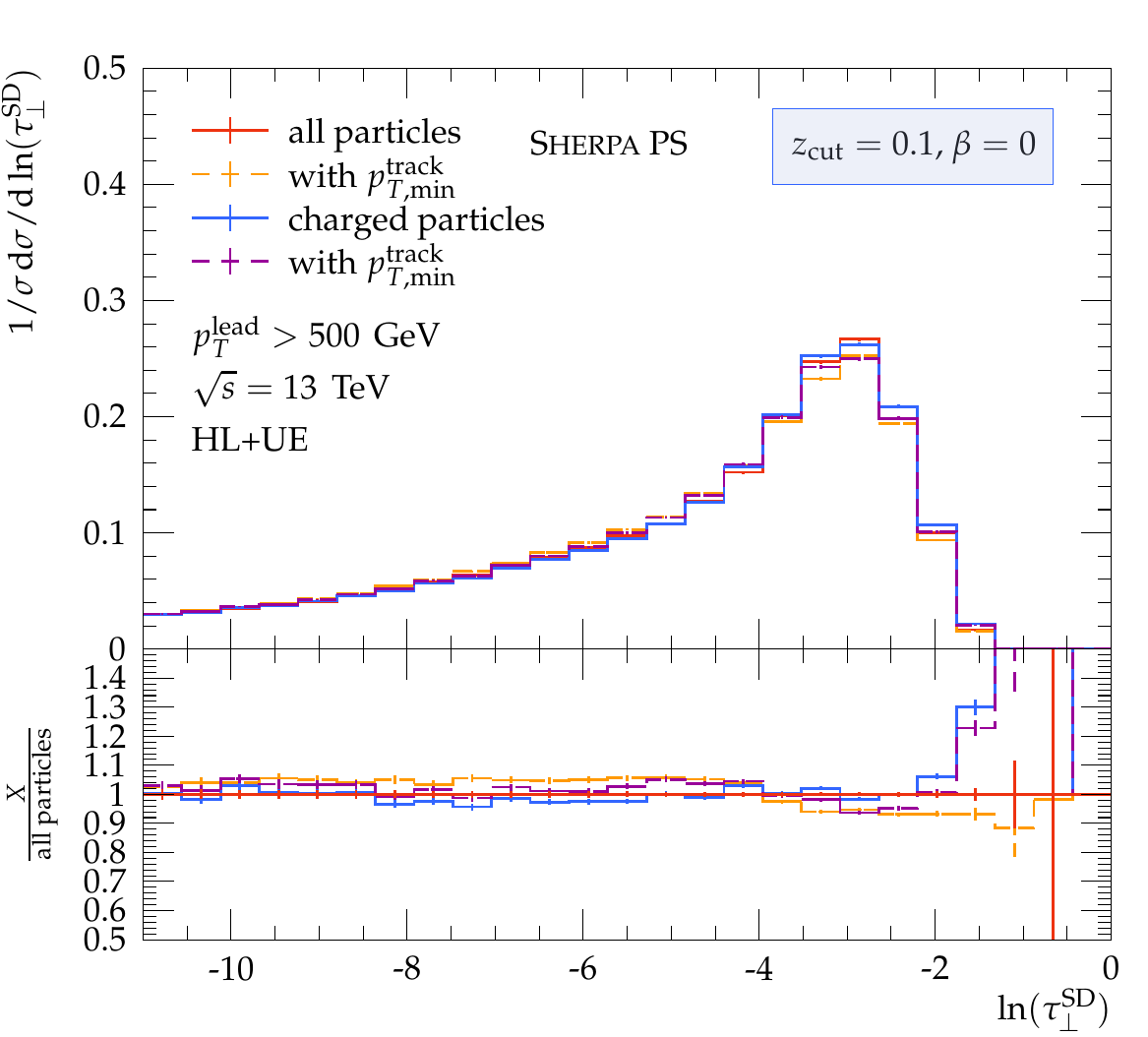}~
	\includegraphics[width=0.32\textwidth]{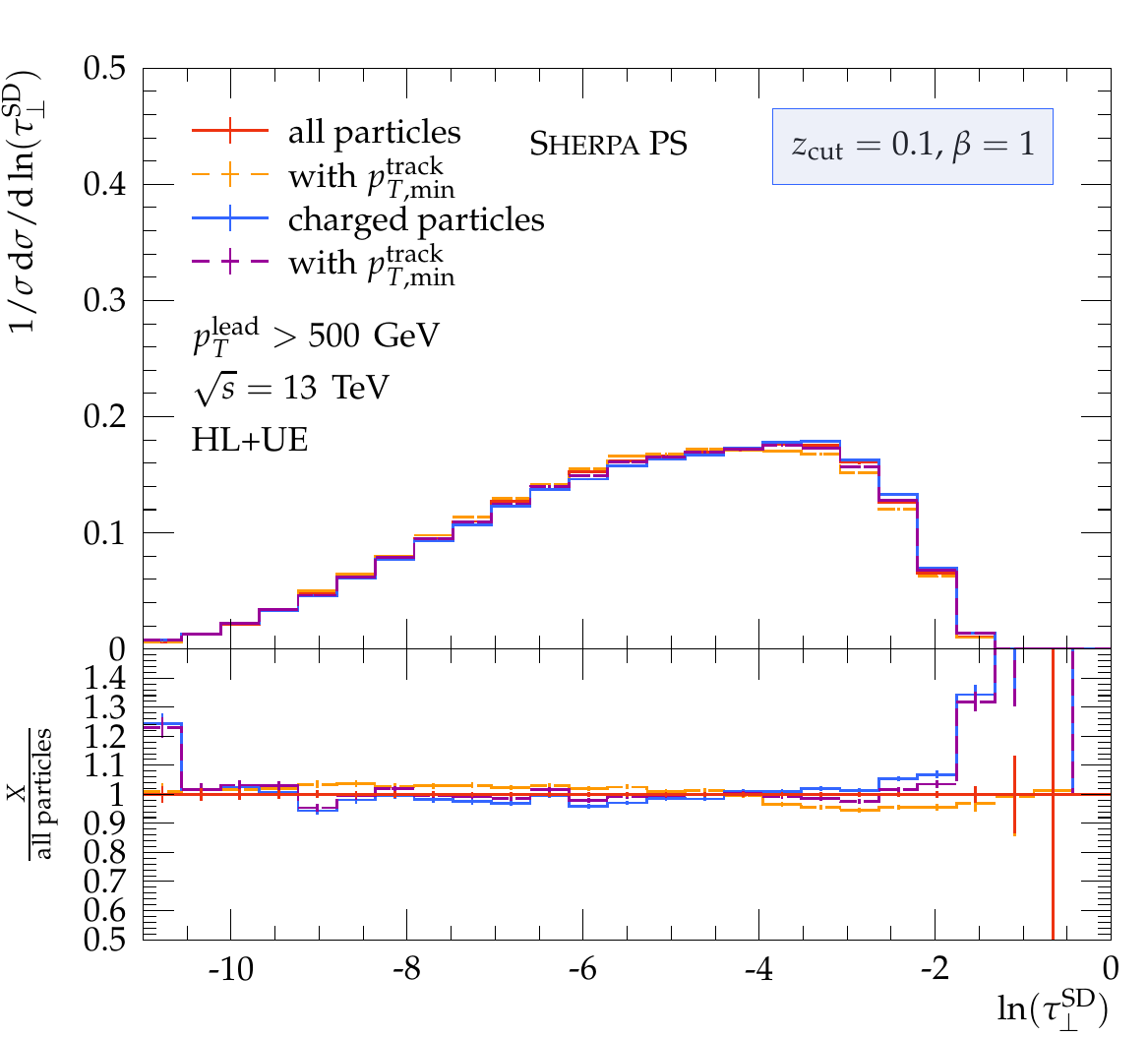}~
	\includegraphics[width=0.32\textwidth]{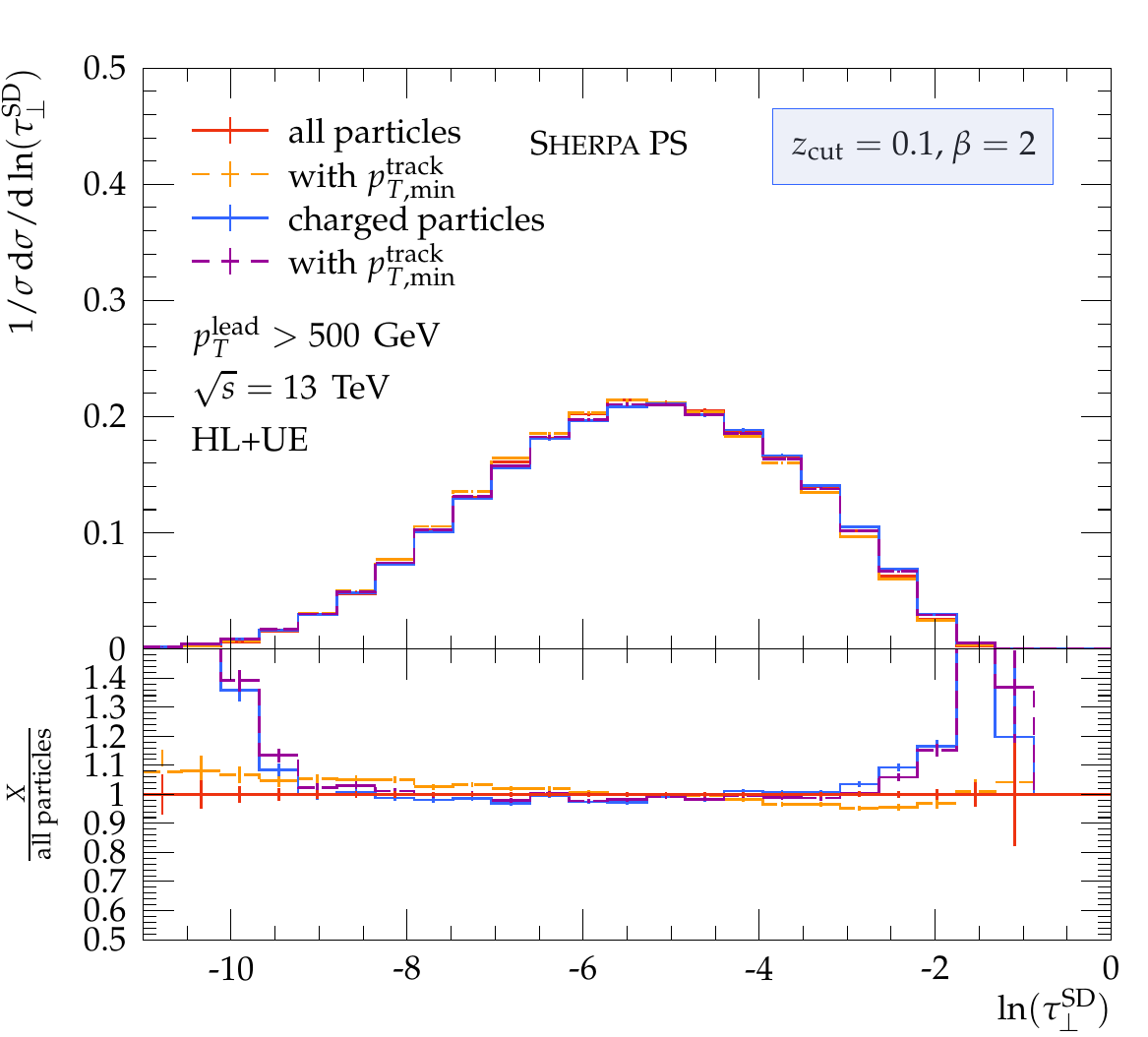}\\
\end{center}
\caption{Hadron-level results for groomed thrust for $\beta\in\{0,1,2\}$ (columns)
  and $\zcut\in\{0.01, 0.02, 0.05, 0.1\}$ (rows) for the $p_{T,\text{min}}=500\;\text{GeV}$ event selection.
  Shown are predictions from a parton-shower based \Sherpa simulation, where the observable gets determined from
  \emph{all} or \emph{charged} final-state particles only. In addition, the effect of a minimal particle
  transverse-momentum cut of $p^{\text{track}}_{T,\text{min}}= 500\;\text{MeV}$ is studied. The lower panels show
  ratios with respect to the \emph{all} particles and \emph{no} track-$p_T$ cut prediction. }
\label{fig:charge-pt_500}
\end{figure}

Here we collect supplementary material for the validation of the matched
resummed predictions, Fig.~\ref{fig:Resum_500_aux}, and parton-level results from \Sherpa,
Fig.~\ref{fig:ResumVSMC_500_aux}, for the $p_{T,\text{min}}$ event selection.
In Fig.~\ref{fig:charge-pt_500} we present additional results for the generator studies
presented in Sec.~\ref{sec:pheno}, for the $p_{T,\text{min}}=500\;\text{GeV}$ dijet
selection.

\FloatBarrier
\clearpage

\bibliographystyle{amsunsrt_modp}
\bibliography{references}

\end{document}